\documentclass[a4paper,12pt]{article}
\pdfoutput=1
\usepackage{epsfig,amsmath,amsfonts,amsthm}
\usepackage[figuresright]{rotating}
\usepackage[left=28mm,right=20mm,top=30mm,bottom=20mm]{geometry}
\usepackage{subfig}
\usepackage{graphicx}
\graphicspath{{SMD_new_figures_20112018/}}
\usepackage{rotating}
\usepackage{booktabs}
\usepackage{moreverb}

\DeclareGraphicsExtensions{.pdf,.png,.jpg}
\usepackage{natbib}
\numberwithin{equation}{section}

\def\var{\hbox{Var}}

\def\max{\hbox{max}}

\providecommand{\keywords}[1]{\textbf{{Keywords}} #1}
\usepackage[nokeyprefix]{refstyle}
\usepackage{varioref}
\usepackage{xr}
\usepackage{hyperref}
\begin{document}

\title{Simulation study of  estimating between-study variance and overall effect in meta-analysis of  standardized mean difference}

\author{Ilyas Bakbergenuly, David C. Hoaglin  and Elena Kulinskaya}

\date{\today}

\maketitle

\abstract
{Methods for random-effects meta-analysis require an estimate of the between-study variance, $\tau^2$. The performance of estimators of $\tau^2$ (measured by bias and coverage) affects their usefulness in assessing heterogeneity of study-level effects, and also the performance of related estimators of the overall effect. For the effect measure standardized mean difference (SMD), we provide the results  from extensive simulations  on  five point estimators of $\tau^2$ (the popular methods of DerSimonian-Laird, restricted maximum likelihood, and Mandel and Paule (MP); the less-familiar method of Jackson; the  new method (KDB) based on the improved approximation to the distribution of the Q statistic by Kulinskaya,  Dollinger and Bj{\o}rkest{\o}l  (2011) ), five interval estimators for $\tau^2$ (profile likelihood, Q-profile, Biggerstaff and Jackson, Jackson, and the new KDB method), six point estimators of the overall effect (the five related to the point estimators of $\tau^2$ and an estimator whose weights use only study-level sample sizes), and eight interval estimators for the overall effect (five based on the point estimators for $\tau^2$; the Hartung-Knapp-Sidik-Jonkman (HKSJ) interval; a modification of HKSJ; and an interval based on the sample-size-weighted estimator).  }


\maketitle
\section{Introduction}\label{sec:Intro}
Meta-analysis is a statistical methodology for combining estimated effects from several studies in order to assess their heterogeneity and obtain an overall estimate. In this paper we focus on meta-analysis of standardized mean difference.  The data and, often, existing tradition determine the choice of outcome measure. In a comparative study with continuous subject-level data for a treatment arm (T) and a control arm (C), the customary outcome measures are the mean difference (MD) and the standardized mean difference (SMD). The Cochrane Handbook \citep[Part 2, Chapter 9]{higgins2011cochrane} points out that the choice of MD over SMD depends on whether  \lq\lq outcome measurements in all studies are made on the same scale.''  However, fields of application have established preferences:  MD in medicine and SMD in social sciences. 

If the studies can be assumed to have the same true effect, a meta-analysis uses a fixed-effect (FE) model (common-effect model) to combine the estimates. Otherwise, the studies' true effects can depart from homogeneity in a variety of ways. Most commonly, a random-effects (RE) model regards those effects as a sample from a distribution and summarizes their heterogeneity via its variance, usually denoted by $\tau^2$.
The between-studies variance, $\tau^2$, has a key role in estimates of the mean of the distribution of random effects; but it is also important as a quantitative indication of heterogeneity \citep{higgins2009re}. In studying estimation for meta-analysis of  SMD, we focus first on $\tau^2$ and then proceed to the overall effect.

\cite{veroniki2015methods} provide a comprehensive overview and recommendations  on methods of estimating $\tau^2$ and its uncertainty. Their review, however, has two important limitations. First, the authors study only \lq\lq methods that can be applied for any type of outcome data.'' However,  the performance of the methods that we study varies widely among effect measures. \cite[Section 6.1]{veroniki2015methods} mention this only in passing as a hypothetical possibility. Second, any review on the topic, such as \cite{veroniki2015methods}, currently can draw on limited empirical information on the comparative performance of the methods. The (short) list of previous simulation studies for SMD is given in Table~\ref{simulations_summary_for_SMD}. From this table, it is clear that only three studies (\cite{viechtbauer2005bias}
 \cite{petropoulou2017comparison} and  \cite{Langan_2018_RSM_1316}) considered and compared  several point estimators of $\tau^2$.

However,   to assess bias of the estimators of heterogeneity variance, \cite{petropoulou2017comparison} use mean absolute error (the only performance measure reported for SMD in their Table S8), which is not a measure of bias; it is the linear counterpart of mean squared error.

\cite{Langan_2018_RSM_1316} studied bias and mean squared error of estimators of $\tau^2$, as well as coverage of confidence intervals for the overall effect, but they used only one value of SMD=0.5 because they believed that the value of SMD does not matter. We show in our simulations that it does matter. Additionally, both \cite{viechtbauer2005bias} and  \cite{Langan_2018_RSM_1316}) consider a very restricted range of $\tau^2$ values.   There appear to be no studies
at all on coverage of $\tau^2$.  Most of the  simulation studies   consider only inverse-variance-based estimation of the overall effect, and only two (\cite{Langan_2018_RSM_1316} and \cite{Hamman2018}) consider t-based confidence intervals for it.

Additionally,  all moment-based point  estimators of $\tau^2$ (i.e., the vast majority of the estimators listed in Table 1), use the inferior  $\chi^2_{K-1}$  approximation to the distribution of Cochran's $Q$ statistic, which does not perform well for small to medium sample sizes   \citep{kulinskaya2011testing}.

To address this gap in information on methods of estimating the heterogeneity variance for  SMD, we use simulation to study four methods recommended by \cite{veroniki2015methods}. These are the well-established methods of \cite{dersimonian1986meta}, restricted maximum likelihood, and \cite{mandel1970interlaboratory}, and the less-familiar method of \cite{jackson2013confidence}. We also include a new method  based on improved approximations to the distribution of the $Q$ statistic for  SMD \citep{kulinskaya2011testing}. We also study coverage of confidence intervals for $\tau^2$ achieved by five methods, including the Q-profile method of \cite{viechtbauer2007confidence}, a Q-profile method based on an improved approximation to the distribution of Cochran's $Q$, and profile-likelihood-based intervals.

For each estimator of $\tau^2$,  we also study bias of the corresponding inverse-variance-weighted estimator of the overall effect. As our work progressed, it became clear that those inverse-variance-weighted estimators generally had unacceptable bias for SMD. Therefore, we added an estimator (SSW) whose weights depend only on the sample sizes of the Treatment and Control arms. We studied the coverage of the confidence intervals associated with the inverse-variance-weighted estimators, and also the HKSJ interval, the HKSJ interval using the improved estimator of $\tau^2$, and the interval centered at SSW and using the improved $\hat\tau^2$ in estimating its variance.

\begin{sidewaystable}[]
	\resizebox{0.8\textwidth}{!}{%
\begin{tabular}{|l|l|l|l|l|l|l|l|l|l|}\hline

	Study &SMD measure             & $\delta$     & $\tau^2$            & n and/or $\bar{n}$                 & K                                                                    & $\hat{\tau}^2$ & Coverage of $\tau^2$ & $\hat{\delta}$ &  Coverage of $\delta$                                               \\\hline
	Viechtbauer 2005  &    Hedges$'$s d       & 0, 0.2 ,0.5 ,0.8   & 0, 0.01 ,0.025, 0.05 ,0.1           & $\bar{n}=20 ,40, 80, 160, 320$                                                     & 5, 10, 20, 40, 80                             & DL                       &                                            & IV                              &                                                 \\
	&           &                 &                                 &                                         $n_i \sim N(\bar n, \bar n/3)$                             &                                            & ML                                           &                                            &                                               &                                                 \\
	&           &                 &                                 &                                       $n_{Ti}=n_{Ci}=n_i/2$                               &                                            & REML                                         &                                            &                                               &                                                 \\
	&           &                 &                                 &                                                                      &                                            & HE                             &                                            &                                               &                                                 \\
	&           &                 &                                 &                                                                      &                                            & HS                     &                                            &                                               &                                                 \\\hline
	Friedrich et al. 2008          &   Hedges$'$s d           & 0.2 ,0.5 ,0.8     & 0, 0.5                           & $n_{T}=n_{C}=10,100$                               & 5, 10, 30                              & DL                       &                                            & IV                              & IV  \\\hline
	Petropoulou and Mavridis 2017             &           &0 \& 0.5             & 0,0.01,0.05,0.5             &    $n_{T}=n_{C} \sim U(20,\;200)$                                          & 5,10,20, 30                              & 20                 &                                            & IV                              & IV  \\
	&           &                 &                                 &                                                                      &                                            & estimators                      &                                            &                                               &                                                 \\
	&           &                 &                                 &                                                                      &                                            & of $\tau^2$&                                            &                                               &                                                 \\
	&           &                 &                                 &                                                                      &                                            & &                                            &                                               &                                                 \\
	&           &                 &                                 &                                                                      &                                            & &                                            &                                               &                                                 \\\hline
	Langan et al. 2018            &Hedges$'$s d  & 0.5             & depends on $I^2$, 0\%,15\% & $n=40$, $n \sim U(40,400),$                           & 2 ,3, 5,                             & REML                                         &                                            & IV                              & IV  \\
	&           &                 &  30\%, 45\%, 60\%,        &                      $n=400$, $n \sim U(2000,4000)$,                          &  10, 20, 30,                                           & CA                           &                                            &                                               & IV+t                 \\
	&           &                 &     75\%, 90\%, 95\%          &       $n=40$+$N \sim U(2000,4000)$                                           &    50, 100                                         & PM or MP                                 &                                            &                                               & IV +HKSJ   \\
	&           &                 &                                 &          $n_T=n_C=n/2$                                                            &                                            & $PM_{CA}$               &                                            &                                               &                                                 \\
	&           &                 &                                 &                                                                      &                                            & $PM_{DL}$           &                                            &                                               &                                                 \\
	&           &                 &                                 &                                                                      &                                            & HM                         &                                            &                                               &                                                 \\
	&           &                 &                                 &                                                                      &                                            & SJ                          &                                            &                                               &                                                 \\
	&           &                 &                                 &                                                                      &                                            & $SJ_{CA}$            &                                            &                                               &                                                 \\\hline
	Lin 2018    & Cohen$'$s d         & 0, 0.2, 0.5, 0.8, 1                 & 0,0.2,0.5 & U(5,10), U(10,20), & 5, 10, 20, 50        & DL                        &                                            & IV                              & IV  \\
	& Hedges$'$s d  &                 &                                 &            U(20,30), U(30,50)                                                 &                                            &                                              &                                            &                                               &           \\
	&  &                          &                                  &                    U(50,100), U(100,500),                                                 &                                            &                                              &                                            &                                               &           \\
	&  &                          &                                  &                    U(500,1000), $n_T=n_C$                                                  &                                            &                                              &                                            &                                               &           \\\hline
	Hamman et al. 2018 &Hedges$'$s d &0,0.1,0.15,0.25&0,0.1,0.5,1,&$\bar{n}=4,6,8,10,12,$&5,10,15,25,35,&REML&&IV& IV +HKSJ     \\
      				   &&0.35,0.5,0.6,0.75,&2.5,5,10&14,16,20,25&45,55,75,100,125&&&SSW(H)&       \\
      				   &&1,1.25,1.5,2.5&&&&&&EW&       \\\hline
Mar\'{\i}n-Mart\'{\i}nez and &&0.2,0.5,0.8&0,0.04,0.08,0.16,0.32 &$\bar{n}=30,50,80,100$&5,10,20,40,100& DL &&FE& \\      				
S\'{a}nchez-Meca 2010 &&& &$n_T=n_C$&&ML &&IV&\\
 &&& &&& &&HS& \\
  \hline   				
\end{tabular}
}
\begin{footnotesize}\caption{Simulation studies on meta-analysis of SMD. \\
Estimators of $\tau^2$: DL - DerSimonian and Laird estimator, ML - Maximum likelihood, REML - Restricted maximum likelihood estimator, HE - Hedges estimator, HS - Hunter-Schmidt estimator, 
CA - Cochran ANOVA, PM or MP - Mandel-Paule estimator, $PM_{CA}$ - two-step Cochran ANOVA, $PM_{DL}$ -two-step DerSimonian-Laird, HM - Hartung-Makambi, SJ - Sidik-Jonkman, $SJ_{CA}$ - alternative Sidik-Jonkman, BM  - Bayes Modal estimator; \\
Estimators of $\delta$:  SSW(H) - sample-size-weighted (Hedges 1982), EW -  equal weights, IV - inverse-variance-weighted, HS -  Hunter and Schmidt (1990) total-sample-size-weighted;\\
Coverage of $\delta$: IV - confidence interval centered at inverse-variance-weighted estimator of  $\delta$ with z quantiles, IV+t - confidence interval centered at inverse-variance-weighted estimator of  $\delta$ with t quantiles, IV+HKSJ - Hartung-Knapp-Sidik-Jonkman confidence interval centered at inverse-variance-weighted estimator of  $\delta$.
}\end{footnotesize}

\end{sidewaystable}\label{simulations_summary_for_SMD}

\section{Study-level estimation of standardized mean difference}\label{sec:EffectSMD}
We assume that each of the $K$ studies in the meta-analysis consists of two arms, Treatment and Control, with sample sizes $n_{iT}$ and $n_{iC}$. The total sample size in Study $i$ is $n_i=n_{iT}+n_{iC}$. We denote the ratio of the control sample size to the total by  $q_i=n_{iC}/n_{i}$.  The subject-level data in each arm are assumed to be normally distributed with means $\mu_{iT}$ and $\mu_{iC}$ and variances $\sigma_{iT}^2$ and $\sigma_{iC}^2$. The sample means are $\bar{x}_{ij}$, and the sample variances are $s^2_{ij}$, for $i = 1,\ldots,K$ and $j = C$ or $T$.

The standardized mean difference effect measure  is
$$\delta_{i}=\frac{\mu_{iT}-\mu_{iC}}{\sigma_{i}}.$$
The plug-in estimator $d_i = (\bar{x}_{iT} - \bar{x}_{iC}) / s_{i}$, known as Cohen's $d$, is biased in small samples, and we do not consider it further.  Instead, we study the unbiased estimator
$${g}_i=J(m_i)\frac{\bar{x}_{iT}-\bar{x}_{iC}}{s_{i}},$$
where $m_{i}=n_{iT}+n_{iC}-2$, and the factor $$J(m)=\frac{\Gamma \left(\frac{m}{2}\right)}{\sqrt{\frac{m}{2}}\Gamma \left( \frac{m - 1}{2} \right)},$$
often approximated by $1 - 3 / (4m - 1)$, corrects for bias
\citep{hedges1983random}. This estimator of $\delta$ is sometimes called Hedges's $g$. The variances in the Treatment and Control arms are usually assumed to be equal. Therefore, $\sigma_i$ is estimated by the square root of the  pooled sample variance
$$s_i^2=\frac{(n_{iT}-1)s_{iT}^2 +(n_{iC}-1)s_{iC}^2}{n_{iT}+n_{iC}-2}.$$
For the variance of ${g}_i$ we use the unbiased estimator
\begin{equation}\label{eq:g_var} v_{i}^2=\frac{n_{iT}+n_{iC}}{n_{iT}n_{iC}}+\left(1-\frac{(m_{i}-2)}
{m_{i}J(m_{i})^2}\right)g^2_{i},\end{equation}
\cite{hedges1983random}.  The sample SMD ${g}_i$ has a scaled non-central $t$-distribution with non-centrality parameter  $[n_iq_i(1-q_i)]^{1/2}\delta_i$ :
\begin{equation}\label{eq:g_dist}\frac{\sqrt{n_iq_i(1-q_i)}}{J(m_i)}{g}_i\sim t_{m_i}( [n_iq_i(1-q_i)]^{1/2}\delta_i).\end{equation}
\cite{cohen1988statistical} categorized values of  $\delta=0.2,\;0.5,\;0.8$ as small, medium, and large effect sizes.  Four studies ( \cite{viechtbauer2005bias}, \cite{Friedrich2008},  \cite{martynez-2010, sanches-2010} and \cite{Lin_2018_PLoSONE_e0204056}) use these values of SMD in their simulations. However, these definitions of ``small,'' ``medium,'' and ``large'' may not be appropriate outside the behavioral sciences.  \cite{Ferguson2009} proposed the values $0.41, \; 1.15,\; 2.70$ as  benchmarks in the social sciences. In an empirical study of 21 ecological meta-analyses by \cite{M-J-2002}, 136 observed values of SMD varied in magnitude from $0.005$ to $3.416$, with mean $0.721$ and $95\%$ confidence interval ($0.622-0.820$).

\section{Standard random-effects model} \label{sec:StdREM}
In meta-analysis, the standard random-effects model assumes that within- and between-study variabilities are accounted for by approximately normal distributions of within- and between-study effects.  For a generic measure of effect,
\begin{equation}\label{eq:standardREM}
\hat{\theta}_{i}\sim N(\theta_{i},{\sigma}_{i}^2)\quad\text{and}\quad \theta_{i}\sim N(\theta,\tau^2),
\end{equation}
resulting in the marginal distribution $\hat{\theta}_{i}\sim N(\theta,\sigma_{i}^2+\tau^2)$. $\hat{\theta}_{i}$ is the estimate of the effect in Study $i$, and its within-study variance is $\sigma_{i}^2$, estimated by $\hat{\sigma}_{i}^2$, $i=1,\ldots,K$.  $\tau^{2}$ is the between-study variance, which is estimated by $\hat{\tau}^2$. The overall effect $\theta$ can be estimated by the weighted mean
\begin{equation}\label{eq:WAverChapter6}
\hat{\theta}_{\mathit{RE}}=\frac{\sum\limits_{i=1}^{K}\hat{w}_{i}(\hat{\tau}^2)\hat{\theta}_{i}}{\sum\limits_{i=1}^{K}\hat{w}_{i}(\hat{\tau}^2)},
\end{equation}
where the $\hat{w}_{i}(\hat{\tau}^2)=(\hat{\sigma}_{i}^2+\hat{\tau}^2)^{-1}$ are inverse-variance weights. The FE estimate $\hat{\theta}$  uses weights $\hat{w}_{i}=\hat{w}_{i}(0)$.

If $w_i = 1/\var(\hat{\theta}_i)$, the variance of the weighted mean of the $\hat{\theta}_i$ is $1/\sum w_{i}$. Thus, many authors estimate the variance of $\hat{\theta}_{\mathit{RE}}$ by $\left[\sum_{i=1}^{K}\hat{w}_{i}(\hat{\tau}^2)\right]^{-1}$.  In practice, however, this estimate may not be satisfactory \cite{sidik2006robust, li1994bias,rukhin2009weighted}.


\section{Point and interval estimation of $\tau^2$ by the Kulinskaya-Dollinger-Bj{\o}rkest{\o}l method  (KDB)}
Because the $\hat{w}_i(\tau^2)$ in Equation~(\ref{eq:WAverChapter6}) involve the $\hat{\sigma}_{i}^2$, $K - 1$ is an adequate approximation for the expected value of  Cochran's $Q$ statistic only for very large sample sizes. As an alternative one can use one of the improved approximations to the expected value of Cochran's $Q$. Corrected Mandel-Paule methods for estimating $\tau^2$ equate Cochran's  $Q$ statistic with the weights $\hat{w}_i(\tau^2)$ to the first moment of an improved approximate null distribution.

More-realistic approximations to the distribution of $Q$  are available for several effect measures. In these approximations the estimates $\hat{\sigma}_{i}^2$ are not treated as equal to the $\sigma_{i}^2$.  For SMD, \cite{kulinskaya2011testing} derived  $O(1/n)$ corrections to moments of $Q$ and suggested using the chi-squared distribution with degrees of freedom equal to the estimate of the corrected first moment to approximate the distribution of $Q$.  \cite{kulinskaya2011testing} give expressions from which it can be calculated, along with a computer program in R.

We propose a new method of estimating $\tau^2$ based on this improved approximation.
Let $E_{KDB}({Q})$ denote the corrected expected value of $Q$.  Then one obtains the KBD estimate of $\tau^2$ by iteratively solving
\begin{equation}
Q(\tau^2)=\sum\limits_{i=1}^{K}\frac{(\theta_{i}-\hat{\theta}_{RE})^{2}}{\hat{\sigma}_{i}^2+\tau^2}=E_{KDB}({Q}).
\end{equation}
We denote the resulting estimator of $\tau^2$ by $\hat{\tau}_{KDB}^2$.

We also propose a new KDB confidence interval for the between-study variance. This  interval for $\tau^2$ combines the Q-profile approach and the improved approximation by \cite{kulinskaya2011testing} (i.e.,  the chi-squared distribution with fractional degrees of freedom based on the corrected first moment of $Q$).

This corrected Q-profile confidence interval can be estimated from the lower and upper quantiles of $F_Q$, the cumulative distribution function for the improved approximation to the distribution of $Q$:
\begin{equation}
Q(\tau_{L}^2)=F_{Q;0.975}\qquad Q(\tau_{U}^2)=F_{Q;0.025}
\end{equation}
The upper and lower confidence limits for $\tau^2$ can be calculated iteratively.

\section{Sample-size-weighted (SSW)  point and interval  estimators  of $\theta$}

In an attempt to avoid the bias in the inverse-variance-weighted estimators, we included a point estimator whose weights depend only on the studies' sample sizes. For this estimator (SSW),
$w_{i} = \tilde{n}_i = n_{iT}n_{iC}/(n_{iT} + n_{iC})$; $\tilde{n}_i$ is the effective sample size in Study $i$. These weights would coincide with the inverse-variance weights
when $\delta=0$. These effective-sample-size-based weights were suggested in \cite [p.110]{hedges1985statistical}.   

The interval estimator corresponding to SSW (SSW KDB) uses the SSW point estimator as its center, and its  half-width equals the estimated standard deviation of SSW under the random-effects model times the critical value from the $t$ distribution on $K - 1$ degrees of freedom.  The estimator of the variance of SSW is
\begin{equation}\label{eq:varianceOfSSW}
\widehat{\var}(\hat{\theta}_{\mathit{SSW}})= \frac{\sum \tilde{n}_i^2 (v_i^2 + \hat{\tau}^2)} {(\sum \tilde{n}_i)^2},
\end{equation}
in which $v_i^2$ comes from Equation~(\ref{eq:g_var})  and $\hat{\tau}^2 = \hat{\tau}_{\mathit{KDB}}^2$.

\section{Simulation study}\label{sec:simsect}
As mentioned in Section~\ref{sec:Intro}, other studies have used simulation to examine estimators of $\tau^2$ or of the overall effect for SMD, but gaps in evidence remain.

Our simulation study for SMD uses $0 \leq \delta \leq 2$ and $0\leq \tau^2\leq 2.5$ as realistic for a range of applications. 

Our simulation study assesses the performance of five methods for point estimation of between-study variance $\tau^2$ (DL, REML, J, MP, and KDB) and five methods of interval estimation of $\tau^2$ (Q-profile-based methods corresponding to DerSimonian-Laird and KDB, the generalized Q-profile intervals of \cite{biggerstaff2008exact} and \cite{jackson2013confidence}, and the profile-likelihood confidence interval based on REML).

We also assess the performance of the point and interval estimators of  $\delta$ in the random-effects model for SMD.

We vary five parameters: the overall true SMD ($\delta$), the between-studies variance ($\tau^2$), the number of studies ($K$), the studies' total sample size ($n$ and $\bar{n}$), and the proportion of observations in the Control arm ($q$). The combinations of parameters are listed in Table~\ref{tab:altdataSMD}.

All simulations use the same numbers of studies  $K = 5, \;10, \;30$ and,  for each combination of parameters, the same vector of total sample sizes $n = (n_{1},\ldots, n_{K})$ and the same proportions of observations in the Control arm $q_i = n_{iC}/n_i = .5, \;.75$ for all $i$. The values of $q$ reflect two situations for the two arms of each study: approximately equal (1:1) and quite unbalanced (1:3). The sample sizes in the Treatment and Control arms are $n_{iT}=\lceil{(1 - q_i)n_{i}}\rceil$ and $n_{iC}=n_{i}-n_{iT}$, $i=1,\ldots,K$.

We study equal and unequal study sizes. For equal study sizes $n_i$ is as small as 20, and for unequal study sizes $n_i$ is as small as 12, in order to examine how the methods perform for the extremely small sample sizes that arise in some areas of application.
In choosing unequal study sizes, we follow a suggestion of \cite{sanches-2000}, who selected study sizes having skewness of 1.464, which they considered typical in behavioral and health sciences. Table~\ref{tab:altdataSMD} gives the details.

The patterns of sample sizes are illustrative; they do not attempt to represent all patterns seen in practice. By using the same patterns of sample sizes for each combination of the other parameters, we avoid the additional variability in the results that would arise from choosing sample sizes at random (e.g., uniformly between 20 and 200).

We use a total of $10,000$ repetitions for each combination of parameters. Thus, the simulation standard error for estimated coverage of $\tau^2$ or $\delta$ at the $95\%$ confidence level is roughly $\sqrt{0.95 \times 0.05/10,000}=0.00218$.

We generate the true effect sizes $\delta_{i}$ from a normal distribution:  $\delta_{i} \sim N(\delta, \tau^2)$. We generate the values of Hedges's estimator ${g}_{i}$ directly from the appropriately scaled non-central $t$-distribution, given by Equation (\ref{eq:g_dist}), and obtain their estimated within-study variances from Equation~(\ref{eq:g_var}).

The simulations were programmed in R version 3.3.2 using the University of East Anglia 140-computer-node High Performance Computing (HPC) Cluster, providing a total of 2560 CPU cores, including parallel processing and large memory resources. For each configuration, we divided the 10,000 replications into 10 parallel sets of 1000 replications.

The structure of the simulations invites an analysis of the results along the lines of a designed experiment, in which the variables are $\tau^2$, $n$, $K$, $q$,  and $\delta$. Most of the variables are crossed, but two have additional structure. Within the two levels of $n$, equal and unequal, the values are nested: $n = 20,\;40,\; 100,\; 250$ and $\bar{n} = 30, \;60, \;100, \;160$.   We approach the analysis of the data from the simulations qualitatively, to identify the variables that substantially affect (or do not affect) the performance of the estimators as a whole and the variables that reveal important differences in performance.  We might hope to describe the estimators' performance one variable at a time, but such ``main effects'' often do not provide an adequate summary: important differences are related to certain combinations of two or more variables.

We use this approach to examine bias and coverage in estimation of $\tau^2$ and bias and coverage in estimation of  $\delta$. Our summaries of results are based on examination of the figures in the corresponding Appendices. Section~\ref{sec:SMDresults} gives brief summaries, and Appendix A contains more detail.


\begin{table}[ht]
	\caption{\label{tab:altdataSMD} \emph{Combinations of parameters in the simulations for SMD}}
	\begin{footnotesize}
		\begin{center}
			\begin{tabular}
				{|l|l|l|l|}
				\hline
				SMD&Equal study sizes& Unequal study sizes&Results in\\
                &&&Appendix\\
				\hline
				$K$ (number of studies)& 5, 10, 30&5, 10, 30&\\
				$n$ or $\bar{n}$  (average (individual) study size,  & 20, 40, 100, 250& 30 (12,16,18,20,84),& \\
				total of the two arms)&30, 50, 60, 70&60 (24,32,36,40,168),&\\
				For  $K=10$ and $K=30$,  the same set of &&100 (64,72,76,80,208),& \\
				unequal study sizes is used 2 or 6 times, respectively.&&160 (124,132,136,140,268)& \\
				$q$ (proportion of each study in the control arm) & 1/2, 3/4&1/2, 3/4&\\
				$\tau^{2}$ (variance of random effect)&0(0.5)2.5&0(0.5)2.5&  A1, A2\\
                $\delta$ (true value of the SMD) &0, 0.2, 0.5, 1, 2&0, 0.2, 0.5, 1, 2&   B1, B2\\
				\hline
			\end{tabular}
		\end{center}
	\end{footnotesize}
\end{table}

\section {Methods of estimation of $\tau^2$ and $\delta$ used  in simulations}
\subsection*{Point estimators of $\tau^2$}
\begin{itemize}
\item DL - method of \cite{dersimonian1986meta}
\item J - method of \cite{jackson2013confidence}
\item KDB - method based on corrected null moment of $Q$ per \cite{kulinskaya2011testing}
\item MP - method of \cite{mandel1970interlaboratory}
\item REML - restricted maximum-likelihood method
\end{itemize}

\subsection*{Interval estimators of $\tau^2$}
\begin{itemize}
\item BJ - method of \cite{biggerstaff2008exact}
\item J - method of \cite{jackson2013confidence}
\item KDB - Q-profile method based on corrected null distribution of $Q$ per \cite{kulinskaya2011testing}
\item PL - profile-likelihood confidence interval based on $\hat{\tau}_{REML}^2$
\item QP - Q-profile confidence interval of \cite{viechtbauer2007confidence}
\end{itemize}

\subsection*{Point estimators of $\delta$}
Inverse-variance-weighted methods with $\tau^2$ estimated by:
\begin{itemize}
\item DL
\item J
\item REML
\item KDB
\item MP
\end{itemize}
and
\begin{itemize}
\item SSW - weighted mean with weights that depend only on studies sample sizes
\end{itemize}

\subsection*{Interval estimators of $\delta$}
Inverse-variance-weighted methods using normal quantiles, with  $\tau^2$ estimated by:
\begin{itemize}
\item DL
\item J
\item KDB
\item MP
\item REML
\end{itemize}
Inverse-variance-weighted methods with modified variance of $\hat{\delta}$ and t-quantiles as in  \cite{hartung2001refined} and \cite{sidik2002simple}
\begin{itemize}
\item HKSJ (DL) -  $\tau^2$ estimated by DL
\item HKSJ KDB -  $\tau^2$ estimated by KDB
\end{itemize}
and
\begin{itemize}
\item SSW KDB - SSW point estimator of $\delta$ with estimated variance given by (\ref{eq:varianceOfSSW}) and t-quantiles
\end{itemize}

\section{Results}\label{sec:SMDresults}

Our full simulation results, comprising $130$ figures, each presenting $12$ combinations of the 4 values of $n$ or $\bar{n}$ and the 3 values of $K$, are provided in Appendices A and B.  A summary  is given below.

\subsection{Bias in estimation of $\tau^2$ (Appendix A1)}

\noindent The five estimators (DL, REML, J, MP, and KDB) have biases whose traces fan out from the same small positive bias at $\tau^2 = 0$. As $\tau^2$ increases, KDB remains positive and increases slowly; MP stays close to 0 (and slightly below); REML stays negative, with a negative slope; J stays negative, with a more-negative slope; and DL becomes increasingly negative, showing noticeable curvature. The value of $\delta$ has little effect on this pattern (except that the bias of DL and J has smaller magnitude when $\delta = 2$).

As $n$ or $\bar{n}$ increases, the traces for KDB, MP, and REML flatten, and their bias is essentially 0 when $n$ or $\bar{n} = 100$. The traces for J and DL become less steep, but substantial bias remains at $n = 250$ (or $\bar{n} = 160$).

As $K$ increases, the trace for KDB flattens somewhat, but the traces for the other estimators become steeper.

The traces for J and DL are slightly less steep when $q = .75$ than when $q = .5$.

In summary,  the patterns of bias indicate a choice among the five estimators of $\tau^2$ (DL, REML, J, MP, and KDB). When $n \leq 40$, MP is closer to unbiased than KDB when $K = 5$, the magnitudes of their biases are roughly equal when $K = 10$, and KDB is closer to unbiased when $K = 30$. When $n \geq 100$, MP, KDB, and REML are nearly unbiased. DL and J seriously underestimate $\tau^2$. The average of MP and KDB should be close to unbiased.
\subsection{Coverage in estimation of $\tau^2$ (Appendix A2)}

\noindent The five estimators (PL, QP, BJ, J, and KDB) share the feature that their coverage decreases as $\tau^2$ increases from 0 to 0.5. At $\tau^2 = 0$ all five have coverage $\geq .95$. KDB is highest (e.g., .99 when $q = .5$ and $n = 20$), but it drops below .95 at $\tau^2 = 0.5$ or $\tau^2 = 1.0$ and remains slightly below .95. BJ is next highest (e.g., .98 when $q = .5$ and $n = 20$), and it remains above .95 (say, .96 to .97) when $K = 5$ and $K = 10$. QP is close to .95 for $\tau^2 \geq 0.5$. PL is between BJ and QP, and it remains above QP when $K = 30$. The trace for BJ behaves quite differently when $K = 30$ than when $K \leq 10$, decreasing steeply and linearly to around .77 at $\tau^2 = 2.5$ (when $q = .5$ and $n = 20$). When $K \leq 10$, J is between BJ and QP;  and when $K = 30$, it also decreases linearly, but less steeply (e.g., to around .92 at $\tau^2 = 2.5$ when $q = .5$ and $n = 20$).

Coverage does not change noticeably as $n$ or $\bar{n}$ increases, when $K \leq 10$.  When $K = 30$, the slopes of BJ and J become slightly less steep, and the traces of the other estimators move closer together and are closer to .95.

Setting aside the behavior of BJ and J when $K = 30$, and of the other estimators when $n= 20$ and $K = 30$, the traces move closer together as $K$ increases.

The slopes of BJ and J when $K = 30$ are less steep when $q = .75$ than when $q = .5$.

As $\delta$ increases, the coverage of QP at $K = 30$ increases slightly; it is substantially closer to .95 when $\delta = 2$.

In summary, all five interval estimators of $\tau^2$ have coverage substantially above .95 when $\tau^2 = 0$. When $\tau^2 \geq 0.5$, QP is generally closest to .95. The unusual behavior of BJ (and, to a lesser extent, J) when $K = 30$ adds to the evidence against it.
\subsection{Bias and mean squared error in estimation of $\delta$ (Appendix B1)}

\noindent When $\delta = 0$, the bias of the six estimators (DL, REML, MP, KDB, J, and SSW) follows a single trace, close to 0, for all values of $\tau^2$. When $\delta > 0$, SSW stays close to 0, and the others shift down, to increasingly negative bias, as $\delta$ increases, and their traces separate. For example, when $\delta = 1$, $q = .5$, $n = 20$, $K = 10$, and $\tau^2 = 2.5$, the bias ranges from $-0.05$ (KDB) to $-0.07$ (DL), and MP, REML, and J (in that order) have intermediate values.

When $\delta \leq 0.5$, bias has little relation to $\tau^2$. When $\delta \geq 1$, however, the bias of the estimators other than SSW (especially DL) becomes increasingly negative as $\tau^2$ increases. (The plot for $n = 20$ and $K = 30$ in \ref{BiasThetaSMD2}  shows an extreme example.)

Where bias is nonzero, increasing $n$ or $\bar{n}$ moves the traces toward (or to) 0, decreasing separation between them. Some plots (e.g., ~\ref{BiasThetaSMD2q75} and \ref{BiasThetaSMD2q75unequal}) show slight evidence of greater separation among traces when sample sizes are unequal and $\delta = 2$.

For the most part, $K$ has little or no effect on bias. Some plots suggest that, where bias is nonzero, separation among traces increases as $K$ increases, especially from $K = 10$ to $K = 30$.

Bias does not differ noticeably between $q = .5$ and $q = .75$.

SSW essentially avoids the bias that we found in the inverse-variance-weighted estimators of $\delta$. To provide an additional measure of its performance (besides coverage, discussed below). we estimated the mean squared error of SSW and the best two inverse-variance-weighted estimators, KDB and MP. Appendix E1 includes figures that plot (versus $\tau^2$) the ratios MSE(SSW)/MSE(KDB) and MSE(SSW)/MSE(MP) for the five values of $\delta$, the two values of $q$, and $n_{i} = 20,\;40,\; 100,\; 250$. For most situations the two ratios are essentially equal and differ little among values of $K$ and $q$. In most situations the traces are essentially flat as $\tau^2$ increases; otherwise, they curve downward as $\tau^2$ approaches 0. As $n$ increases, the ratios approach 1. For example, when $\delta = 0$ and $K = 5$, they decrease from around 1.1 when $n = 20$ to nearly 1.0 when $n = 250$. As $\delta$ increases ($\geq 0.5$), the ratios at small $\tau^2$ decrease. This pattern is first noticeable when $\delta = 0.5$ and $n = 20$ and $K = 30$; and as $\delta$ increases, it becomes more pronounced at that combination of $n$ and $K$ and extends to larger $n$ (with $K = 30$) and to $n \leq 40$ and $K = 10$. When $\delta = 2$, $q = .5$, $n = 20$, and $K = 30$, the traces for the two ratios are separate and curve up from around 0.55 at $\tau^2 = 0$ to slightly $< 1$ when $\tau^2 = 2.5$. The patterns are similar for $q = .75$.

In summary, the bias of SSW is close to 0, and the other five estimators (DL, REML, J, MP, and KDB), which use inverse-variance weights, have greater (and negative) bias, amounting to 5 -- 10\% when sample sizes are small and $\delta \geq 1$. This bias increases as $\tau^2$ increases.  SSW usually has slightly greater mean squared error than KDB and MP when $n$ is small, but its MSE can be substantially smaller, especially for small $\tau^2$.

\subsection{Coverage in estimation of $\delta$ (Appendix B2)}

\noindent Coverage of the estimators that rely on inverse-variance weights and normal critical values (DL, REML, MP, KDB, and J) is influenced most by $\tau^2$ ($= 0$ versus $> 0$) and $K$. At $\tau^2 = 0$ their coverage is around .97, but at $\tau^2 = 0.5$ (when $q = .5$ and $n = 20$) it is mostly below .95: .90 to .91 when $K = 5$, .92 to.94 when $K = 10$, and .93 to .95+ when $K = 30$. As $\tau^2$ increases, their coverage either is flat (REML, MP, and KDB) or decreases (DL, J). DL almost always has the lowest coverage, and the gap between it and J widens as $K$ increases.

Except for the effect of $K$ on SSW KDB at $\tau^2 = 0$ (above .99 when $K = 5$, decreasing to .97 when $K = 30$) and below-nominal coverage of HKSJ and HKSJ KDB in a region whose definition involves mainly $\delta$, $n$ (or $\bar{n}$), $K$, and $\tau^2$, the coverage of SSW and the HKSJ-type estimators is close to .95 for all $K$ and $\tau^2$. The challenging situations in that region generally involve $\delta = 1$ or $\delta = 2$, the smaller $n$ or $\bar{n}$, $K = 10$ or $K = 30$, and the smaller $\tau^2$. For example, when $\delta = 2$ and $n = 20$ or $\bar{n} = 30$ and $K = 30$, coverage can be as low as .84 when $\tau^2 = 0$.

When $\tau^2 \geq 0.5$, coverage of DL, REML, MP, KDB, and J increases as $K$ increases, usually staying below .95.

The effect of $\delta$ on coverage is slight except for some situations involving $\delta = 2$. When $n = 20$ and $K = 30$, all of the estimators except SSW KDB have low coverage at $\tau^2 = 0$, ranging from $< .82$ to .86. Their traces rise as $\tau^2$ increases; when $\tau^2 = 2.5$, KDB and HKSJ KDB are almost .94, and DL is .86 (up from .84). The pattern is similar when $n = 40$ and $K = 30$, but much reduced.

When $\tau^2 \geq 0.5$, coverage \textit{decreases} slightly as $n$ or $\bar{n}$ increases (except for SSW and the two HKSJ-type estimators); coverage is somewhat lower when sample sizes are unequal.

The plots show at most slight differences between $q = .5$ and $q = .75$.

In summary, except when $\delta = 2$ and $K = 30$, HKSJ and HKSJ KDB have coverage closest to .95; they differ little, and departures from .95 (toward lower coverage) are seldom serious. SSW KDB is rather conservative when $K=5$ and for other $K$ when $\tau^2=0$. Otherwise it provides reliable, albeit slightly conservative, coverage.  When $\delta = 2$ and $K = 30$, SSW KDB is the best alternative.  All of the estimators that use inverse-variance weights and critical values from the normal distribution (DL,REML, J, MP, and KDB) often have coverage substantially below .95.

\section{Discussion: Practical implications for meta-analysis}

The results of our simulations for SMD give a rather disappointing picture. In brief:\\
Because the study-level effects and their variances are related (cf.  Equation~(\ref{eq:g_var}) for SMD), the performance of all statistical methods depends on the effect measures, estimates of overall effects are biased, and coverage of confidence intervals is too low, especially for small sample sizes.

The conventional wisdom is that these deficiencies do not matter, as meta-analysis usually deals with studies that are ``large,'' so all these little problems are automatically resolved.  Unfortunately, this is not true, even in medical meta-analyses; in Issue 4 of the Cochrane Database 2004, the maximum study size was $63$ or less in $25\%$ of meta-analyses with $K\geq 3$ that used SMD as an effect measure, and less than $110$ in $50\%$ of them (our own analysis). 
 We have not surveyed typical study sizes in psychology, but  \cite{sanches-2010}, promoting MA in  psychological research, use an example with 24 studies in which the smallest study size is $12$ and the largest is $121$. 
In ecology, typical sample sizes are between 4 and 25 \citep{Hamman2018}. An effect-measure-specific estimator of $\tau^2$, such as KDB for SMD, can reduce inherent biases.

Arguably, the main purpose of a meta-analysis is to provide point and interval estimates of an overall effect.

Usually, after estimating the between-study variance $\tau^2$, inverse-variance weights are used in estimating the overall effect (and, often, its variance). This approach relies on the theoretical result that, for known variances, and given unbiased estimates $\hat{\theta}_i$, it yields a Uniformly Minimum-Variance Unbiased Estimate (UMVUE) of $\theta$. In practice, however, the true within-study variances are unknown, and use of the estimated variances makes the inverse-variance-weighted estimate of the overall effect biased.

Consumers routinely expect point estimates to have no (or small) bias and CIs to have (close to) nominal coverage. Thus, the IV-weighted approach is unsatisfactory because, in general, it cannot produce an unbiased estimate of an overall effect.


A pragmatic approach to unbiased estimation of $\theta$ uses weights that do not involve estimated variances of study-level estimates, for example, weights proportional to the study sizes $n_i$.  \cite{hunter1990methods} and \cite{Shuster-2010}, among others, have proposed such weights, and  \cite{martynez-2010} and \cite{Hamman2018} have studied the method's performance by simulation for SMD.
We prefer to use weights proportional  to an effective sample size,  $\tilde{n}_i=n_{iT}n_{iC}/n_i$; these are the optimal inverse-variance weights for SMD when $\delta=0$ and $\tau^2=0$.
Thus, the overall effect is estimated by $\hat{\theta}_{\mathit{SSW}} = \sum \tilde{n}_i\hat{\theta}_i / \sum \tilde{n}_i$, and its variance is estimated by Equation~(\ref{eq:varianceOfSSW}). \cite{Hamman2018} use weights proposed by \cite{Hedges1982}, which differ slightly for very small sample sizes.

A good estimator of $\tau^2$, such as MP or KDB, can be used as $\hat{\tau}^2$. Further, confidence intervals for $\delta$ centered at $\hat{\delta}_{\mathit{SSW}}$ with $\hat{\tau}_{\mathit{KDB}}^2$ in Equation~(\ref{eq:varianceOfSSW}) can be used.

This approach based on SSW requires further study.  For example, in the confidence intervals we have used critical values from the $t$-distribution on $K - 1$ degrees of freedom, but we have not yet examined the actual sampling distribution of SSW.  The raw material for such an examination is readily available: For each situation in our simulations, each of the $10,000$ replications yields an observation on the sampling distribution of SSW.

\section*{Funding}
The work by E. Kulinskaya was supported by the Economic and Social Research Council [grant number ES/L011859/1].
\section*{Appendices}
\begin{itemize}
	\item Appendix A. SMD: Plots for bias and coverage of $\tau^2$.
	\item Appendix B. SMD: Plots for bias and coverage of standardized mean difference $\delta$.
\end{itemize}

\bibliographystyle{plainnat}
\bibliography{MD_SMD_Bib_19Nov18}%

\clearpage

\section{Appendices}
\subsection*{A1. Bias of point estimators of $\hat{\tau}^2$ for $\hat{\tau}^2=0.0(0.5)2.5$.}
For bias of $\hat{\tau}^2$, each figure corresponds to a value of $\delta (= 0, 0.5, 1, 1.5, 2 , 2.5)$, a value of $q (= .5, .75)$, and a set of values of $n$ (= 20, 40, 100, 250 or 30, 50, 60, 70) or $\bar{n} (= 30, 60, 100, 160)$.\\
Each figure contains a panel (with $\tau^2$ on the horizontal axis) for each combination of n (or $\bar{n}$) and $K (=5, 10, 30)$.\\
The point estimators of $\tau^2$ are
\begin{itemize}
	\item DL (DerSimonian-Laird)
	\item REML (restricted maximum likelihood)
	\item MP (Mandel-Paule)
	\item KDB (improved moment estimator based on Kulinskaya, Dollinger and  Bj{\o}rkest{\o}l (2011))
	\item J (Jackson)
\end{itemize}

\clearpage

\renewcommand{\thefigure}{A1.\arabic{figure}}

\begin{figure}[t]
	\centering
	\includegraphics[scale=0.33]{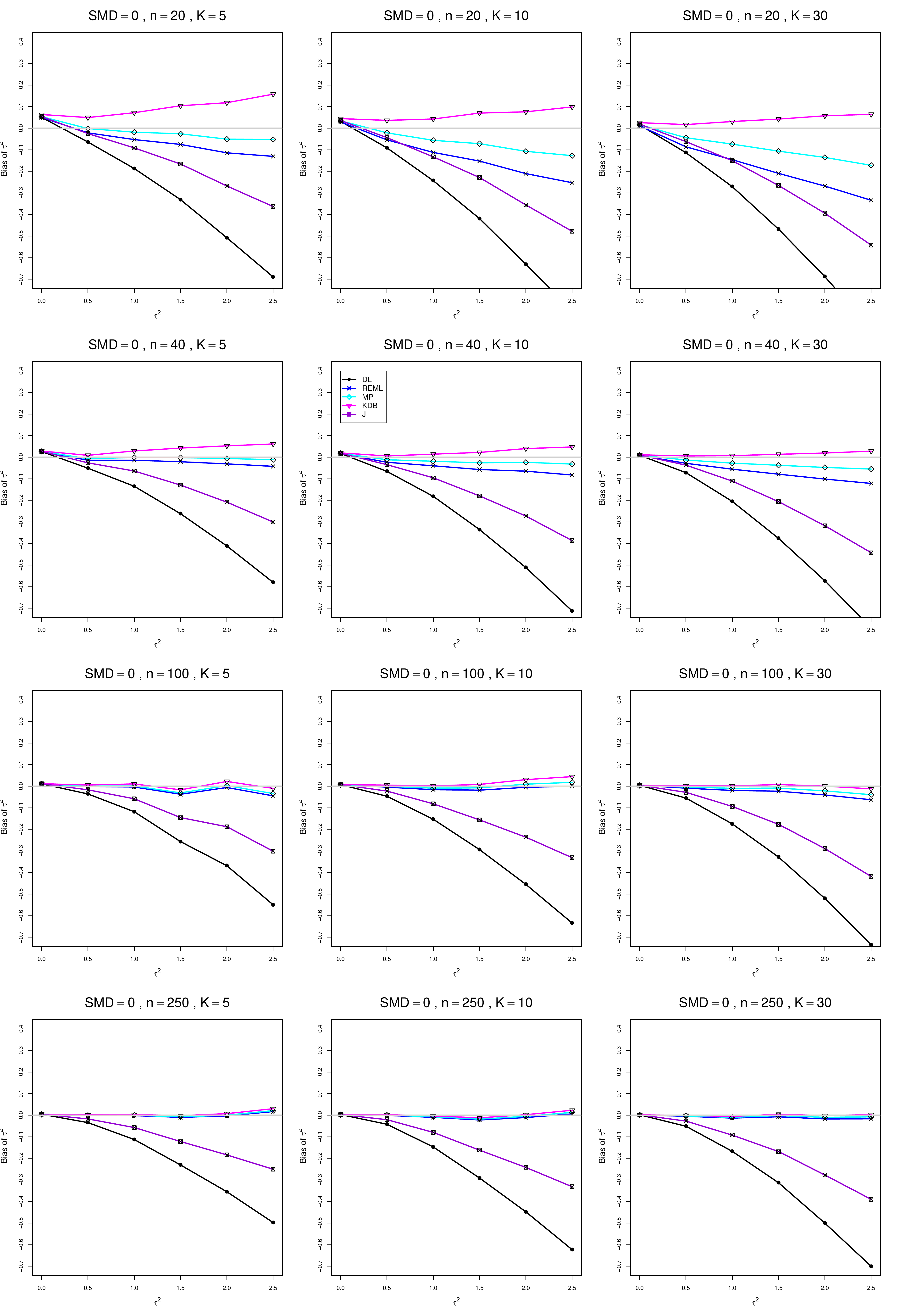}
	\caption{Bias of the estimation of  between-studies variance $\tau^2$ for $\delta=0$, $q=0.5$, $n=20,\;40,\;100,\;250$.
		\label{BiasTauSMD0}}
\end{figure}

\begin{figure}[t]
	\centering
	\includegraphics[scale=0.33]{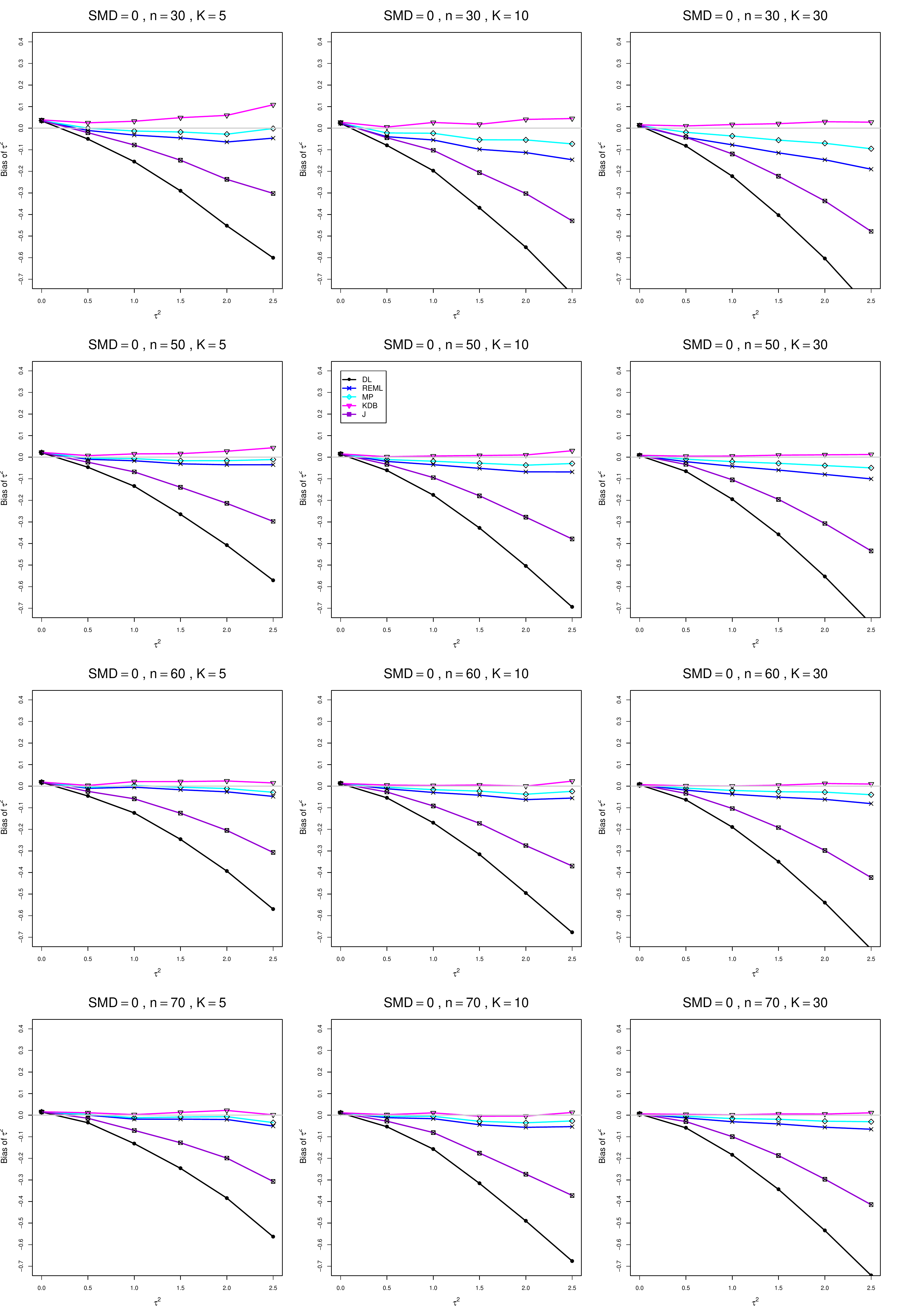}
	\caption{Bias of the estimation of  between-studies variance $\tau^2$ for $\delta=0$, $q=0.5$, $n=30,\;50,\;60,\;70$.
		\label{BiasTauSMD0small}}
\end{figure}

\begin{figure}[t]
	\centering
	\includegraphics[scale=0.33]{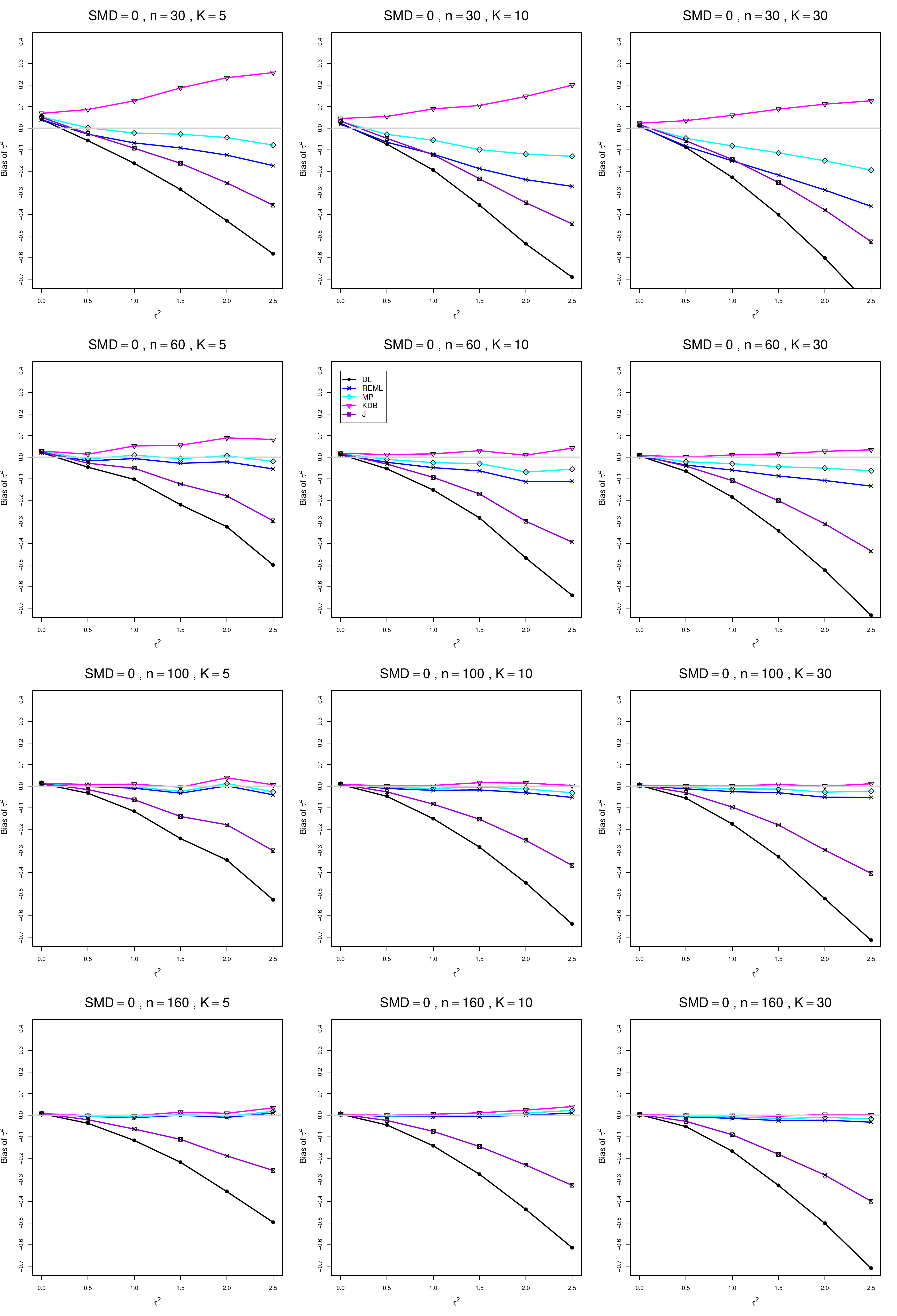}
	\caption{Bias of the estimation of  between-studies variance $\tau^2$ for $\delta=0$, $q=0.5$,  unequal sample sizes with
		$\bar{n}=30,\; 60,\;100,\;160$.
		\label{BiasTauSMD0unequal}}
\end{figure}

\begin{figure}[t]
	\centering
	\includegraphics[scale=0.3]{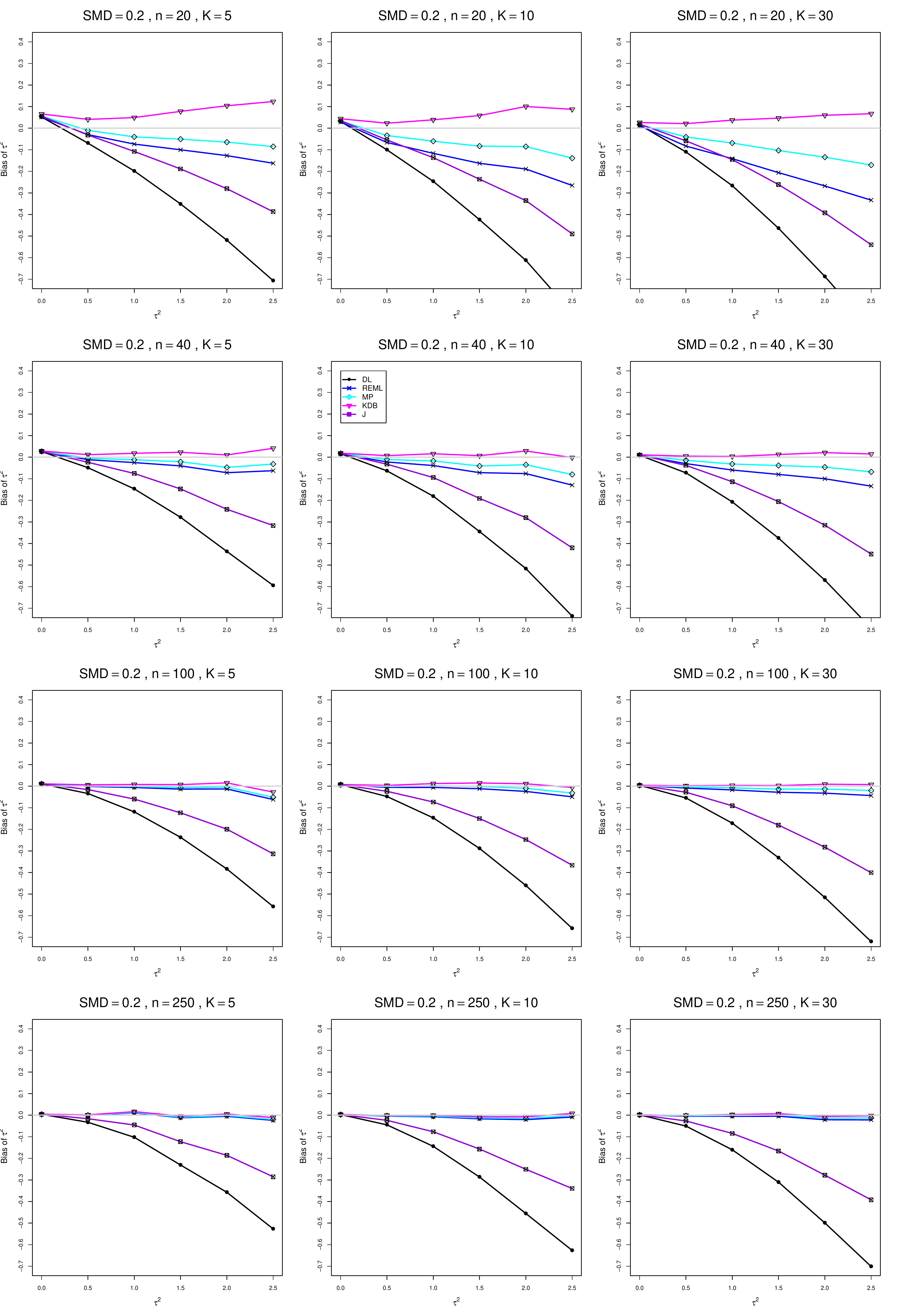}
	\caption{Bias of the estimation of  between-studies variance $\tau^2$ for $\delta=0.2$, $q=0.5$, $n=20,\;40,\;100,\;250$.
		\label{BiasTauSMD02}}
\end{figure}

\begin{figure}[t]
	\centering
	\includegraphics[scale=0.3]{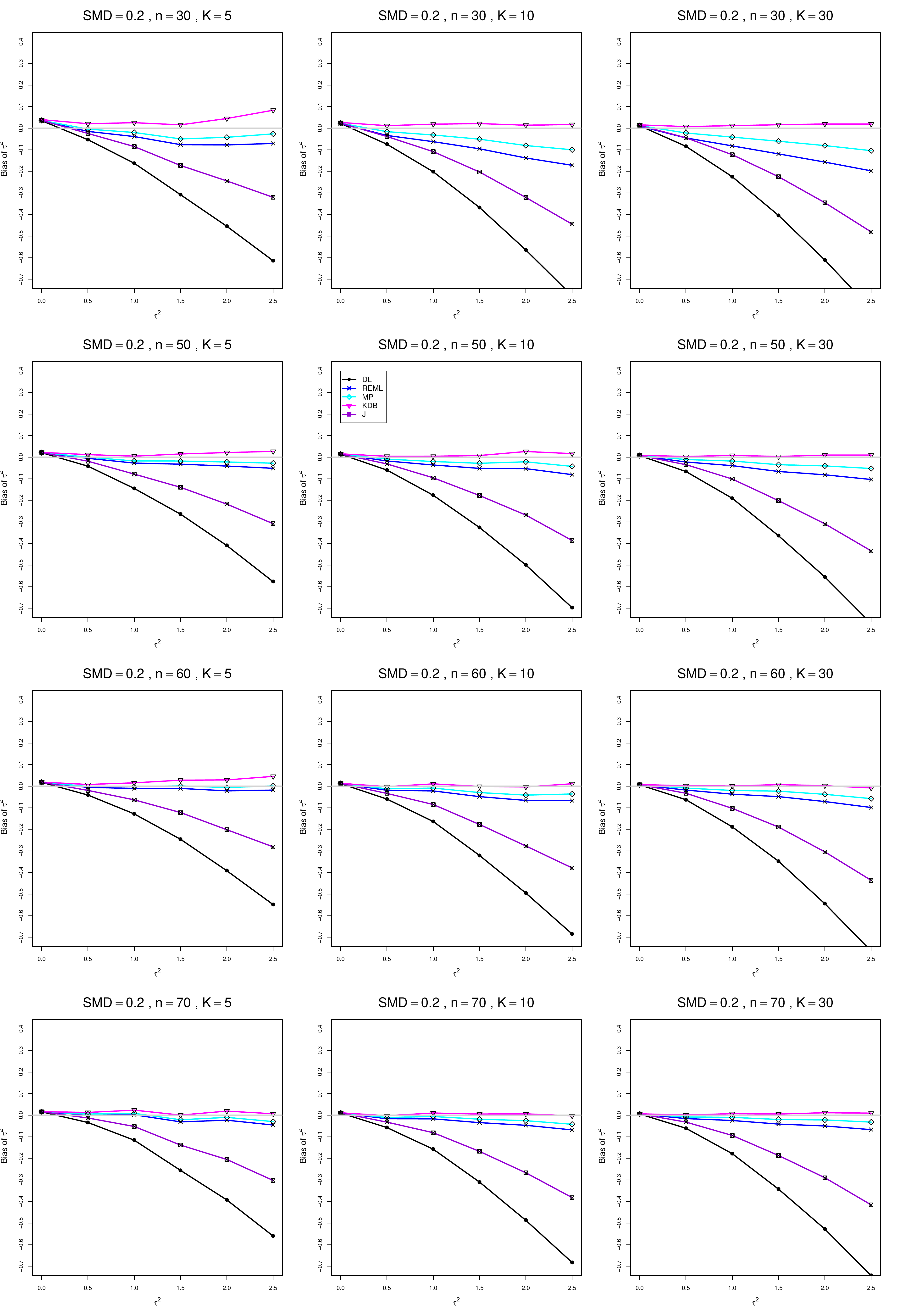}
	\caption{Bias of the estimation of  between-studies variance $\tau^2$ for $\delta=0.2$, $q=0.5$, $n=30,\;50,\;60,\;70$.
		\label{BiasTauSMD02small}}
\end{figure}

\begin{figure}[t]
	\centering
	\includegraphics[scale=0.33]{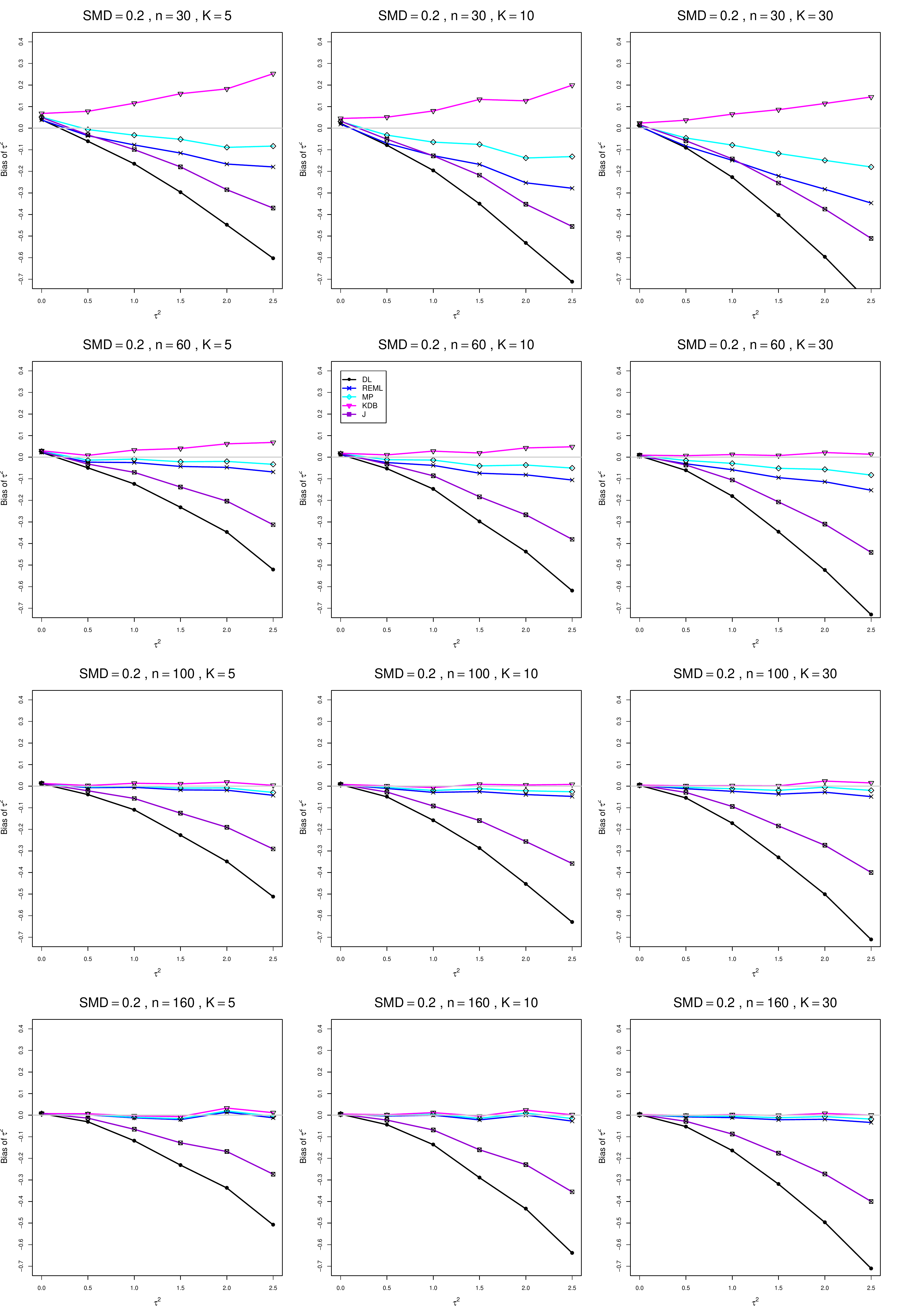}
	\caption{Bias of the estimation of  between-studies variance $\tau^2$ for $\delta=0.2$, $q=0.5$,  unequal sample sizes with
		$\bar{n}=30,\; 60,\;100,\;160$.
		\label{BiasTauSMD02unequal}}
\end{figure}

\begin{figure}[t]
	\centering
	\includegraphics[scale=0.3]{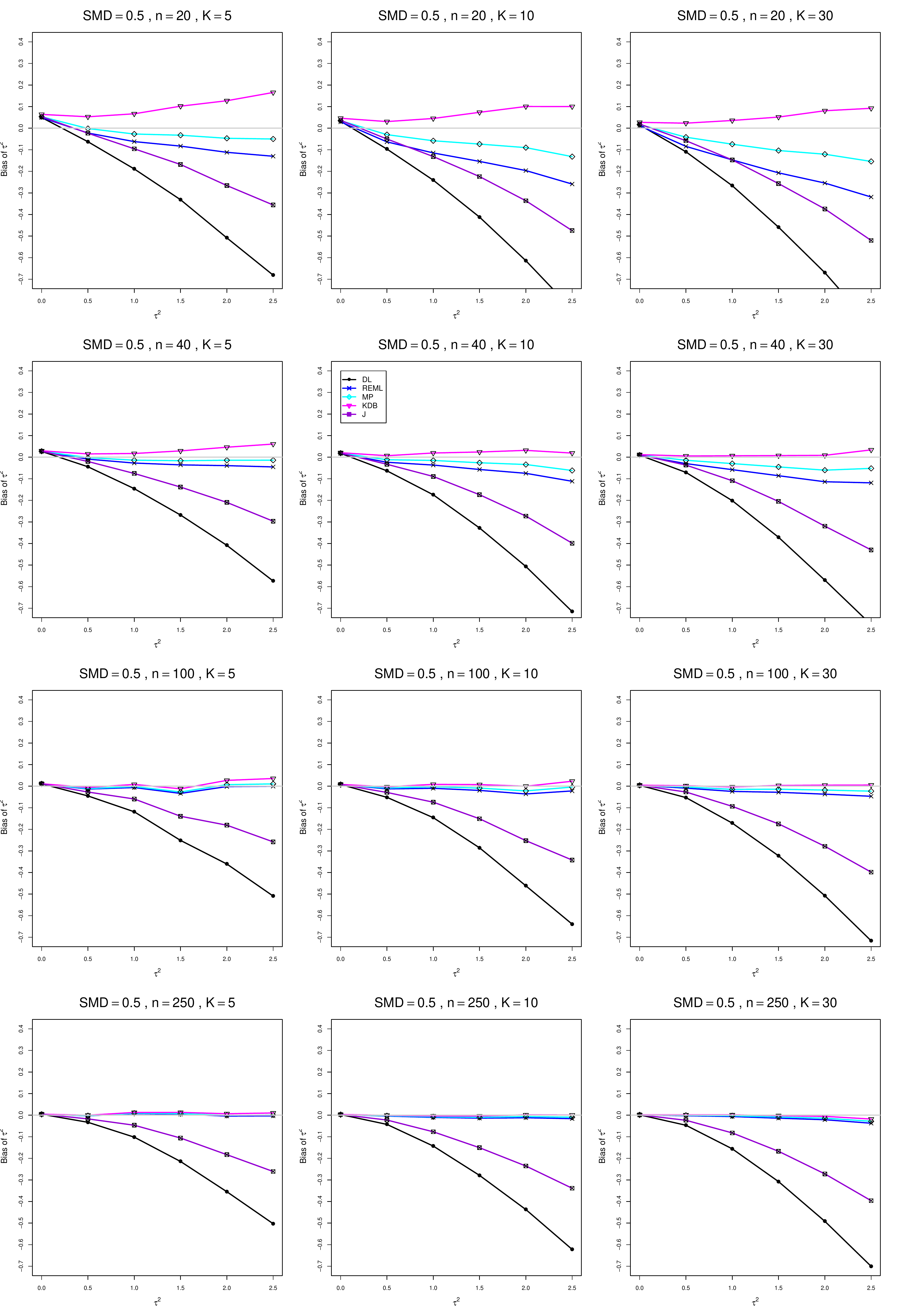}
	\caption{Bias of the estimation of  between-studies variance $\tau^2$ for $\delta=0.5$, $q=0.5$, $n=20,\;40,\;100,\;250$.
		\label{BiasTauSMD05}}
\end{figure}

\begin{figure}[t]
	\centering
	\includegraphics[scale=0.3]{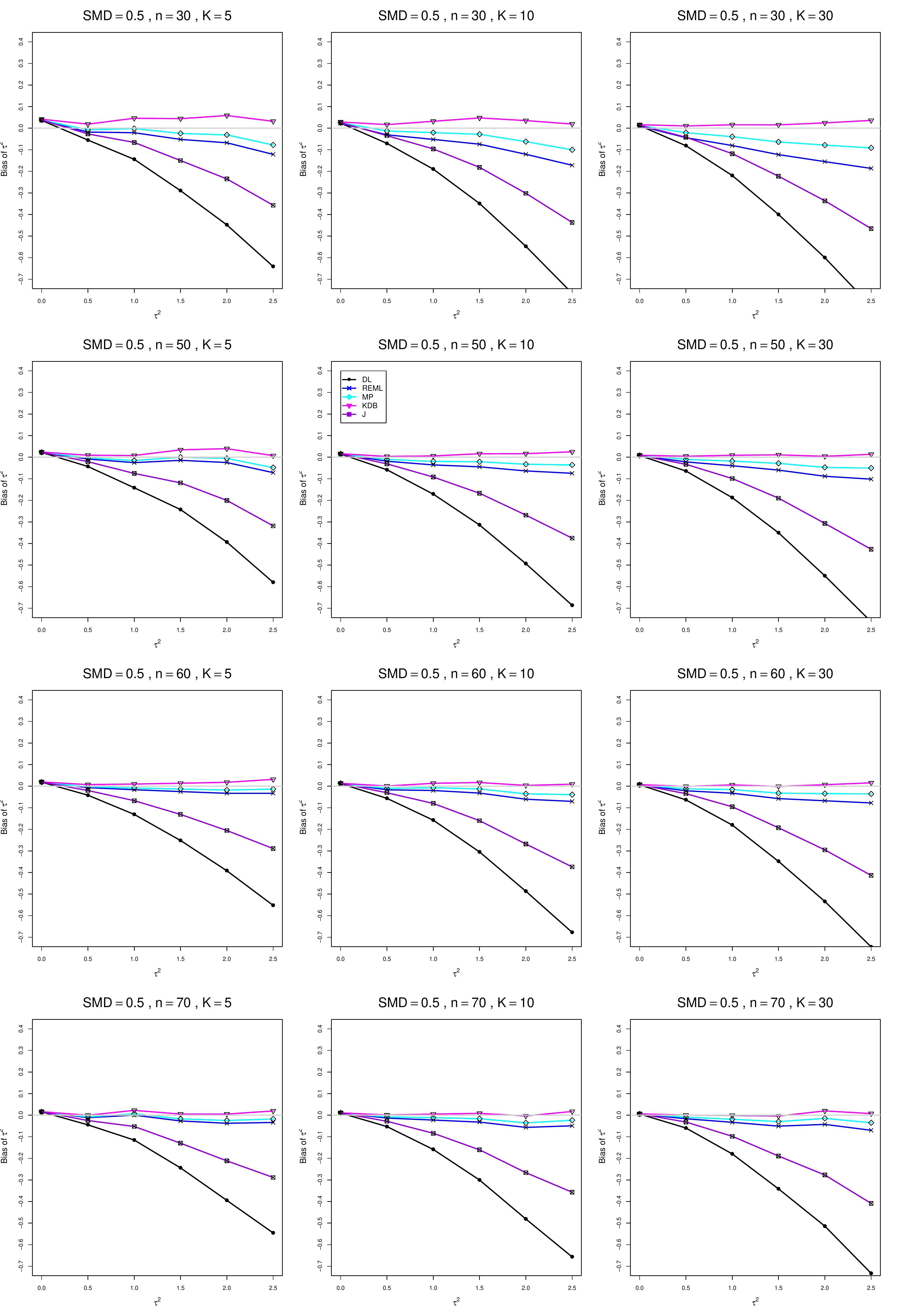}
	\caption{Bias of the estimation of  between-studies variance $\tau^2$ for $\delta=0.5$, $q=0.5$, $n=30,\;50,\;60,\;70$.
		\label{BiasTauSMD05small}}
\end{figure}

\begin{figure}[t]
	\centering
	\includegraphics[scale=0.33]{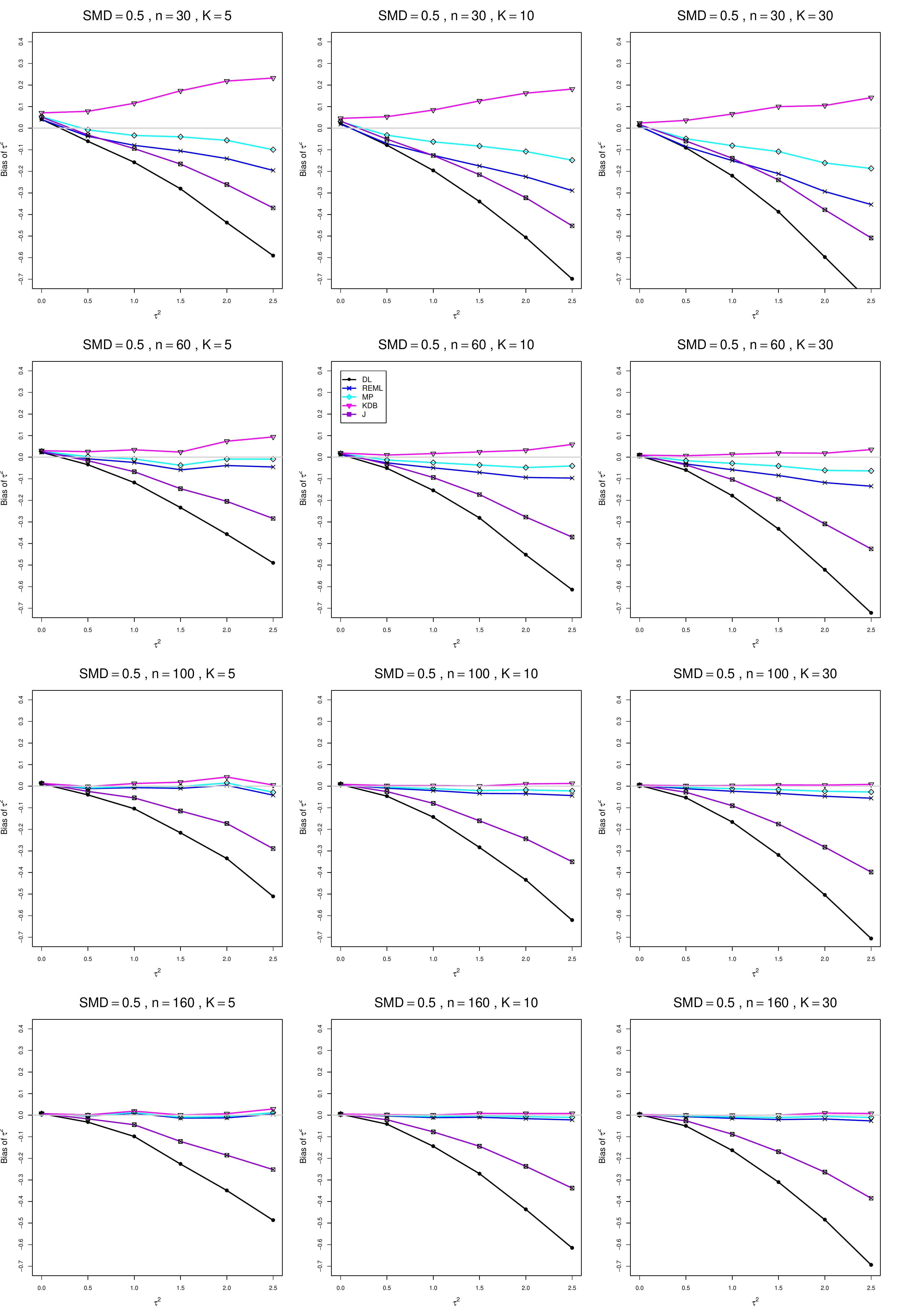}
	\caption{Bias of the estimation of  between-studies variance $\tau^2$ for $\delta=0.5$, $q=0.5$,  unequal sample sizes with
		$\bar{n}=30,\; 60,\;100,\;160$.
		\label{BiasTauSMD05unequal}}
\end{figure}

\begin{figure}[t]
	\centering
	\includegraphics[scale=0.3]{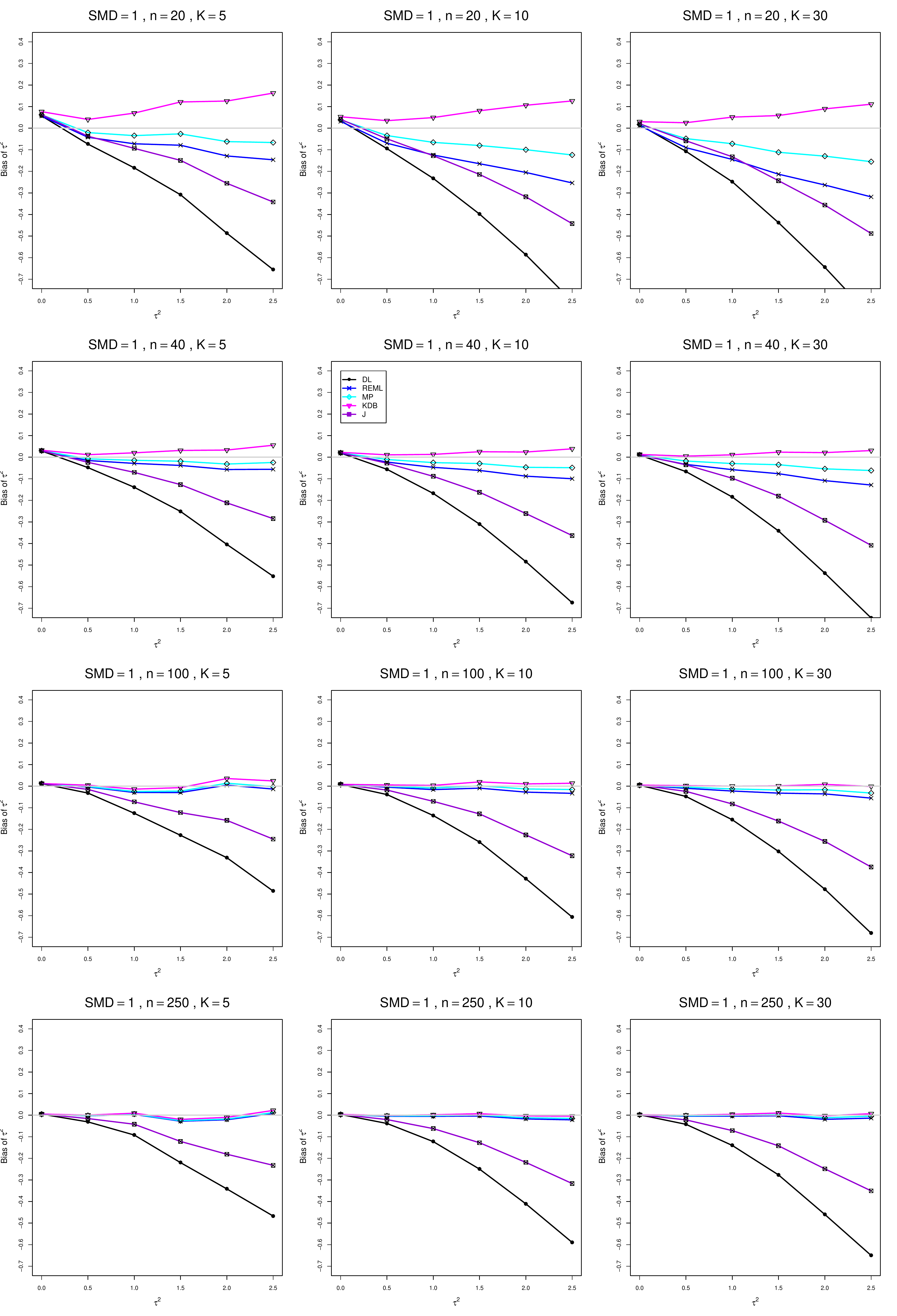}
	\caption{Bias of the estimation of  between-studies variance $\tau^2$ for $\delta=1$, $q=0.5$, $n=20,\;40,\;100,\;250$.
		\label{BiasTauSMD1}}
\end{figure}

\begin{figure}[t]
	\centering
	\includegraphics[scale=0.3]{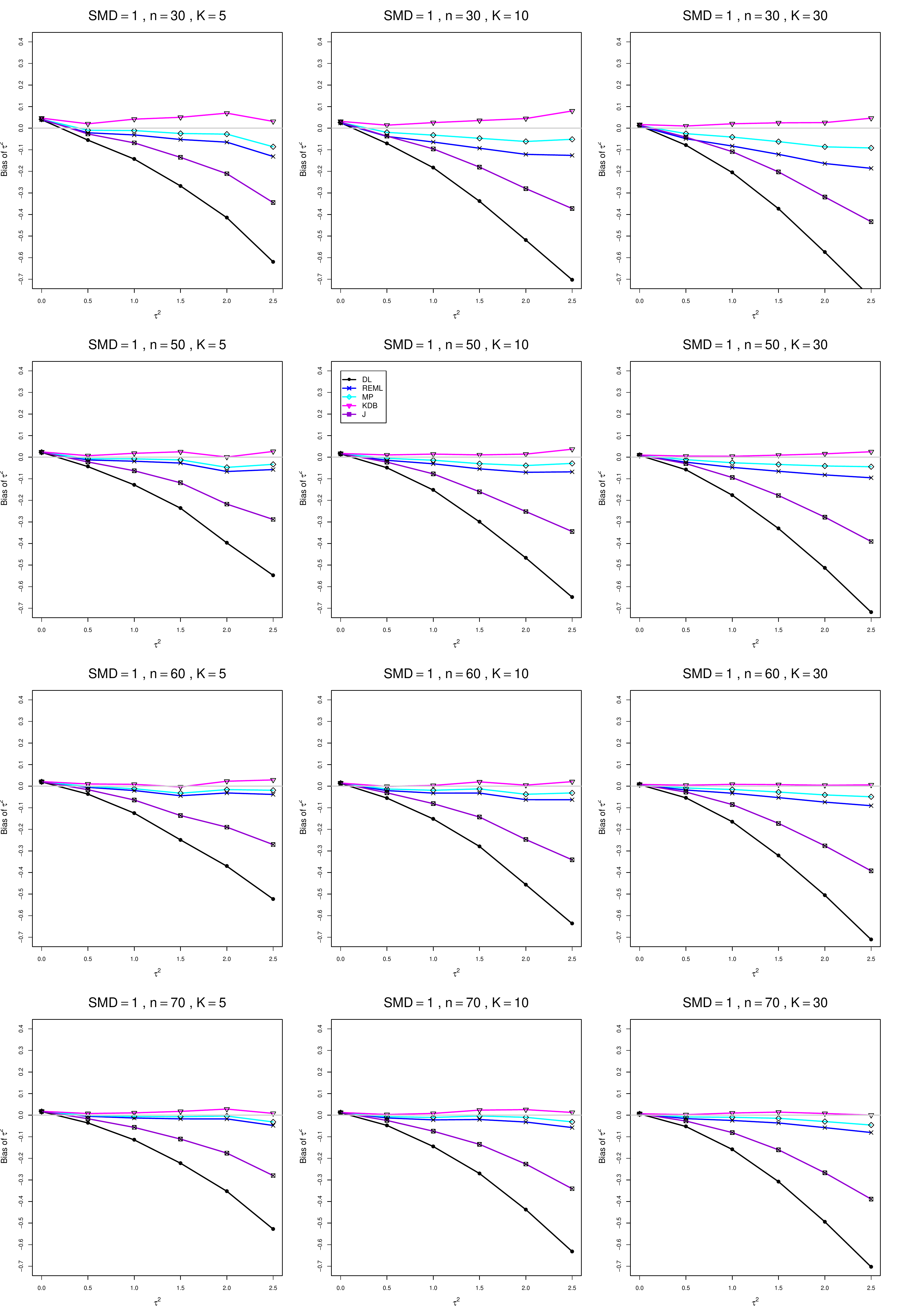}
	\caption{Bias of the estimation of  between-studies variance $\tau^2$ for $\delta=1$, $q=0.5$, $n=30,\;50,\;60,\;70$.
		\label{BiasTauSMD1small}}
\end{figure}

\begin{figure}[t]
	\centering
	\includegraphics[scale=0.33]{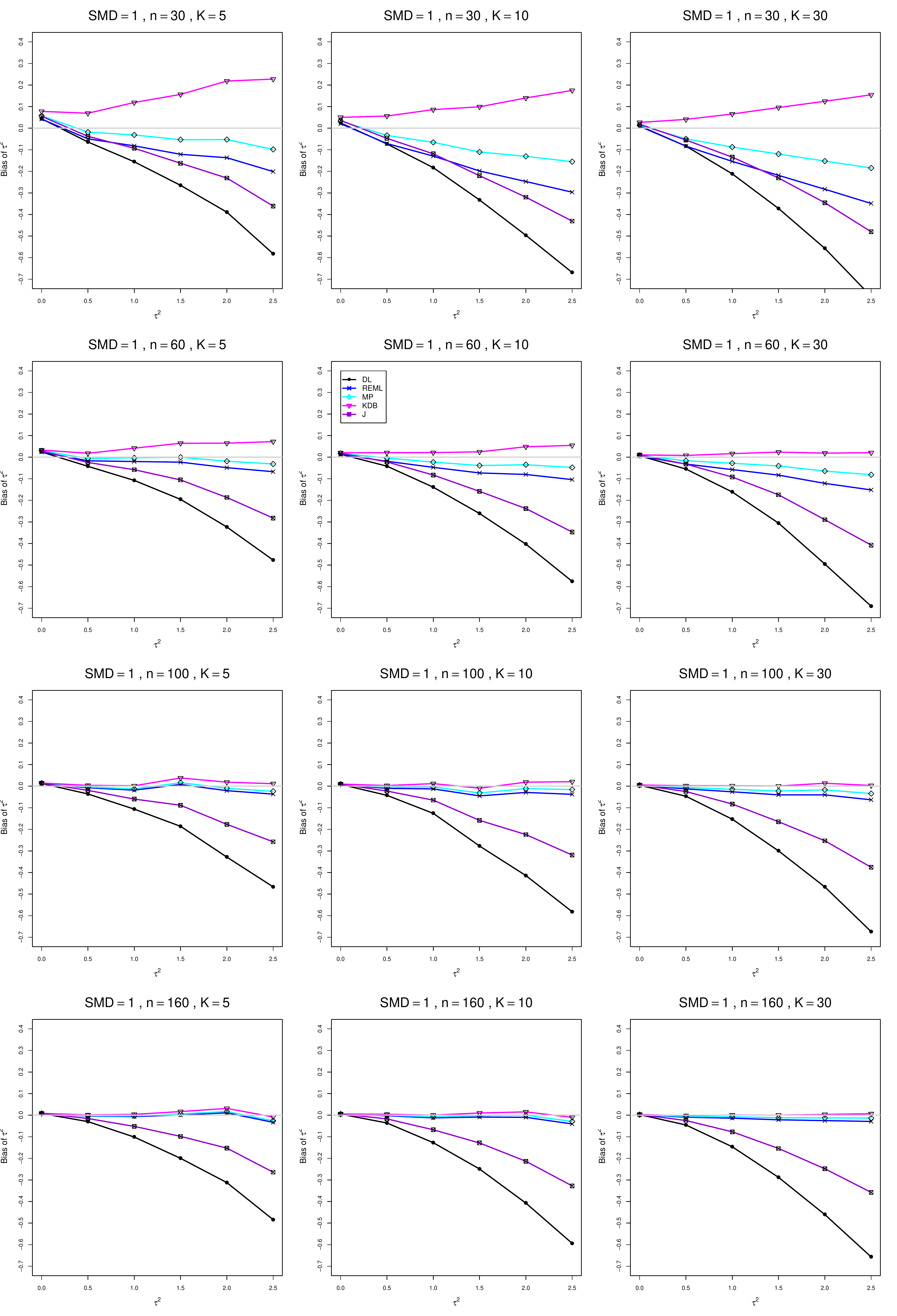}
	\caption{Bias of the estimation of  between-studies variance $\tau^2$ for $\delta=1$, $q=0.5$,  unequal sample sizes with
		$\bar{n}=30,\; 60,\;100,\;160$.
		\label{BiasTauSMD1unequal}}
\end{figure}

\begin{figure}[t]
	\centering
	\includegraphics[scale=0.3]{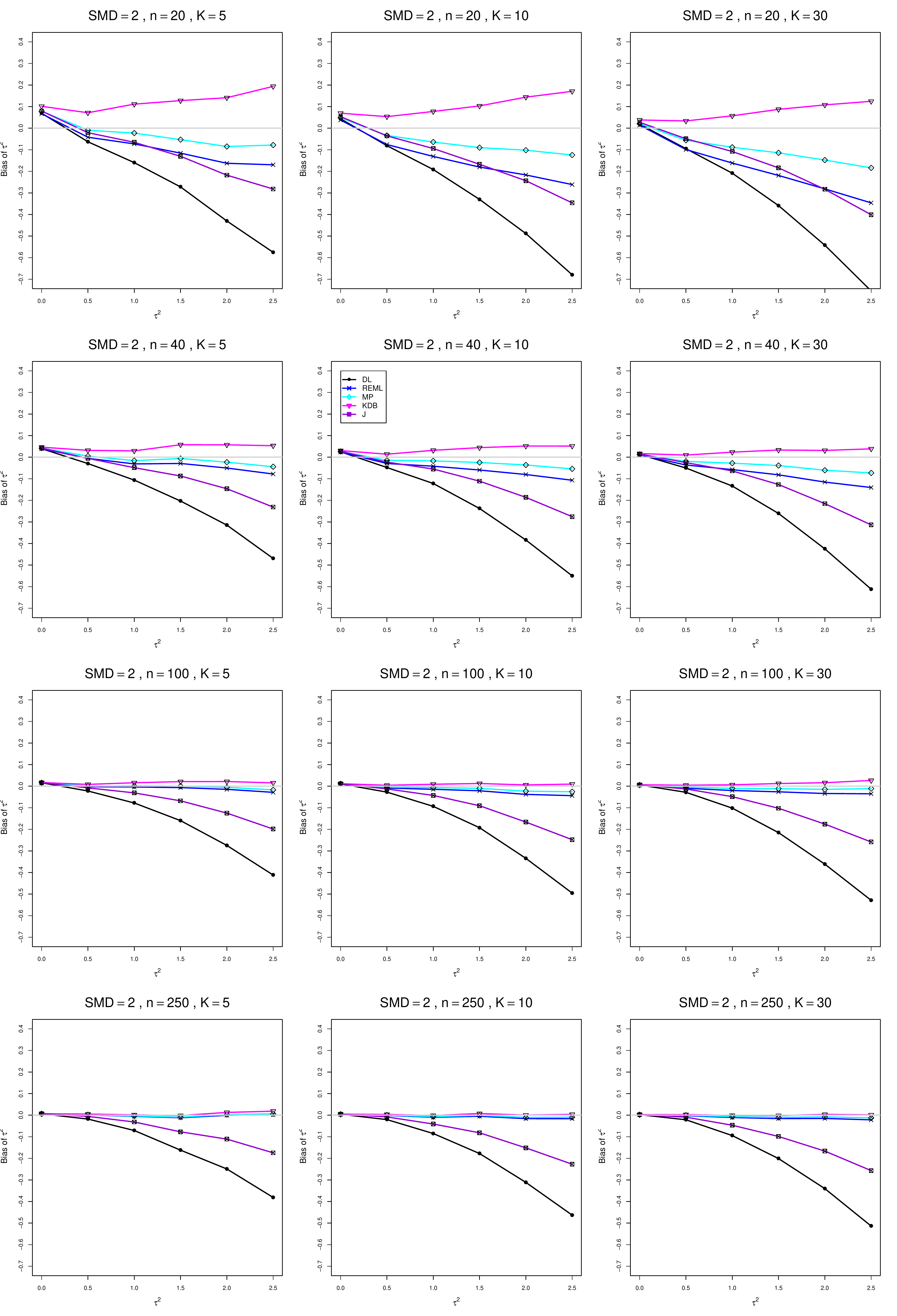}
	\caption{Bias of the estimation of  between-studies variance $\tau^2$ for $\delta=2$, $q=0.5$, $n=20,\;40,\;100,\;250$.
		\label{BiasTauSMD2}}
\end{figure}

\begin{figure}[t]
	\centering
	\includegraphics[scale=0.3]{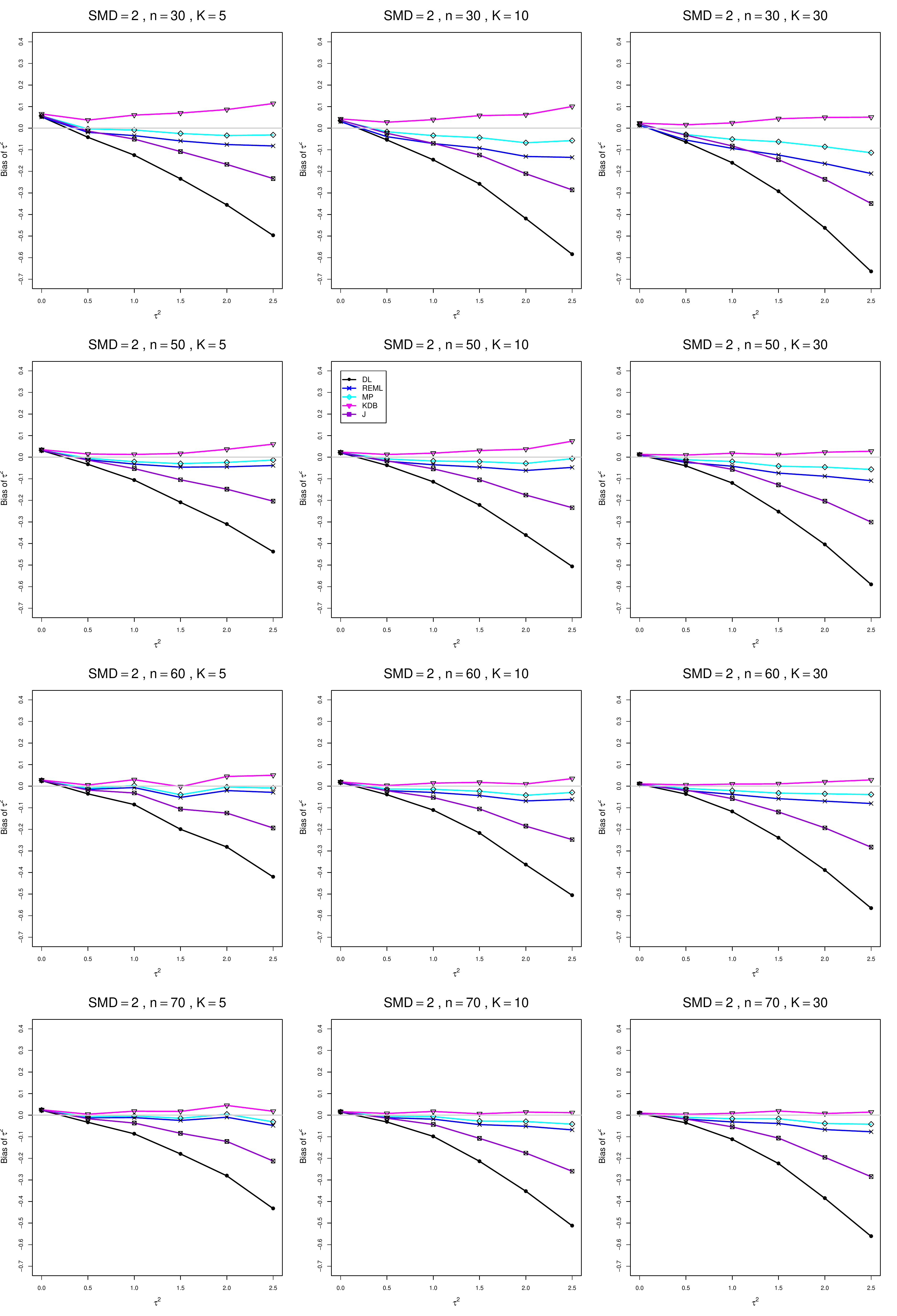}
	\caption{Bias of the estimation of  between-studies variance $\tau^2$ for $\delta=2$, $q=0.5$, $n=30,\;50,\;60,\;70$.
		\label{BiasTauSMD2small}}
\end{figure}

\begin{figure}[t]
	\centering
	\includegraphics[scale=0.33]{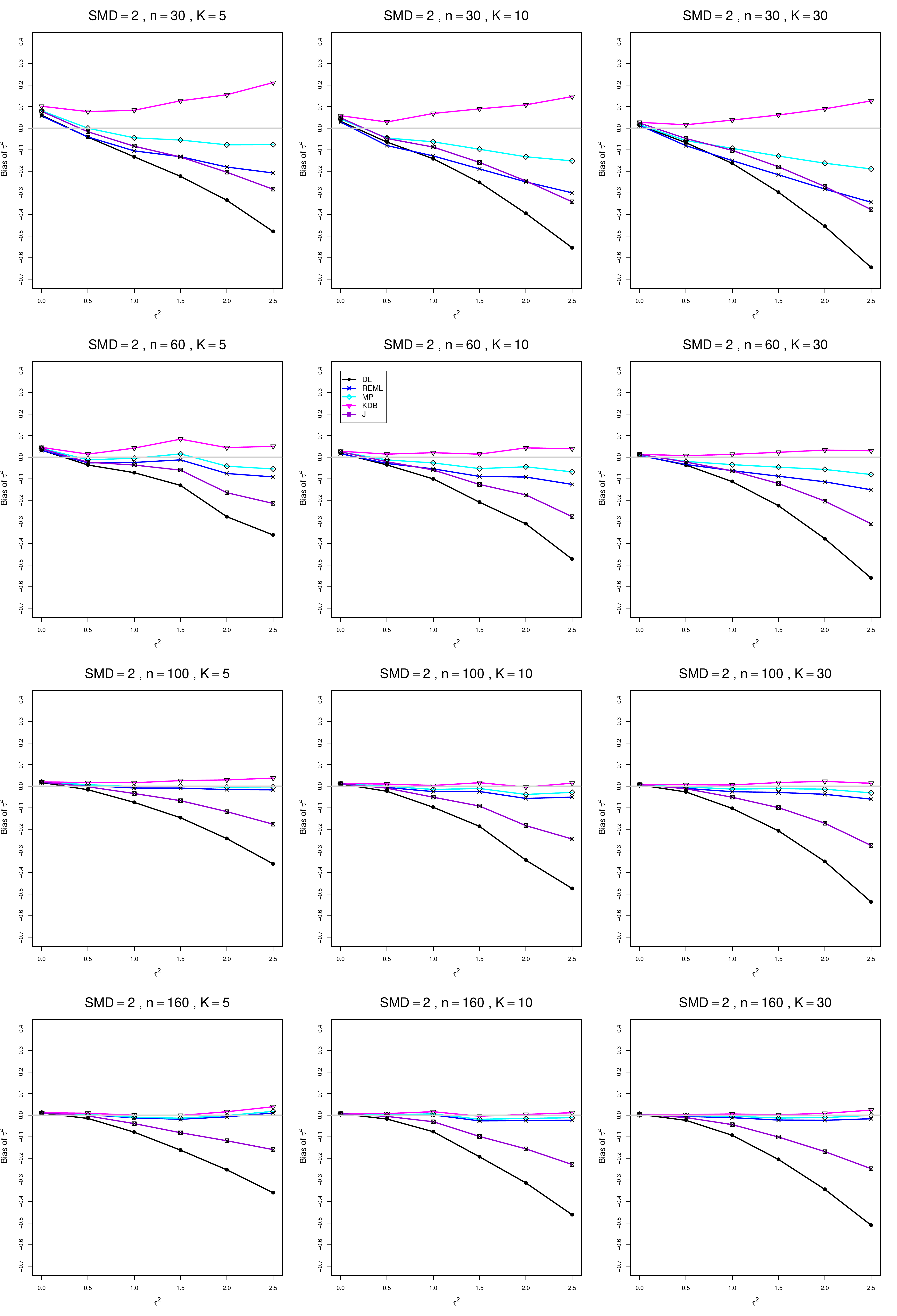}
	\caption{Bias of the estimation of  between-studies variance $\tau^2$ for $\delta=2$, $q=0.5$,  unequal sample sizes with
		$\bar{n}=30,\; 60,\;100,\;160$.
		\label{BiasTauSMD2unequal}}
\end{figure}
\begin{figure}[t]
	\centering
	\includegraphics[scale=0.33]{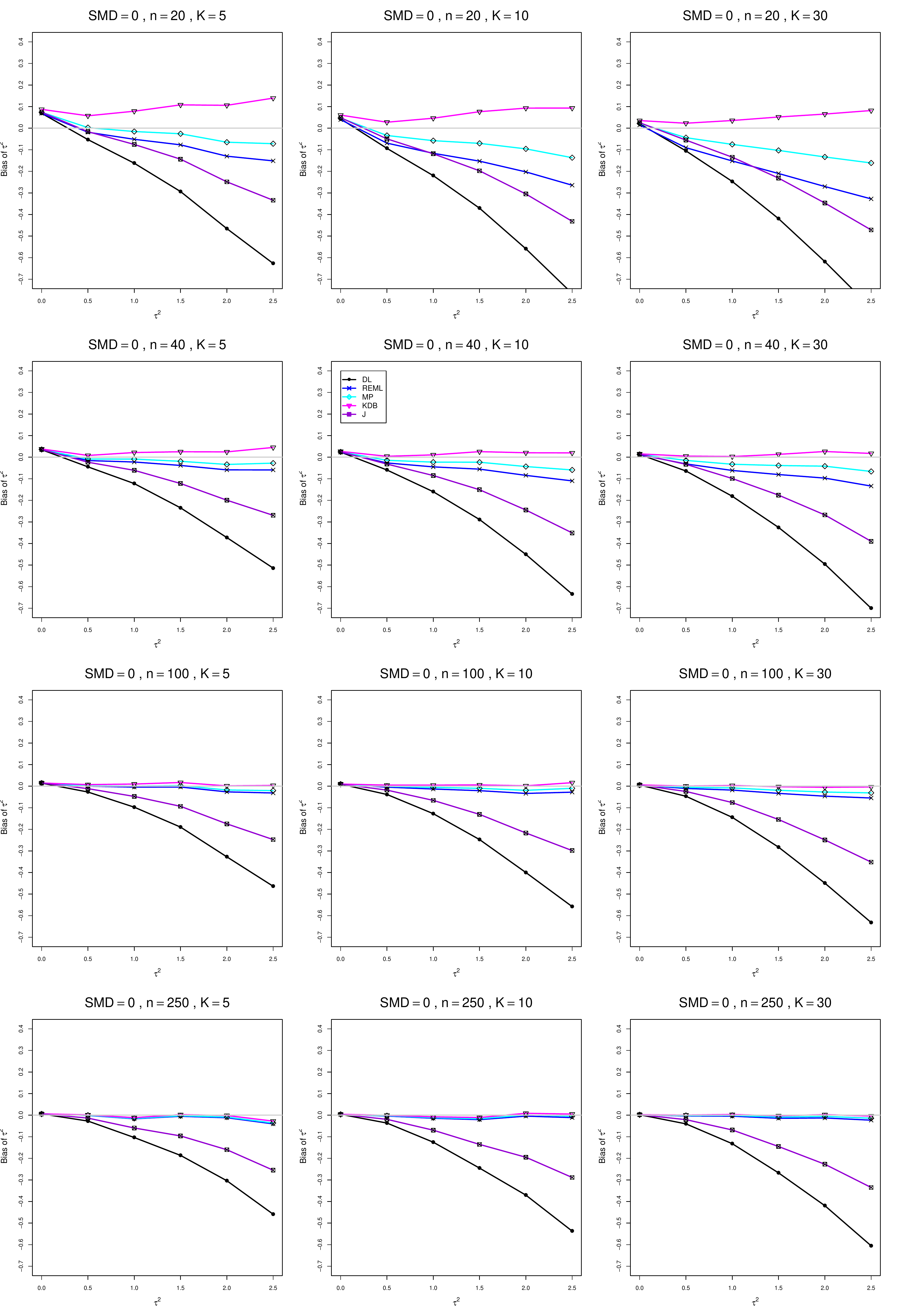}
	\caption{Bias of the estimation of  between-studies variance $\tau^2$ for $\delta=0$, $q=0.75$, $n=20,\;40,\;100,\;250$.
		\label{BiasTauSMD0q75}}
\end{figure}

\begin{figure}[t]
	\centering
	\includegraphics[scale=0.33]{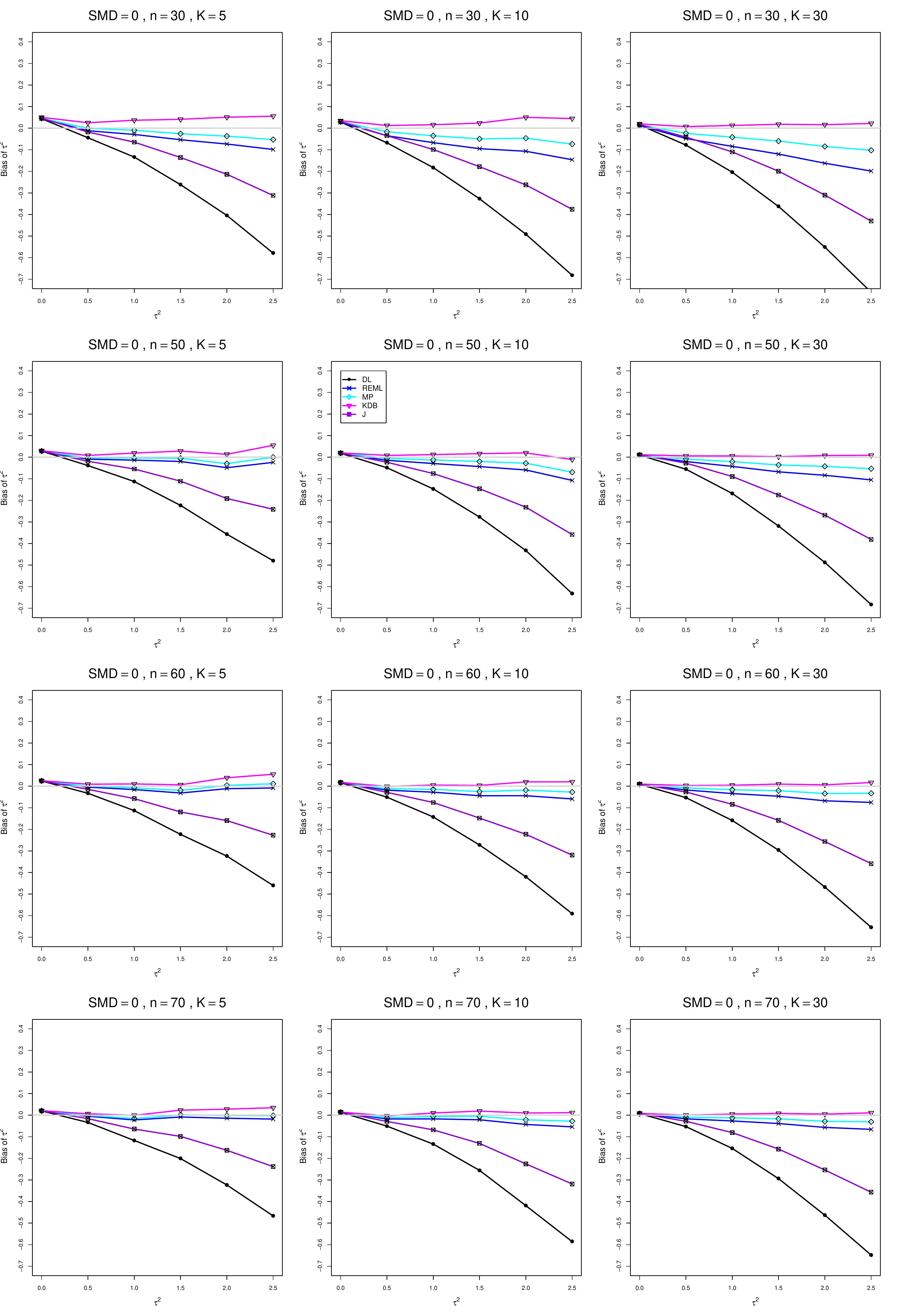}
	\caption{Bias of the estimation of  between-studies variance $\tau^2$ for $\delta=0$, $q=0.75$, $n=30,\;50,\;60,\;70$.
		\label{BiasTauSMD0q75small}}
\end{figure}

\begin{figure}[t]
	\centering
	\includegraphics[scale=0.33]{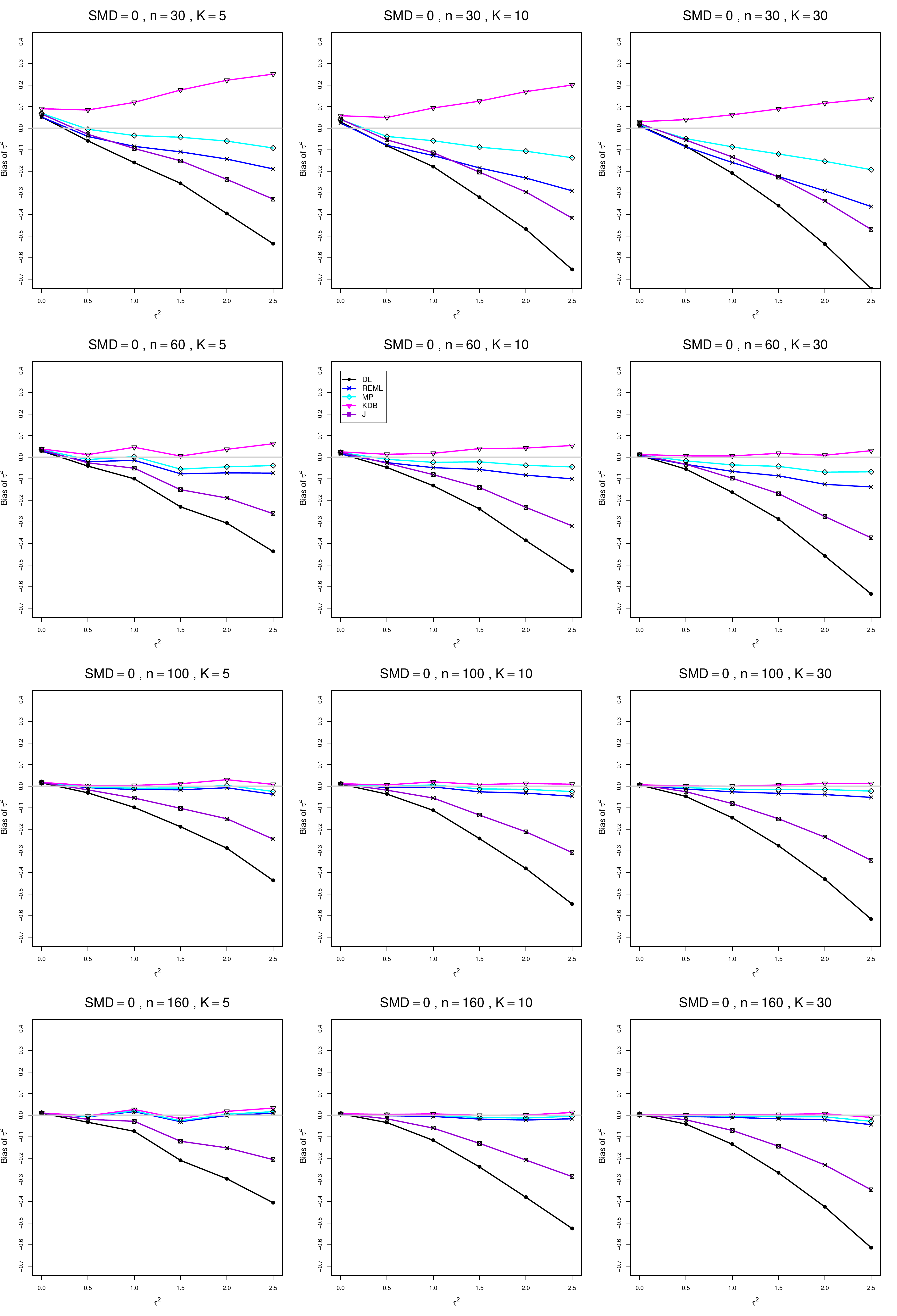}
	\caption{Bias of the estimation of  between-studies variance $\tau^2$ for $\delta=0$, $q=0.75$,  unequal sample sizes with
		$\bar{n}=30,\; 60,\;100,\;160$.
		\label{BiasTauSMD0q75unequal}}
\end{figure}
\clearpage
\begin{figure}[t]
	\centering
	\includegraphics[scale=0.3]{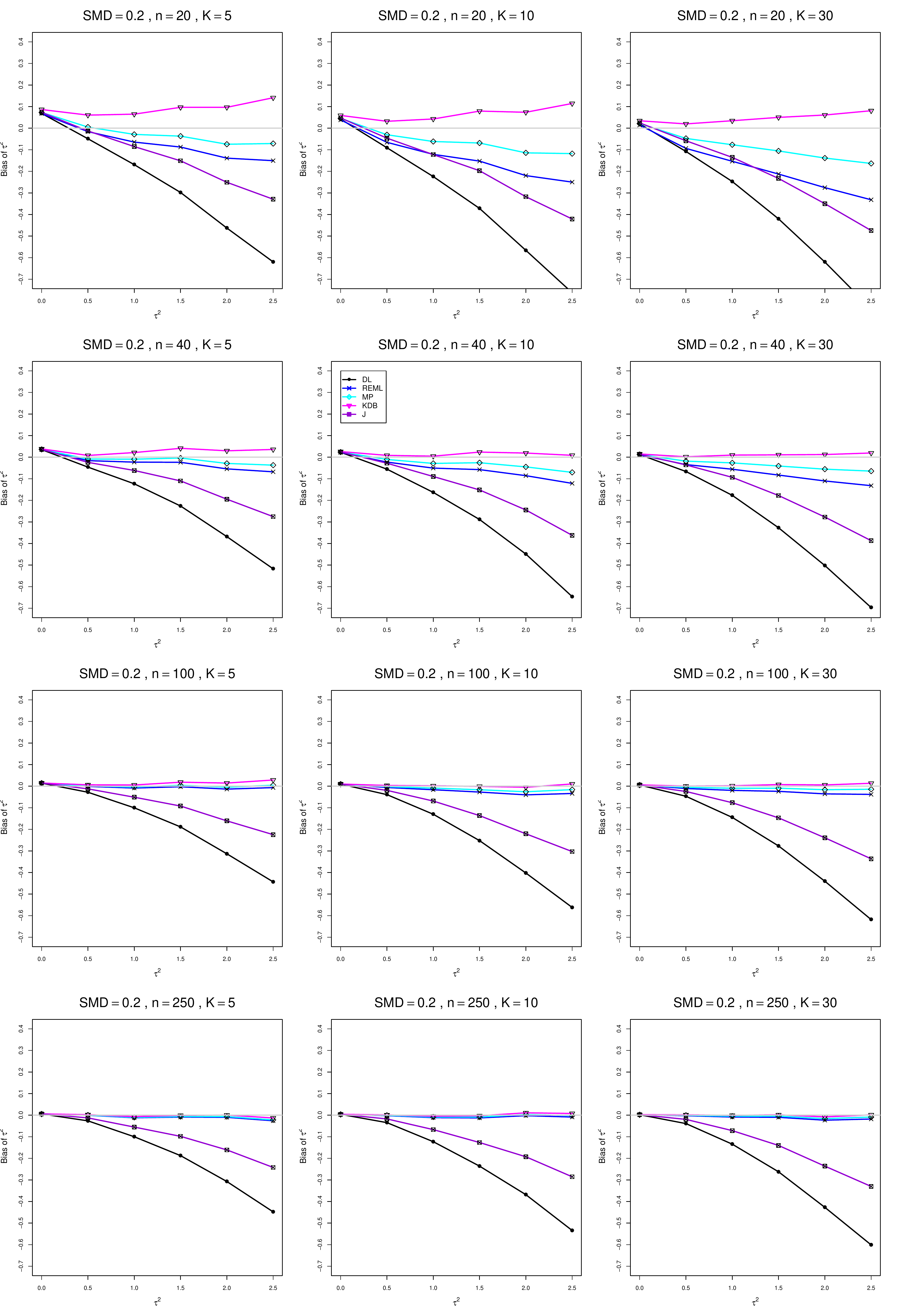}
	\caption{Bias of the estimation of  between-studies variance $\tau^2$ for $\delta=0.2$, $q=0.75$, $n=20,\;40,\;100,\;250$.
		\label{BiasTauSMD02q75}}
\end{figure}

\begin{figure}[t]
	\centering
	\includegraphics[scale=0.3]{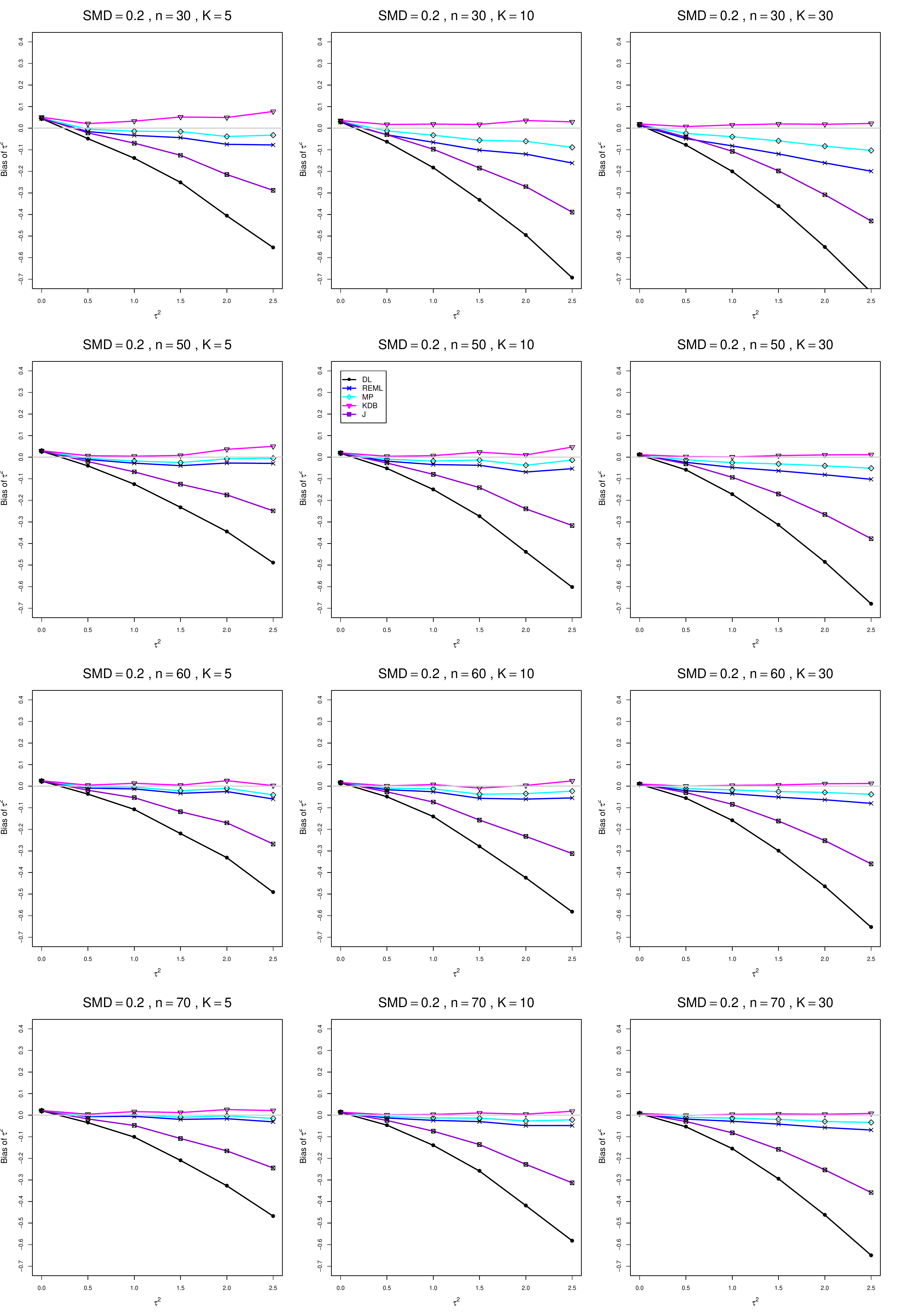}
	\caption{Bias of the estimation of  between-studies variance $\tau^2$ for $\delta=0.2$, $q=0.75$, $n=30,\;50,\;60,\;70$.
		\label{BiasTauSMD02q75small}}
\end{figure}

\begin{figure}[t]
	\centering
	\includegraphics[scale=0.33]{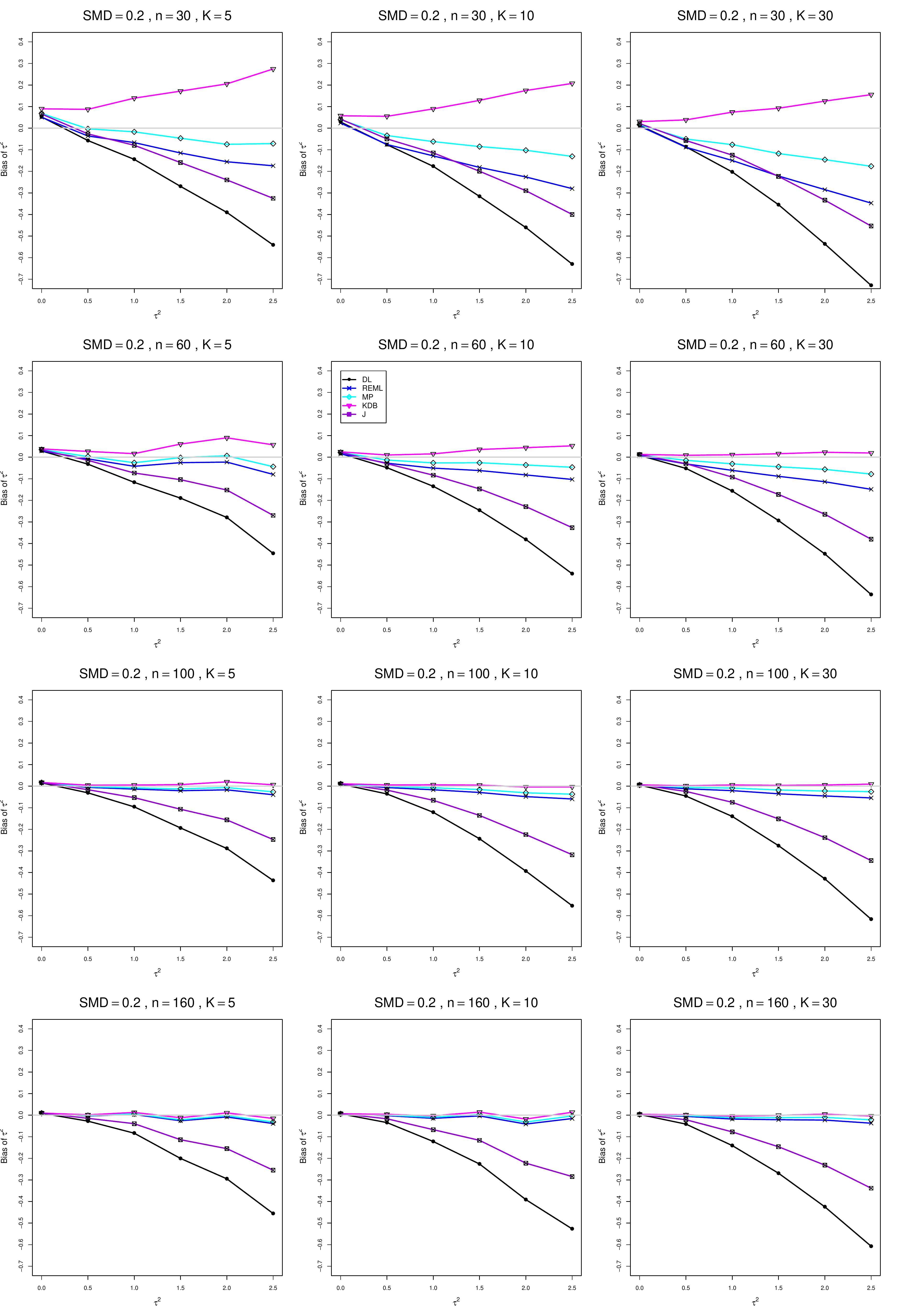}
	\caption{Bias of the estimation of  between-studies variance $\tau^2$ for $\delta=0.2$, $q=0.75$,  unequal sample sizes with
		$\bar{n}=30,\; 60,\;100,\;160$.
		\label{BiasTauSMD02q75unequal}}
\end{figure}

\begin{figure}[t]
	\centering
	\includegraphics[scale=0.3]{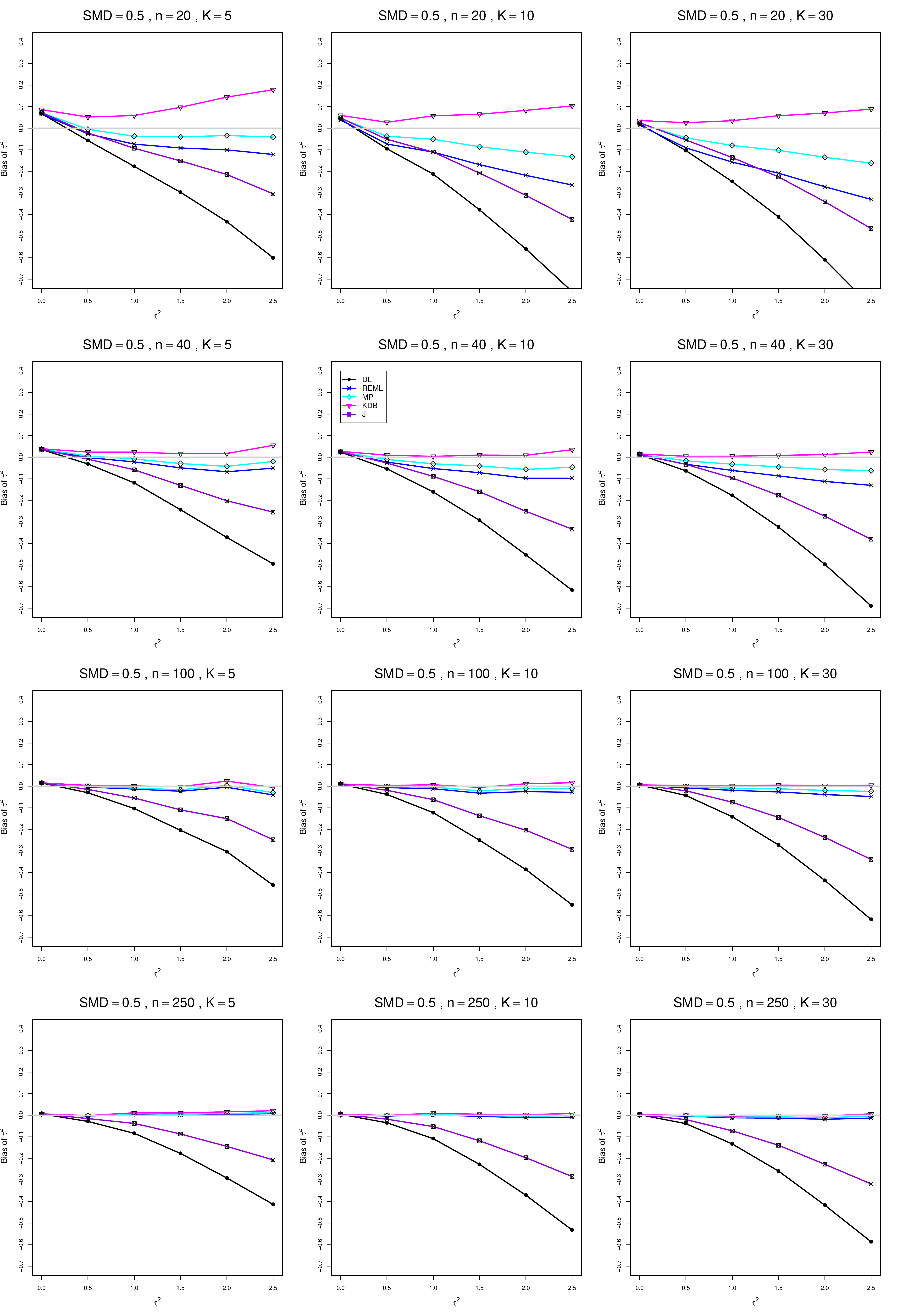}
	\caption{Bias of the estimation of  between-studies variance $\tau^2$ for $\delta=0.5$, $q=0.75$, $n=20,\;40,\;100,\;250$.
		\label{BiasTauSMD05q75}}
\end{figure}

\begin{figure}[t]
	\centering
	\includegraphics[scale=0.3]{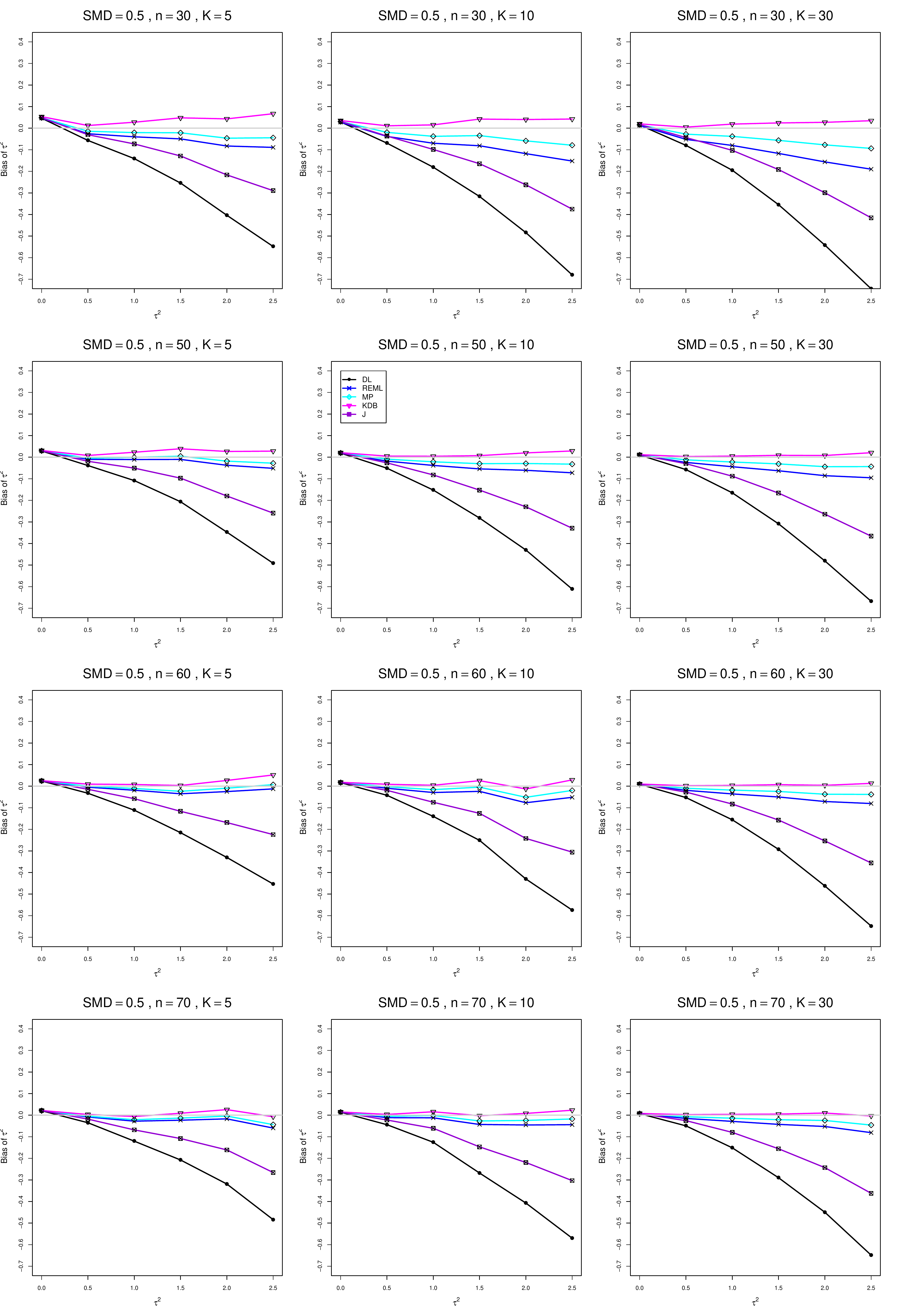}
	\caption{Bias of the estimation of  between-studies variance $\tau^2$ for $\delta=0.5$, $q=0.75$, $n=30,\;50,\;60,\;70$.
		\label{BiasTauSMD05q75small}}
\end{figure}

\begin{figure}[t]
	\centering
	\includegraphics[scale=0.33]{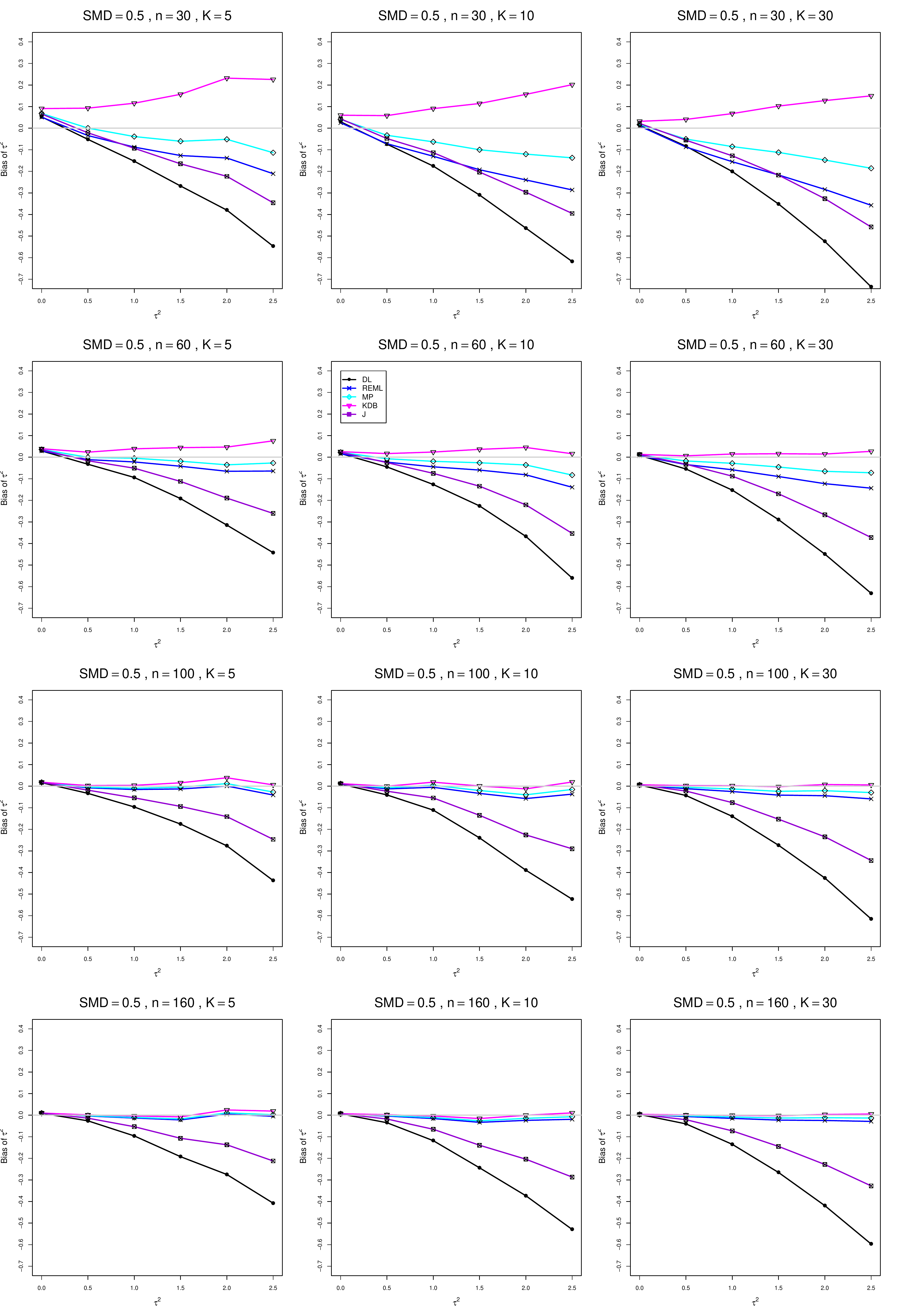}
	\caption{Bias of the estimation of  between-studies variance $\tau^2$ for $\delta=0.5$, $q=0.75$,  unequal sample sizes with
		$\bar{n}=30,\; 60,\;100,\;160$.
		\label{BiasTauSMD05q75unequal}}
\end{figure}

\begin{figure}[t]
	\centering
	\includegraphics[scale=0.3]{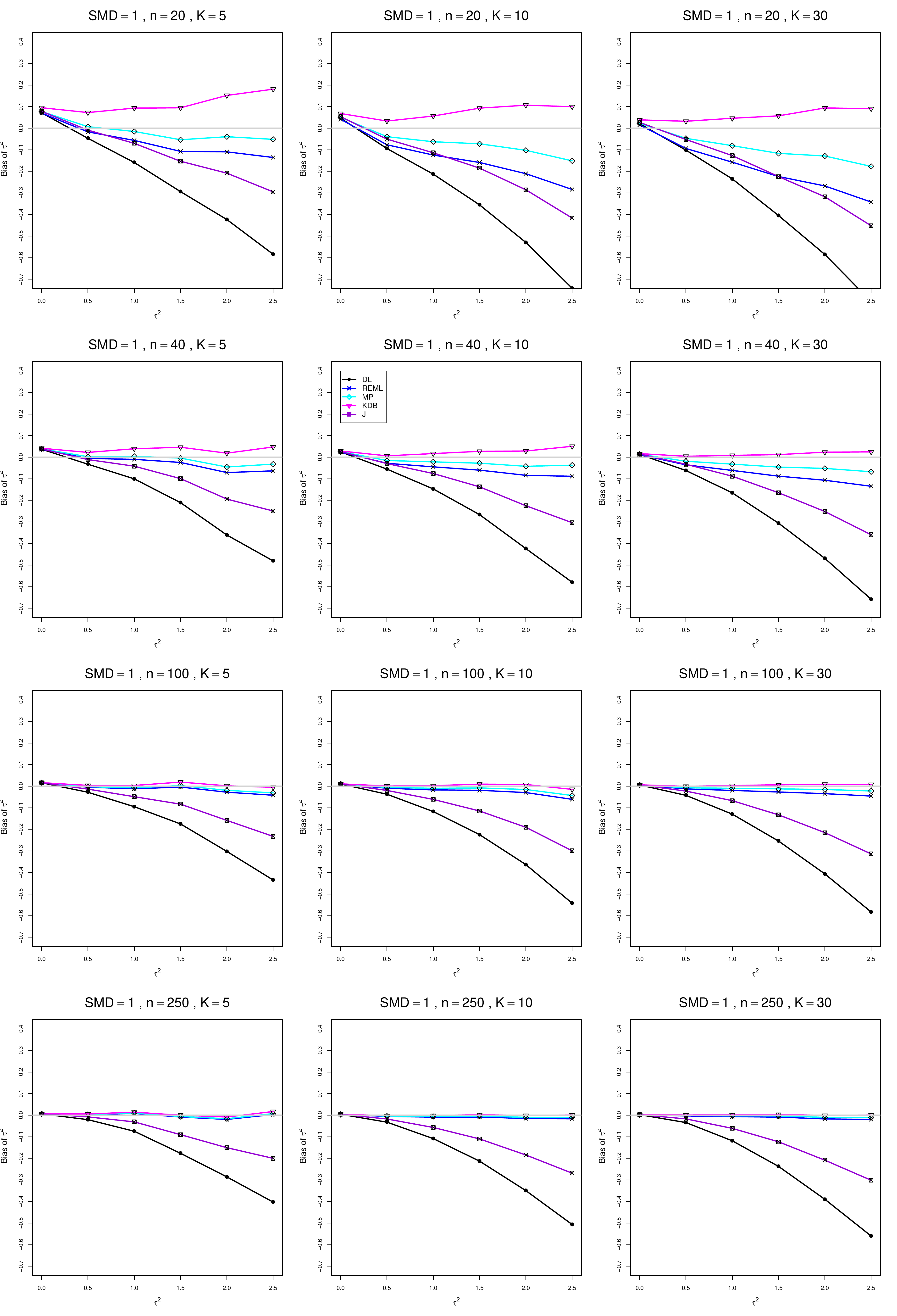}
	\caption{Bias of the estimation of  between-studies variance $\tau^2$ for $\delta=1$, $q=0.75$, $n=20,\;40,\;100,\;250$.
		\label{BiasTauSMD1q75}}
\end{figure}

\begin{figure}[t]
	\centering
	\includegraphics[scale=0.3]{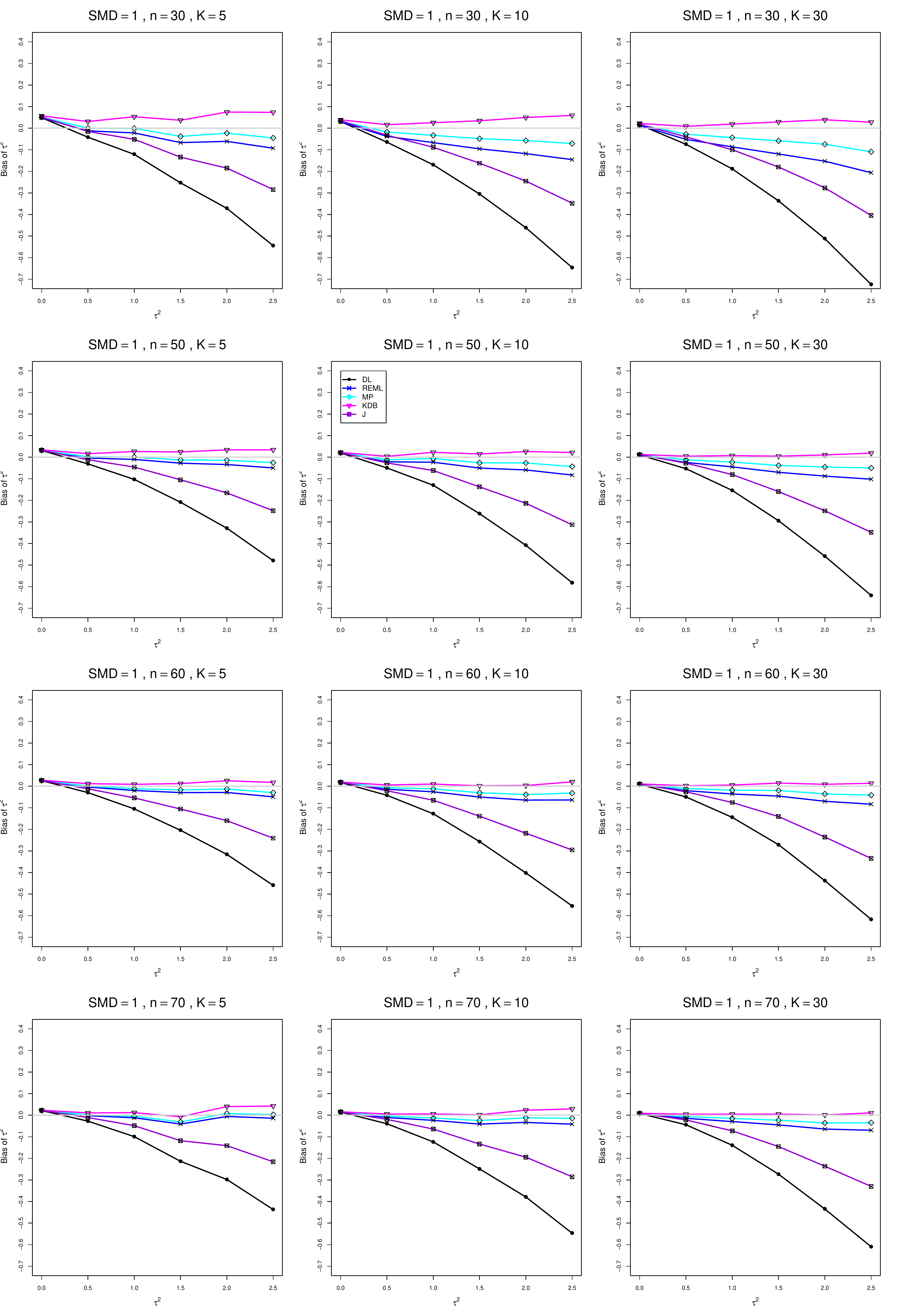}
	\caption{Bias of the estimation of  between-studies variance $\tau^2$ for $\delta=1$, $q=0.75$, $n=30,\;50,\;60,\;70$.
		\label{BiasTauSMD1q75small}}
\end{figure}

\begin{figure}[t]
	\centering
	\includegraphics[scale=0.33]{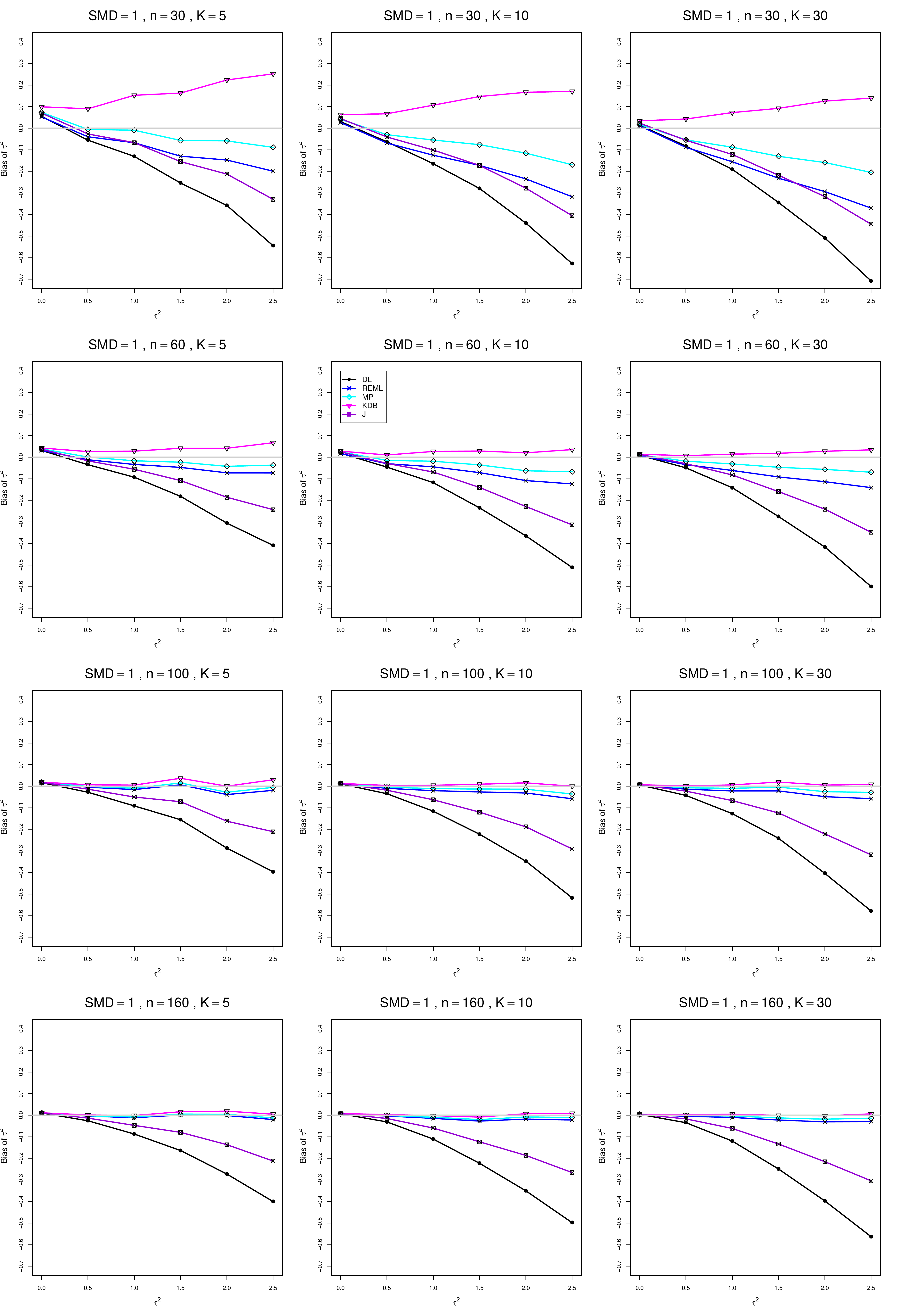}
	\caption{Bias of the estimation of  between-studies variance $\tau^2$ for $\delta=1$, $q=0.75$,  unequal sample sizes with
		$\bar{n}=30,\; 60,\;100,\;160$.
		\label{BiasTauSMD1q75unequal}}
\end{figure}

\begin{figure}[t]
	\centering
	\includegraphics[scale=0.3]{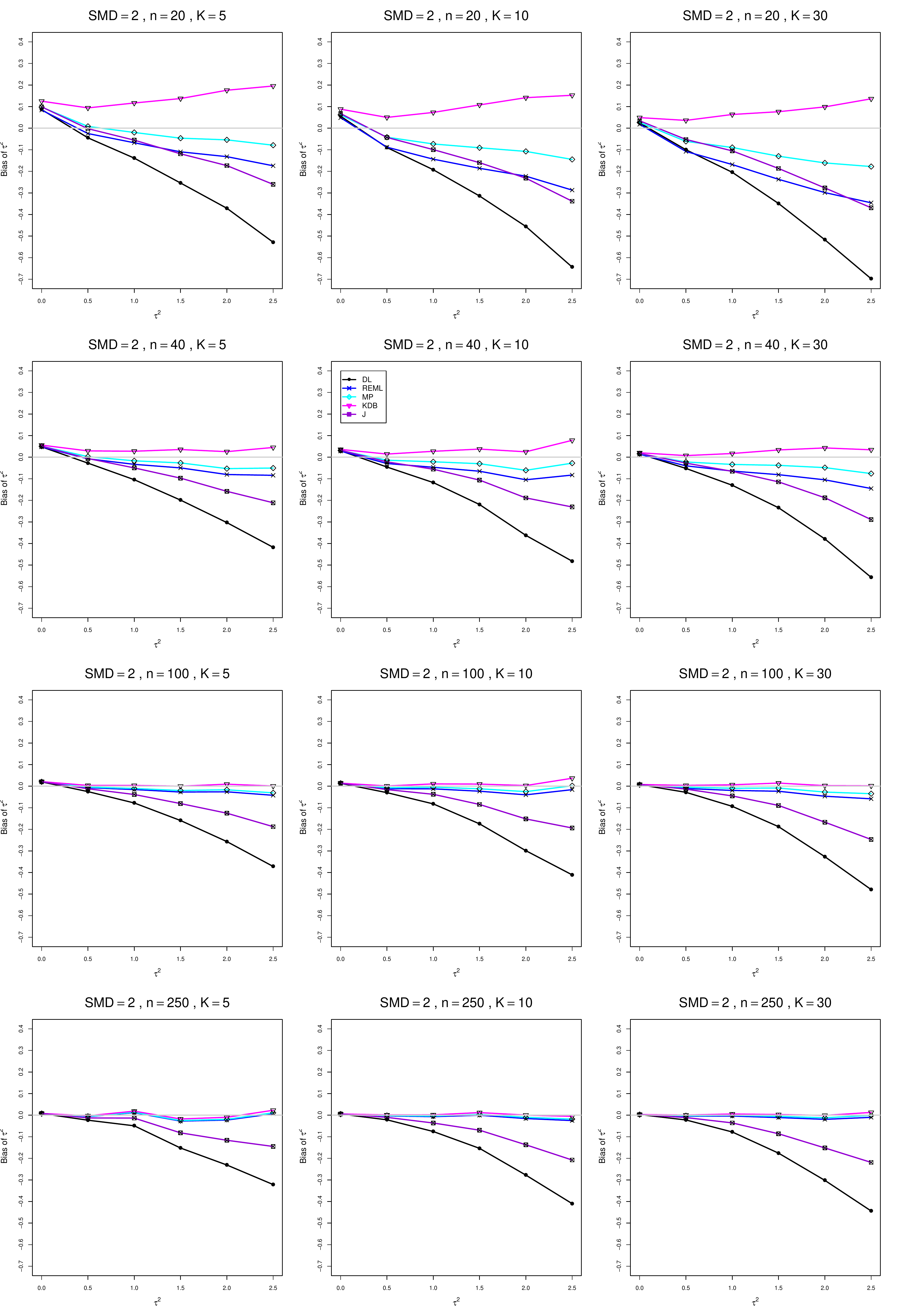}
	\caption{Bias of the estimation of  between-studies variance $\tau^2$ for $\delta=2$, $q=0.75$, $n=20,\;40,\;100,\;250$.
		\label{BiasTauSMD2q75}}
\end{figure}

\begin{figure}[t]
	\centering
	\includegraphics[scale=0.3]{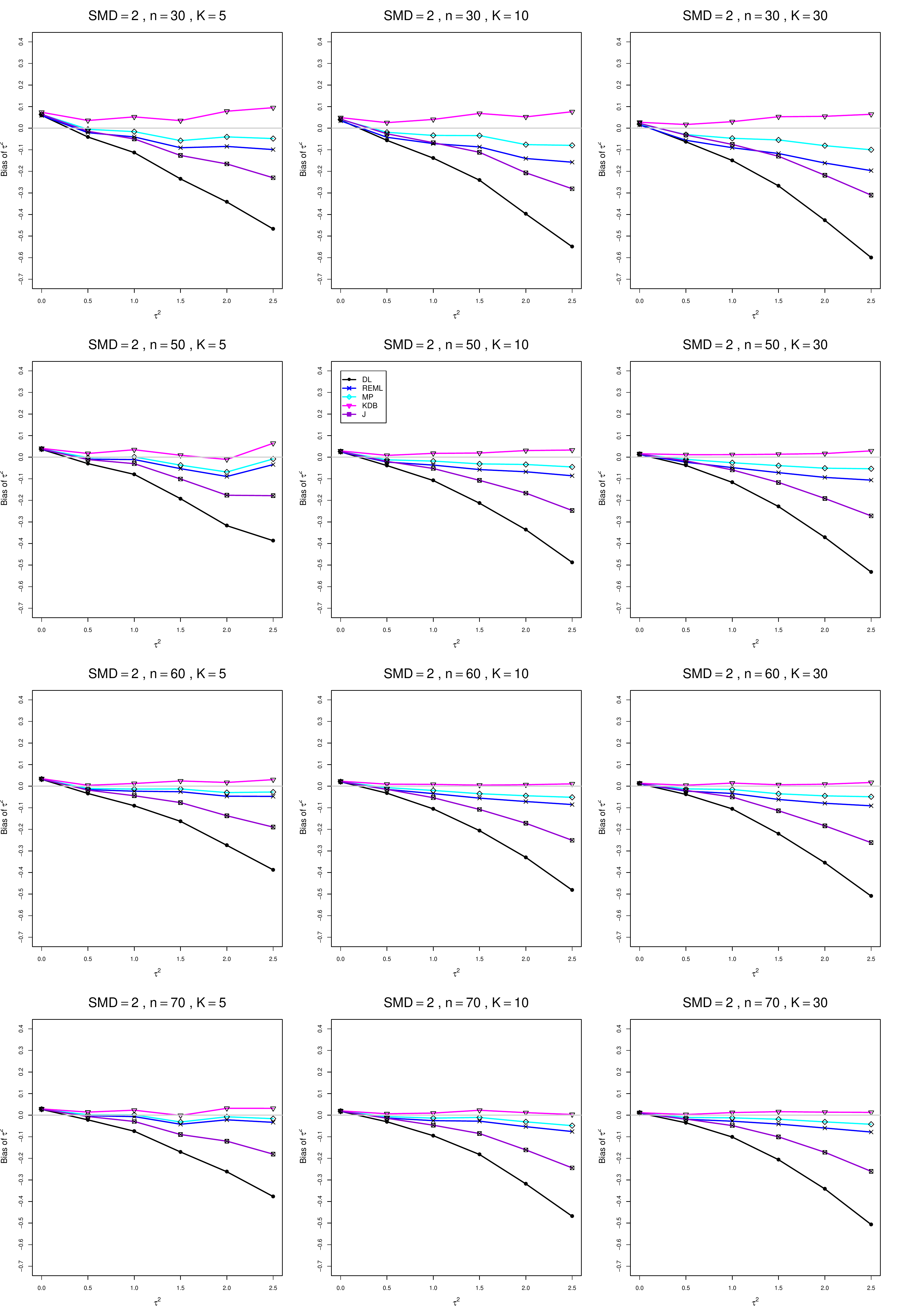}
	\caption{Bias of the estimation of  between-studies variance $\tau^2$ for $\delta=2$, $q=0.75$, $n=30,\;50,\;60,\;70$.
		\label{BiasTauSMD2q75small}}
\end{figure}

\begin{figure}[t]
	\centering
	\includegraphics[scale=0.33]{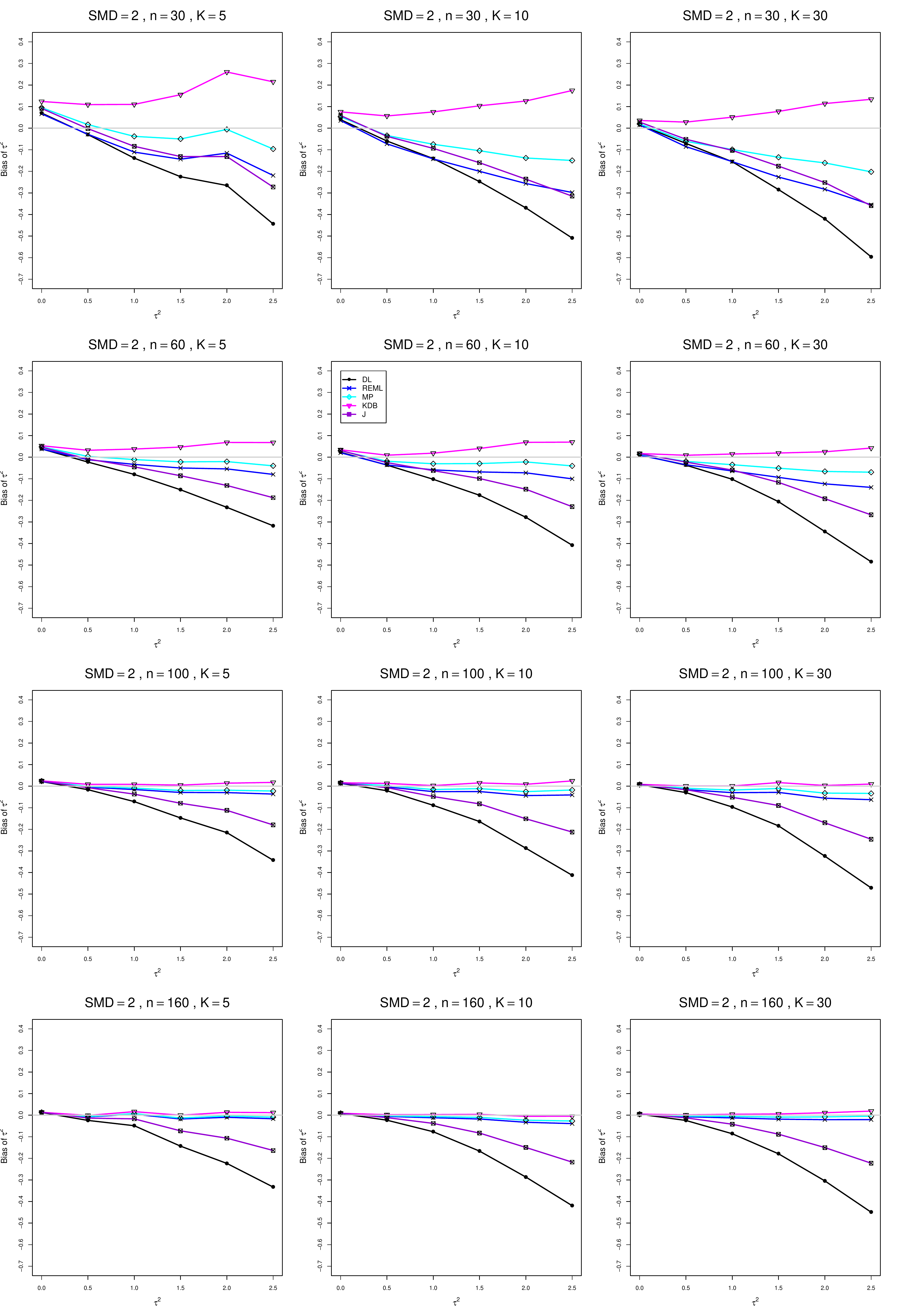}
	\caption{Bias of the estimation of  between-studies variance $\tau^2$ for $\delta=2$, $q=0.75$,  unequal sample sizes with
		$\bar{n}=30,\; 60,\;100,\;160$.
		\label{BiasTauSMD2q75unequal}}
\end{figure}
\clearpage
\setcounter{figure}{0}
\renewcommand{\thefigure}{A2.\arabic{figure}}
\section*{A2. Coverage of of interval estimators of $\hat{\tau}^2$ for $\hat{\tau}^2=0.0(0.5)2.5$.}
For coverage of $\hat{\tau}^2$, each figure corresponds to a value of $\delta (= 0, 0.5, 1, 1.5, 2 , 2.5)$, a value of $q (= .5, .75)$, and a set of values of $n$ (= 20, 40, 100, 250 or 30, 50, 60, 70) or $\bar{n} (= 30, 60, 100, 160)$.\\
Each figure contains a panel (with $\tau^2$ on the horizontal axis) for each combination of n (or $\bar{n}$) and $K (=5, 10, 30)$.\\
The interval estimators of $\tau^2$ are
\begin{itemize}
	\item QP (Q-profile confidence interval)
	\item BJ (Biggerstaff and Jackson interval )
	\item PL (Profile likelihood interval)
	\item KDB ( KDB - improved Q-profile method  based on Kulinskaya, Dollinger and  Bj{\o}rkest{\o}l (2011))
	\item J (Jackson's interval)
\end{itemize}

\begin{figure}[t] \centering
	\includegraphics[scale=0.35]{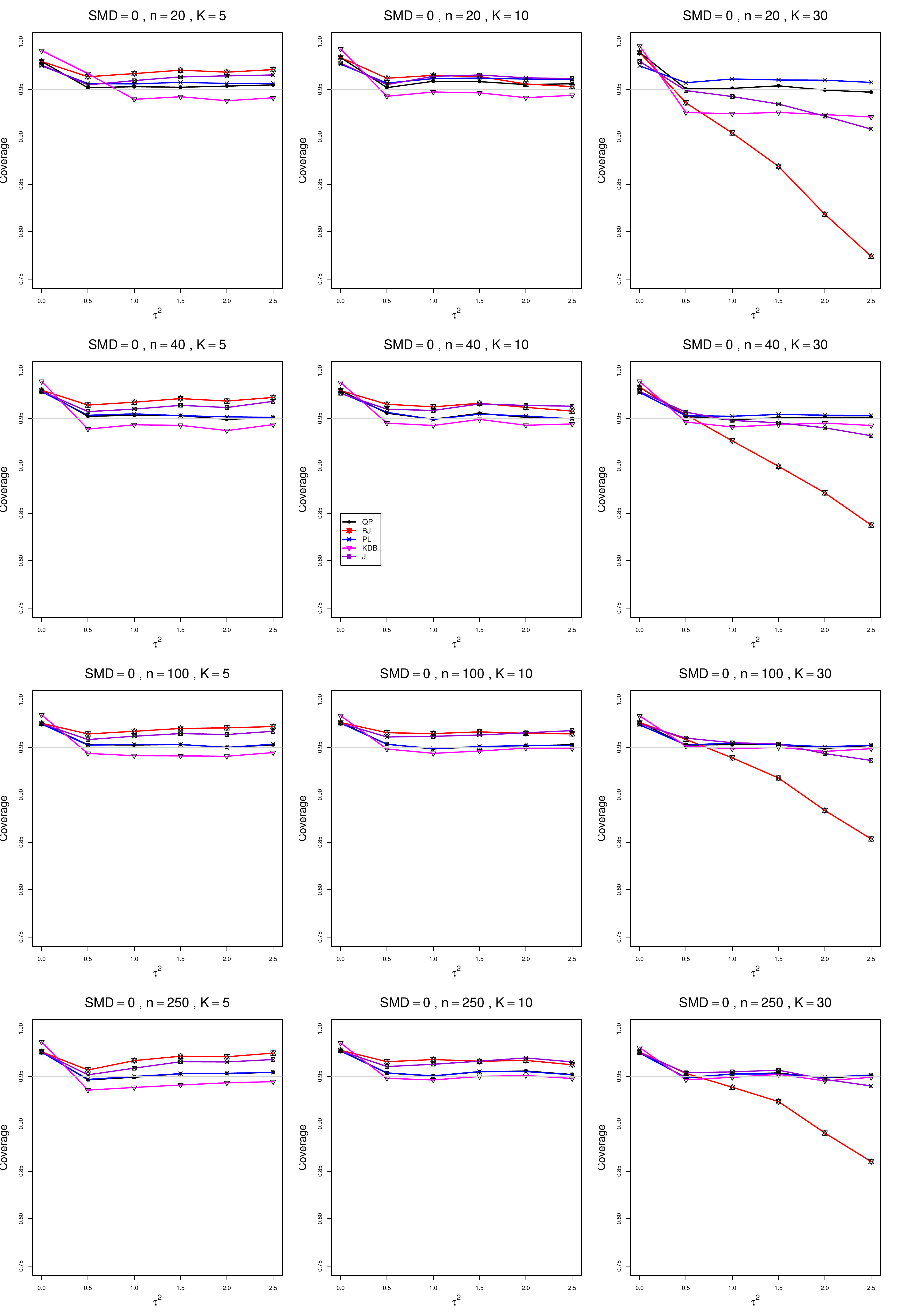}
	\caption{Coverage at  the nominal confidence level of $0.95$ of the  between-studies variance $\tau^2$ for $\delta=0$, $q=0.5$, $n=20,\;40,\;100,\;250$.
		\label{CovTauSMD0}}
\end{figure}

\begin{figure}[t]\centering
	\includegraphics[scale=0.35]{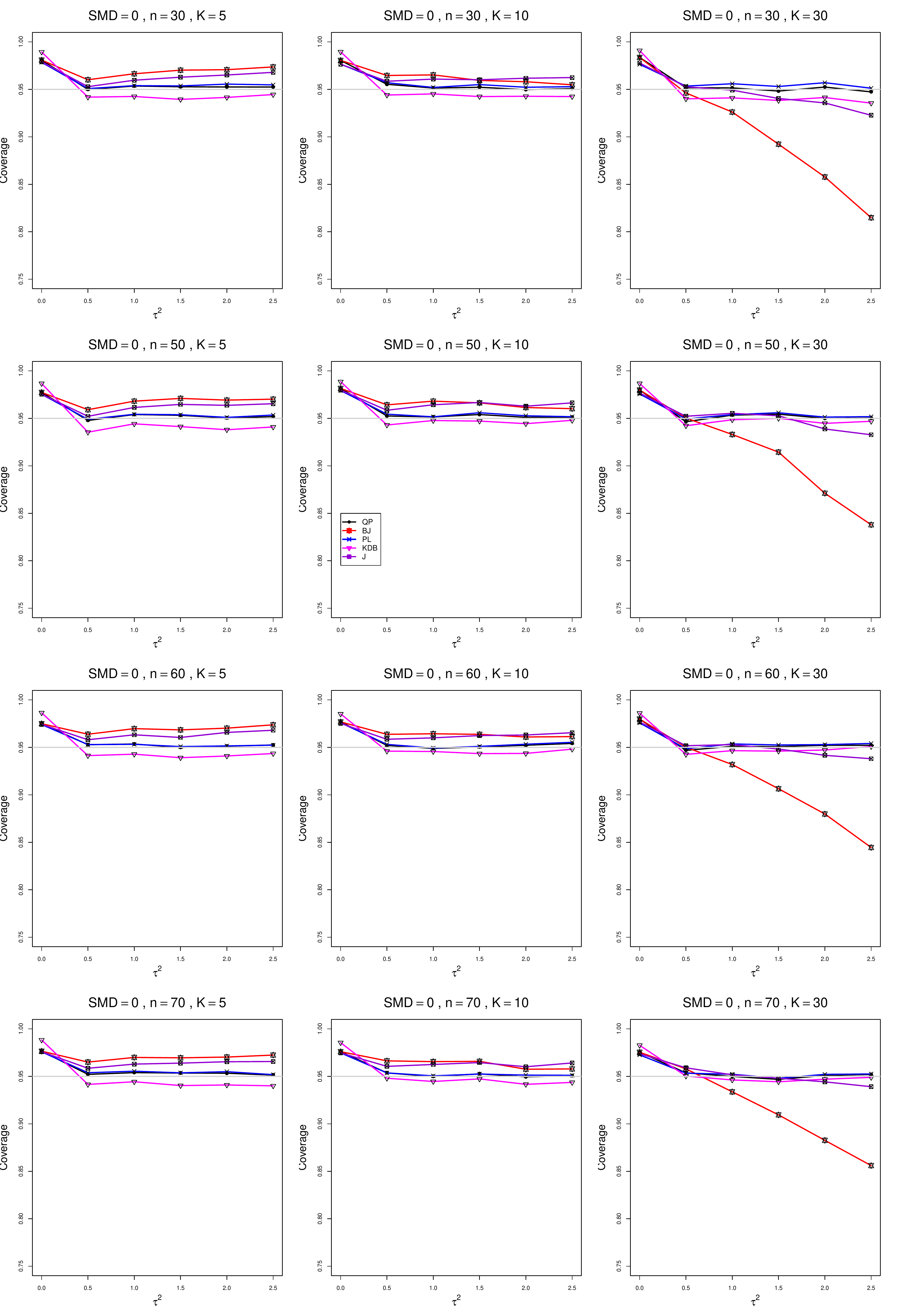}
	\caption{Coverage at  the nominal confidence level of $0.95$ of the  between-studies variance $\tau^2$ for $\delta=0$, $q=0.5$, $n=30,\;50,\;60,\;70$.
		\label{CovTauSMD0small}}
\end{figure}

\begin{figure}[t]\centering
	\includegraphics[scale=0.35]{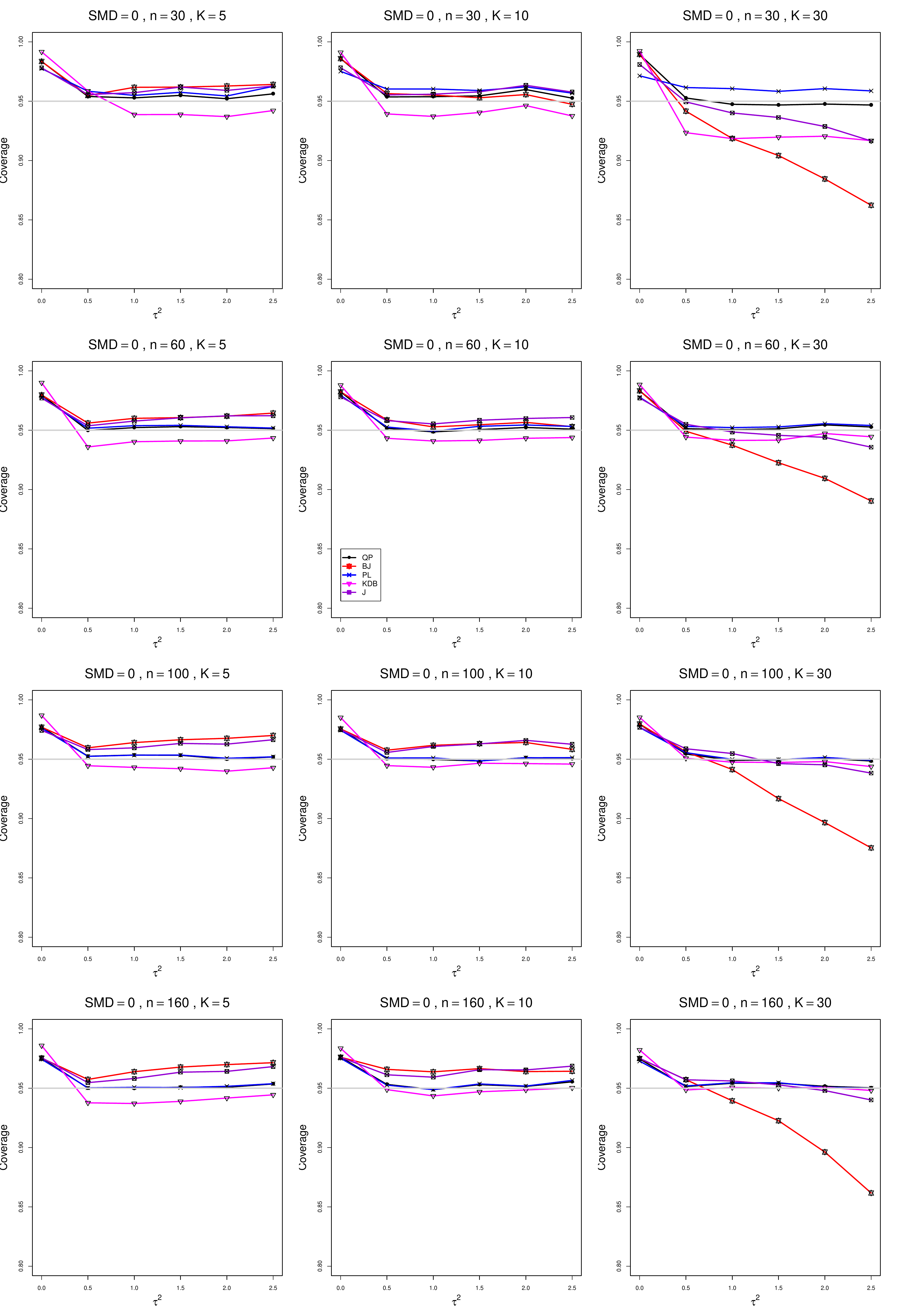}
	\caption{Coverage at  the nominal confidence level of $0.95$ of the  between-studies variance $\tau^2$ for $\delta=0$, $q=0.5$,  unequal sample sizes with $\bar{n}=30,\; 60,\;100,\;160$.
		\label{CovTauSMD0unequal}}
\end{figure}

\begin{figure}[t]\centering
	\includegraphics[scale=0.35]{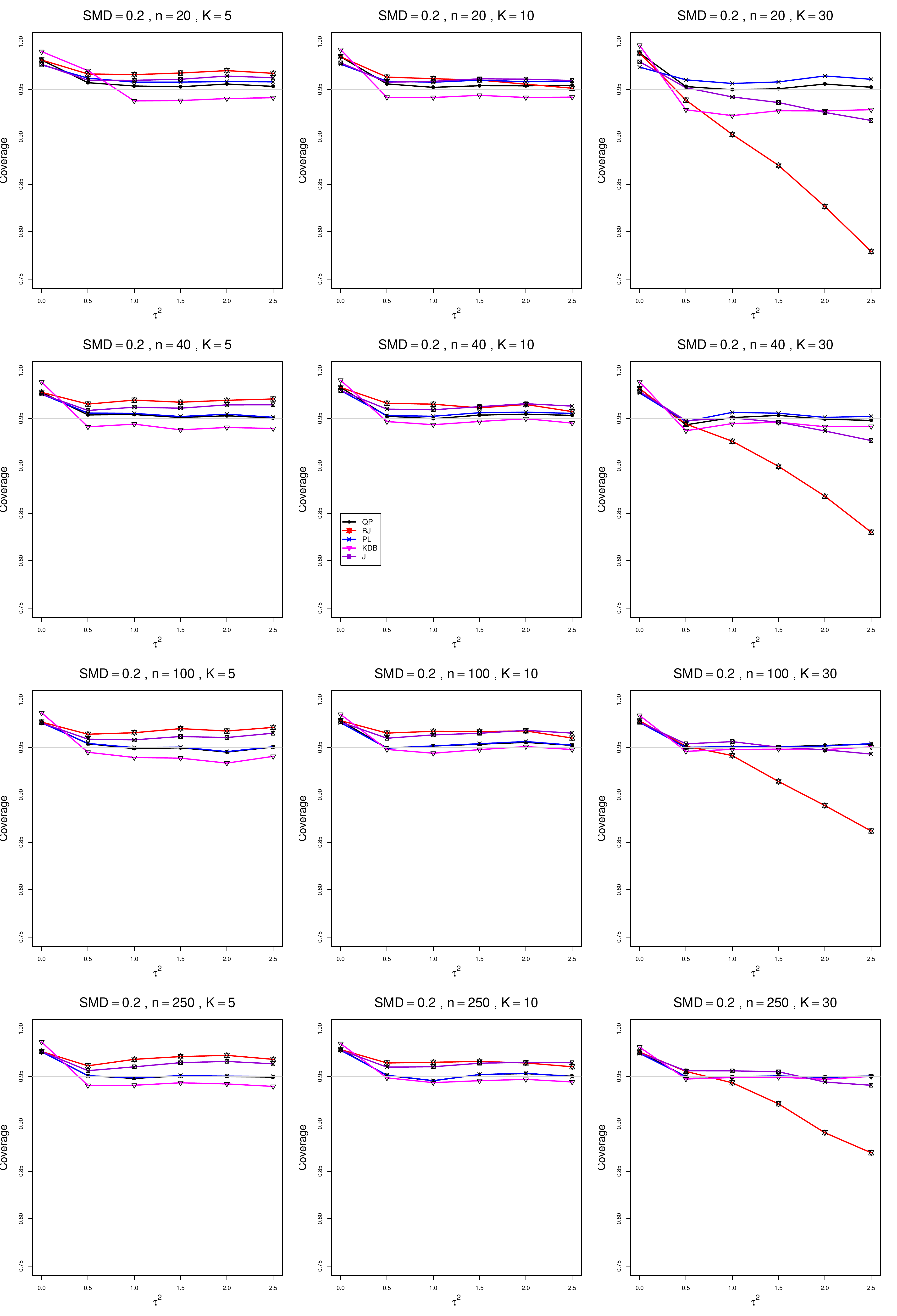}
	\caption{Coverage at  the nominal confidence level of $0.95$ of the  between-studies variance $\tau^2$ for $\delta=0.2$, $q=0.5$, $n=20,\;40,\;100,\;250$.
		\label{CovTauSMD02}}
\end{figure}

\begin{figure}[t]\centering
	\includegraphics[scale=0.35]{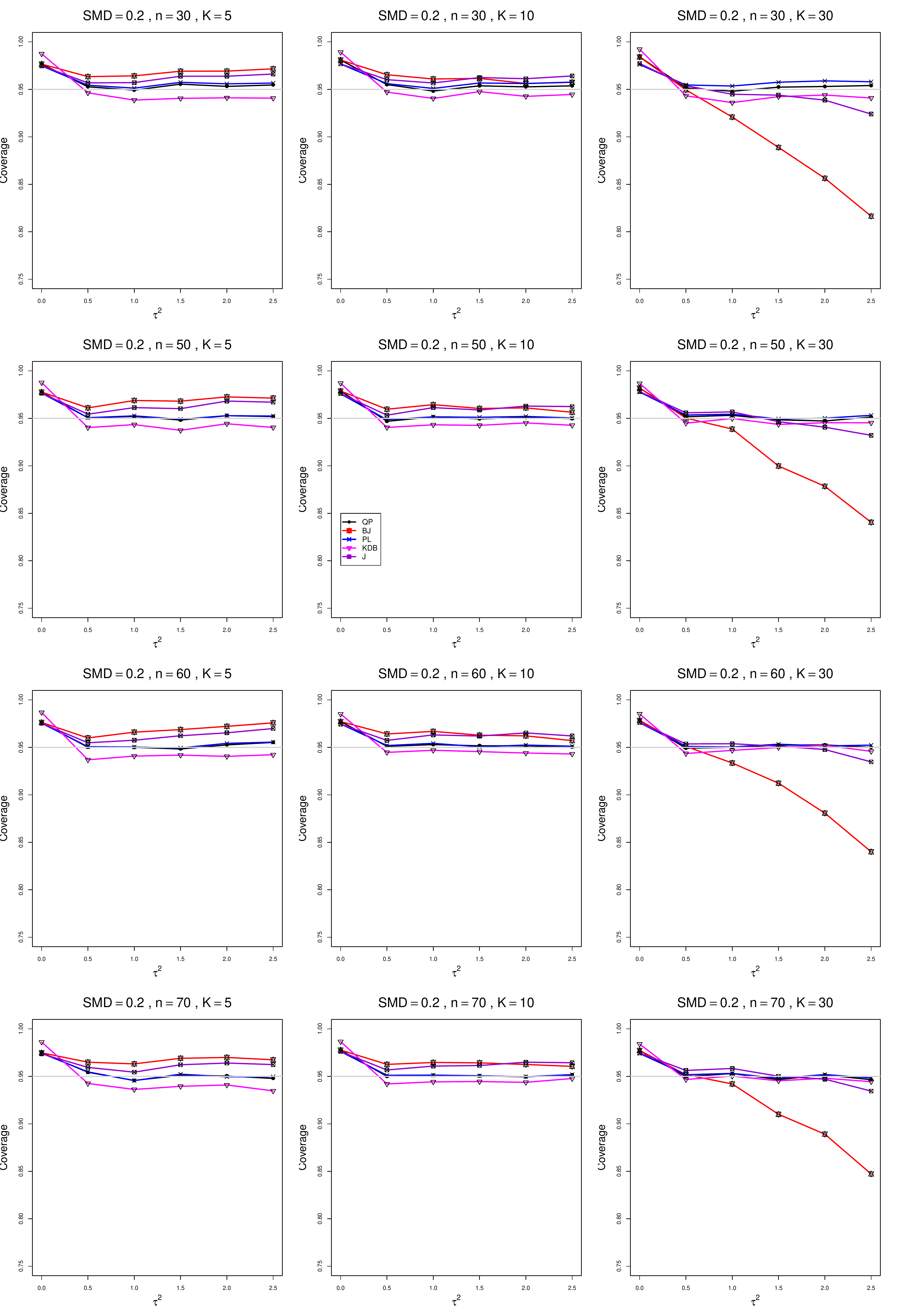}
	\caption{Coverage at  the nominal confidence level of $0.95$ of the  between-studies variance $\tau^2$ for $\delta=0.2$, $q=0.5$, $n=30,\;50,\;60,\;70$.
		\label{CovTauSMD02small}}
\end{figure}

\begin{figure}[t]\centering
	\includegraphics[scale=0.35]{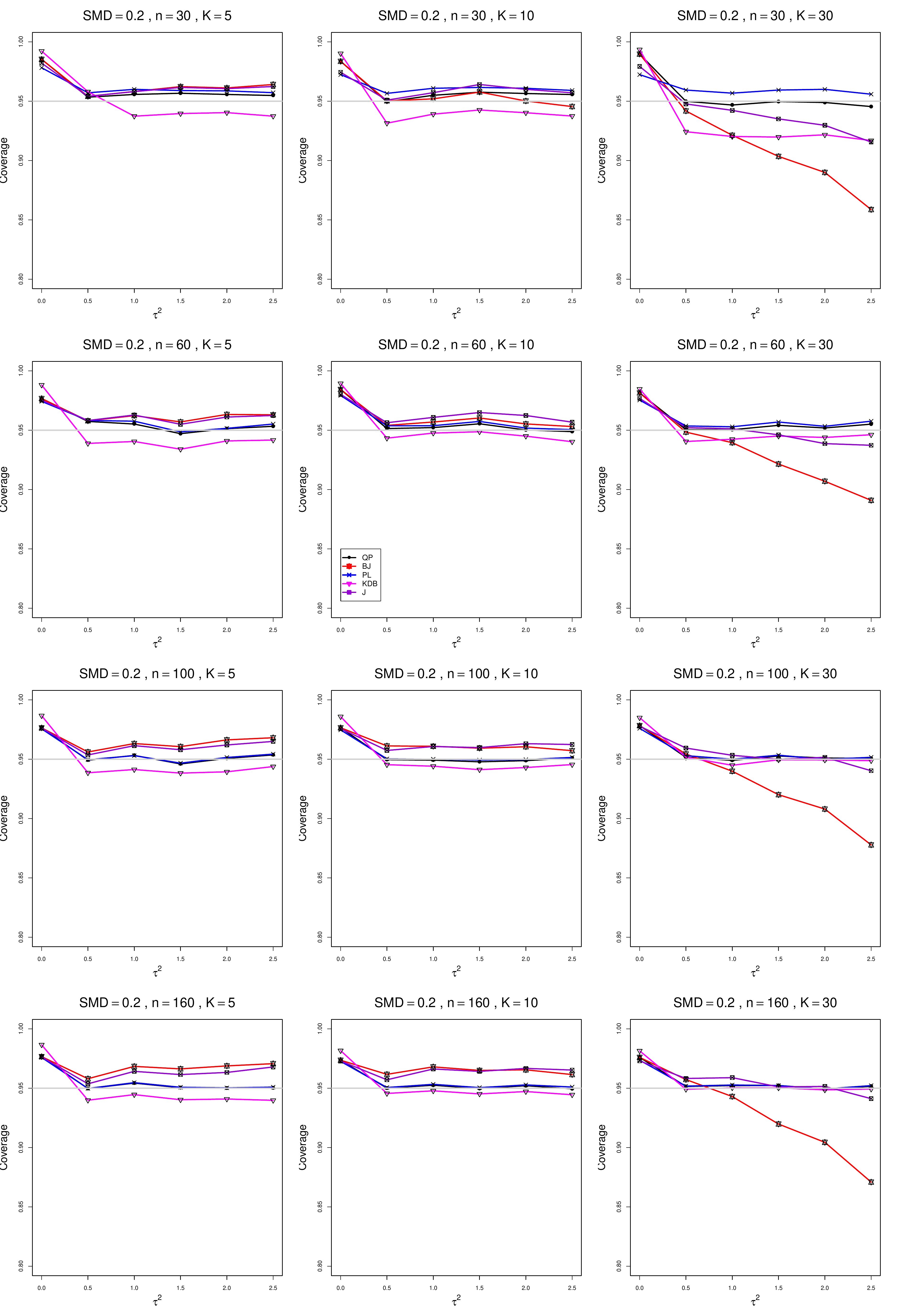}
	\caption{Coverage at  the nominal confidence level of $0.95$ of the  between-studies variance $\tau^2$ for $\delta=0.2$, $q=0.5$,  unequal sample sizes with $\bar{n}=30,\; 60,\;100,\;160$.
		\label{CovTauSMD02unequal}}
\end{figure}

\begin{figure}[t]\centering
	\includegraphics[scale=0.35]{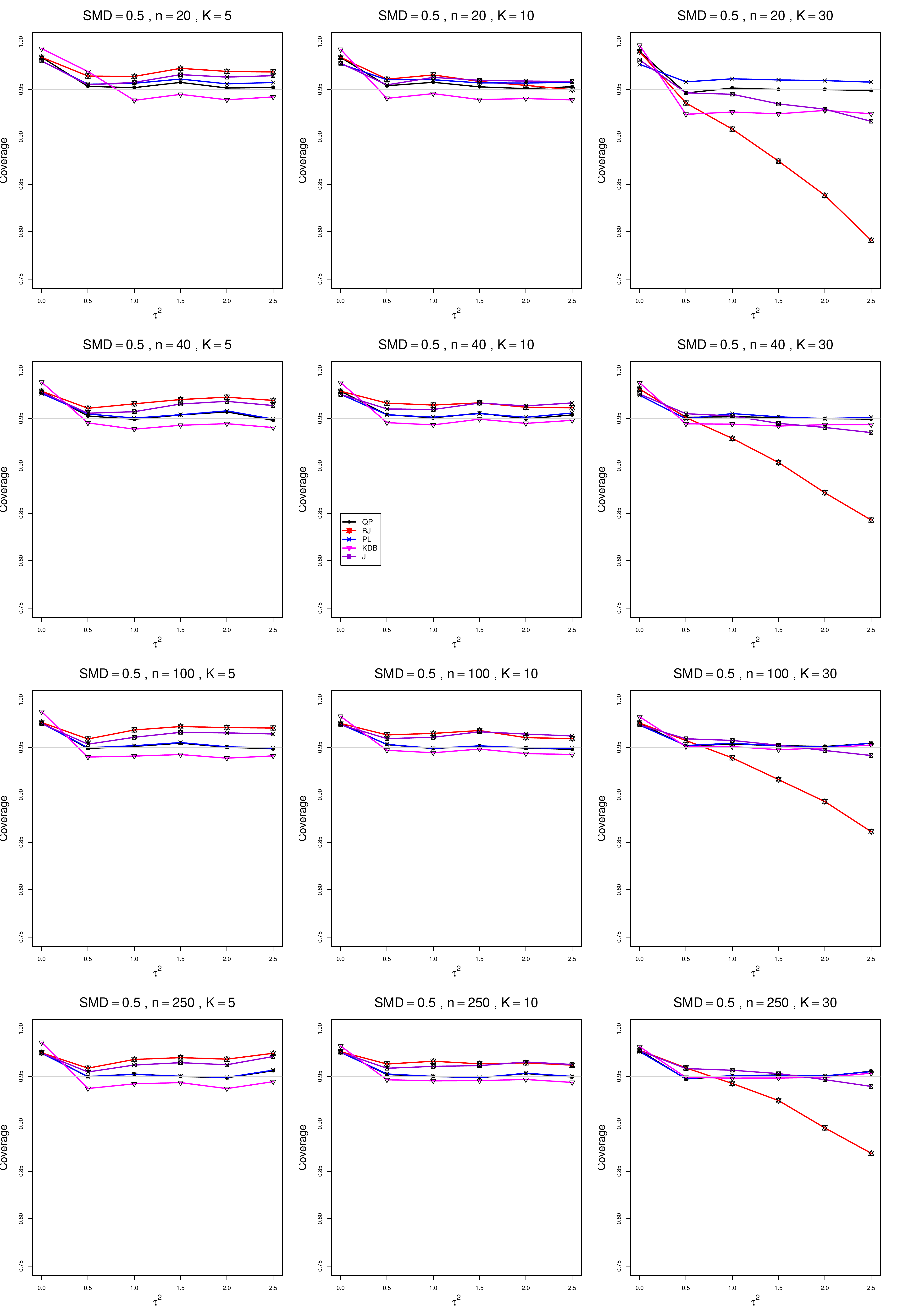}
	\caption{Coverage at  the nominal confidence level of $0.95$ of the  between-studies variance $\tau^2$ for $\delta=0.5$, $q=0.5$, $n=20,\;40,\;100,\;250$.
		\label{CovTauSMD05}}
\end{figure}

\begin{figure}[t]\centering
	\includegraphics[scale=0.35]{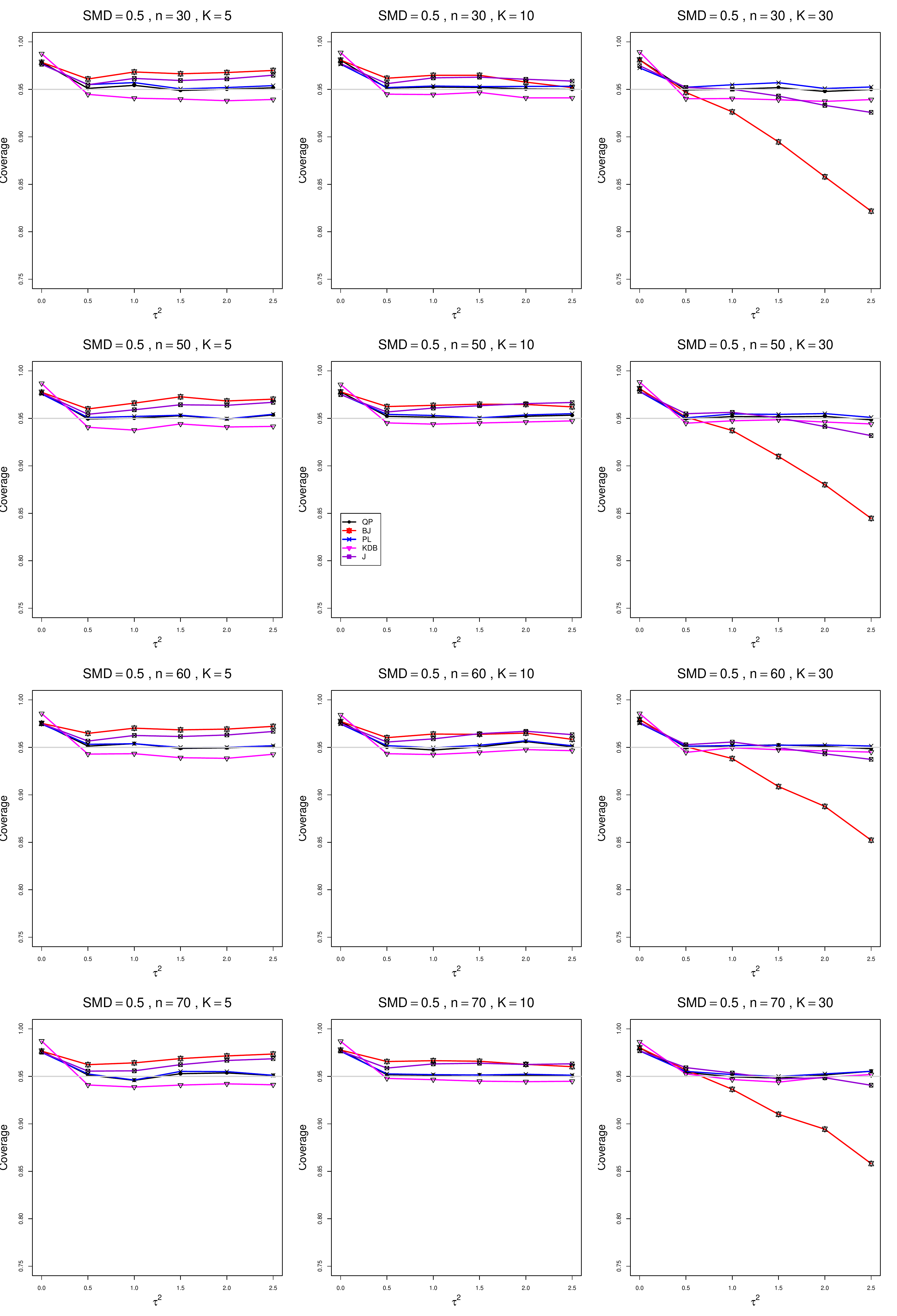}
	\caption{Coverage at  the nominal confidence level of $0.95$ of the  between-studies variance $\tau^2$ for $\delta=0.5$, $q=0.5$, $n=30,\;50,\;60,\;70$.
		\label{CovTauSMD05small}}
\end{figure}

\begin{figure}[t]\centering
	\includegraphics[scale=0.35]{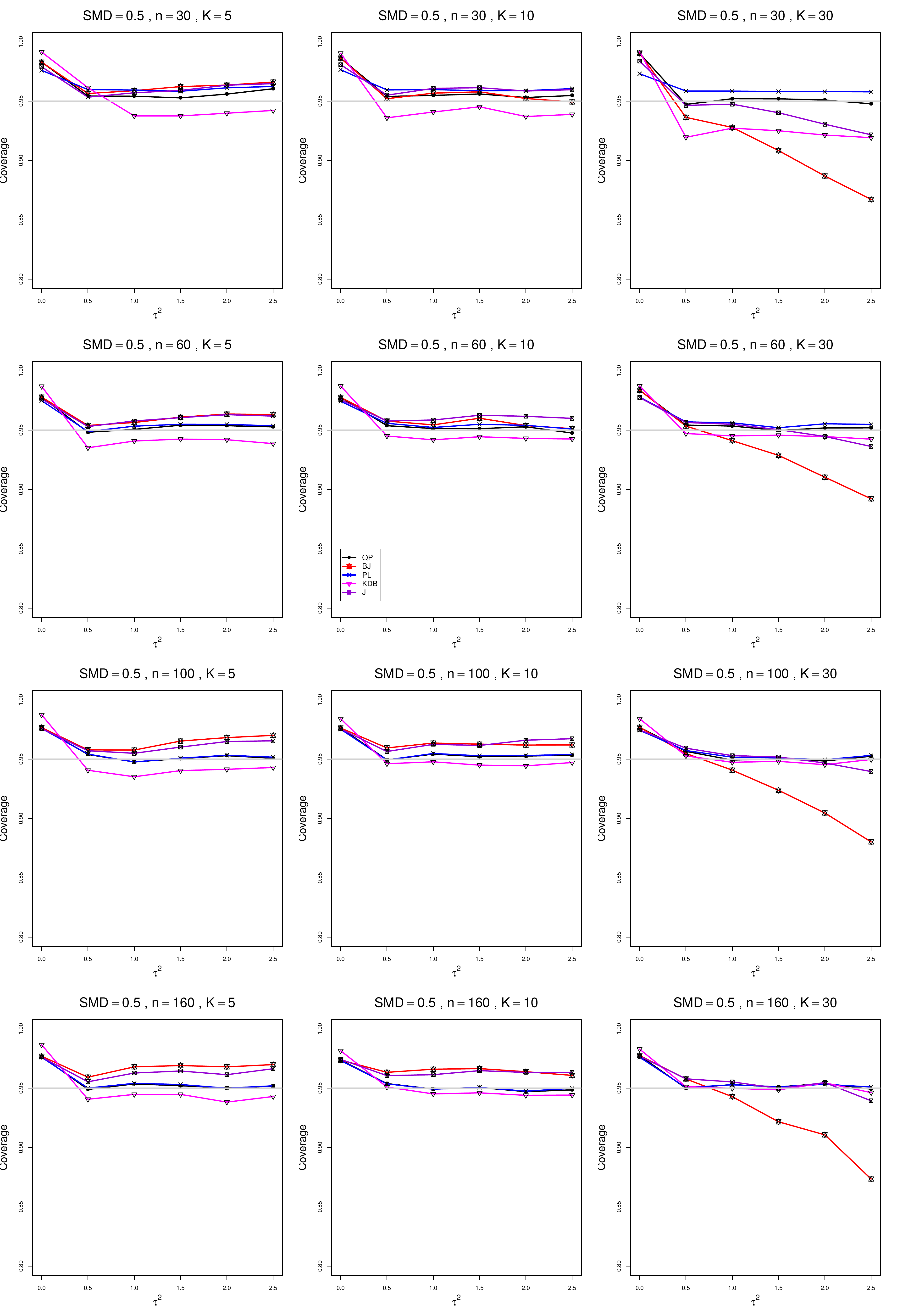}
	\caption{Coverage at  the nominal confidence level of $0.95$ of the  between-studies variance $\tau^2$ for $\delta=0.5$, $q=0.5$,  unequal sample sizes with $\bar{n}=30,\; 60,\;100,\;160$.
		\label{CovTauSMD05unequal}}
\end{figure}

\begin{figure}[t]\centering
	\includegraphics[scale=0.35]{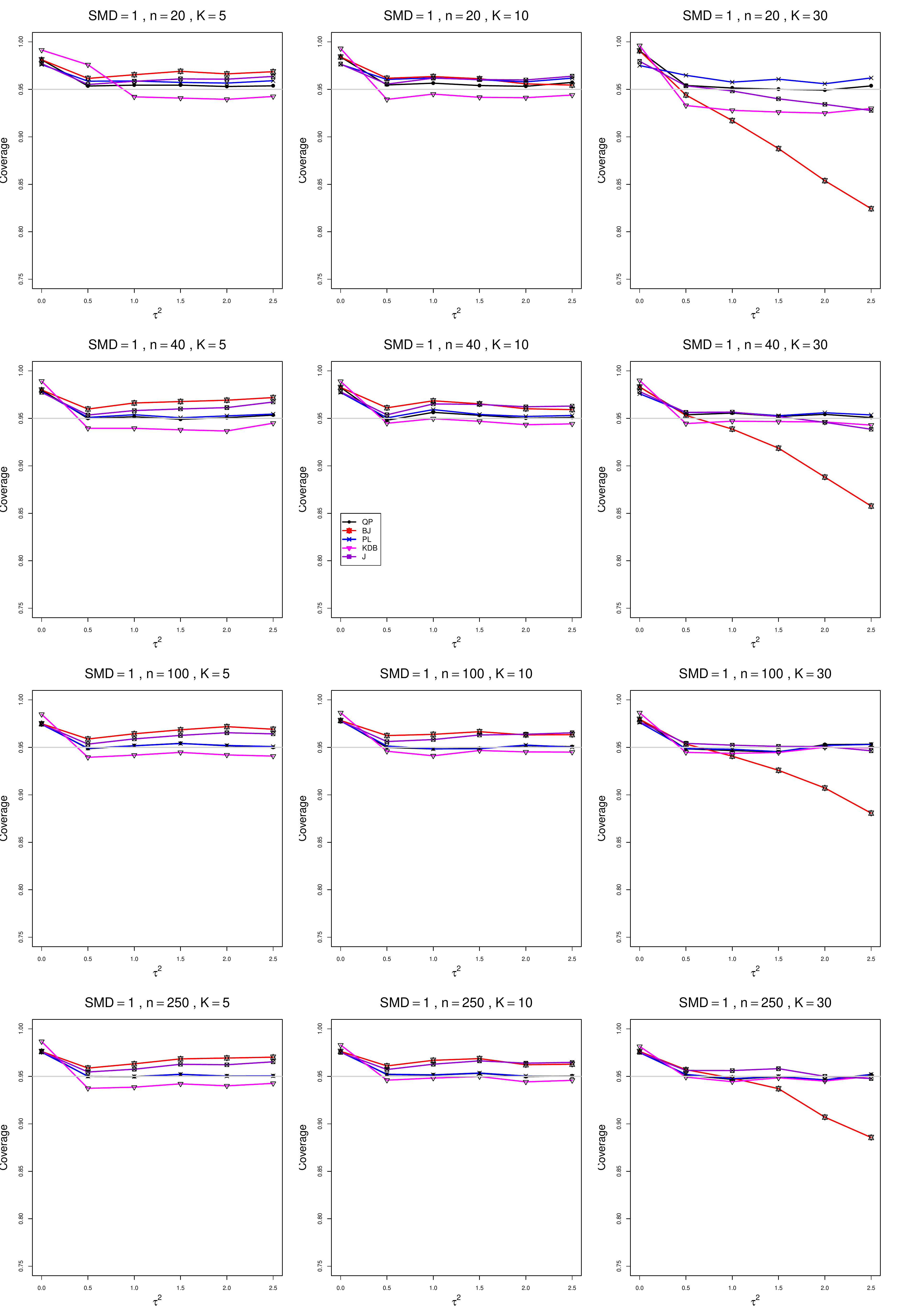}
	\caption{Coverage at  the nominal confidence level of $0.95$ of the  between-studies variance $\tau^2$ for $\delta=1$, $q=0.5$, $n=20,\;40,\;100,\;250$ .
		\label{CovTauSMD1}}
\end{figure}

\begin{figure}[t]\centering
	\includegraphics[scale=0.35]{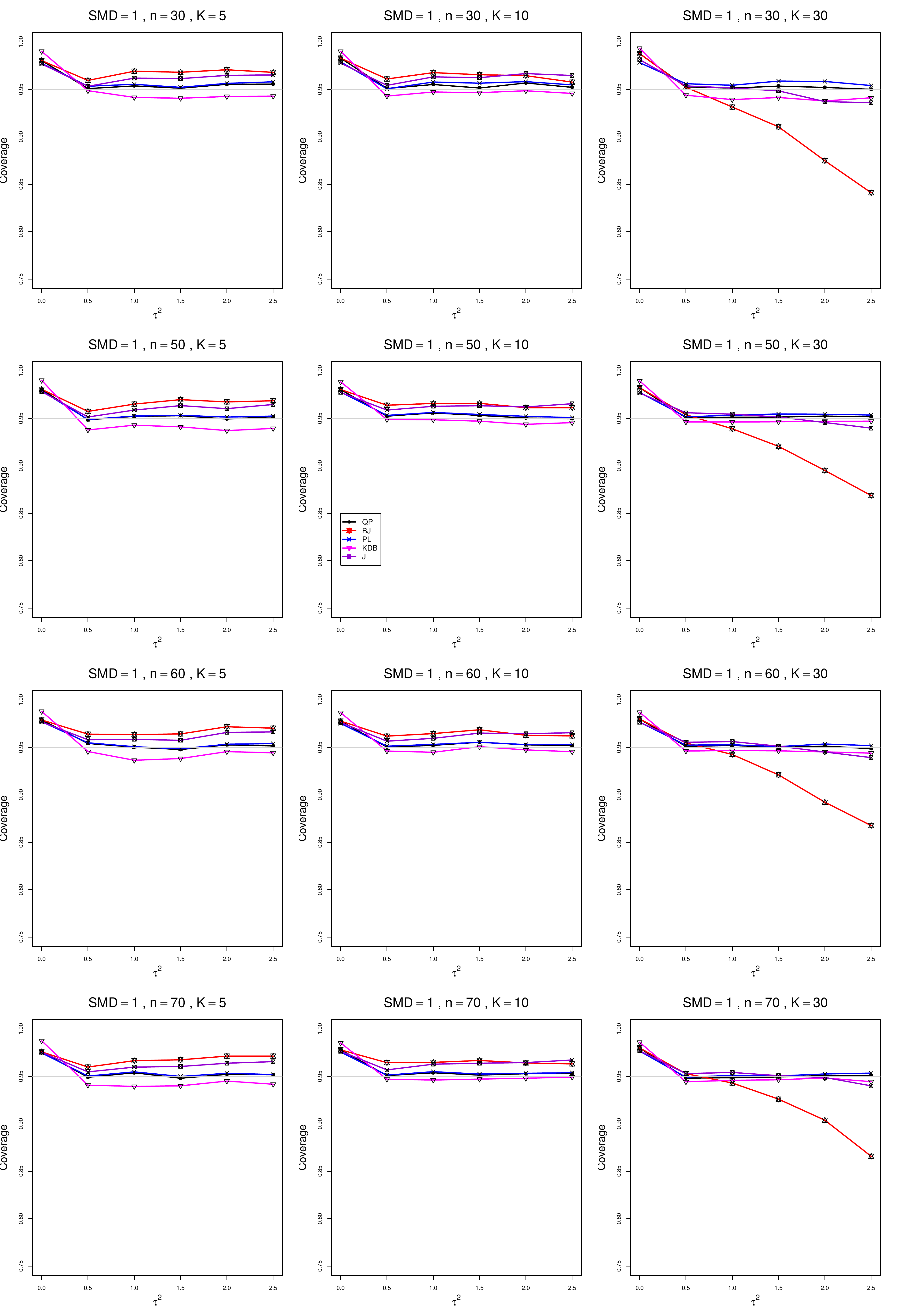}
	\caption{Coverage at  the nominal confidence level of $0.95$ of the  between-studies variance $\tau^2$ for $\delta=1$, $q=0.5$, $n=30,\;50,\;60,\;70$.
		\label{CovTauSMD1small}}
\end{figure}

\begin{figure}[t]\centering
	\includegraphics[scale=0.35]{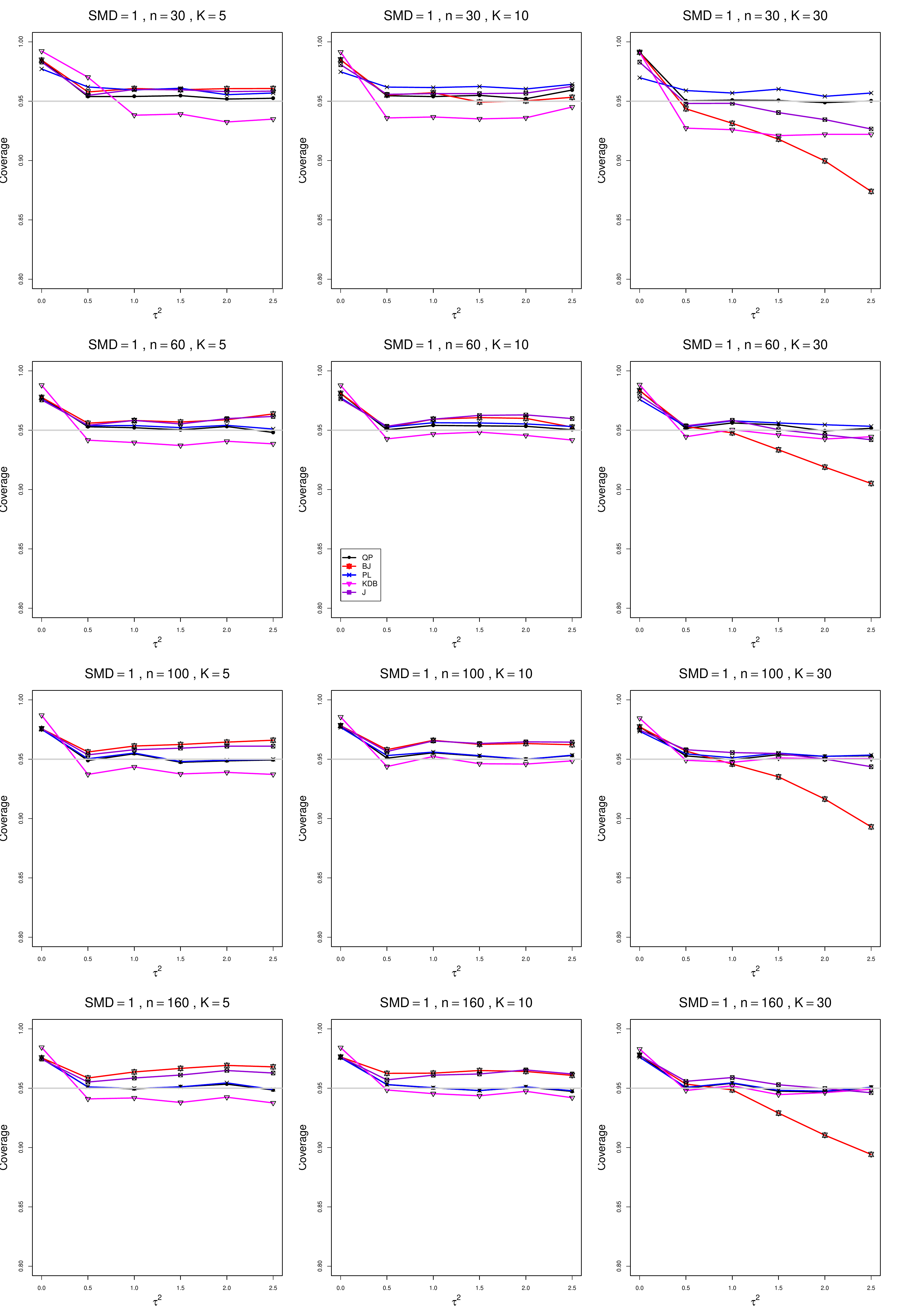}
	\caption{Coverage at  the nominal confidence level of $0.95$ of the  between-studies variance $\tau^2$ for $\delta=1$, $q=0.5$,  unequal sample sizes with $\bar{n}=30,\; 60,\;100,\;160$.
		\label{CovTauSMD1unequal}}
\end{figure}

\begin{figure}[t]\centering
	\includegraphics[scale=0.35]{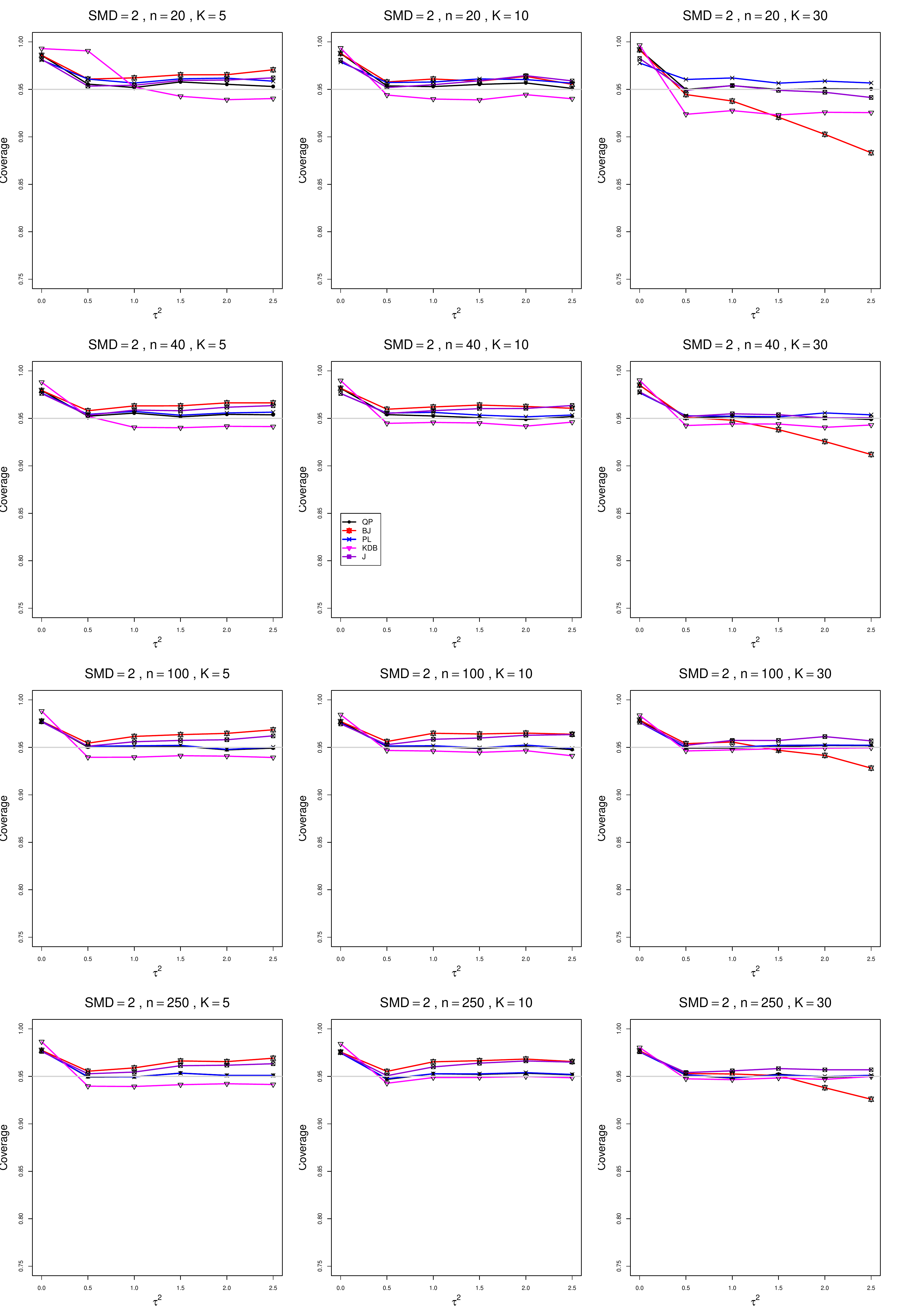}
	\caption{Coverage at  the nominal confidence level of $0.95$ of the  between-studies variance $\tau^2$ for $\delta=2$, $q=0.5$, $n=20,\;40,\;100,\;250$.
		\label{CovTauSMD2}}
\end{figure}

\begin{figure}[t]\centering
	\includegraphics[scale=0.35]{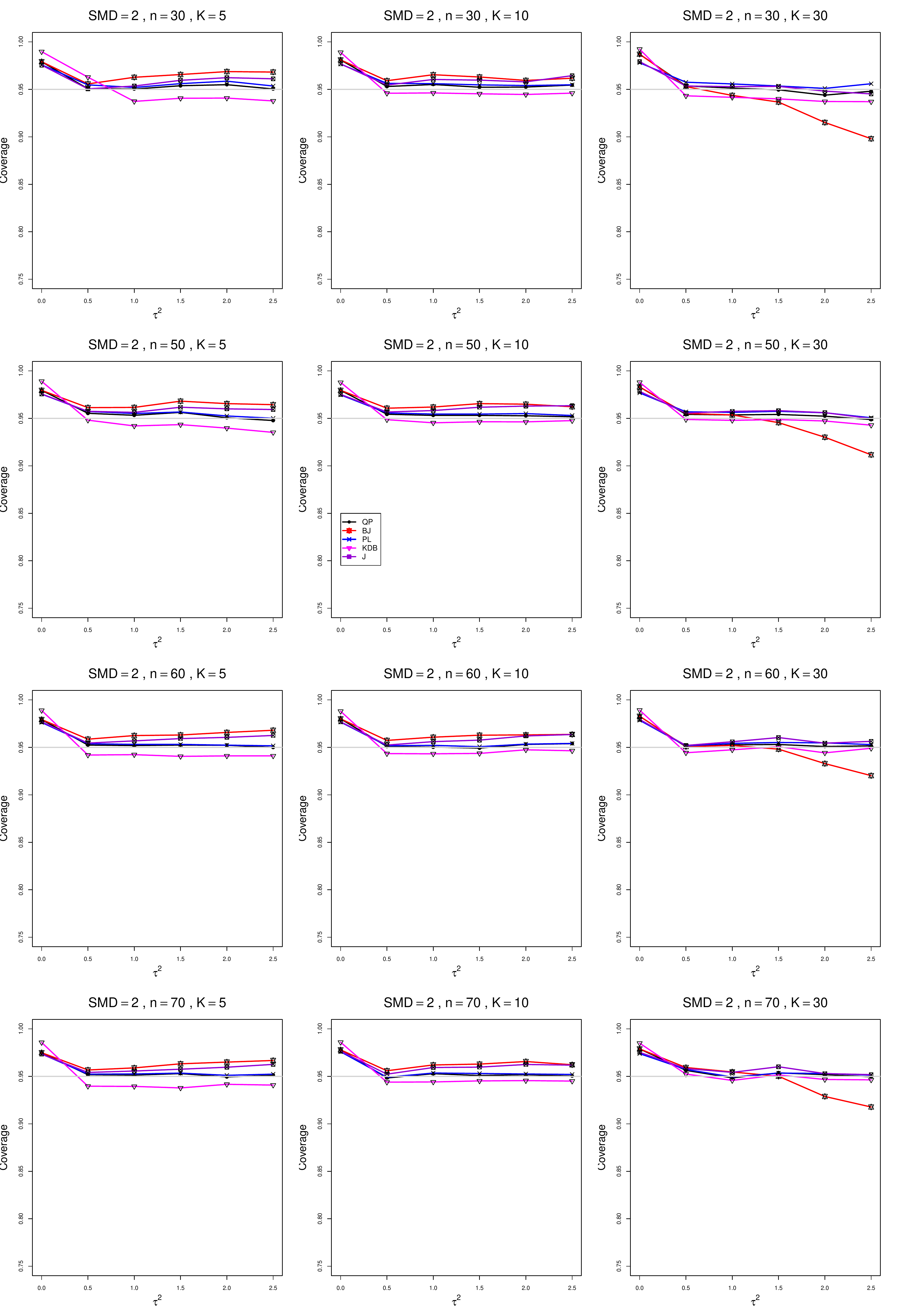}
	\caption{Coverage at  the nominal confidence level of $0.95$ of the  between-studies variance $\tau^2$ for $\delta=2$, $q=0.5$, $n=30,\;50,\;60,\;70$.
		\label{CovTauSMD2small}}
\end{figure}

\begin{figure}[t]\centering
	\includegraphics[scale=0.35]{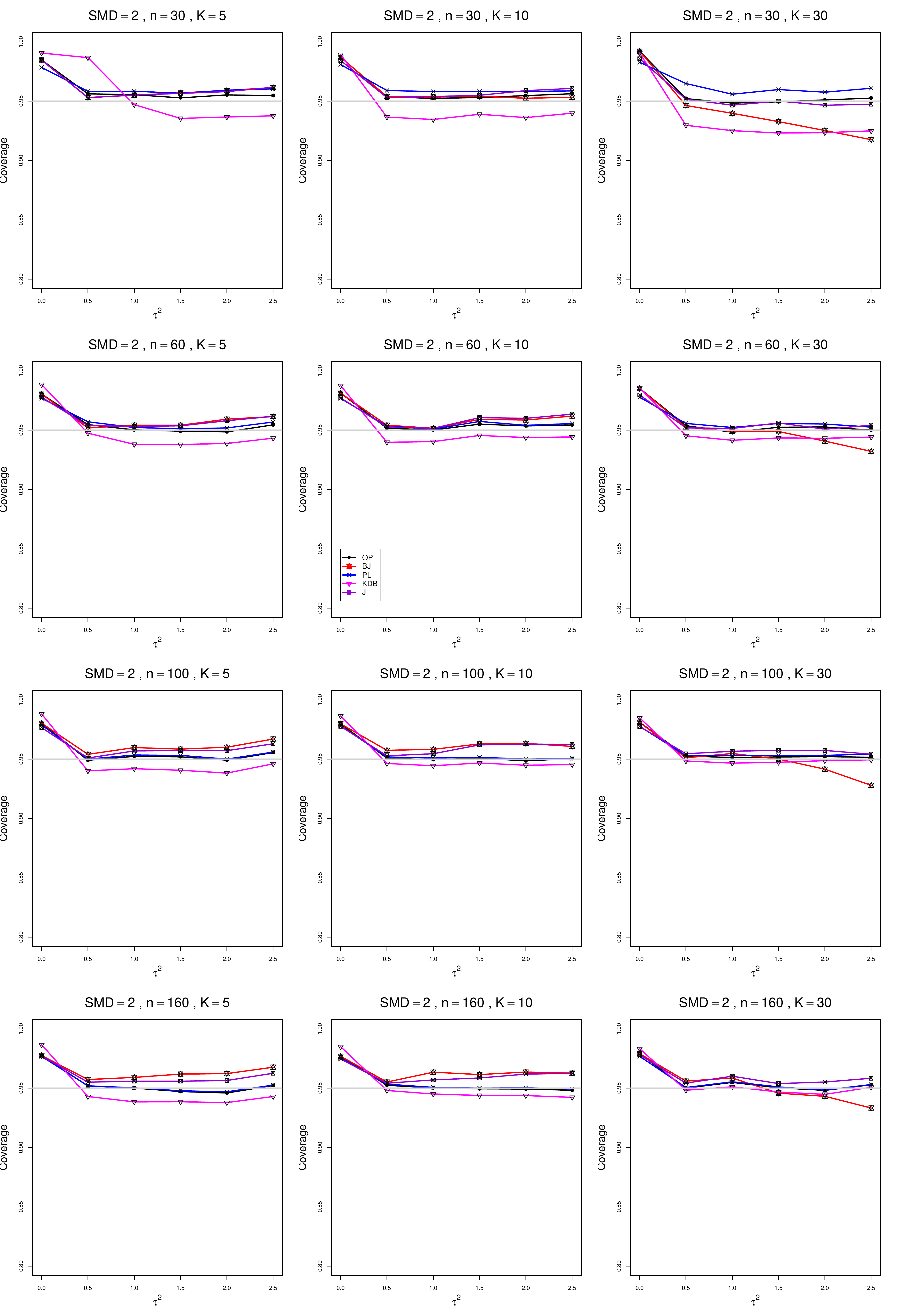}
	\caption{Coverage at  the nominal confidence level of $0.95$ of the  between-studies variance $\tau^2$ for $\delta=2$, $q=0.5$,  unequal sample sizes with $\bar{n}=30,\; 60,\;100,\;160$.
		\label{CovTauSMD2unequal}}
\end{figure}


\begin{figure}[t]
	\centering
	\includegraphics[scale=0.35]{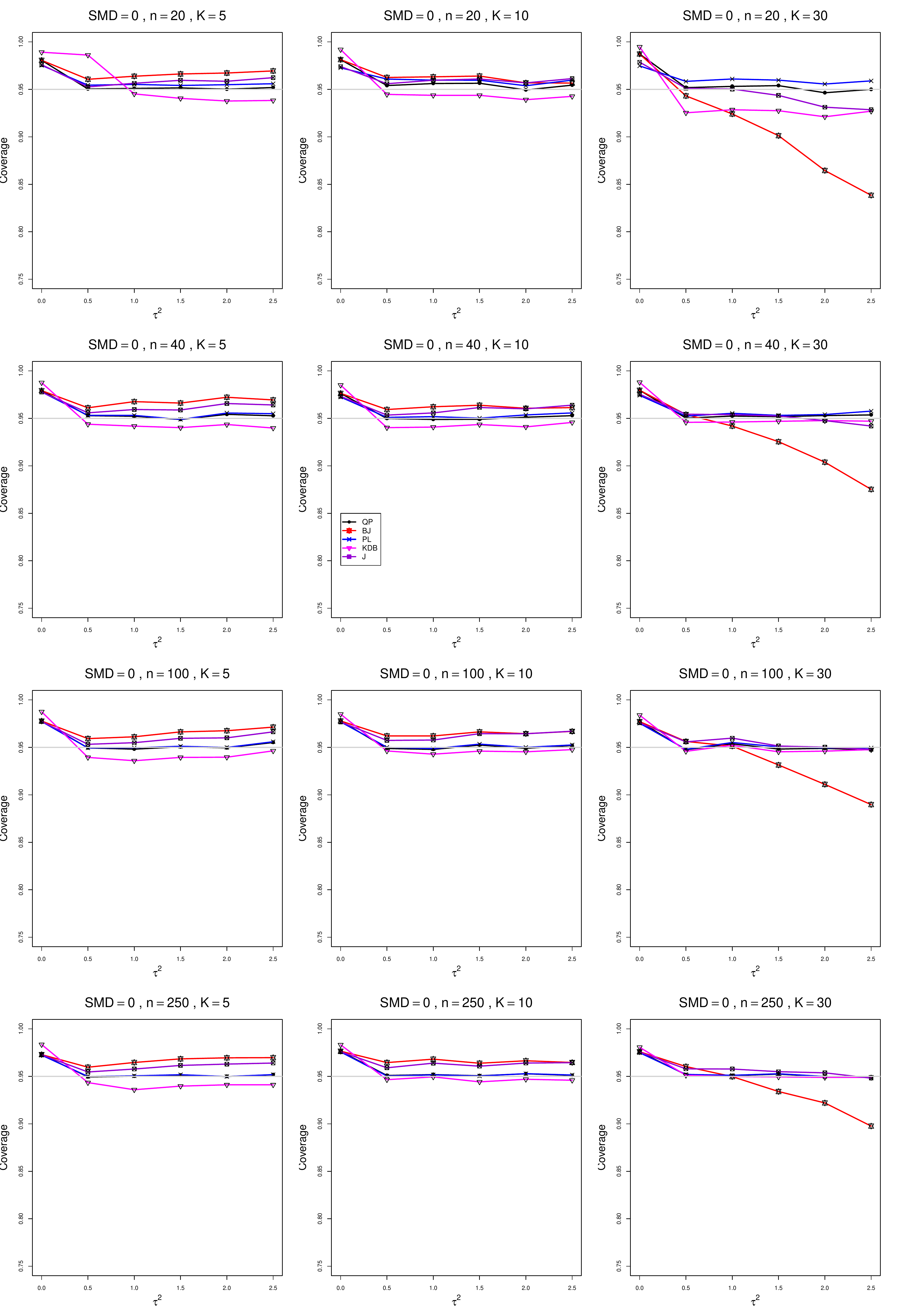}
	\caption{Coverage at  the nominal confidence level of $0.95$ of the  between-studies variance $\tau^2$ for $\delta=0$, $q=0.75$, $n=20,\;40,\;100,\;250$.
		\label{CovTauSMD0q75}}
\end{figure}

\begin{figure}[t]
	\centering
	\includegraphics[scale=0.35]{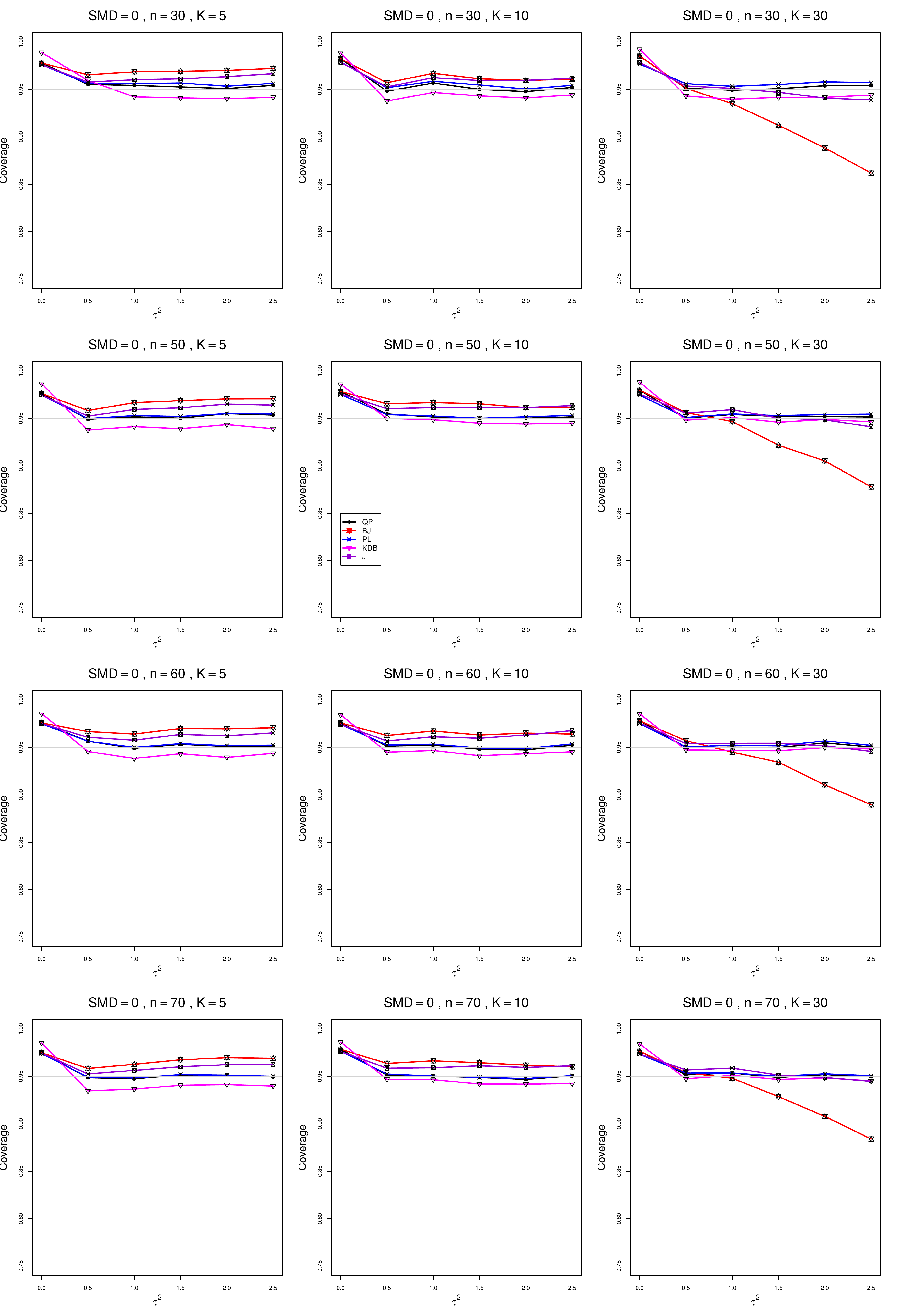}
	\caption{Coverage at  the nominal confidence level of $0.95$ of the  between-studies variance $\tau^2$ for $\delta=0$, $q=0.75$, $n=30,\;50,\;60,\;70$.
		\label{CovTauSMD0q75small}}
\end{figure}

\begin{figure}[t]
	\centering
	\includegraphics[scale=0.35]{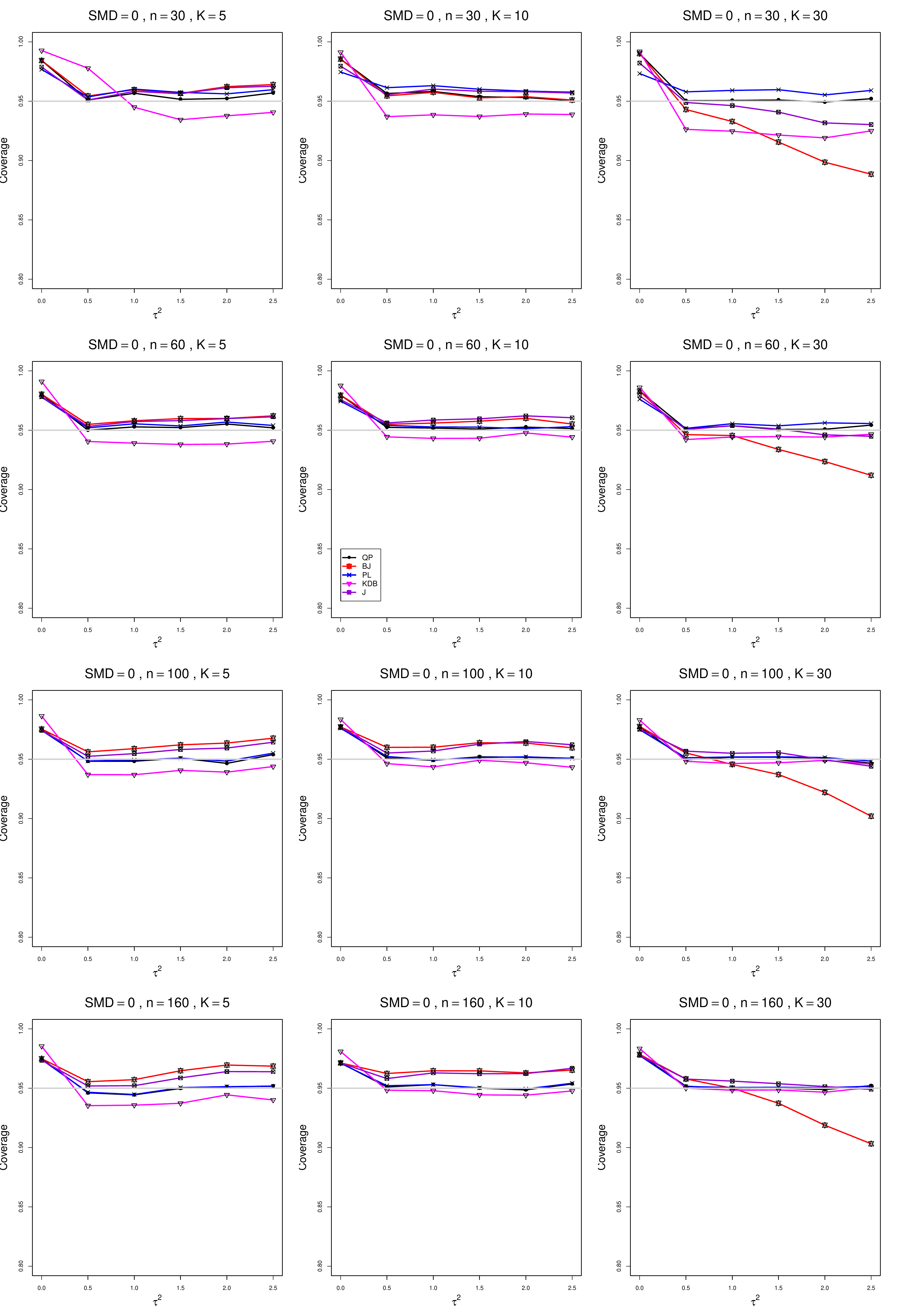}
	\caption{Coverage at  the nominal confidence level of $0.95$ of the  between-studies variance $\tau^2$ for $\delta=0$, $q=0.75$, unequal sample sizes with $\bar{n}=30,\; 60,\;100,\;160$.
		\label{CovTauSMD0q75unequal}}
\end{figure}

\clearpage
\begin{figure}[t]
	\centering
	\includegraphics[scale=0.35]{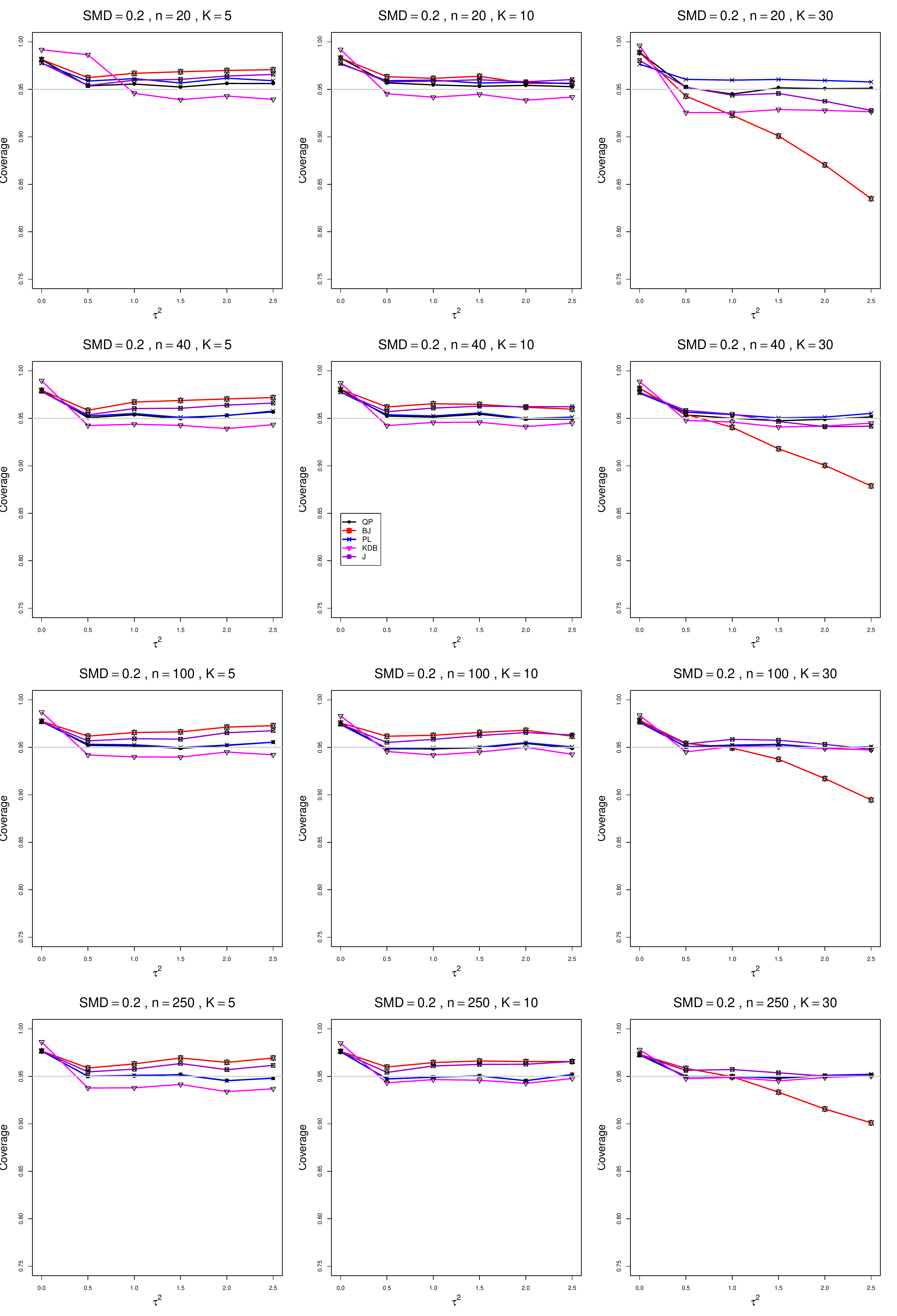}
	\caption{Coverage at  the nominal confidence level of $0.95$ of the  between-studies variance $\tau^2$ for $\delta=0.2$, $q=0.75$, $n=20,\;40,\;100,\;250$.
		\label{CovTauSMD02q75}}
\end{figure}

\begin{figure}[t]
	\centering
	\includegraphics[scale=0.35]{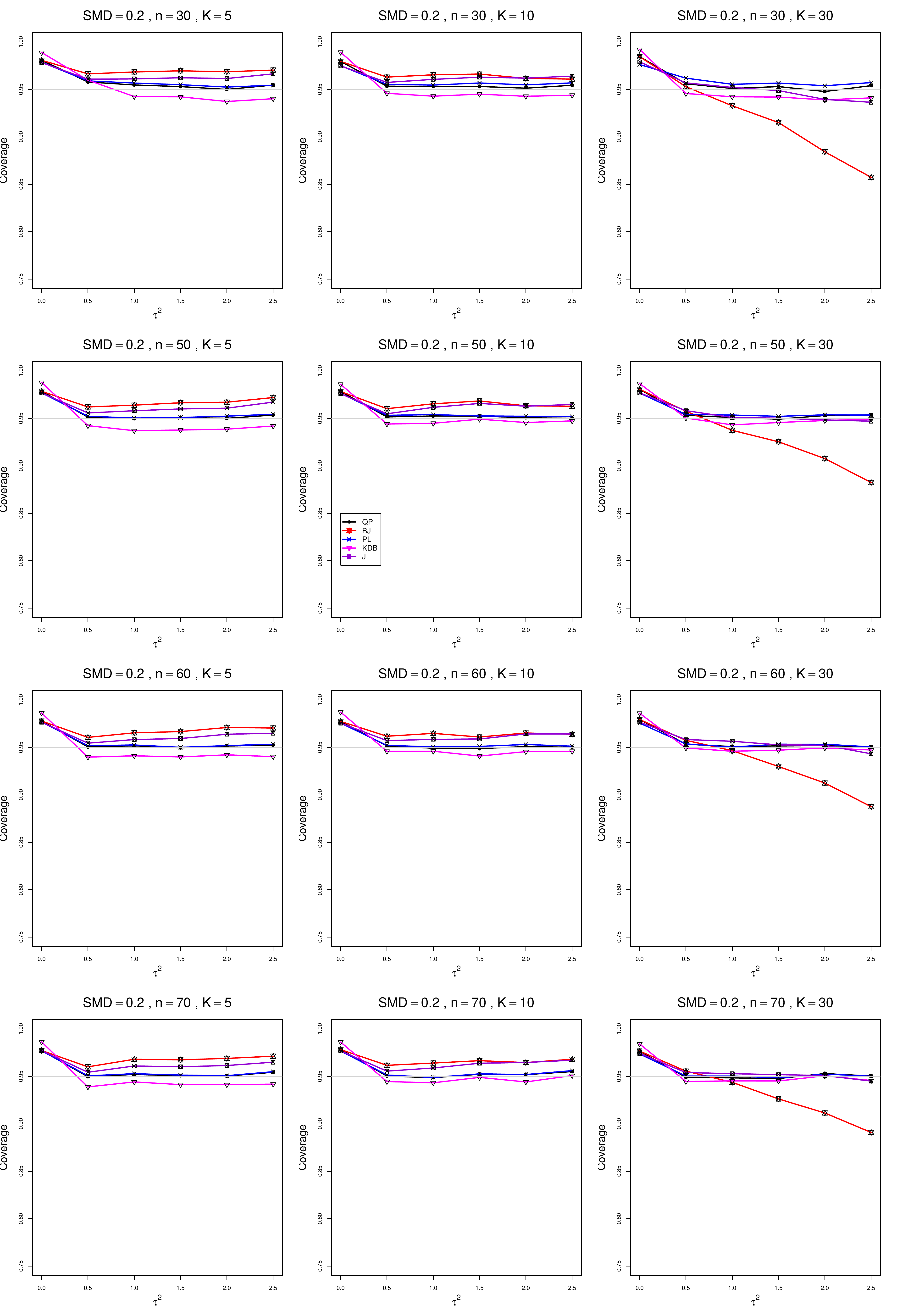}
	\caption{Coverage at  the nominal confidence level of $0.95$ of the  between-studies variance $\tau^2$ for $\delta=0.2$, $q=0.75$, $n=30,\;50,\;60,\;70$.
		\label{CovTauSMD02q75small}}
\end{figure}


\begin{figure}[t]\centering
	\includegraphics[scale=0.35]{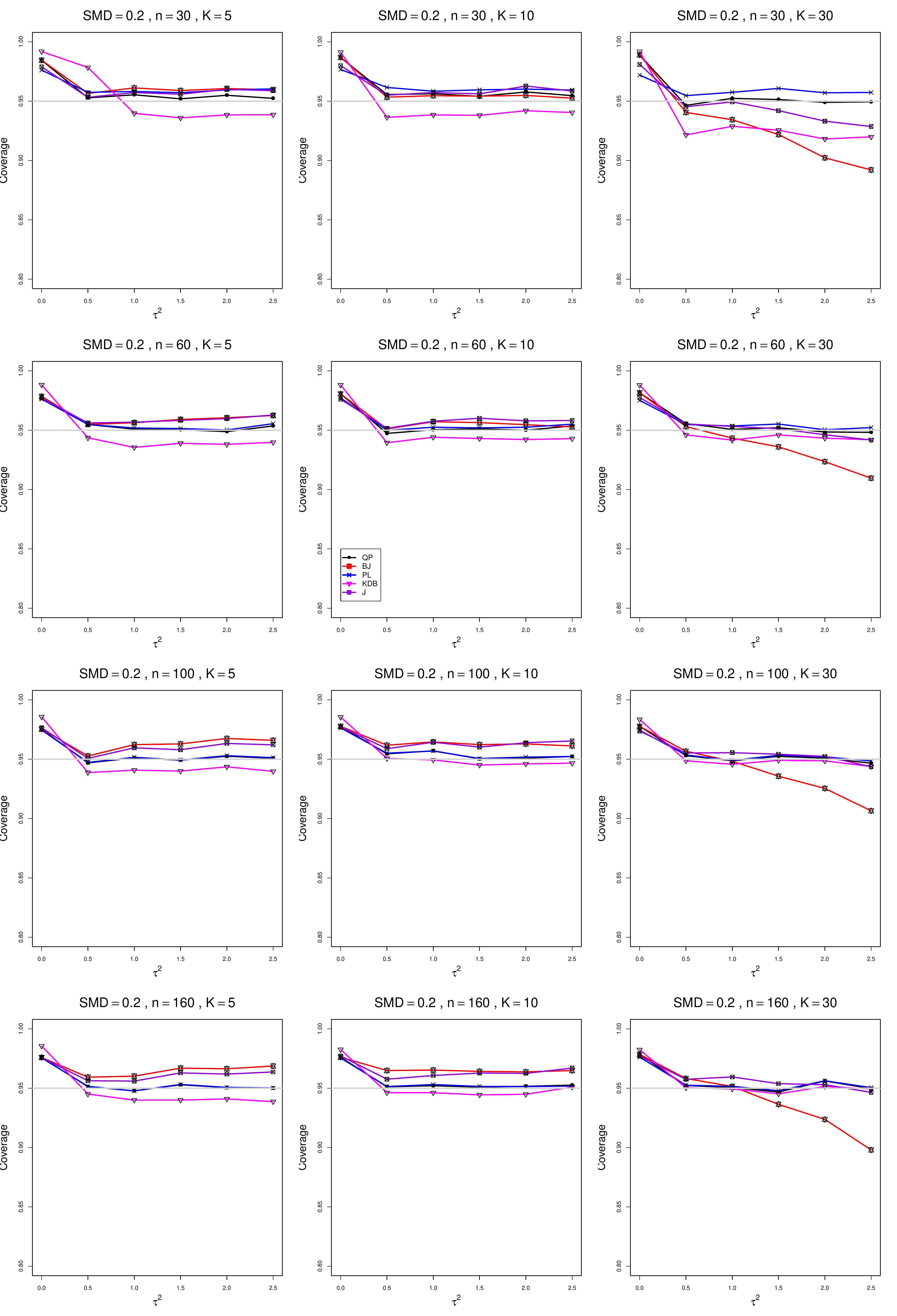}
	\caption{Coverage at  the nominal confidence level of $0.95$ of the  between-studies variance $\tau^2$ for $\delta=0.2$, $q=0.75$, unequal sample sizes with $\bar{n}=30,\; 60,\;100,\;160$.
		\label{CovTauSMD02q75unequal}}
\end{figure}

\begin{figure}[t]
	\centering
	\includegraphics[scale=0.35]{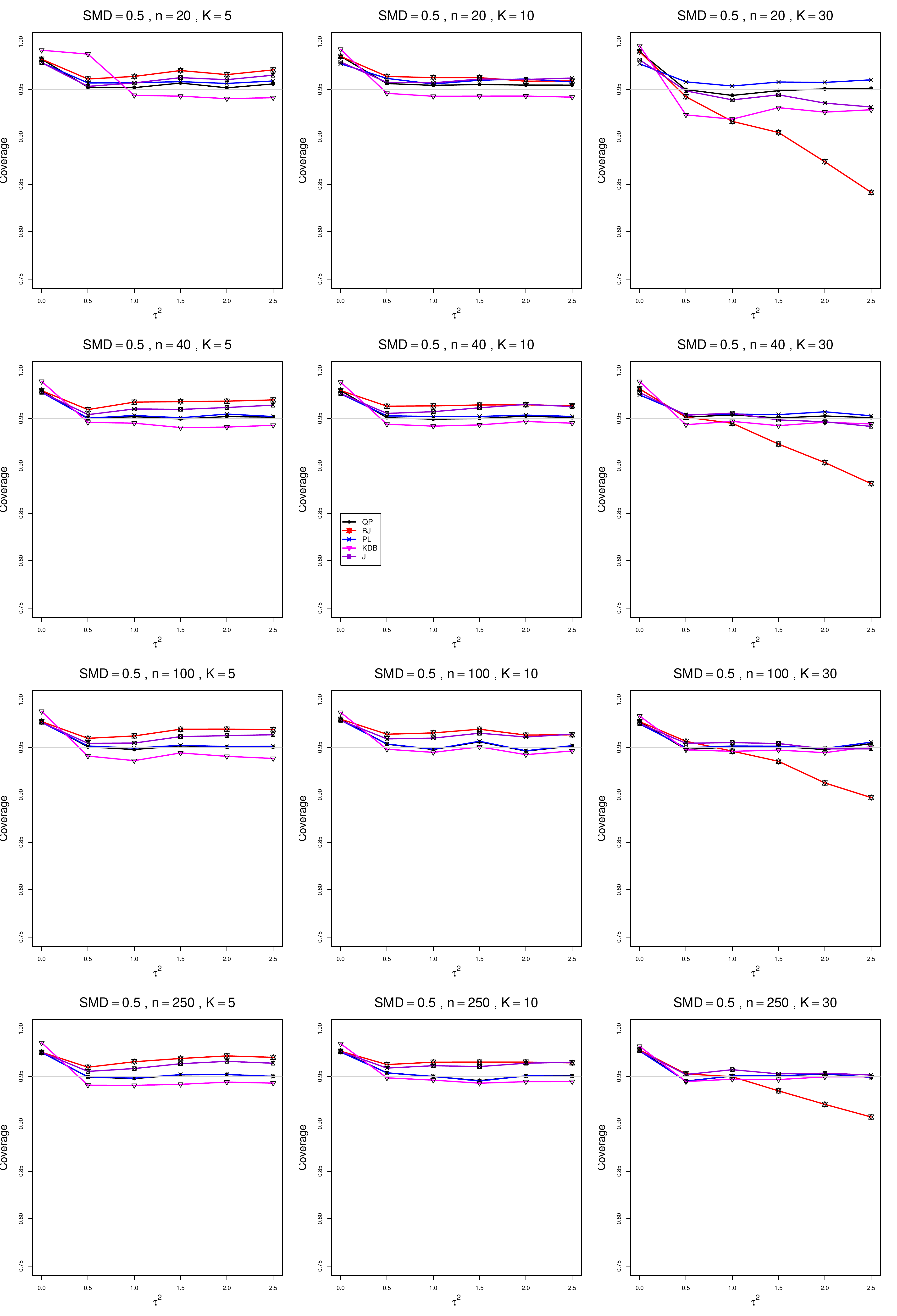}
	\caption{Coverage at  the nominal confidence level of $0.95$ of the  between-studies variance $\tau^2$ for $\delta=0.5$, $q=0.75$, $n=20,\;40,\;100,\;250$.
		\label{CovTauSMD05q75}}
\end{figure}

\begin{figure}[t]
	\centering
	\includegraphics[scale=0.35]{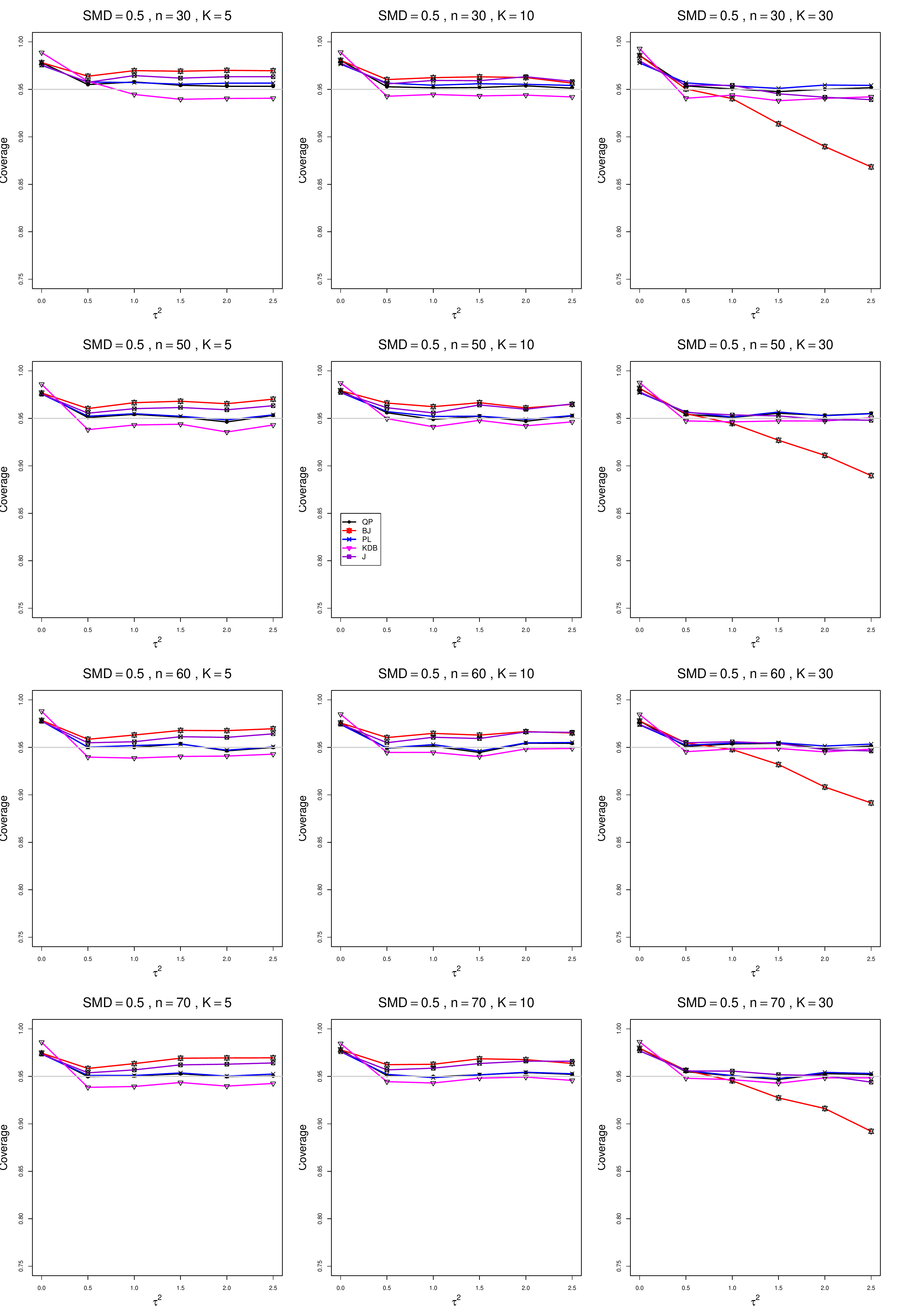}
	\caption{Coverage at  the nominal confidence level of $0.95$ of the  between-studies variance $\tau^2$ for $\delta=0.5$, $q=0.75$, $n=30,\;50,\;60,\;70$.
		\label{CovTauSMD05q75small}}
\end{figure}

\begin{figure}[t]
	\centering
	\includegraphics[scale=0.35]{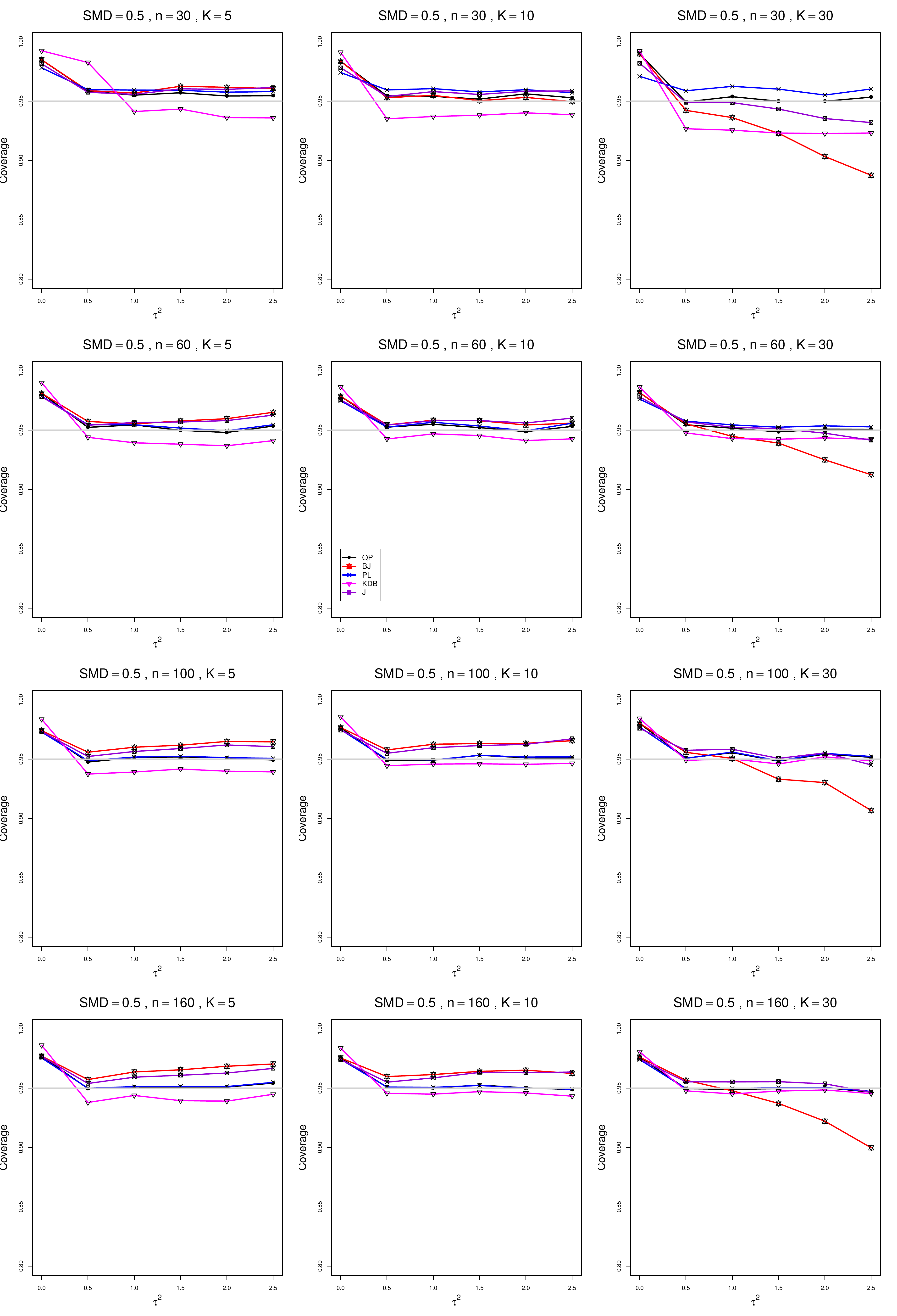}
	\caption{Coverage at  the nominal confidence level of $0.95$ of the  between-studies variance $\tau^2$ for $\delta=0.5$, $q=0.75$, unequal sample sizes with $\bar{n}=30,\; 60,\;100,\;160$.
		\label{CovTauSMD0.5q75unequal}}
\end{figure}

\begin{figure}[t]
	\centering
	\includegraphics[scale=0.35]{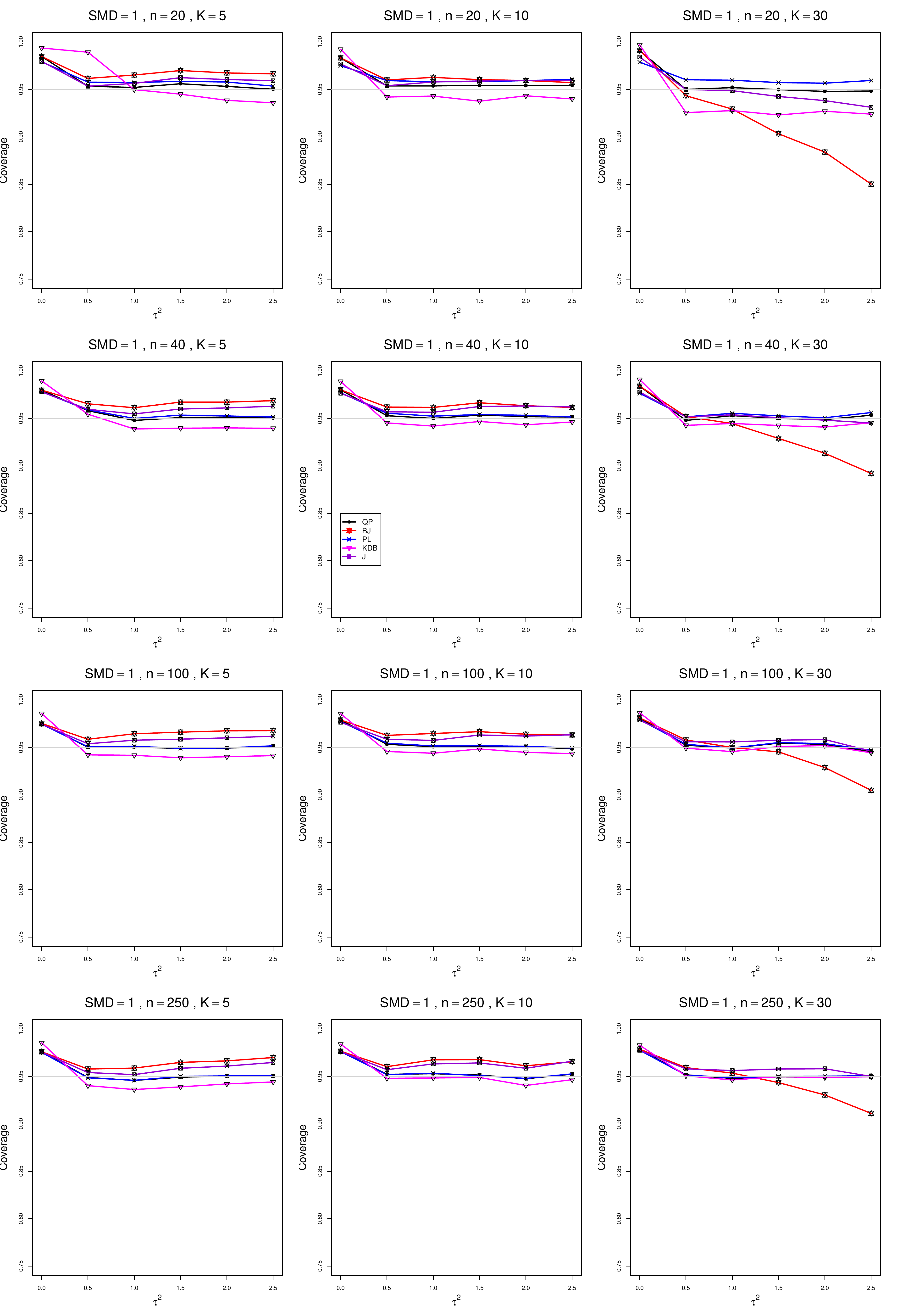}
	\caption{Coverage at  the nominal confidence level of $0.95$ of the  between-studies variance $\tau^2$ for $\delta=1$, $q=0.75$, $n=20,\;40,\;100,\;250$.
		\label{CovTauSMD1q75}}
\end{figure}

\begin{figure}[t]
	\centering
	\includegraphics[scale=0.35]{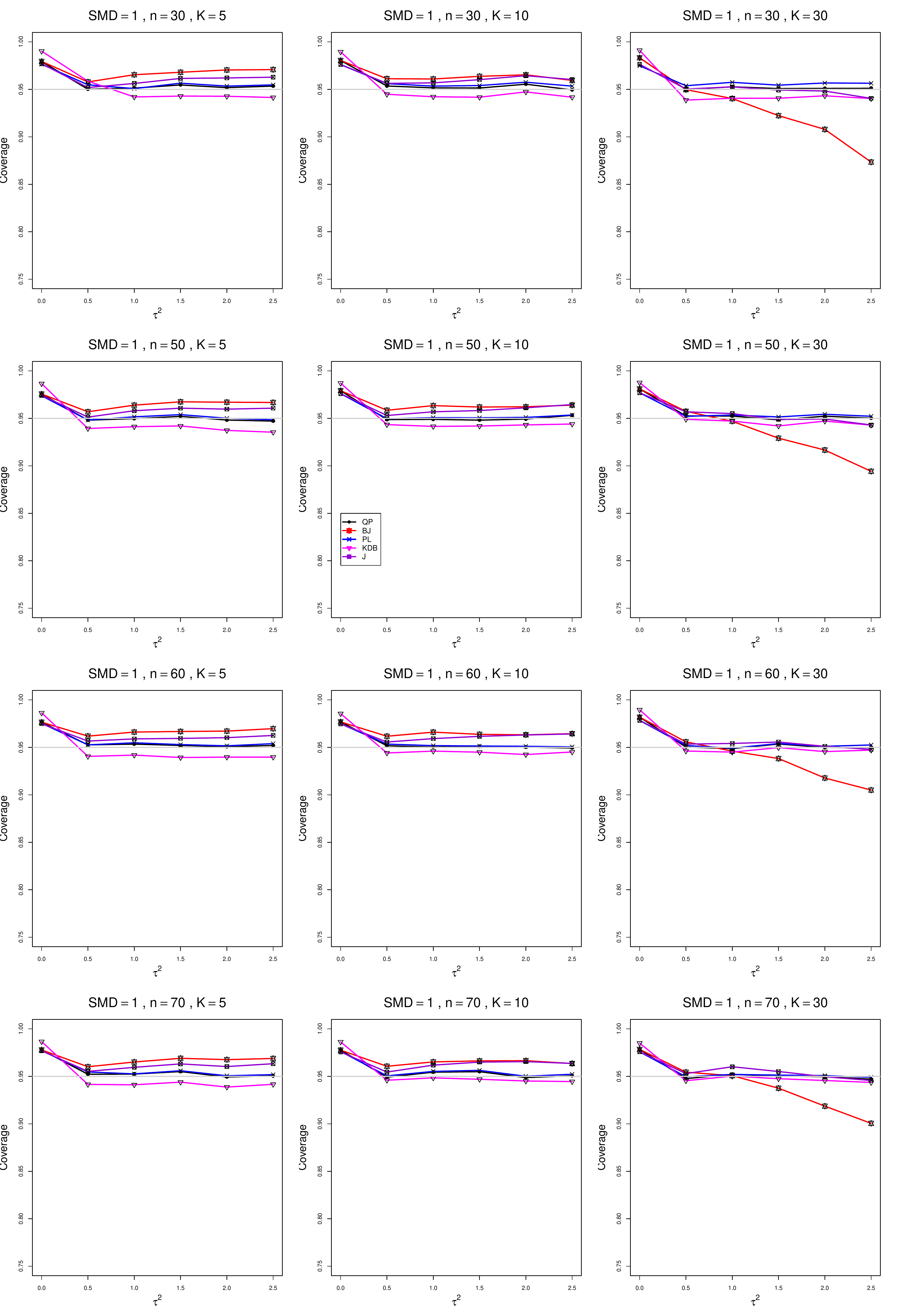}
	\caption{Coverage at  the nominal confidence level of $0.95$ of the  between-studies variance $\tau^2$ for $\delta=1$, $q=0.75$, $n=30,\;50,\;60,\;70$.
		\label{CovTauSMD1q75small}}
\end{figure}

\begin{figure}[t]
	\centering
	\includegraphics[scale=0.35]{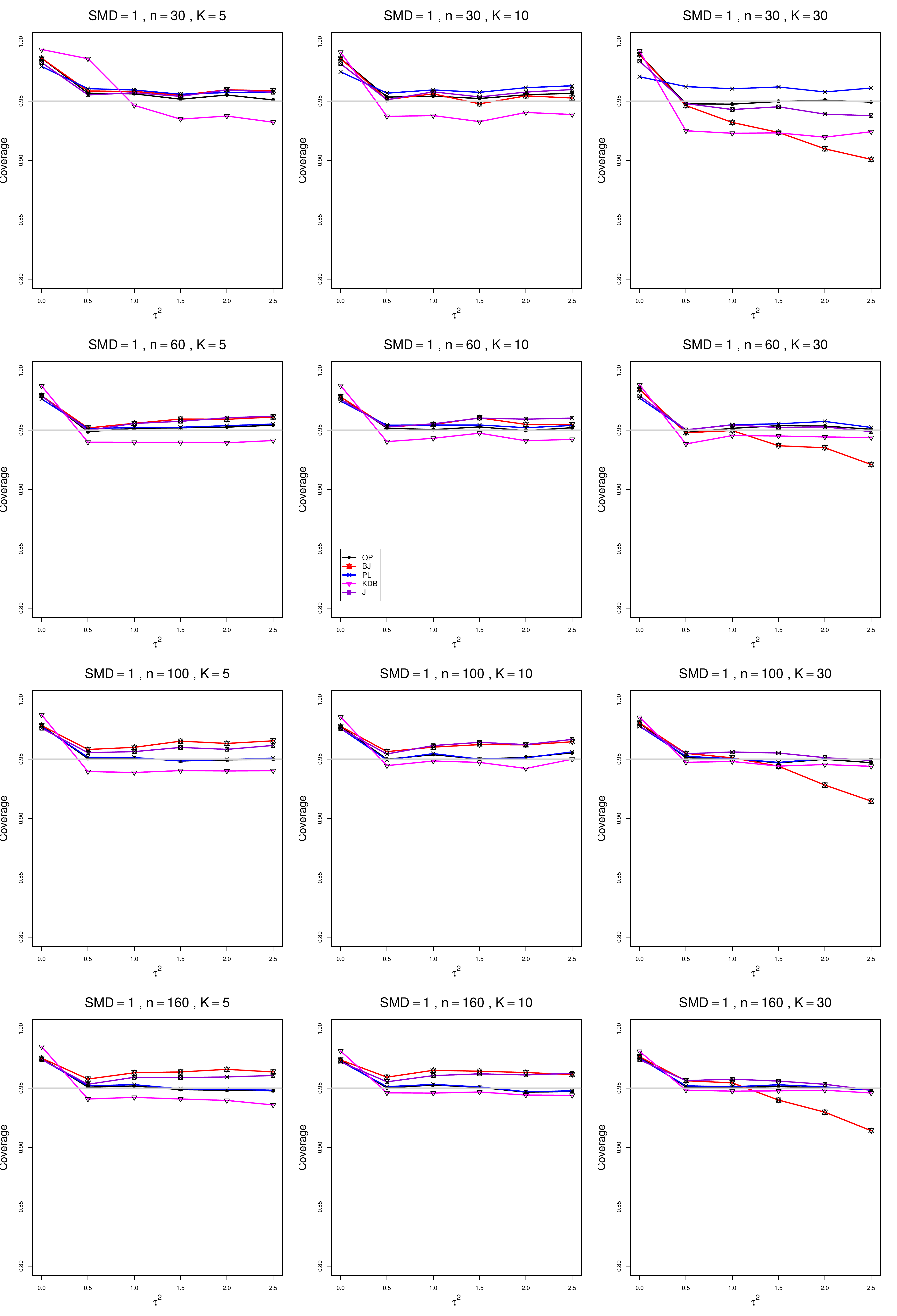}
	\caption{Coverage at  the nominal confidence level of $0.95$ of the  between-studies variance $\tau^2$ for $\delta=1$, $q=0.75$, unequal sample sizes with $\bar{n}=30,\; 60,\;100,\;160$.
		\label{CovTauSMD1q75unequal}}
\end{figure}

\begin{figure}[t]
	\centering
	\includegraphics[scale=0.35]{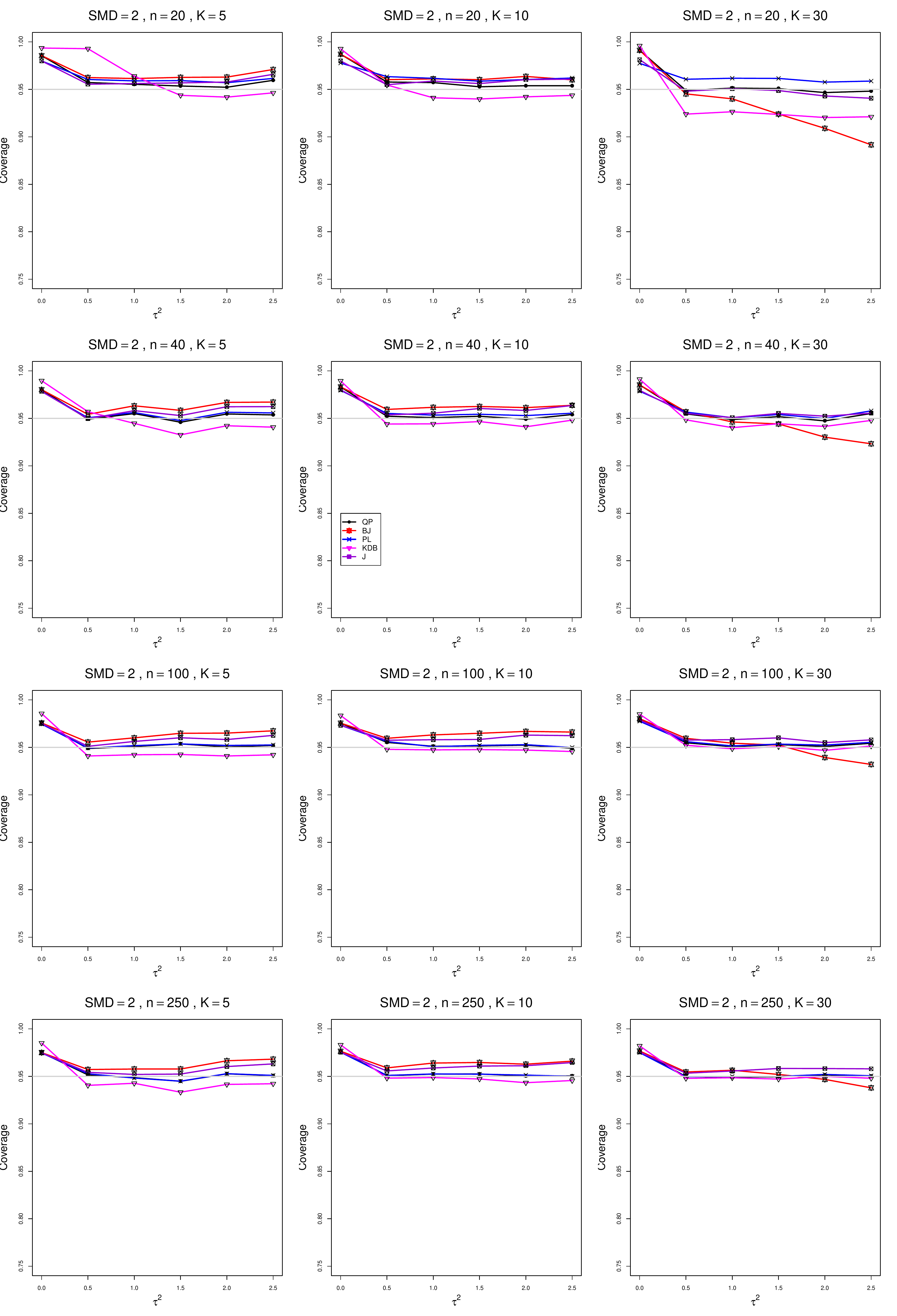}
	\caption{Coverage at  the nominal confidence level of $0.95$ of the  between-studies variance $\tau^2$ for $\delta=2$, $q=0.75$, $n=20,\;40,\;100,\;250$.
		\label{CovTauSMD2q75}}
\end{figure}

\begin{figure}[t]
	\centering
	\includegraphics[scale=0.35]{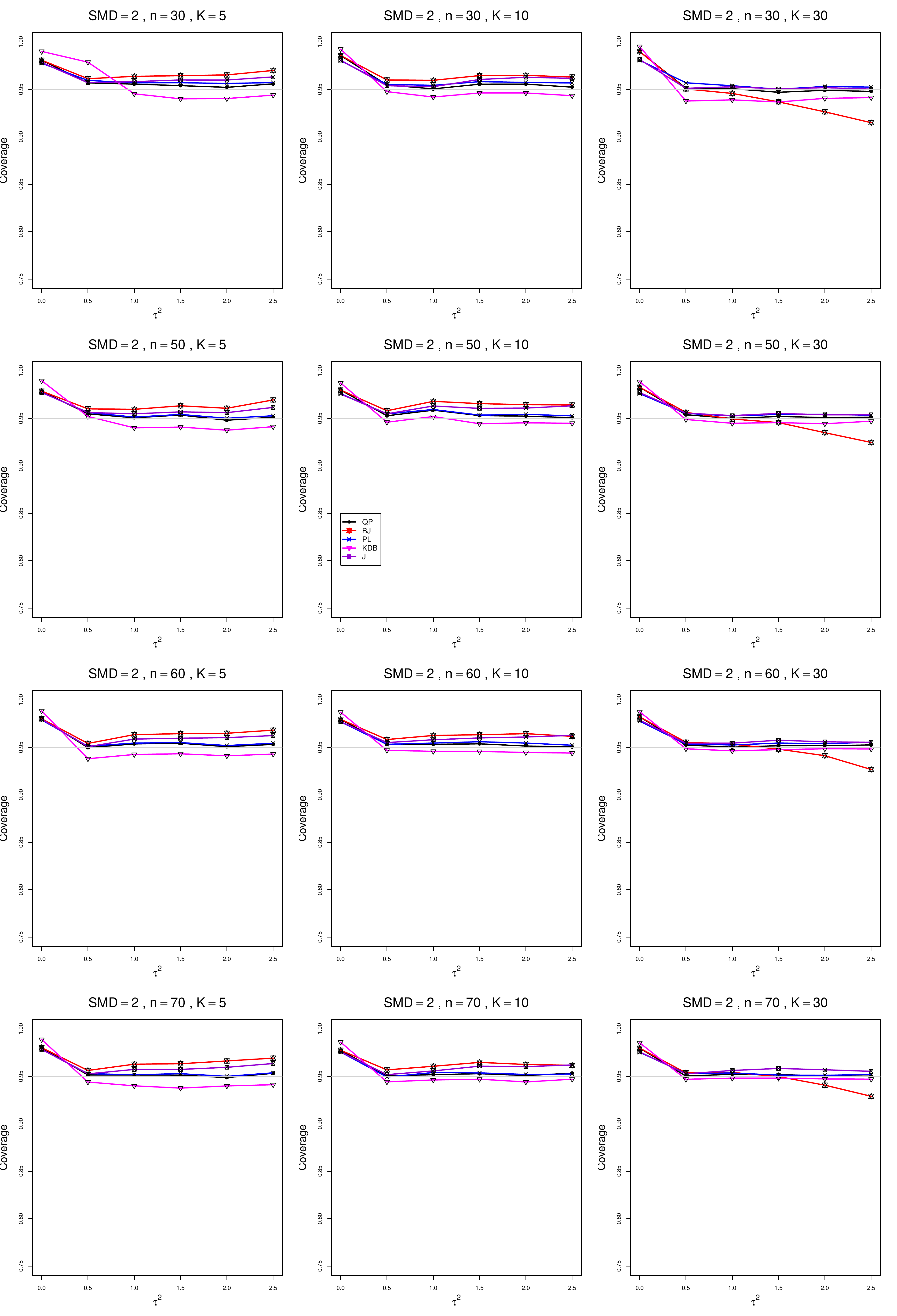}
	\caption{Coverage at  the nominal confidence level of $0.95$ of the  between-studies variance $\tau^2$ for $\delta=2$, $q=0.75$, $n=30,\;50,\;60,\;70$.
		\label{CovTauSMD2q75small}}
\end{figure}

\begin{figure}[t]
	\centering
	\includegraphics[scale=0.35]{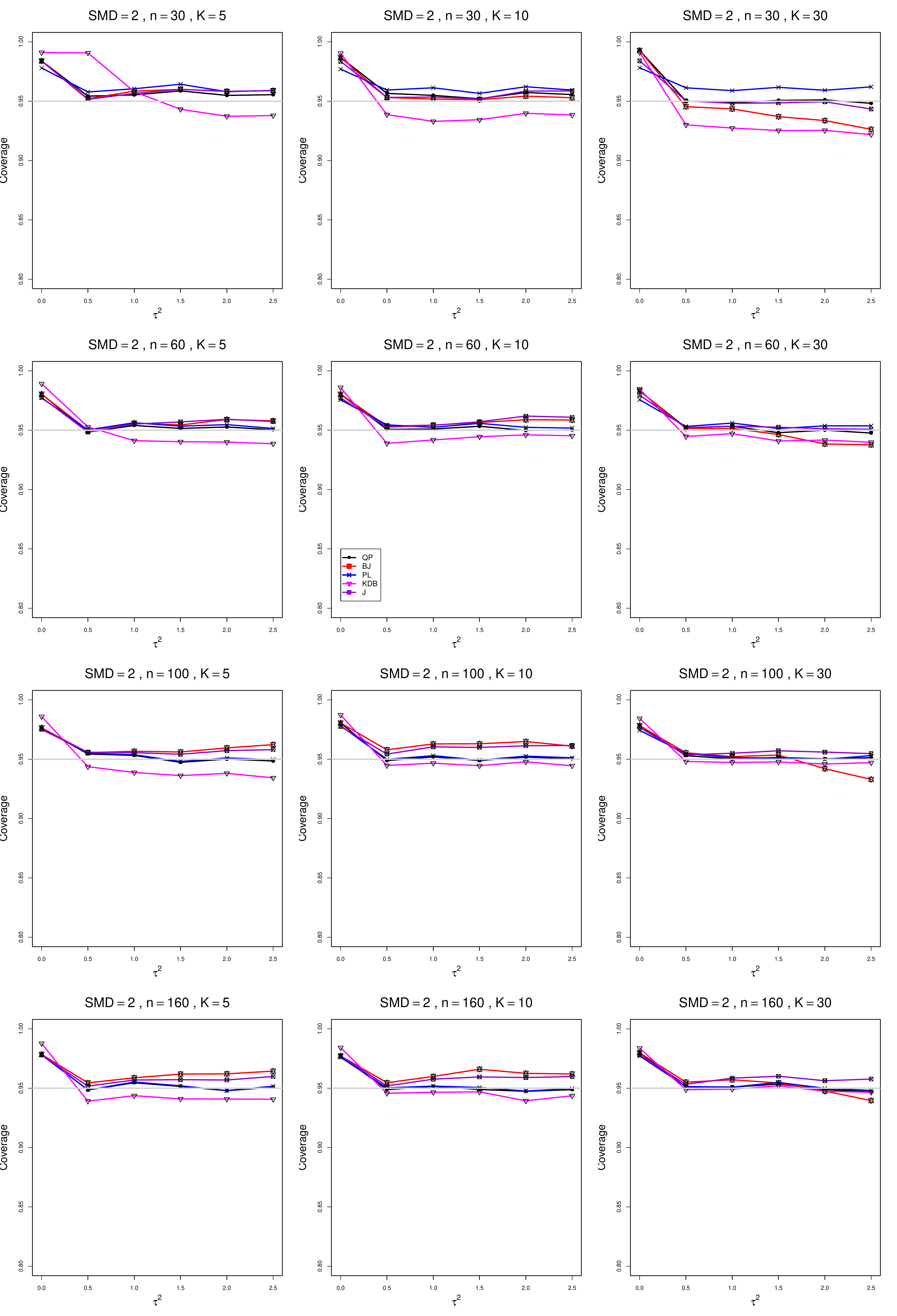}
	\caption{Coverage at  the nominal confidence level of $0.95$ of the  between-studies variance $\tau^2$ for $\delta=2$, $q=0.75$, unequal sample sizes with $\bar{n}=30,\; 60,\;100,\;160$.
		\label{CovTauSMD2
			q75unequal}}
\end{figure}


\clearpage

\section*{B1. Bias and mean squared error of point estimators of $\hat{\delta}$ for $\hat{\tau}^2=0.0(0.5)2.5$.}
For bias of $\hat{\delta}$, each figure corresponds to a value of $\delta (= 0, 0.5, 1, 1.5, 2 , 2.5)$, a value of $q (= .5, .75)$, and a set of values of $n$ (= 20, 40, 100, 250 or 30, 50, 60, 70) or $\bar{n} (= 30, 60, 100, 160)$.\\
Figures for mean squared error (expressed as the ratio of the MSE of SSW to the MSEs of the inverse-variance-weighted estimators that use the MP or KDB estimator of $\tau^2$) use the above values of $\delta$ and q but only n = 20, 40, 100, 250.\\
Each figure contains a panel (with $\tau^2$ on the horizontal axis) for each combination of n (or $\bar{n}$) and $K (=5, 10, 30)$.\\
The point estimators of $\delta$ are
\begin{itemize}
	\item DL (DerSimonian-Laird)
	\item REML (restricted maximum likelihood)
	\item MP (Mandel-Paule)
	\item KDB (improved moment estimator based on Kulinskaya, Dollinger and  Bj{\o}rkest{\o}l (2011))
	\item J (Jackson)
	\item SSW (sample-size weighted)
\end{itemize}

\clearpage
\setcounter{figure}{0}
\renewcommand{\thefigure}{B1.\arabic{figure}}
\begin{figure}[t]\centering
	\includegraphics[scale=0.35]{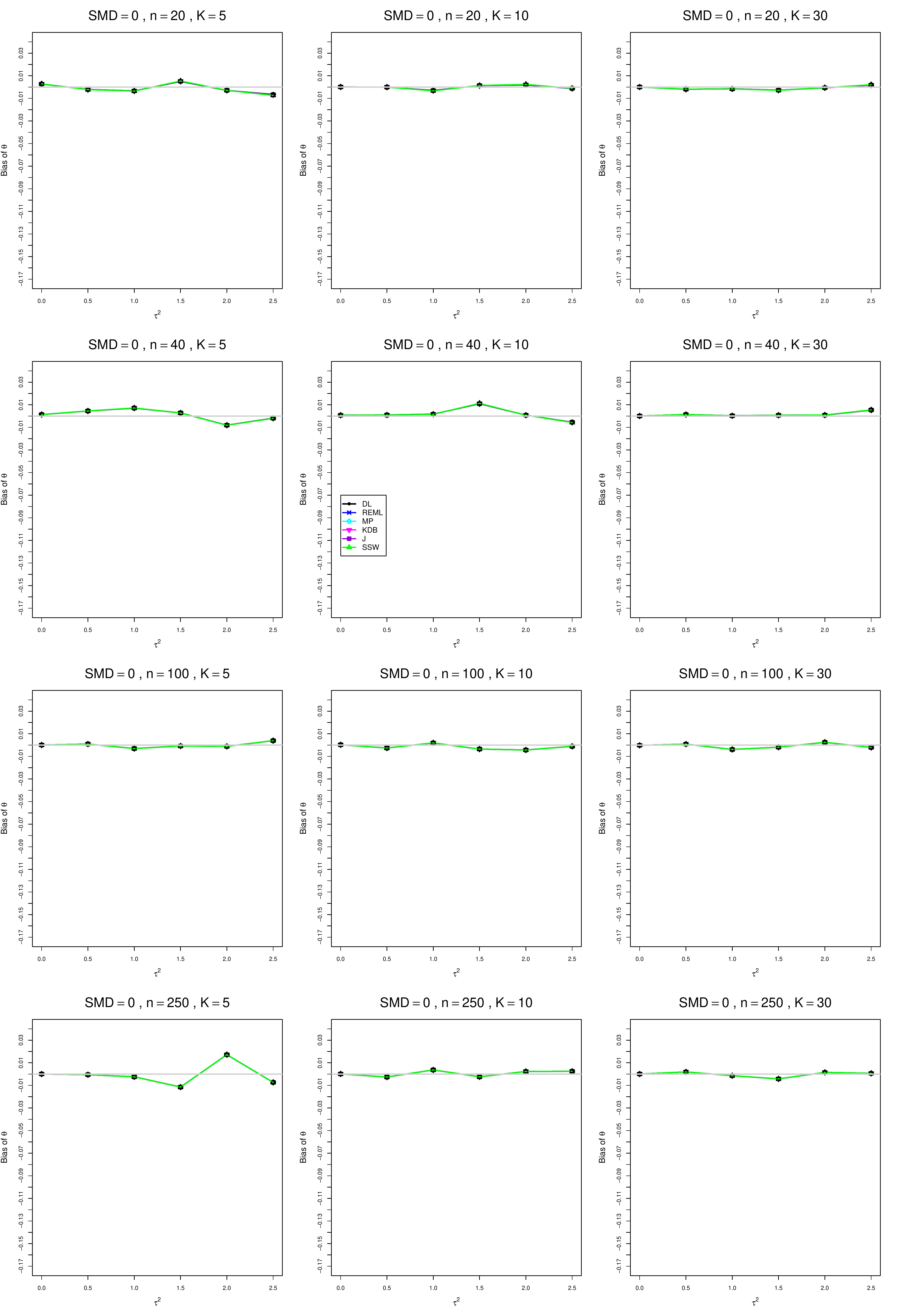}
	\caption{Bias of inverse-variance estimator of $\delta=0$, for $q=0.5$, $n=20,\;40,\;100,\;250$.
		\label{BiasThetaSMD0}}
\end{figure}

\begin{figure}[t]\centering
	\includegraphics[scale=0.35]{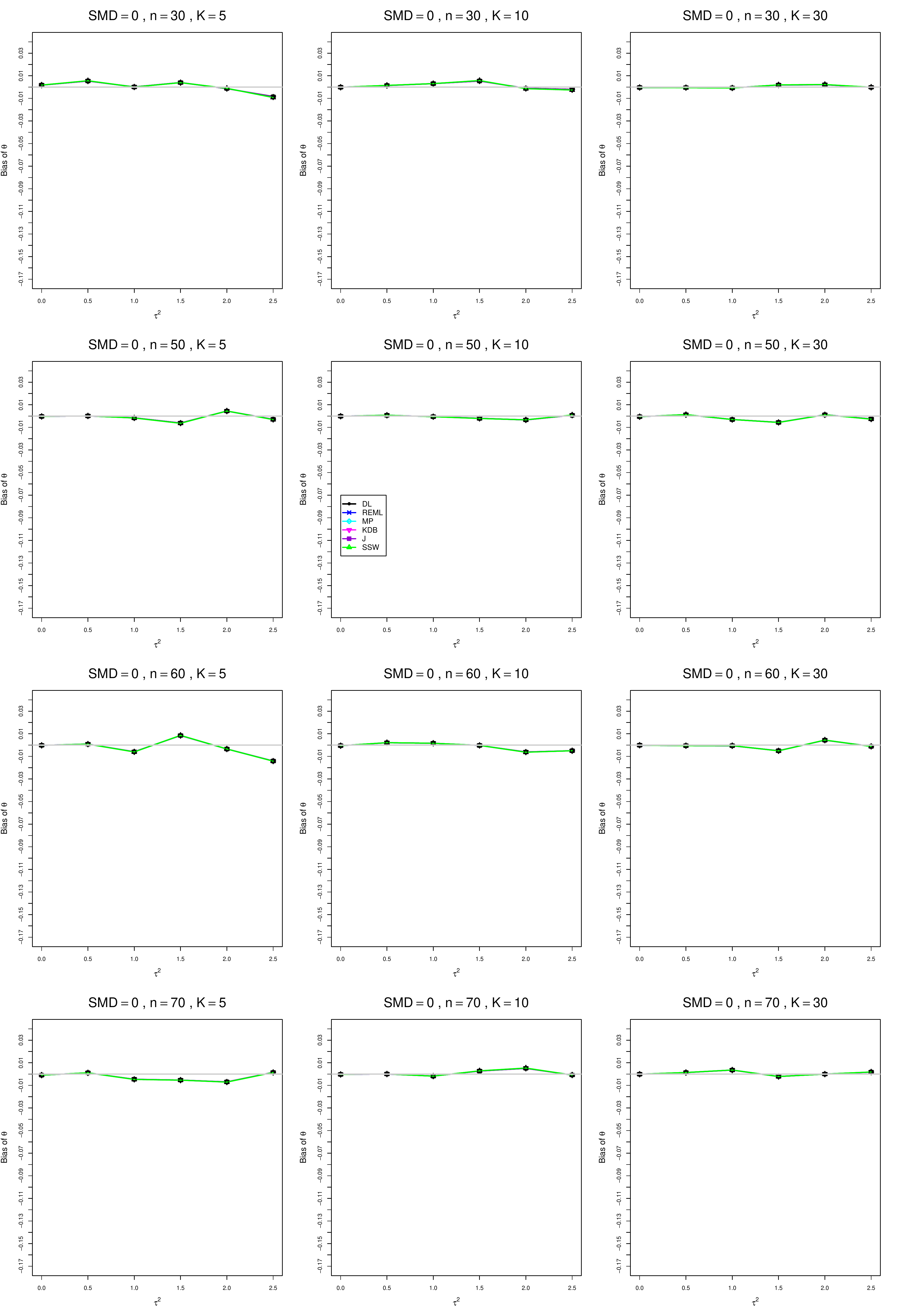}
	\caption{Bias of inverse-variance estimator of $\delta=0$, for $q=0.5$, $n=30,\;50,\;60,\;70$.
		\label{BiasThetaSMD0small}}
\end{figure}

\begin{figure}[t]\centering
	\includegraphics[scale=0.35]{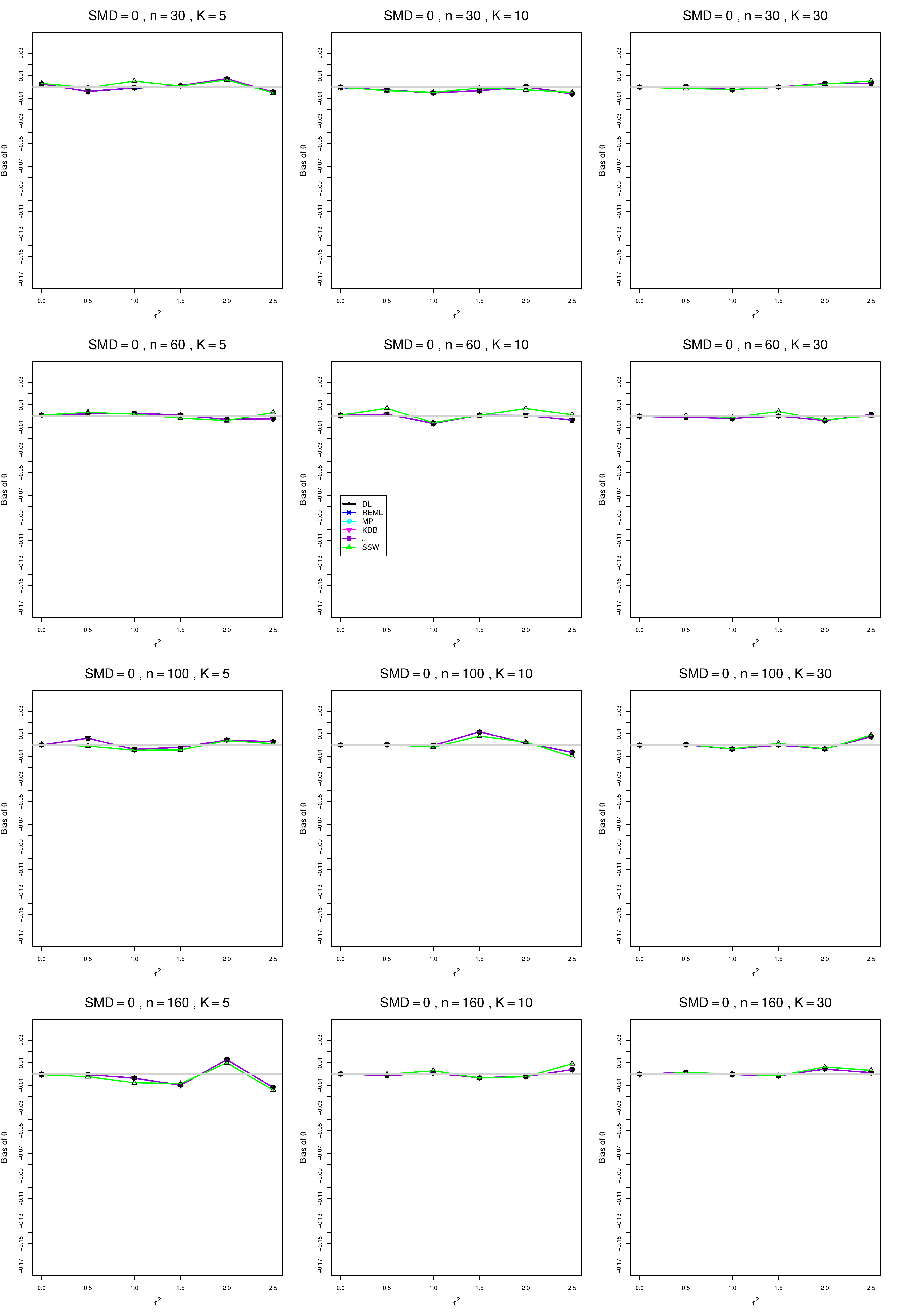}
	\caption{Bias of inverse-variance estimator of $\delta=0$, for $q=0.5$, unequal sample sizes with $\bar{n}=30,\; 60,\;100,\;160$.
		\label{BiasThetaSMD0unequal}}
\end{figure}

\begin{figure}[t]\centering
	\includegraphics[scale=0.35]{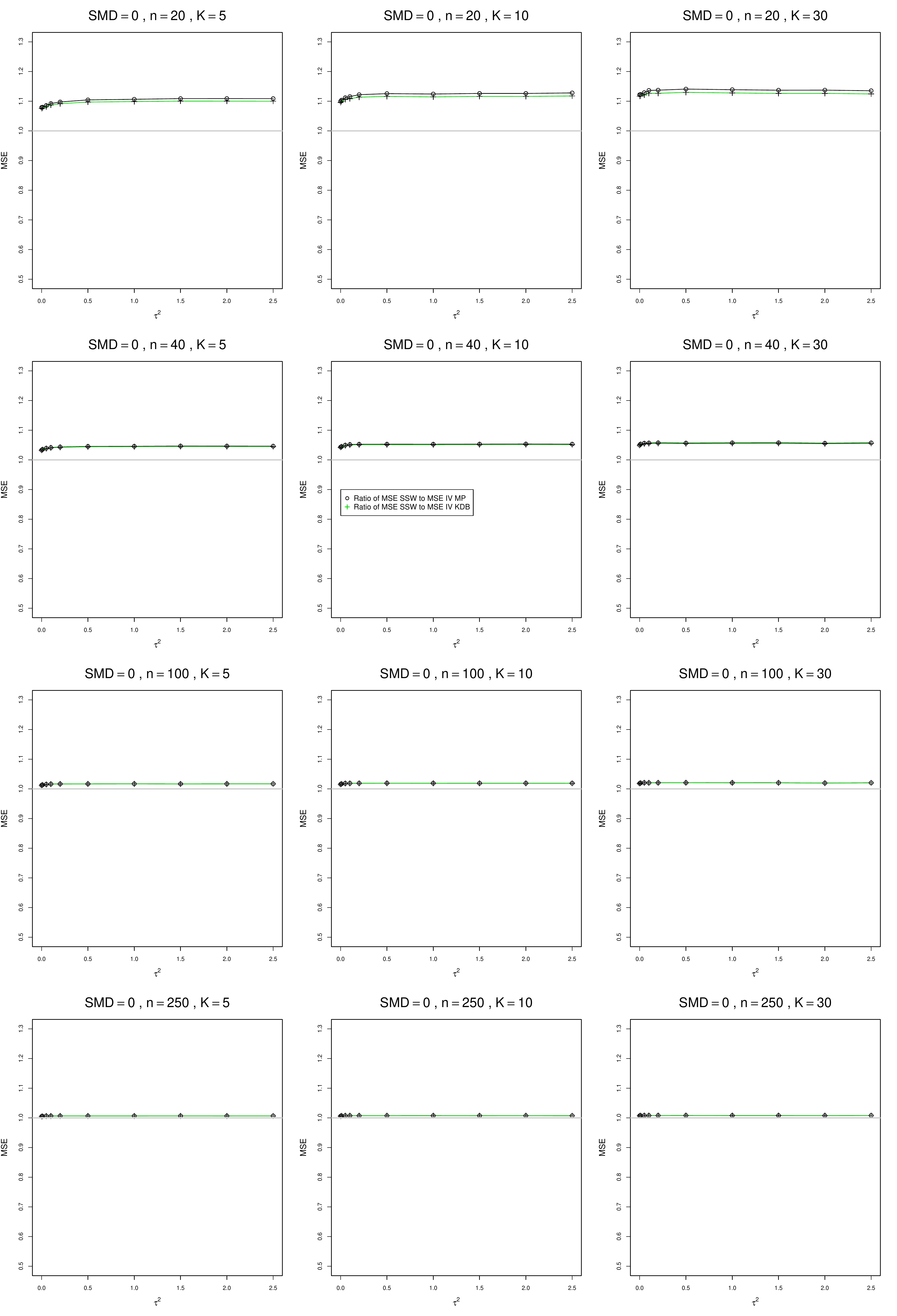}
	\caption{Ratio of mean squared errors of the fixed-weights to mean squared errors of inverse-variance estimator for $\delta=0$, $q=0.5$, $n=20,\;40,\;100,\;250$.
		\label{RatioOfMSEwithSMD0fromMPandCMP}}
\end{figure}

\clearpage

\begin{figure}[t]\centering
	\includegraphics[scale=0.35]{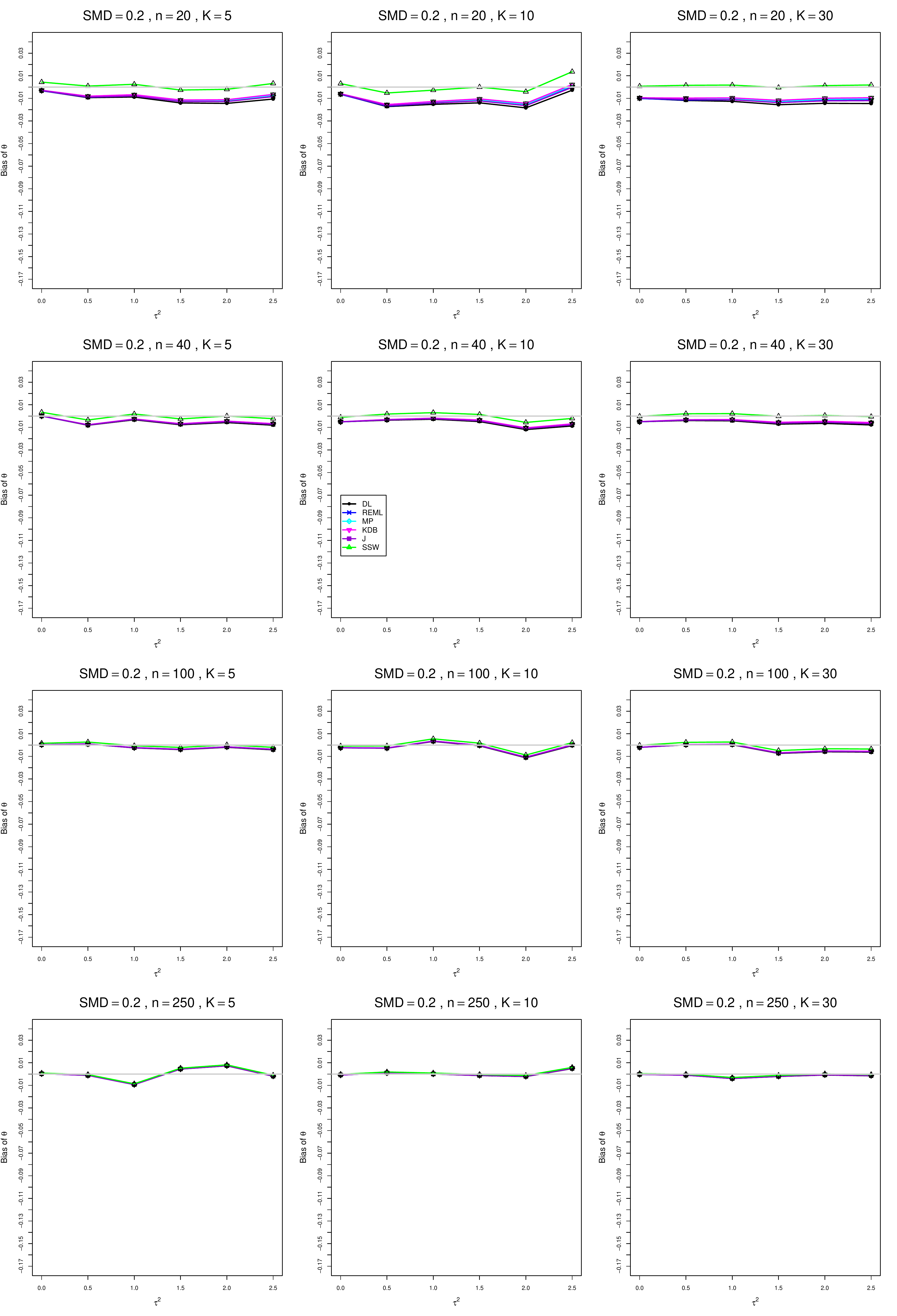}
	\caption{Bias of inverse-variance estimator of $\delta=0.2$, for $q=0.5$, $n=20,\;40,\;100,\;250$.
		\label{BiasThetaSMD02}}
\end{figure}

\begin{figure}[t]\centering
	\includegraphics[scale=0.35]{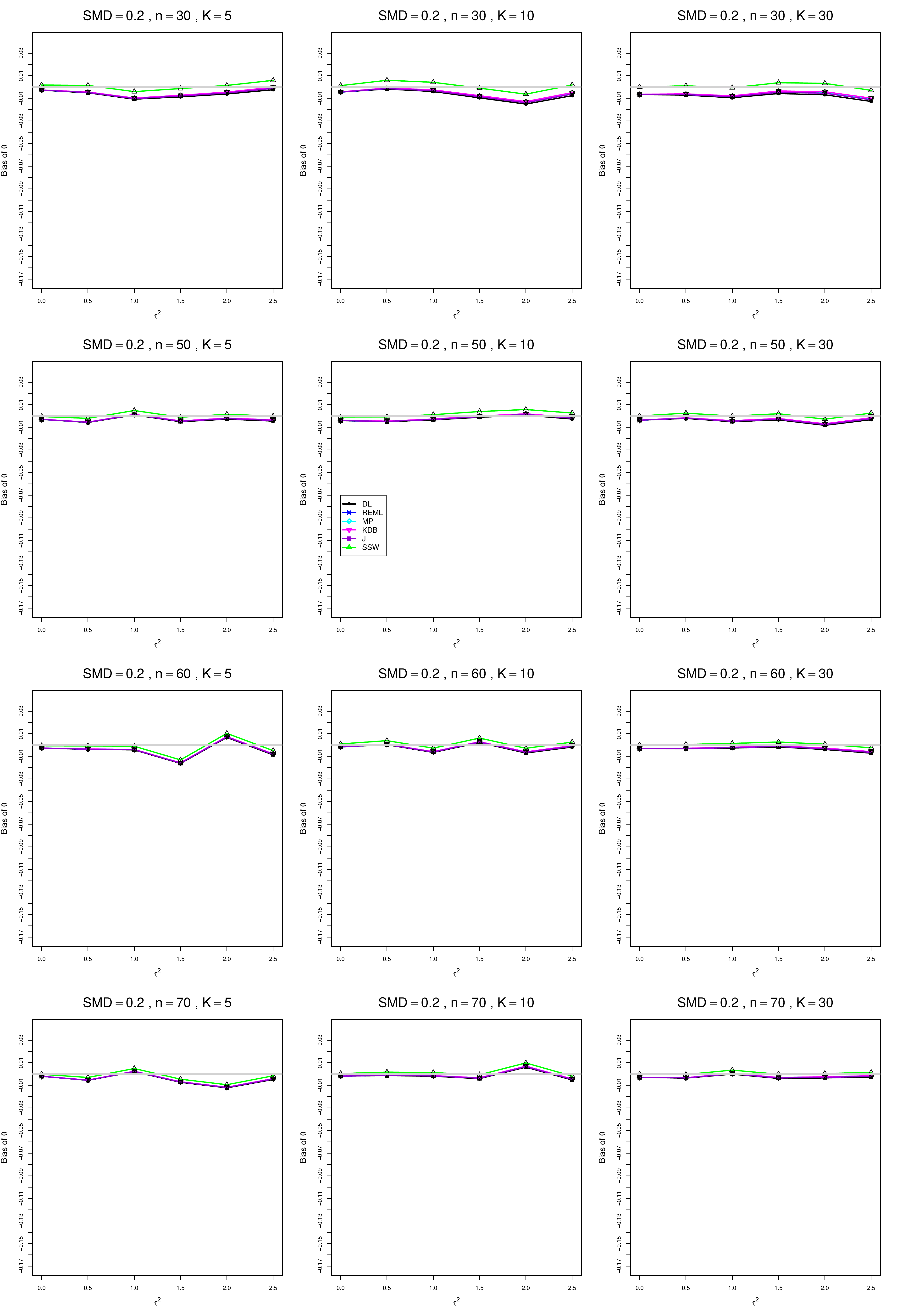}
	\caption{Bias of inverse-variance estimator of $\delta=0.2$, for $q=0.5$, $n=30,\;50,\;60,\;70$.
		\label{BiasThetaSMD02small}}
\end{figure}

\begin{figure}[t]\centering
	\includegraphics[scale=0.35]{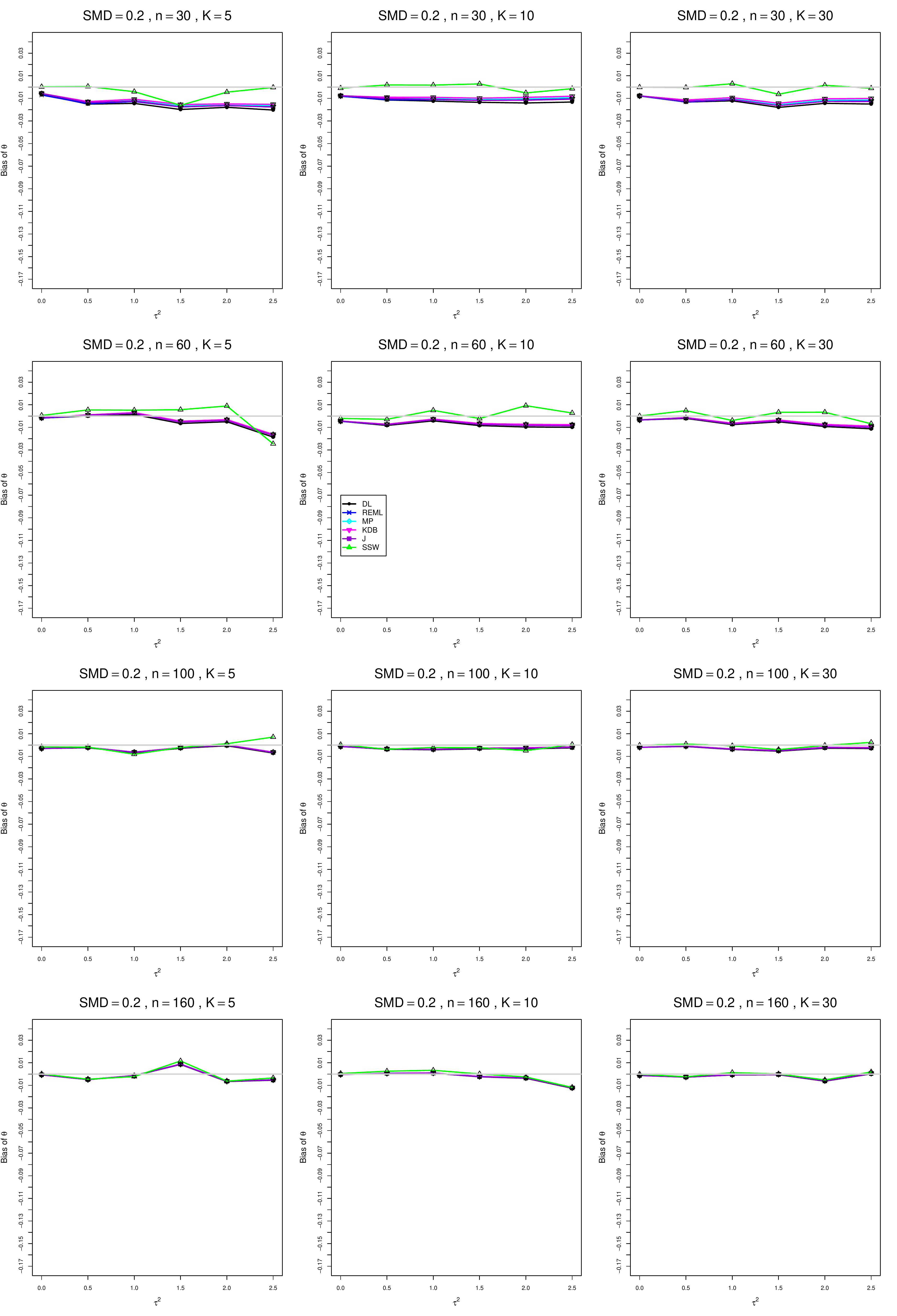}
	\caption{Bias of inverse-variance estimator of $\delta=0.2$, for $q=0.5$, unequal sample sizes with $\bar{n}=30,\; 60,\;100,\;160$.
		\label{BiasThetaSMD02unequal}}
\end{figure}

\clearpage
\begin{figure}[t]\centering
	\includegraphics[scale=0.35]{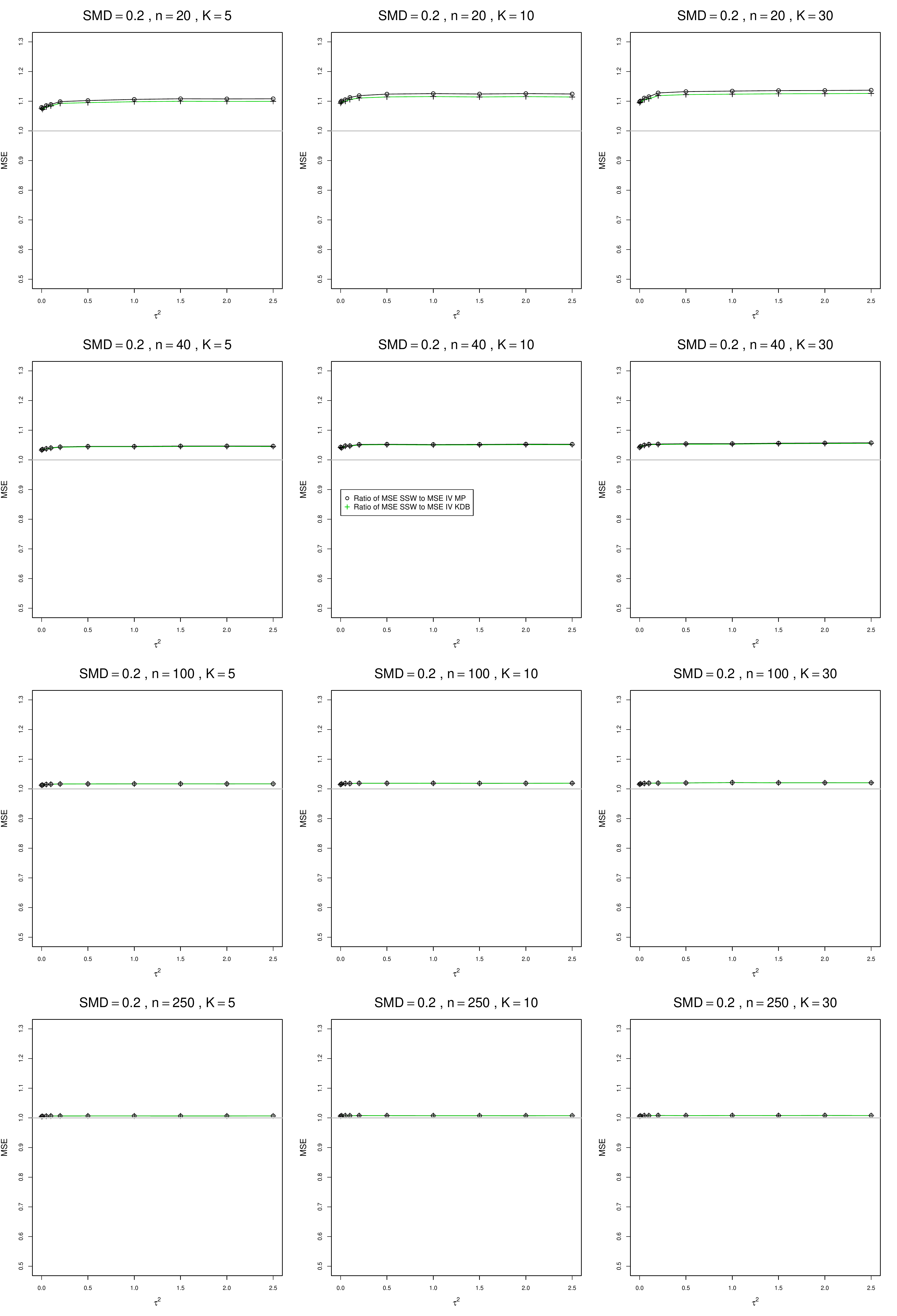}
	\caption{Ratio of mean squared errors of the fixed-weights to mean squared errors of inverse-variance estimator for $\delta=0.2$, $q=0.5$, $n=20,\;40,\;100,\;250$.
		\label{RatioOfMSEwithSMD02fromMPandCMP}}
\end{figure}

\begin{figure}[t]\centering
	\includegraphics[scale=0.35]{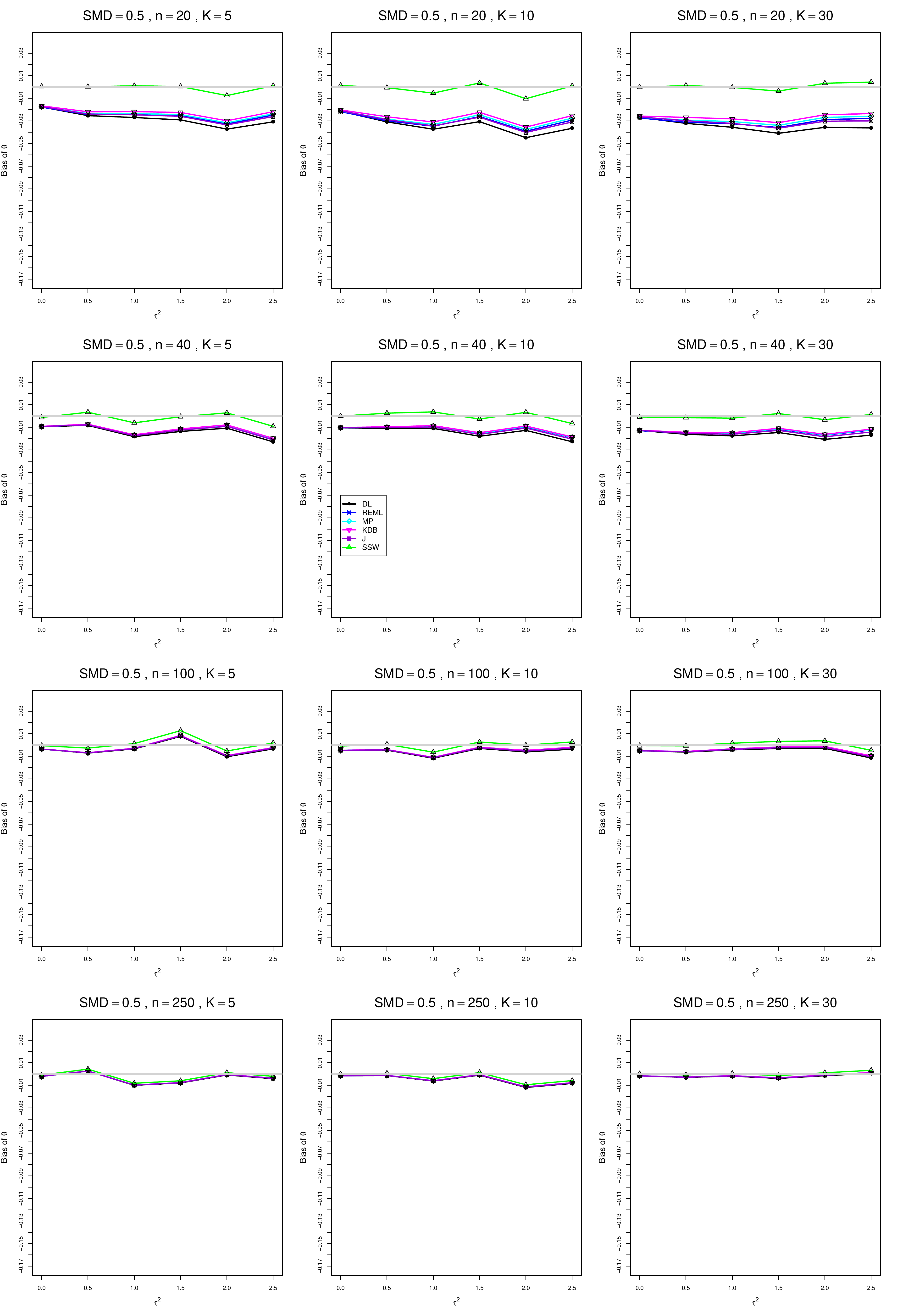}
	\caption{Bias of inverse-variance estimator of $\delta=0.5$, for $q=0.5$, $n=20,\;40,\;100,\;250$.
		\label{BiasThetaSMD05}}
\end{figure}

\begin{figure}[t]\centering
	\includegraphics[scale=0.35]{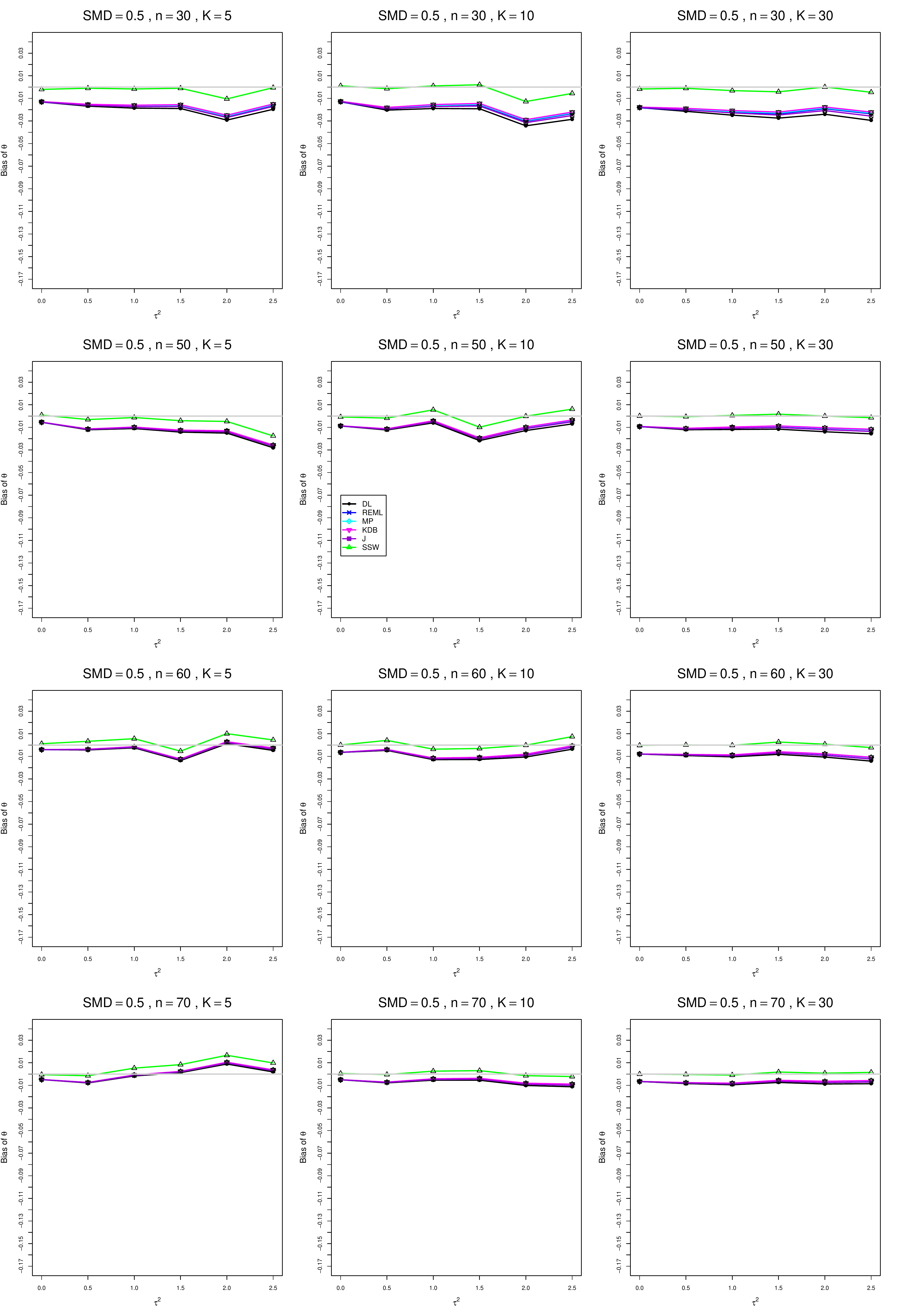}
	\caption{Bias of inverse-variance estimator of $\delta=0.5$, for $q=0.5$, $n=30,\;50,\;60,\;70$.
		\label{BiasThetaSMD05small}}
\end{figure}

\begin{figure}[t]\centering
	\includegraphics[scale=0.35]{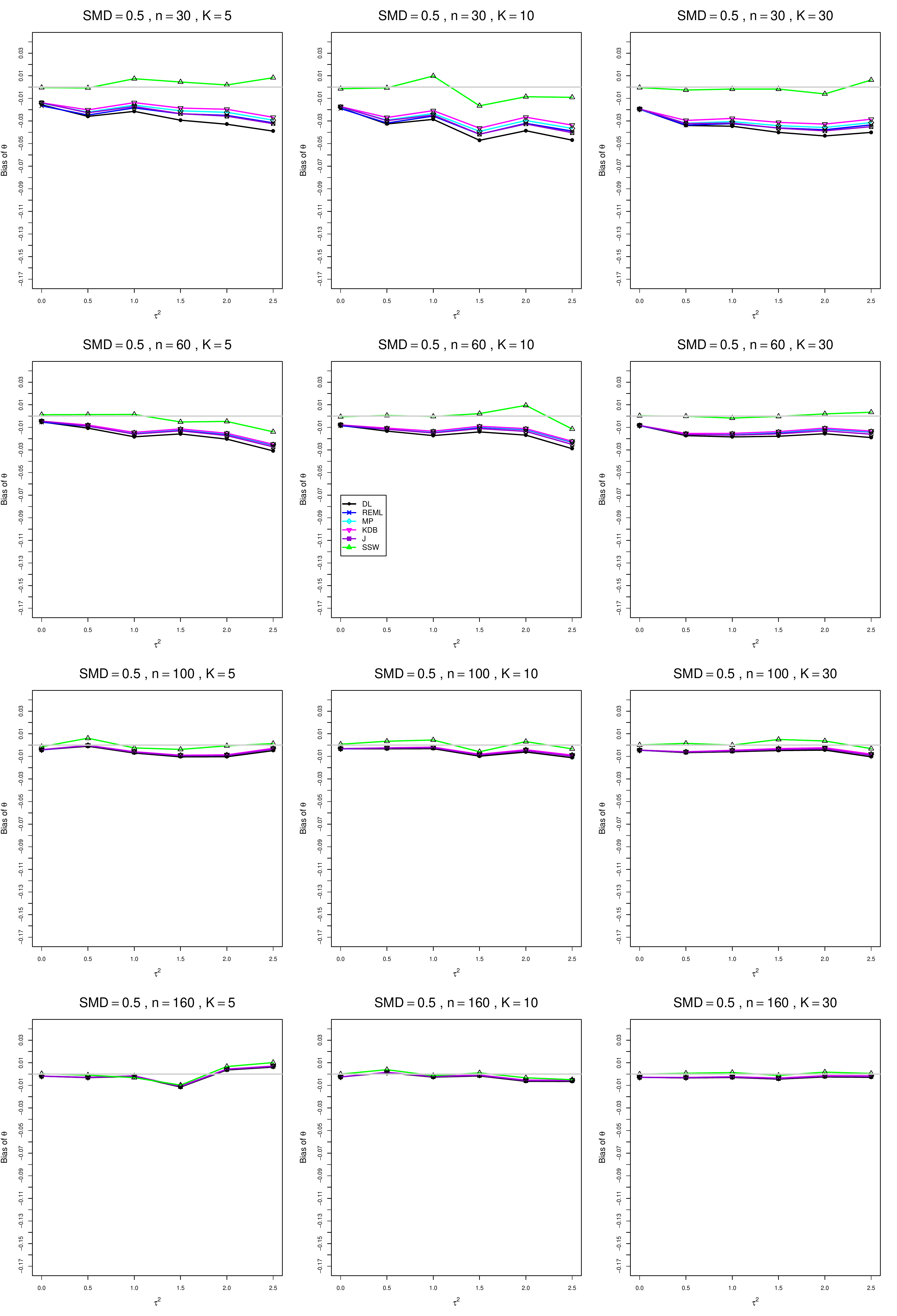}
	\caption{Bias of inverse-variance estimator of $\delta=0.5$, for $q=0.5$, unequal sample sizes with $\bar{n}=30,\; 60,\;100,\;160$.
		\label{BiasThetaSMD05unequal}}
\end{figure}

\begin{figure}[t]\centering
	\includegraphics[scale=0.35]{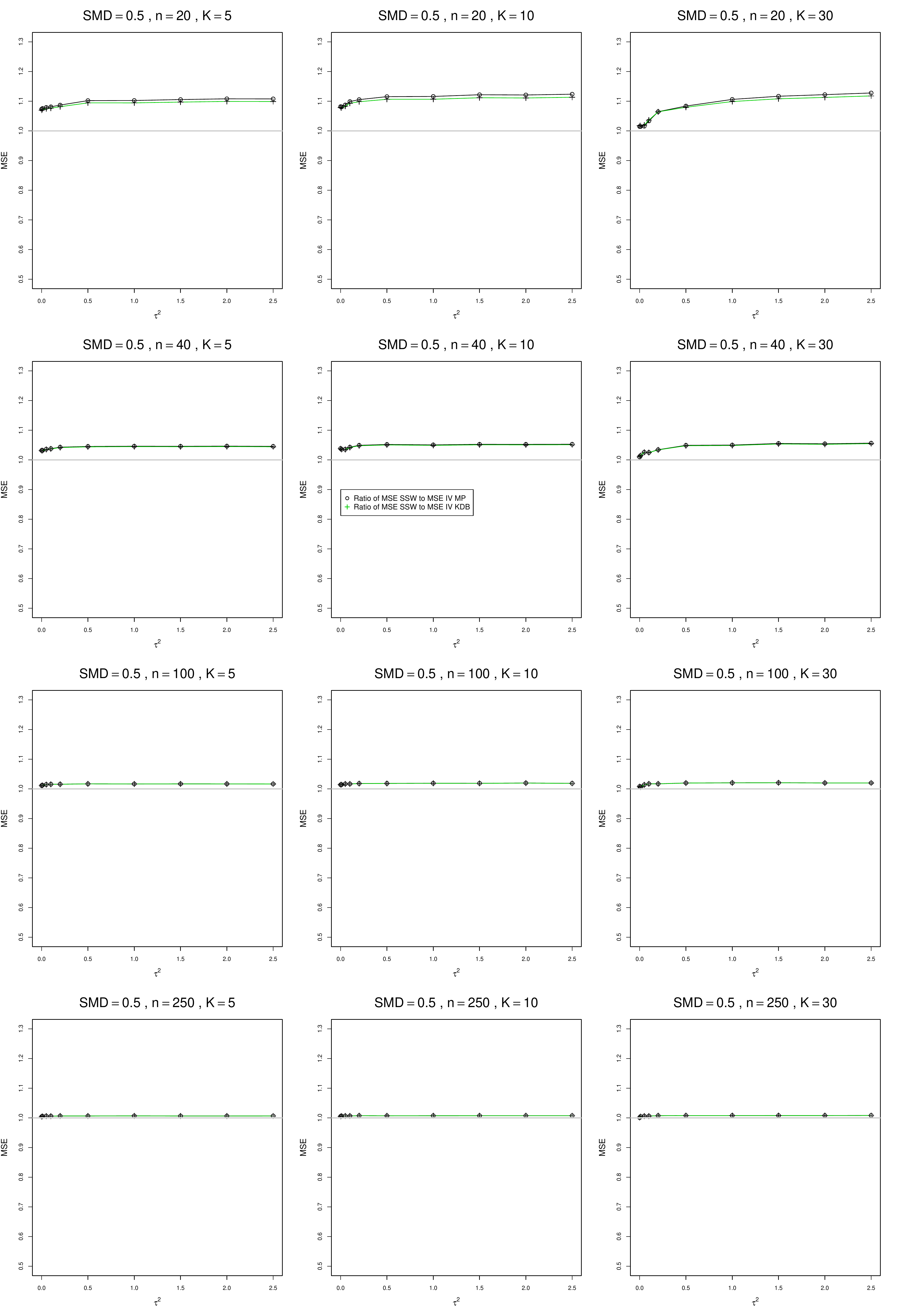}
	\caption{Ratio of mean squared errors of the fixed-weights to mean squared errors of inverse-variance estimator for $\delta=0.5$, $q=0.5$, $n=20,\;40,\;100,\;250$.
		\label{RatioOfMSEwithSMD05fromMPandCMP}}
\end{figure}

\begin{figure}[t]\centering
	\includegraphics[scale=0.35]{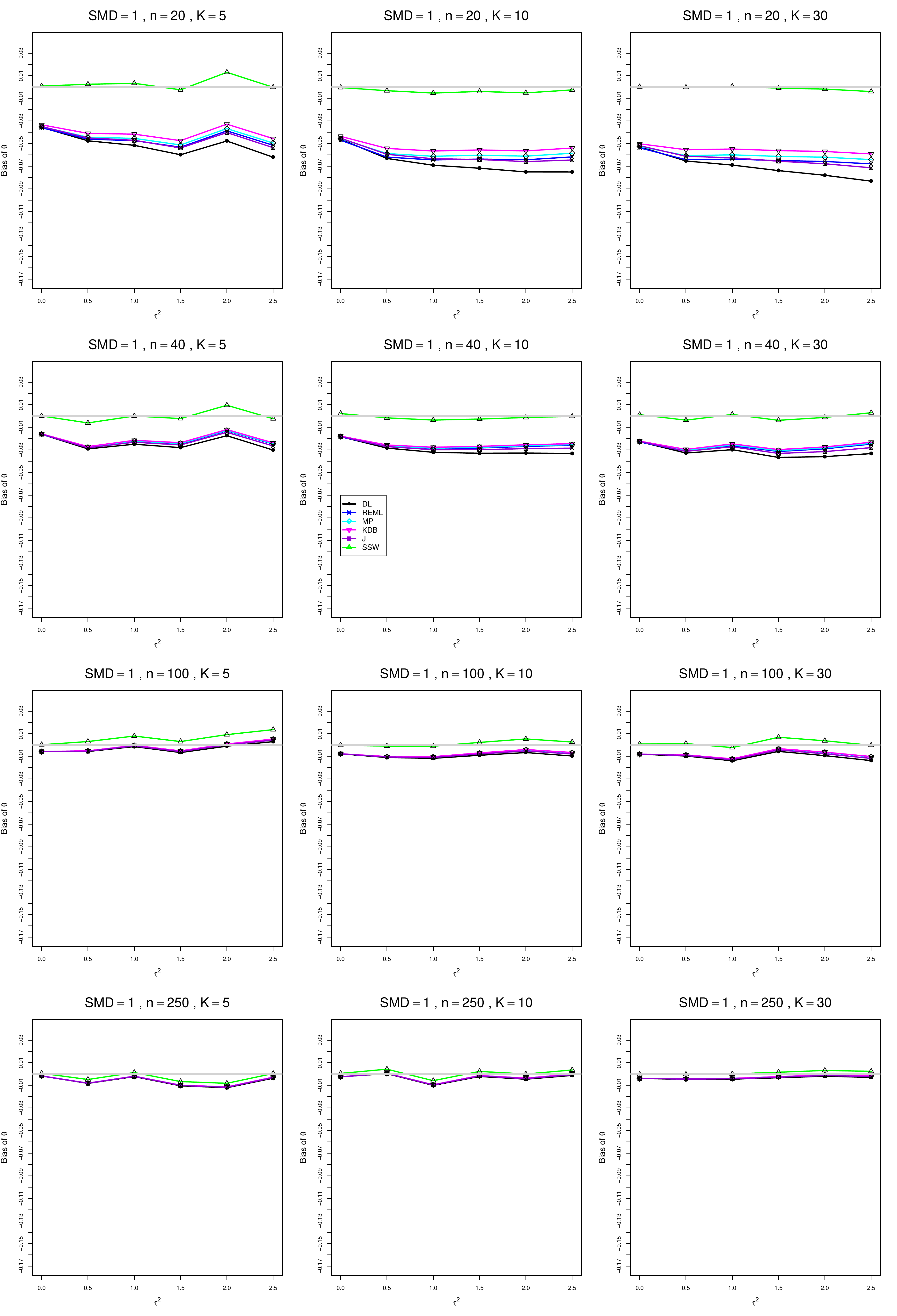}
	\caption{Bias of inverse-variance estimator of $\delta=1$, for $q=0.5$, $n=20,\;40,\;100,\;250$.
		\label{BiasThetaSMD1}}
\end{figure}

\begin{figure}[t]\centering
	\includegraphics[scale=0.35]{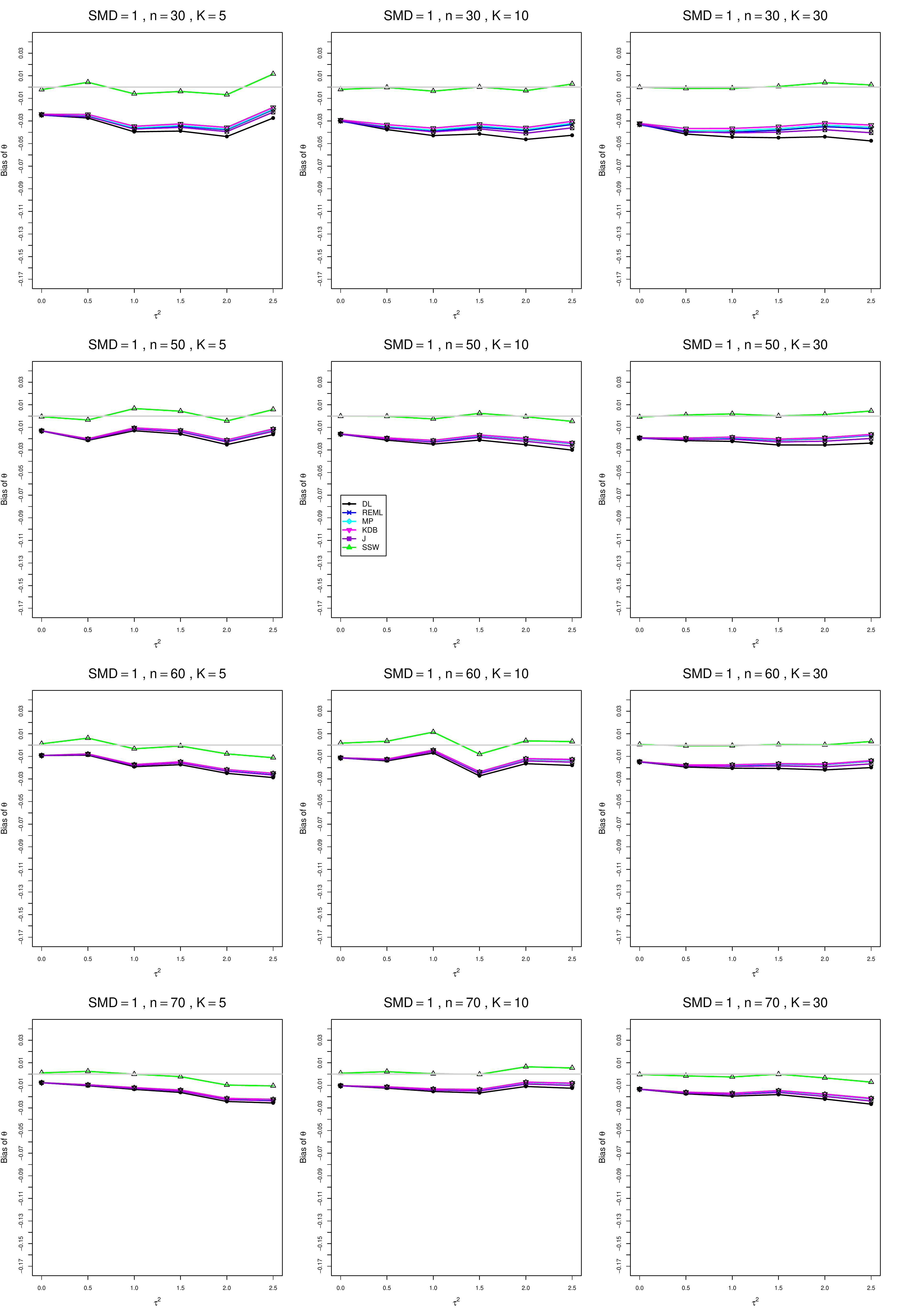}
	\caption{Bias of inverse-variance estimator of $\delta=1$, for $q=0.5$, $n=30,\;50,\;60,\;70$.
		\label{BiasThetaSMD1small}}
\end{figure}

\begin{figure}[t]\centering
	\includegraphics[scale=0.35]{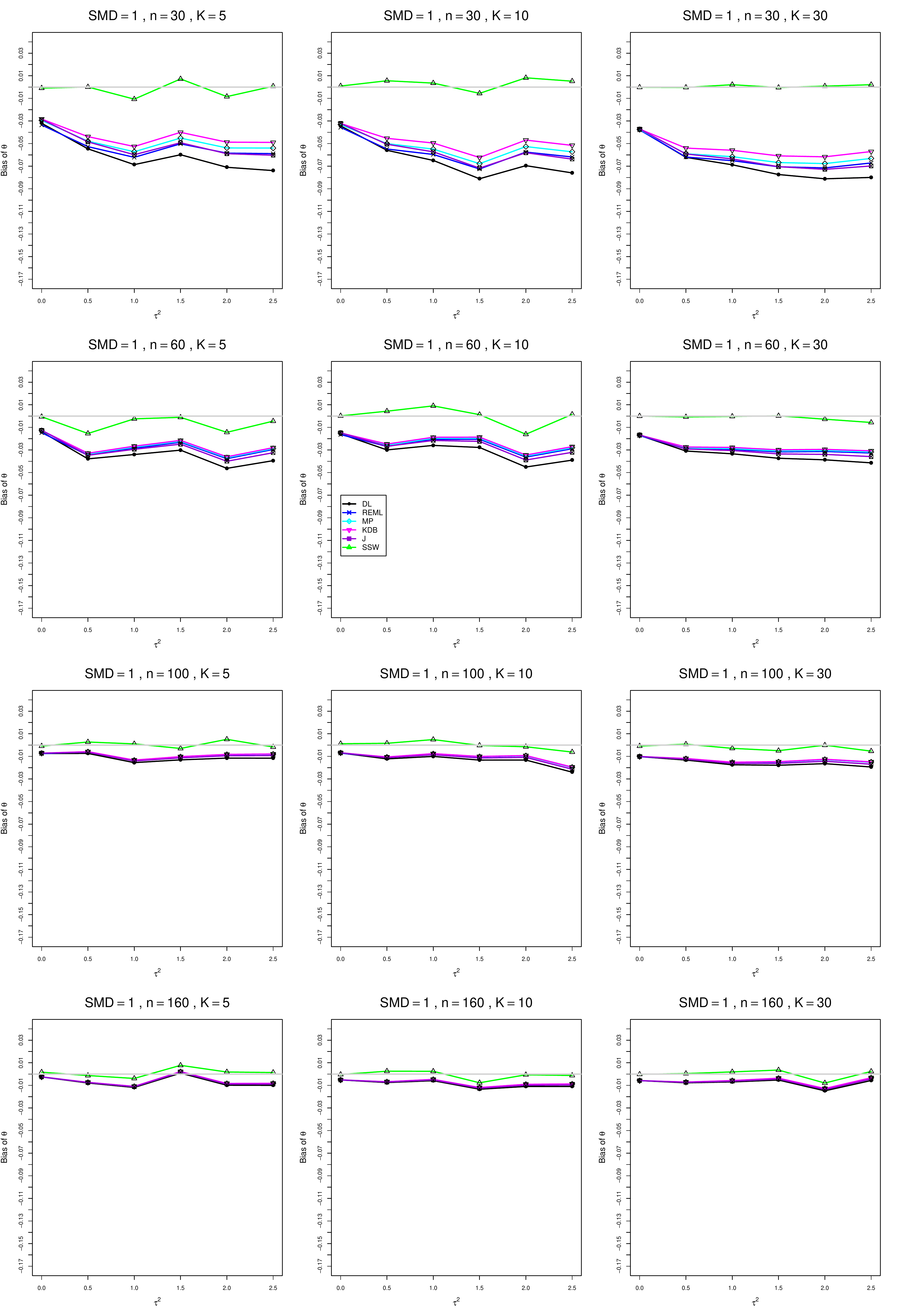}
	\caption{Bias of inverse-variance estimator of $\delta=1$, for $q=0.5$, unequal sample sizes with $\bar{n}=30,\; 60,\;100,\;160$.
		\label{BiasThetaSMD1unequal}}
\end{figure}

\begin{figure}[t]\centering
	\includegraphics[scale=0.35]{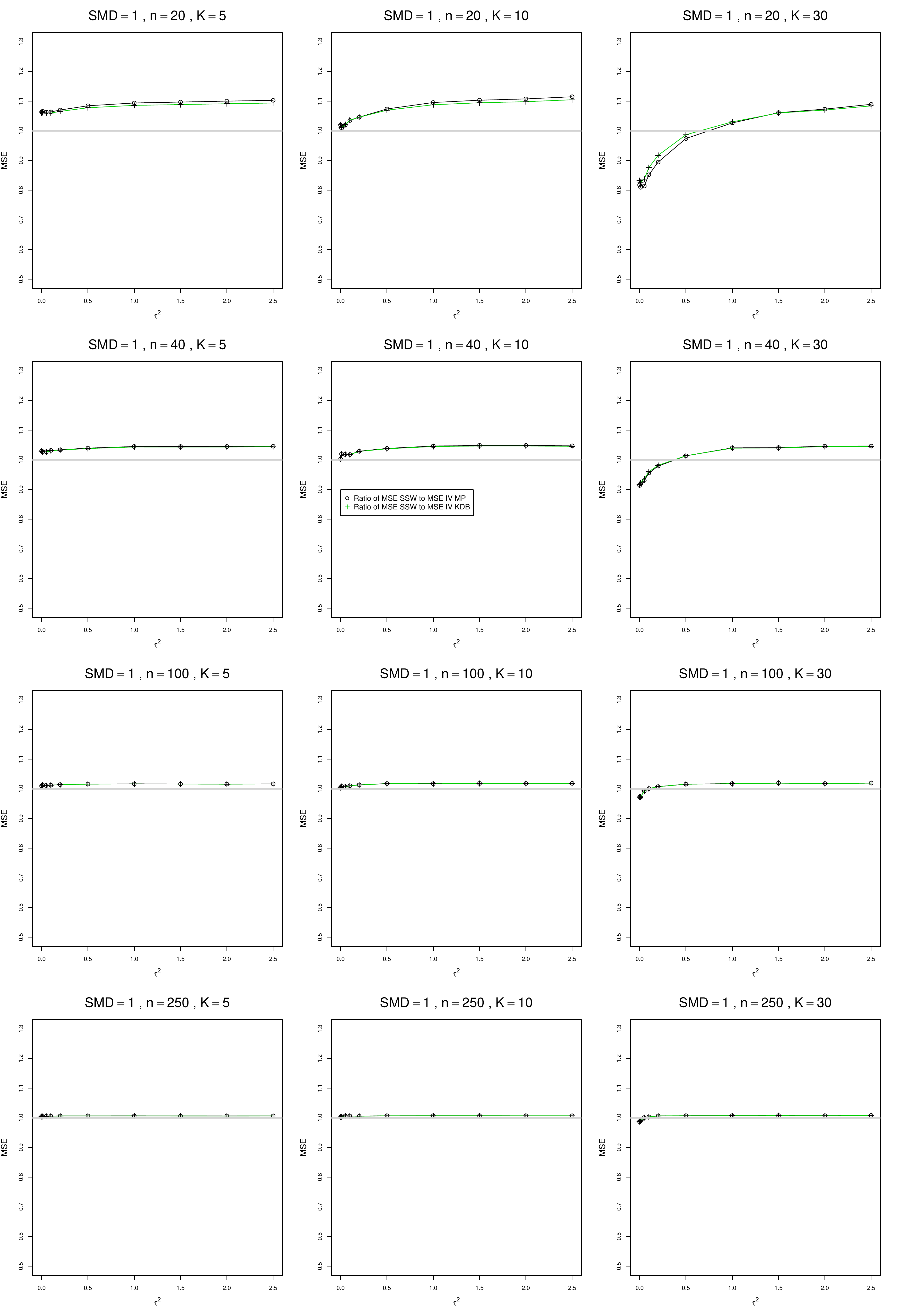}
	\caption{Ratio of mean squared errors of the fixed-weights to mean squared errors of inverse-variance estimator for $\delta=1$, for $q=0.5$, $n=20,\;40,\;100,\;250$. 
		\label{RatioOfMSEwithSMD1fromMPandCMP}}
\end{figure}

\begin{figure}[t]\centering
	\includegraphics[scale=0.35]{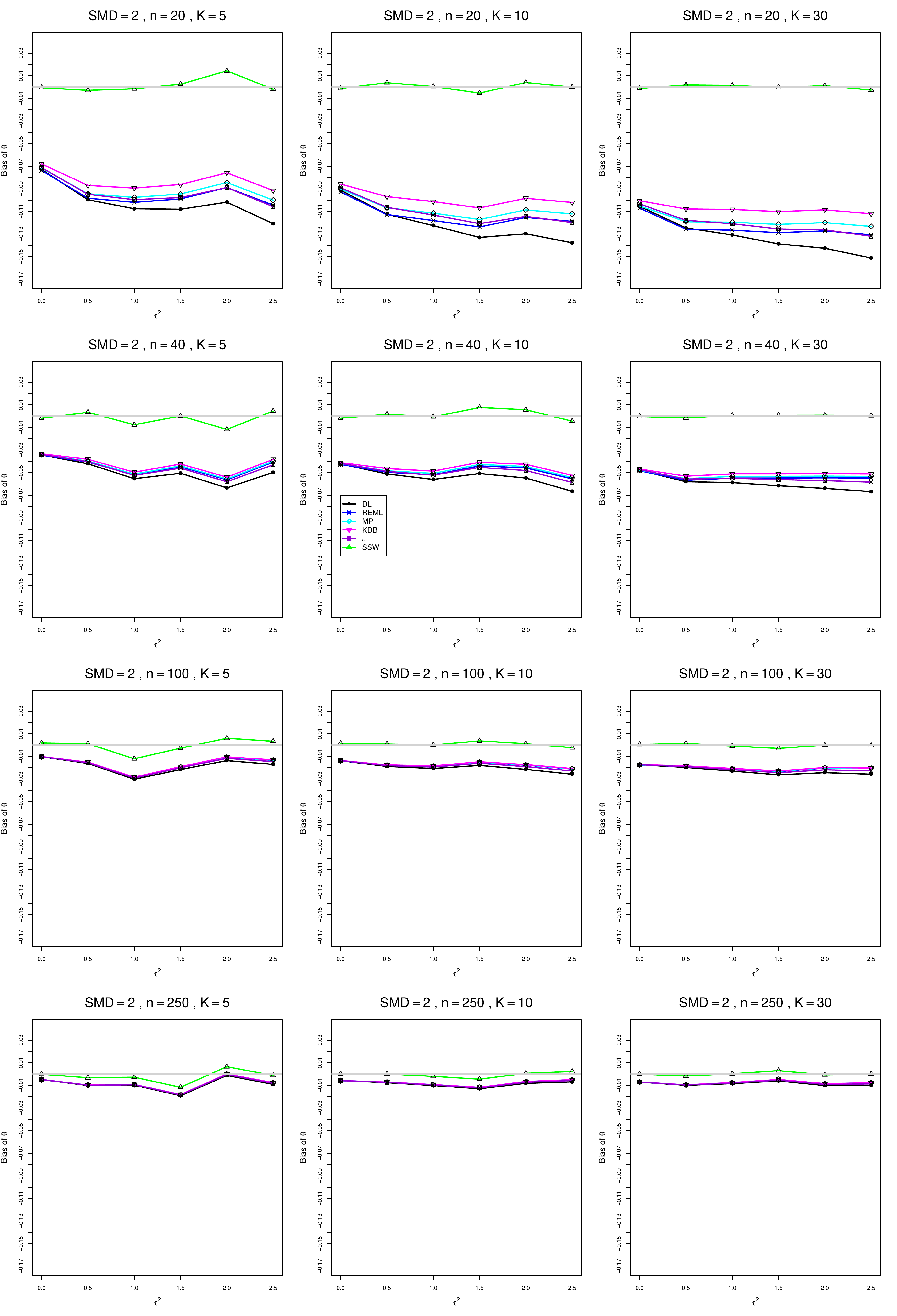}
	\caption{Bias of inverse-variance estimator of $\delta=2$, for $q=0.5$, $n=20,\;40,\;100,\;250$.
		\label{BiasThetaSMD2}}
\end{figure}

\begin{figure}[t]\centering
	\includegraphics[scale=0.35]{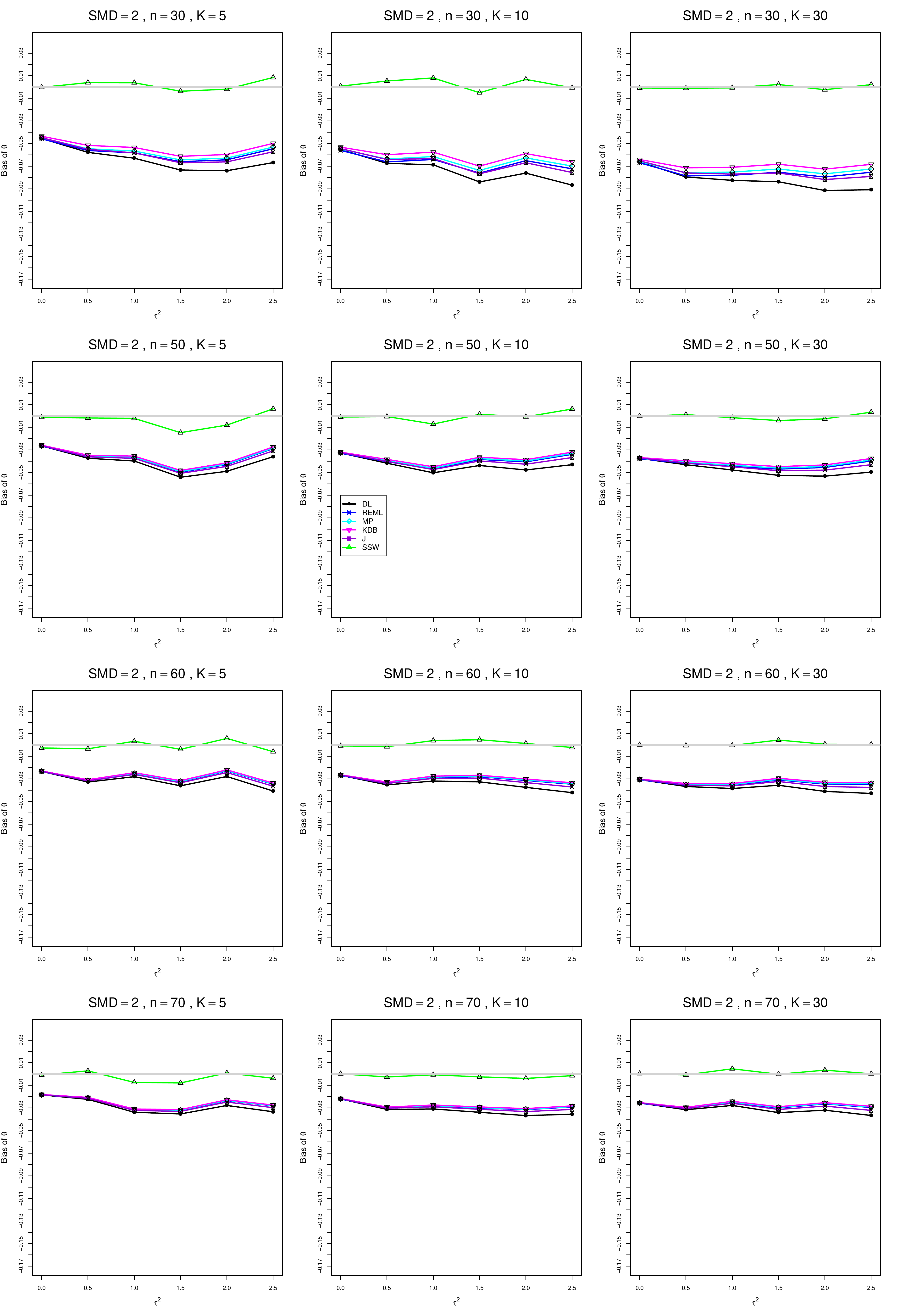}
	\caption{Bias of inverse-variance estimator of $\delta=2$, for $q=0.5$, $n=30,\;50,\;60,\;70$.
		\label{BiasThetaSMD2small}}
\end{figure}

\begin{figure}[t]\centering
	\includegraphics[scale=0.35]{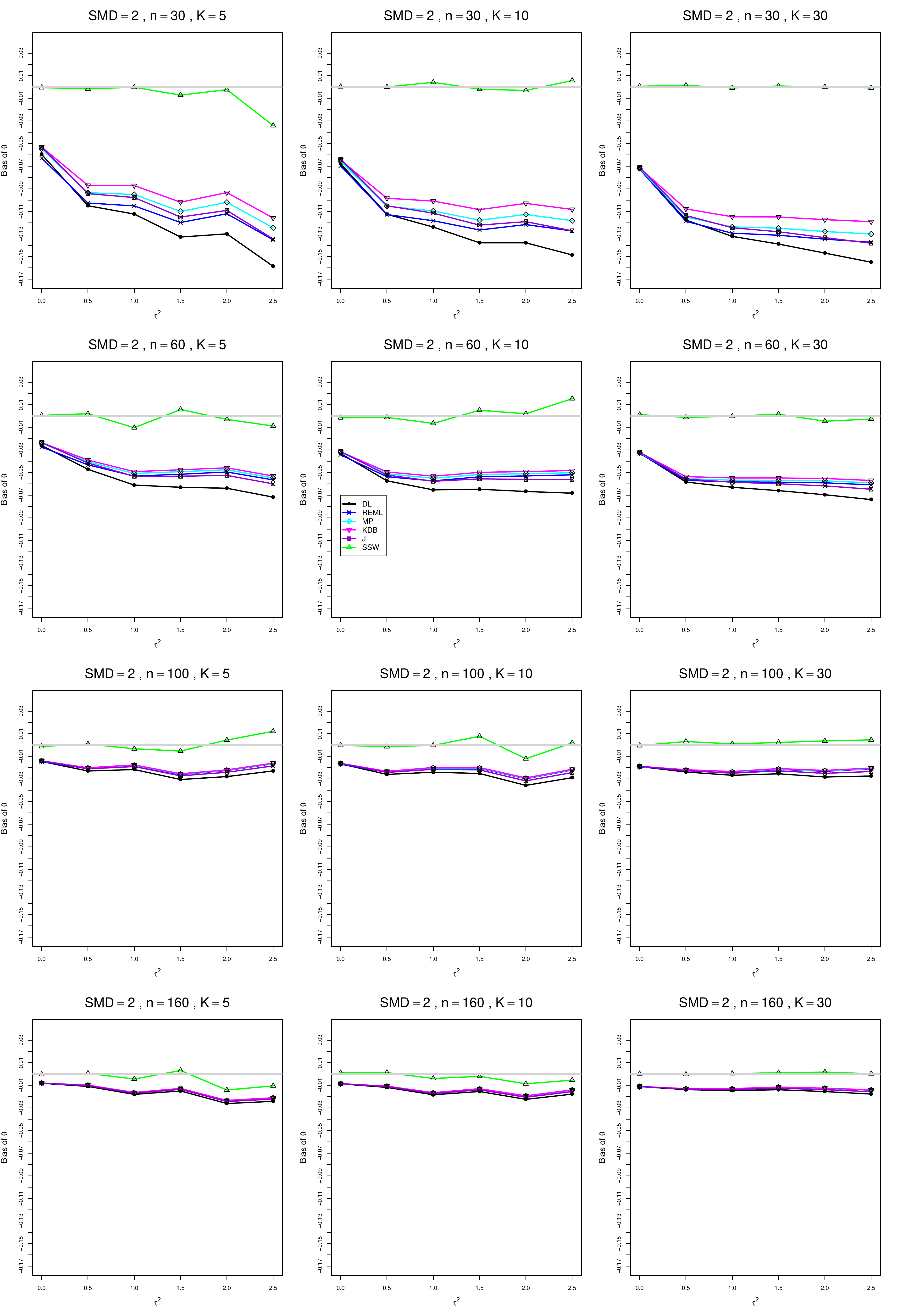}
	\caption{Bias of inverse-variance estimator of $\delta=2$, for $q=0.5$, unequal sample sizes with $\bar{n}=30,\; 60,\;100,\;160$.
		\label{BiasThetaSMD2unequal}}
\end{figure}

\begin{figure}[t]\centering
	\includegraphics[scale=0.35]{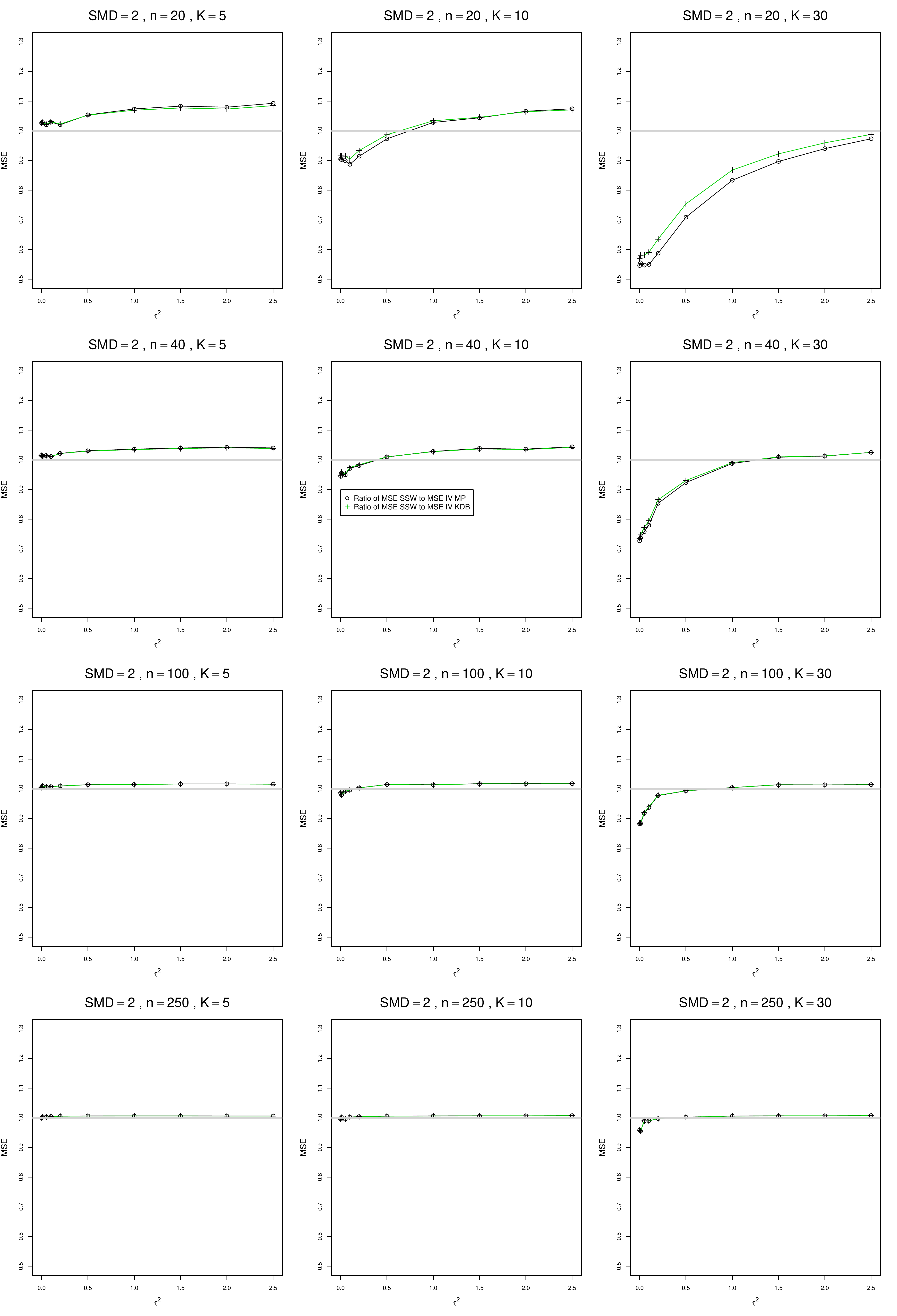}
	\caption{Ratio of mean squared errors of the fixed-weights to mean squared errors of inverse-variance estimator for $\delta=2$, for $q=0.5$, $n=20,\;40,\;100,\;250$. 
		\label{RatioOfMSEwithSMD2fromMPandCMP}}
\end{figure}

\begin{figure}[t]\centering
	\includegraphics[scale=0.35]{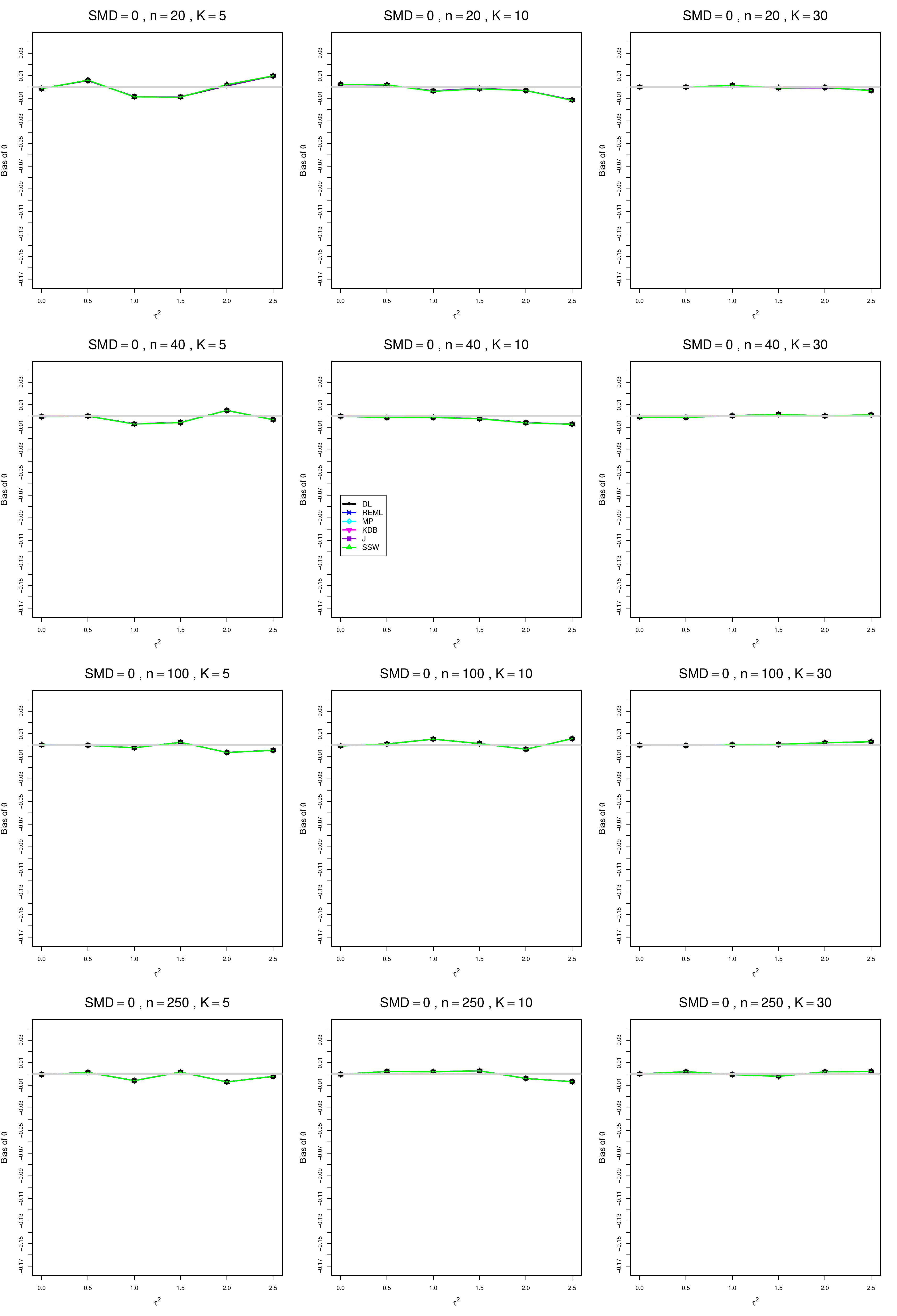}
	\caption{Bias of inverse-variance estimator of $\delta=0$, for $q=0.75$, $n=20,\;40,\;100,\;250$.
		\label{BiasThetaSMD0q75}}
\end{figure}

\begin{figure}[t]\centering
	\includegraphics[scale=0.35]{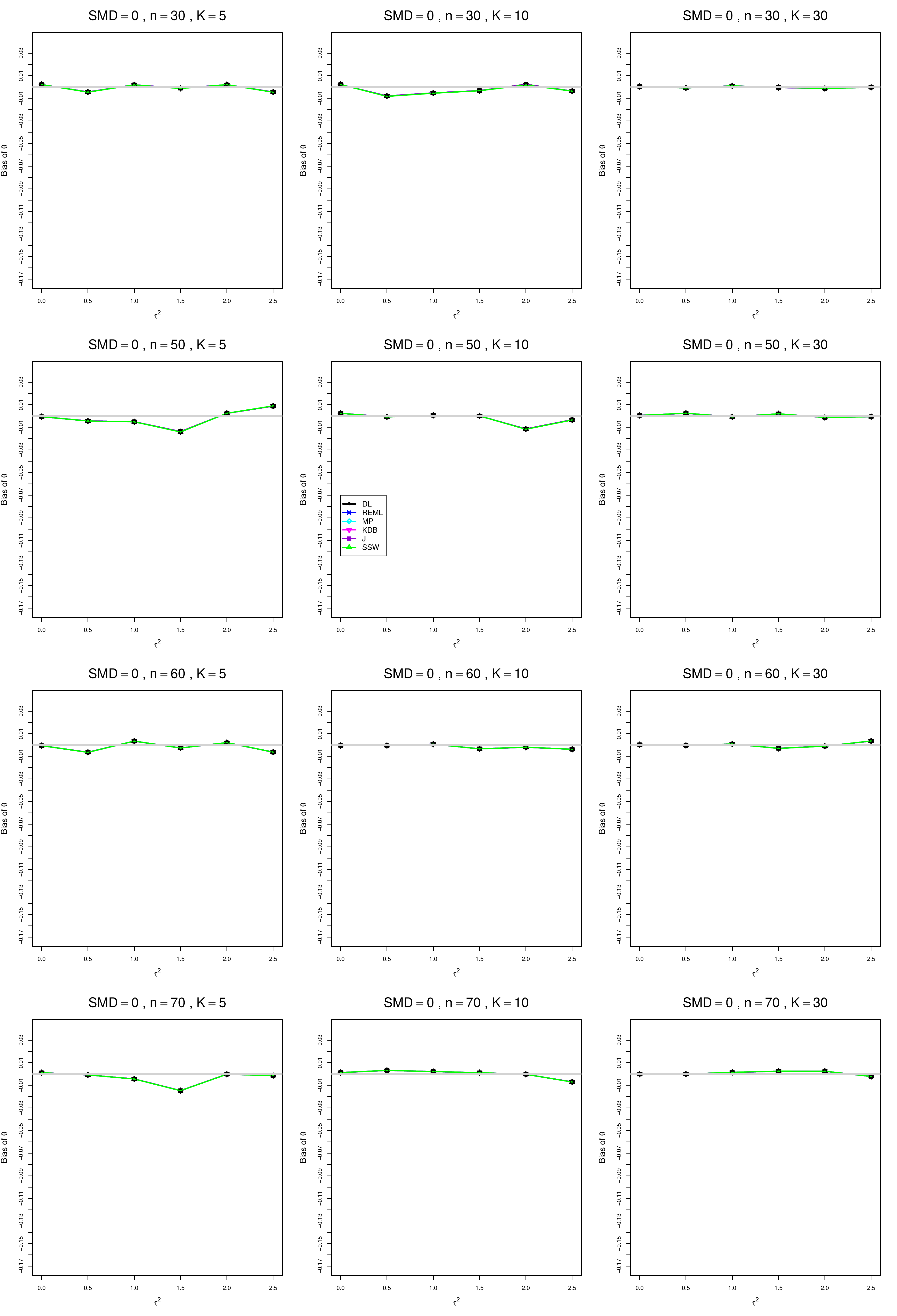}
	\caption{Bias of inverse-variance estimator of $\delta=0$, for $q=0.75$, $n=30,\;50,\;60,\;70$.
		\label{BiasThetaSMD0q75small}}
\end{figure}

\clearpage
\begin{figure}[t]\centering
	\includegraphics[scale=0.35]{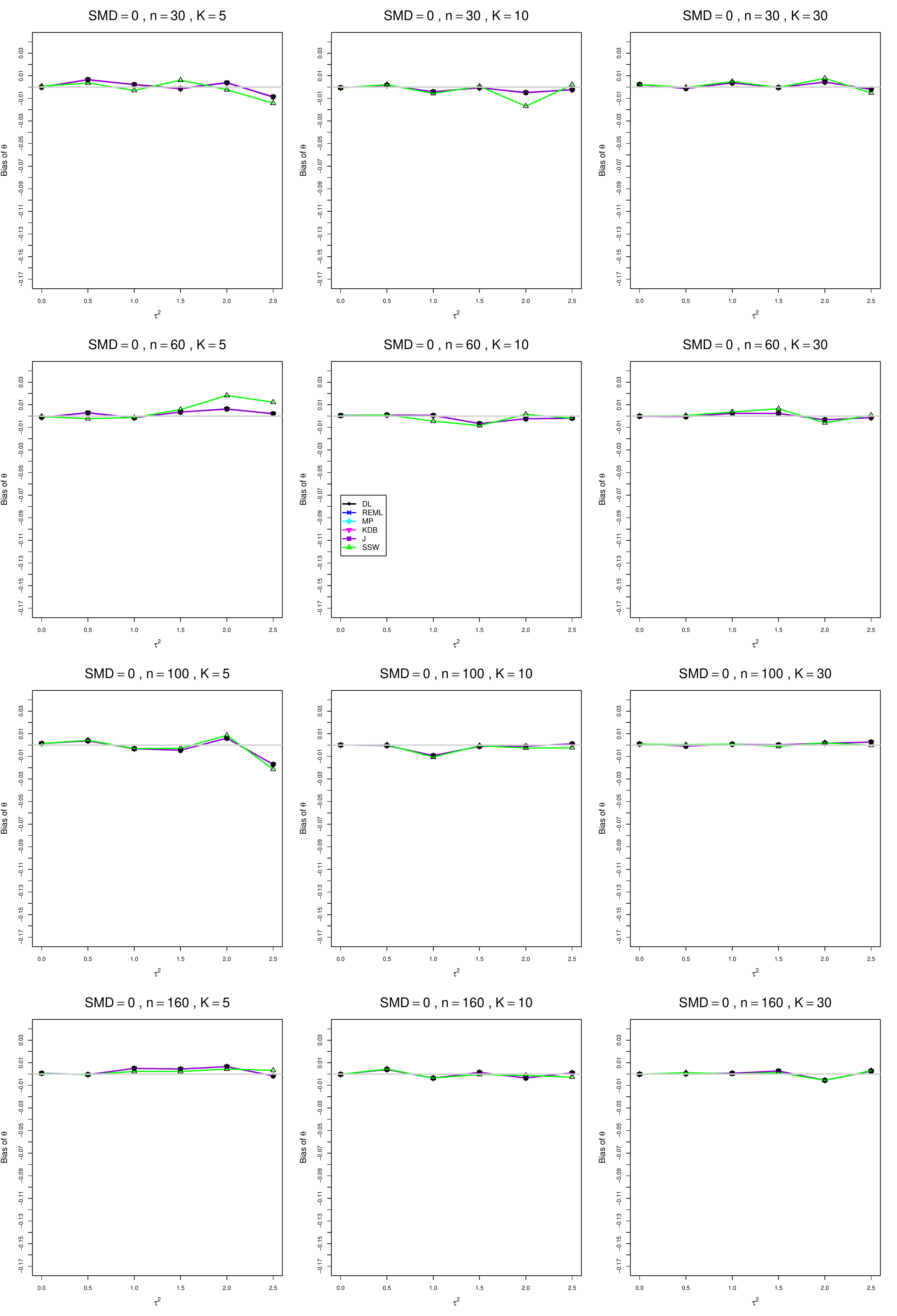}
	\caption{Bias of inverse-variance estimator of $\delta=0$, for  $q=0.75$, unequal sample sizes with $\bar{n}=30,\; 60,\;100,\;160$.
		\label{BiasThetaSMD0q75unequal}}
\end{figure}

\begin{figure}[t]\centering
	\includegraphics[scale=0.35]{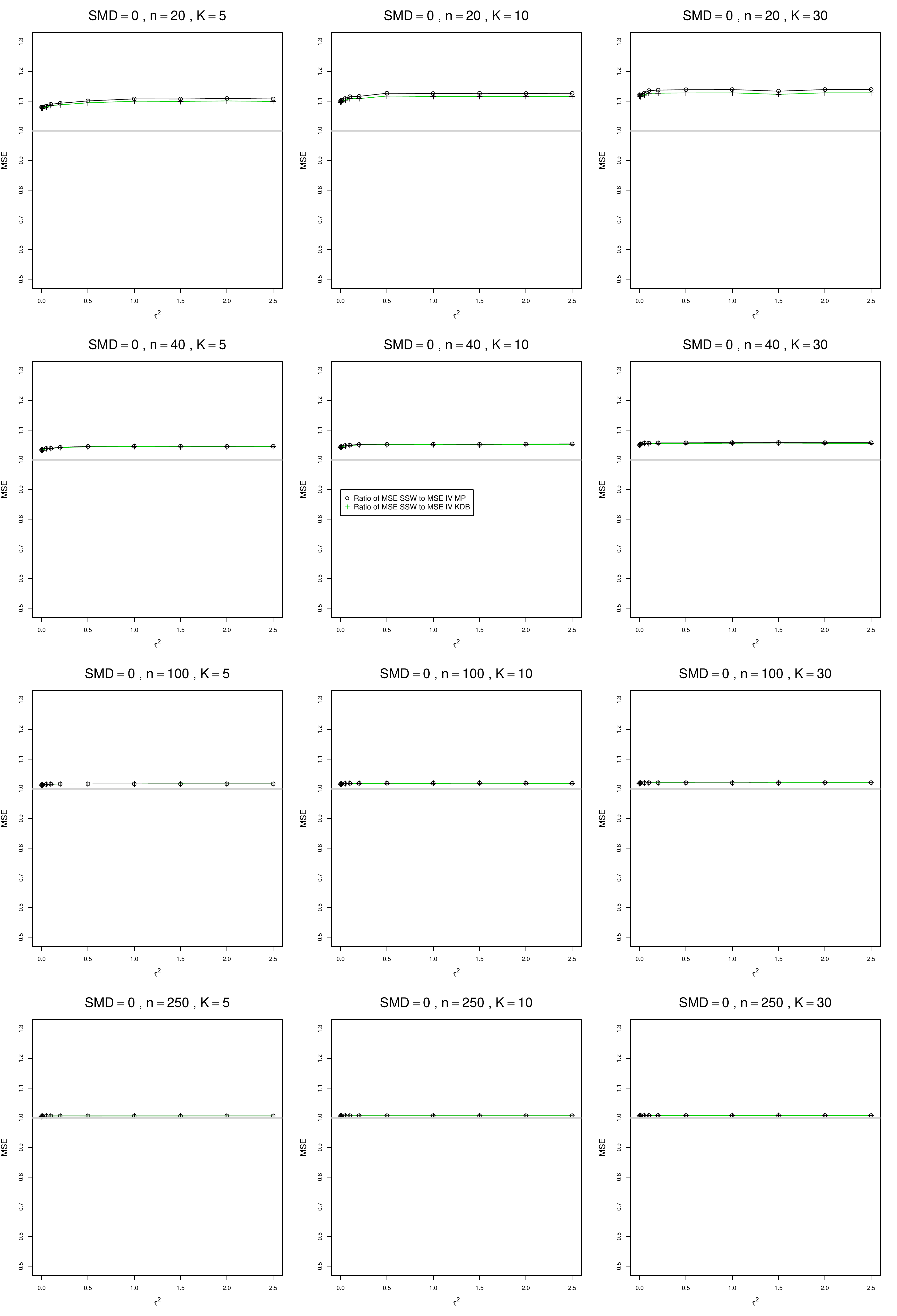}
	\caption{Ratio of mean squared errors of the fixed-weights to mean squared errors of inverse-variance estimator for $\delta=0$, for $q=0.75$, $n=20,\;40,\;100,\;250$.
		\label{RatioOfMSEwithSMD0q075fromMPandCMP}}
\end{figure}

\begin{figure}[t]\centering
	\includegraphics[scale=0.35]{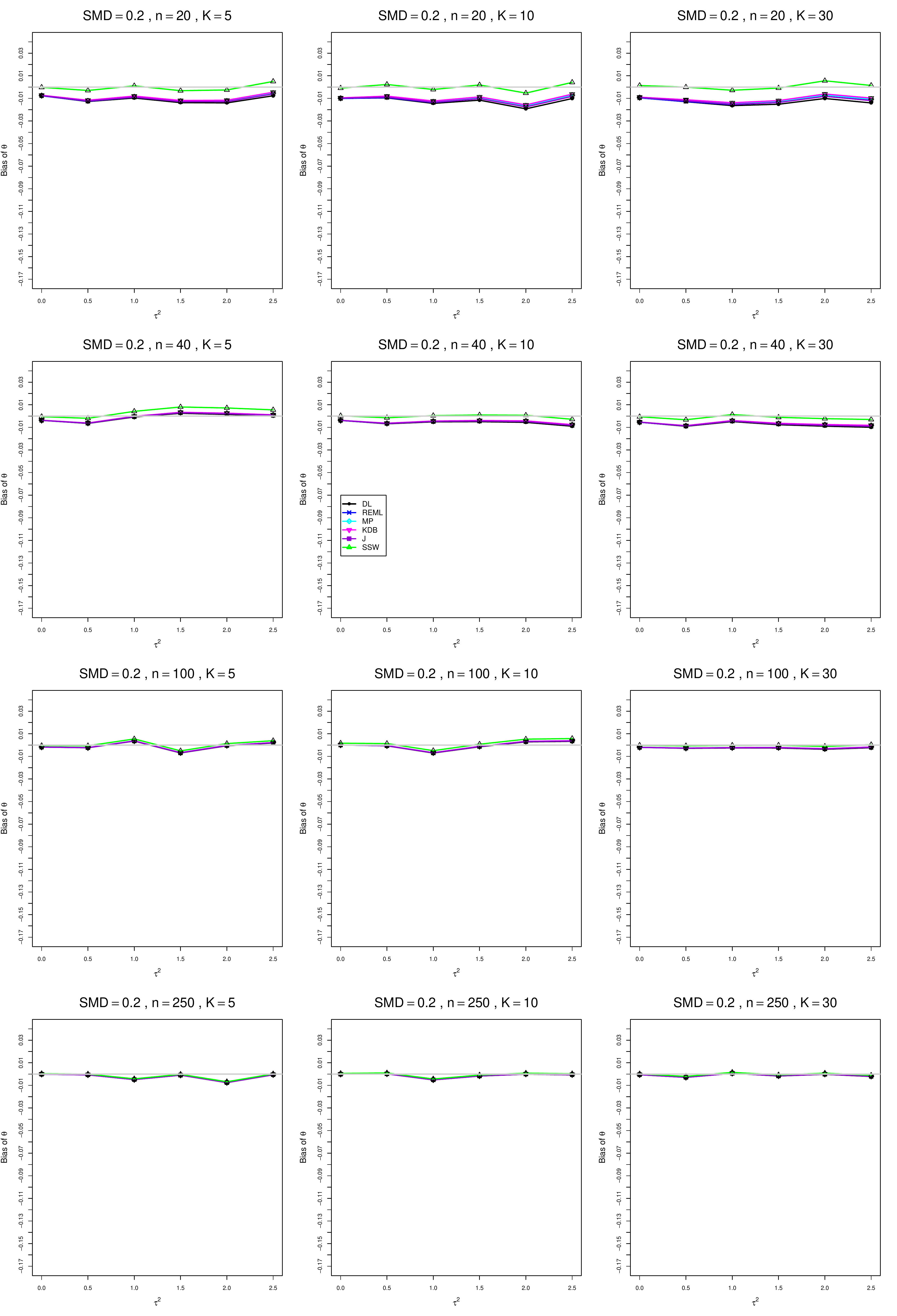}
	\caption{Bias of inverse-variance estimator of $\delta=0.2$, for $q=0.75$, $n=20,\;40,\;100,\;250$.
		\label{BiasThetaSMD02q75}}
\end{figure}

\begin{figure}[t]\centering
	\includegraphics[scale=0.35]{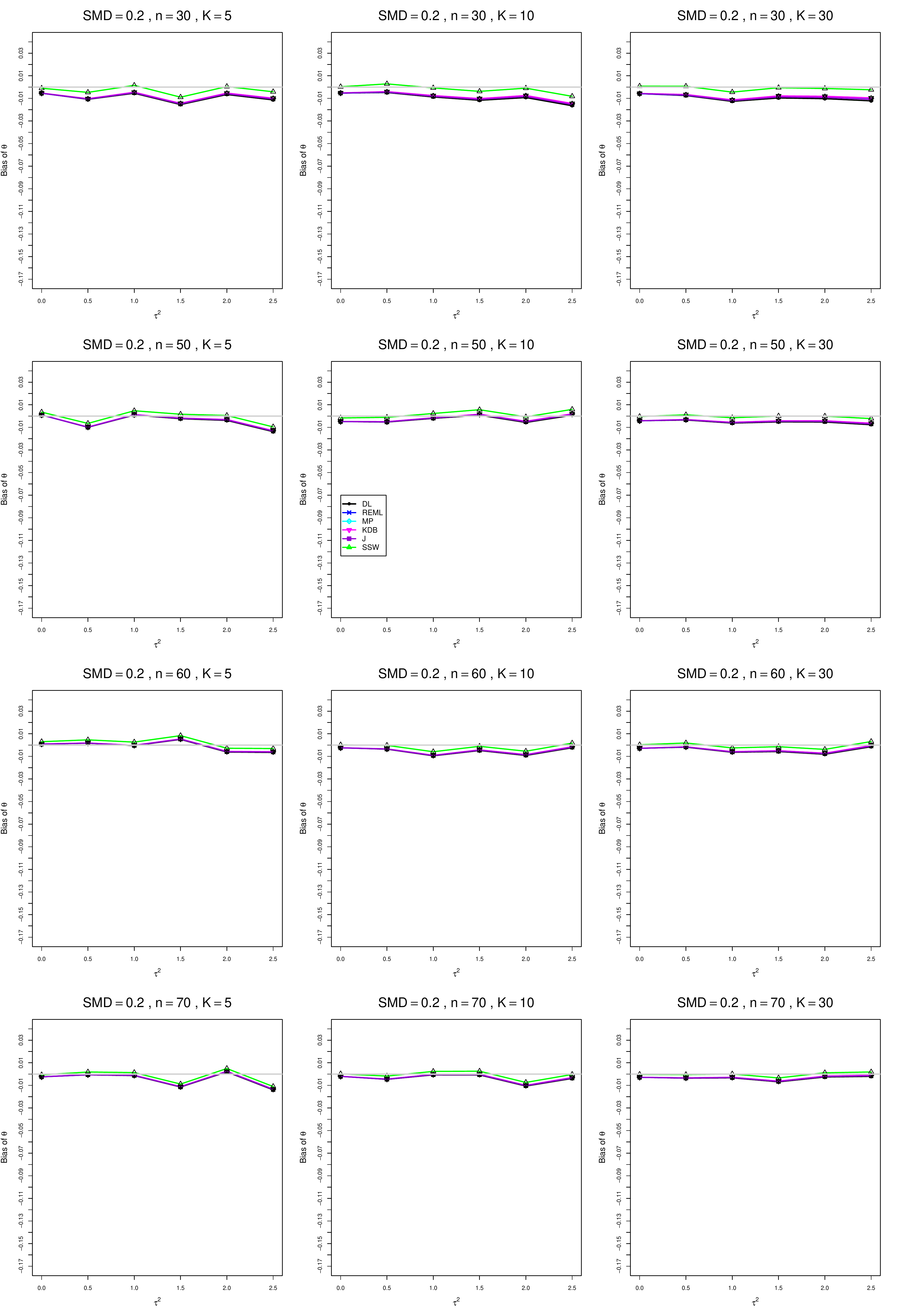}
	\caption{Bias of inverse-variance estimator of $\delta=0.2$, for $q=0.75$, $n=30,\;50,\;60,\;70$.
		\label{BiasThetaSMD02q75small}}
\end{figure}

\begin{figure}[t]\centering
	\includegraphics[scale=0.35]{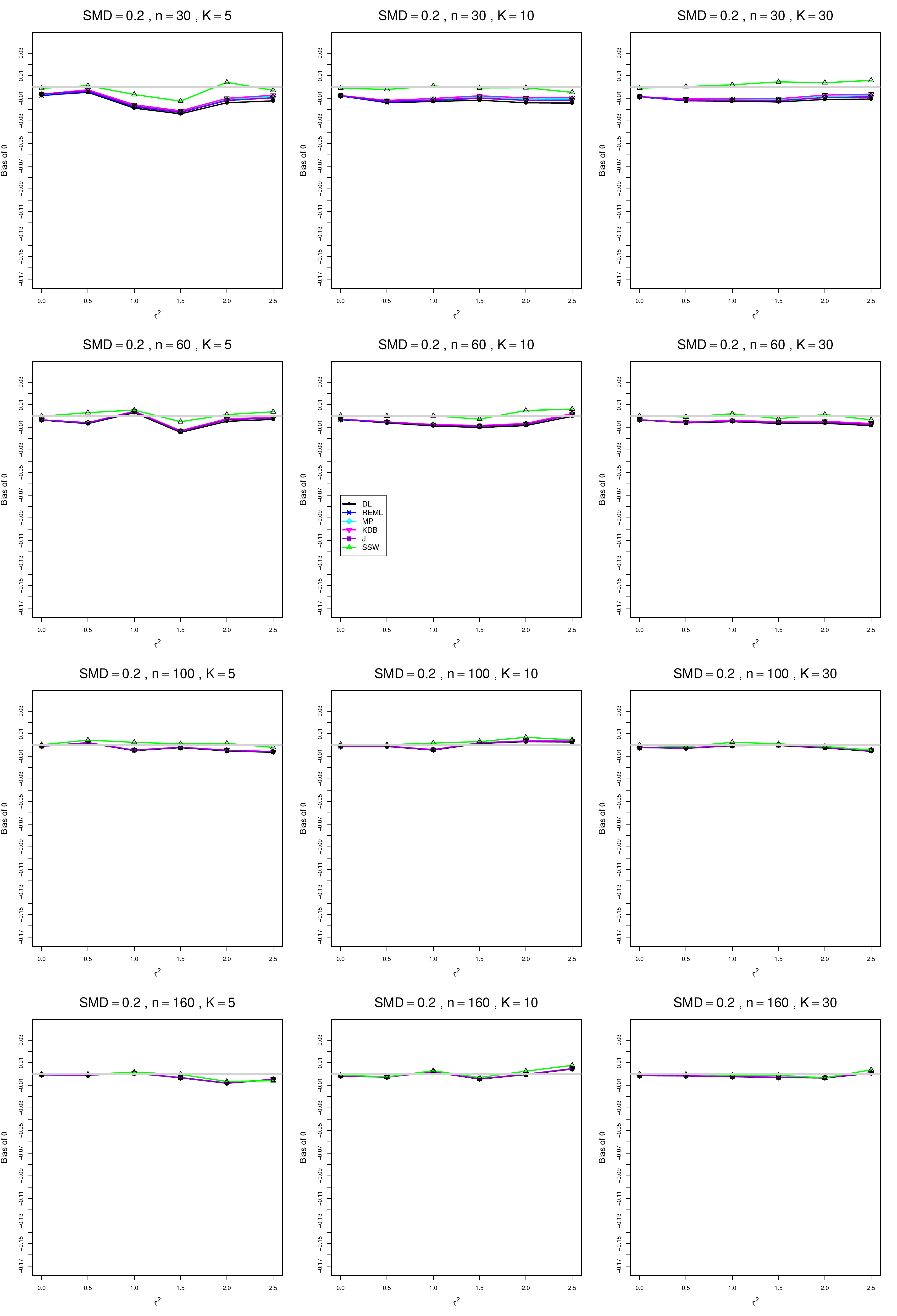}
	\caption{Bias of inverse-variance estimator of $\delta=0.2$, for  $q=0.75$, unequal sample sizes with $\bar{n}=30,\; 60,\;100,\;160$.
		\label{BiasThetaSMD02q75unequal}}
\end{figure}

\begin{figure}[t]\centering
	\includegraphics[scale=0.35]{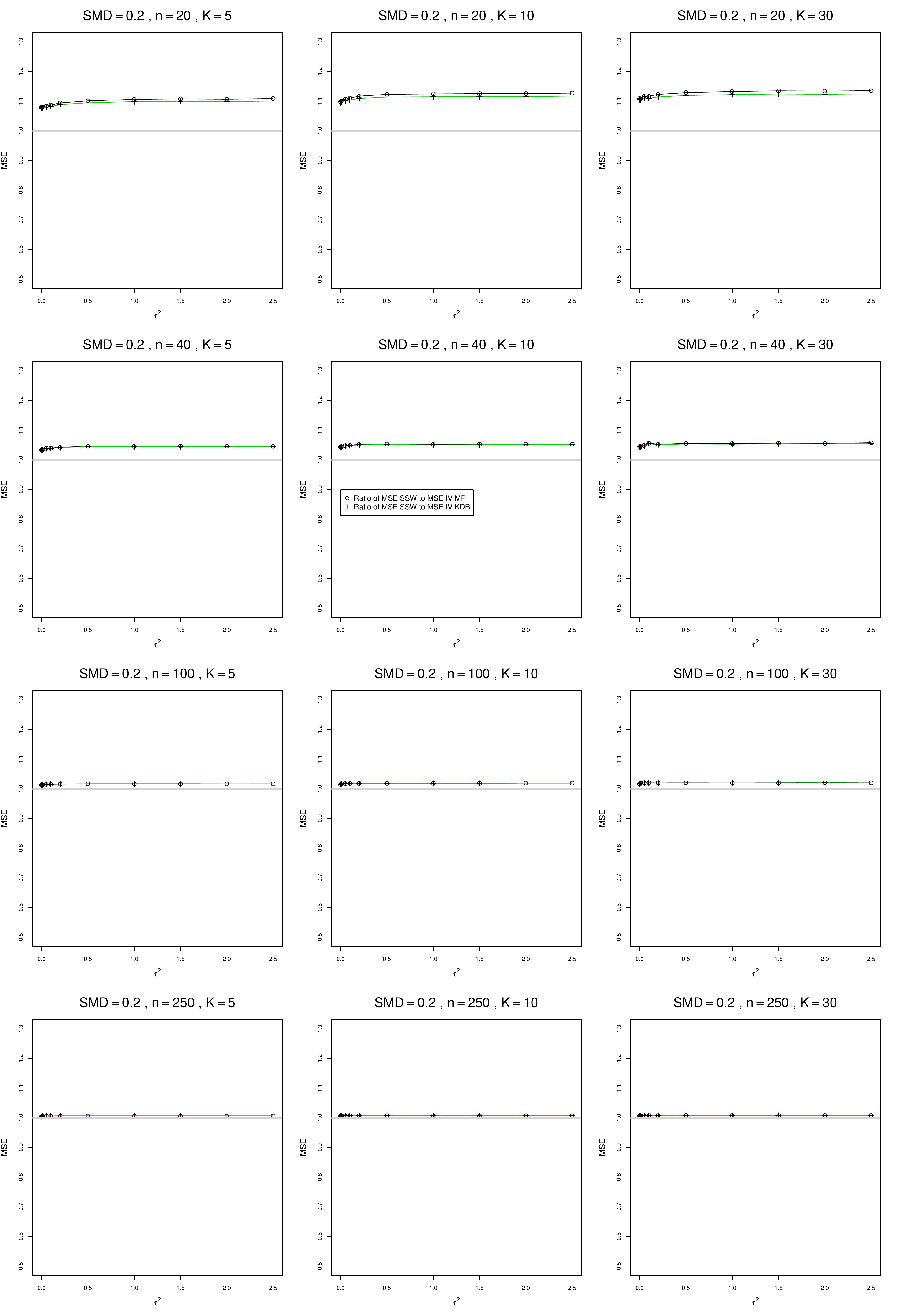}
	\caption{Ratio of mean squared errors of the fixed-weights to mean squared errors of inverse-variance estimator for $\delta=0.2$, for $q=0.75$, $n=20,\;40,\;100,\;250$. 
		\label{RatioOfMSEwithSMD02q075fromMPandCMP}}
\end{figure}

\begin{figure}[t]\centering
	\includegraphics[scale=0.35]{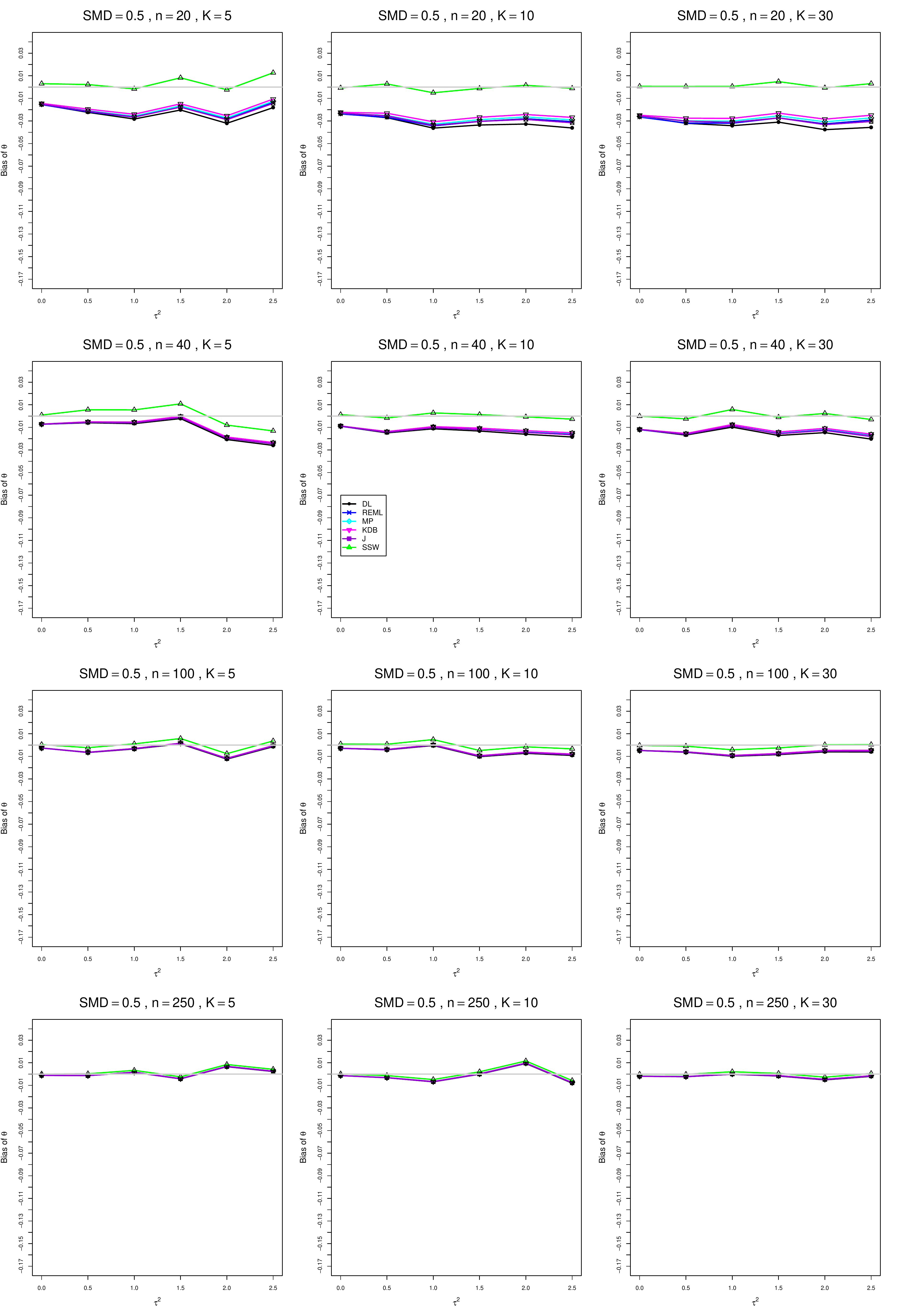}
	\caption{Bias of inverse-variance estimator of $\delta=0.5$, for $q=0.75$, $n=20,\;40,\;100,\;250$.
		\label{BiasThetaSMD05q75}}
\end{figure}

\begin{figure}[t]\centering
	\includegraphics[scale=0.35]{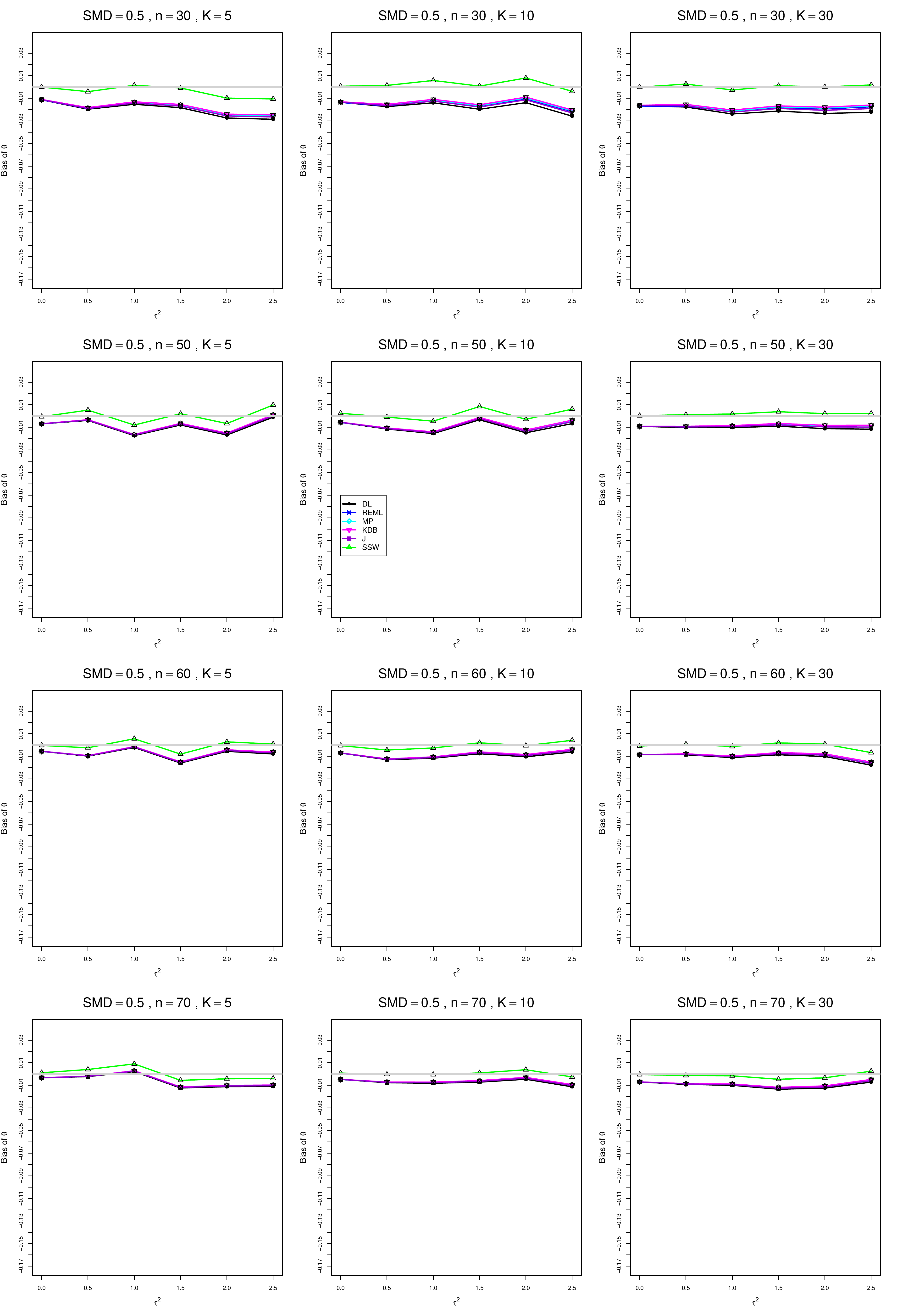}
	\caption{Bias of inverse-variance estimator of $\delta=0.5$, for $q=0.75$, $n=30,\;50,\;60,\;70$.
		\label{BiasThetaSMD05q75small}}
\end{figure}

\begin{figure}[t]\centering
	\includegraphics[scale=0.35]{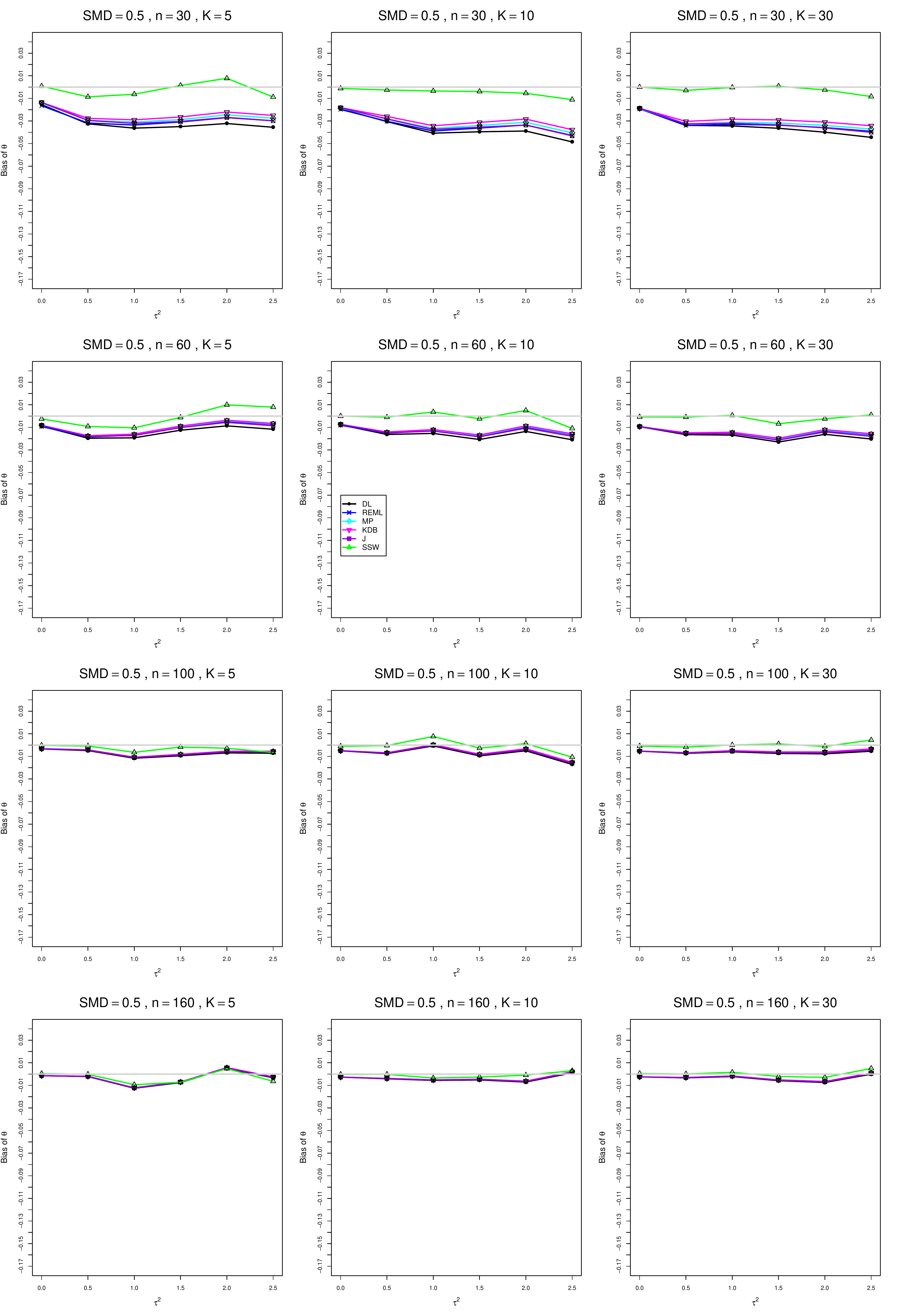}
	\caption{Bias of inverse-variance estimator of $\delta=0.5$, for  $q=0.75$, unequal sample sizes with $\bar{n}=30,\; 60,\;100,\;160$.
		\label{BiasThetaSMD05q75unequal}}
\end{figure}

\begin{figure}[t]\centering
	\includegraphics[scale=0.35]{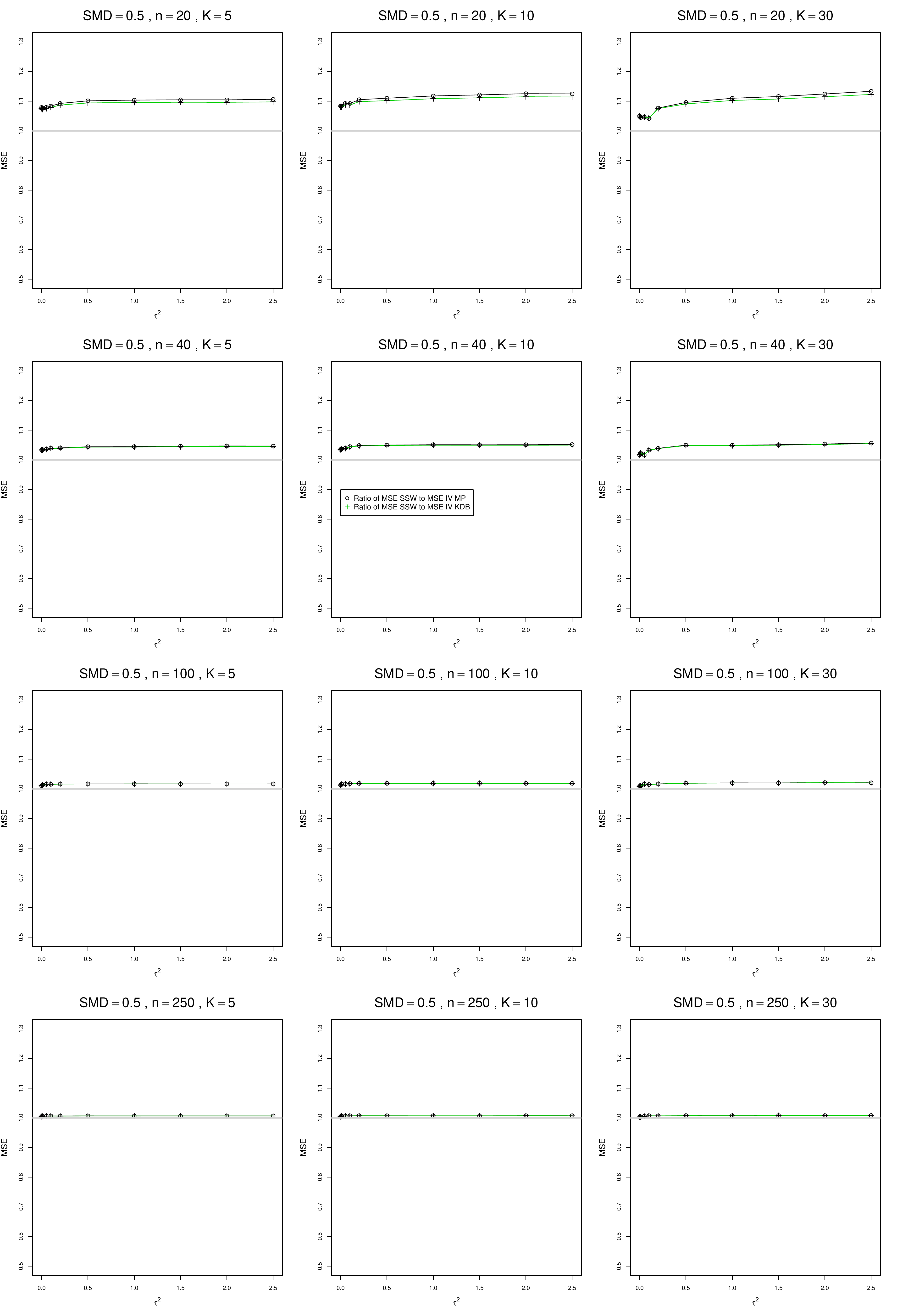}
	\caption{Ratio of mean squared errors of the fixed-weights to mean squared errors of inverse-variance estimator for $\delta=0.5$, for $q=0.75$, $n=20,\;40,\;100,\;250$.
		\label{RatioOfMSEwithSMD05q075fromMPandCMP}}
\end{figure}

\begin{figure}[t]\centering
	\includegraphics[scale=0.35]{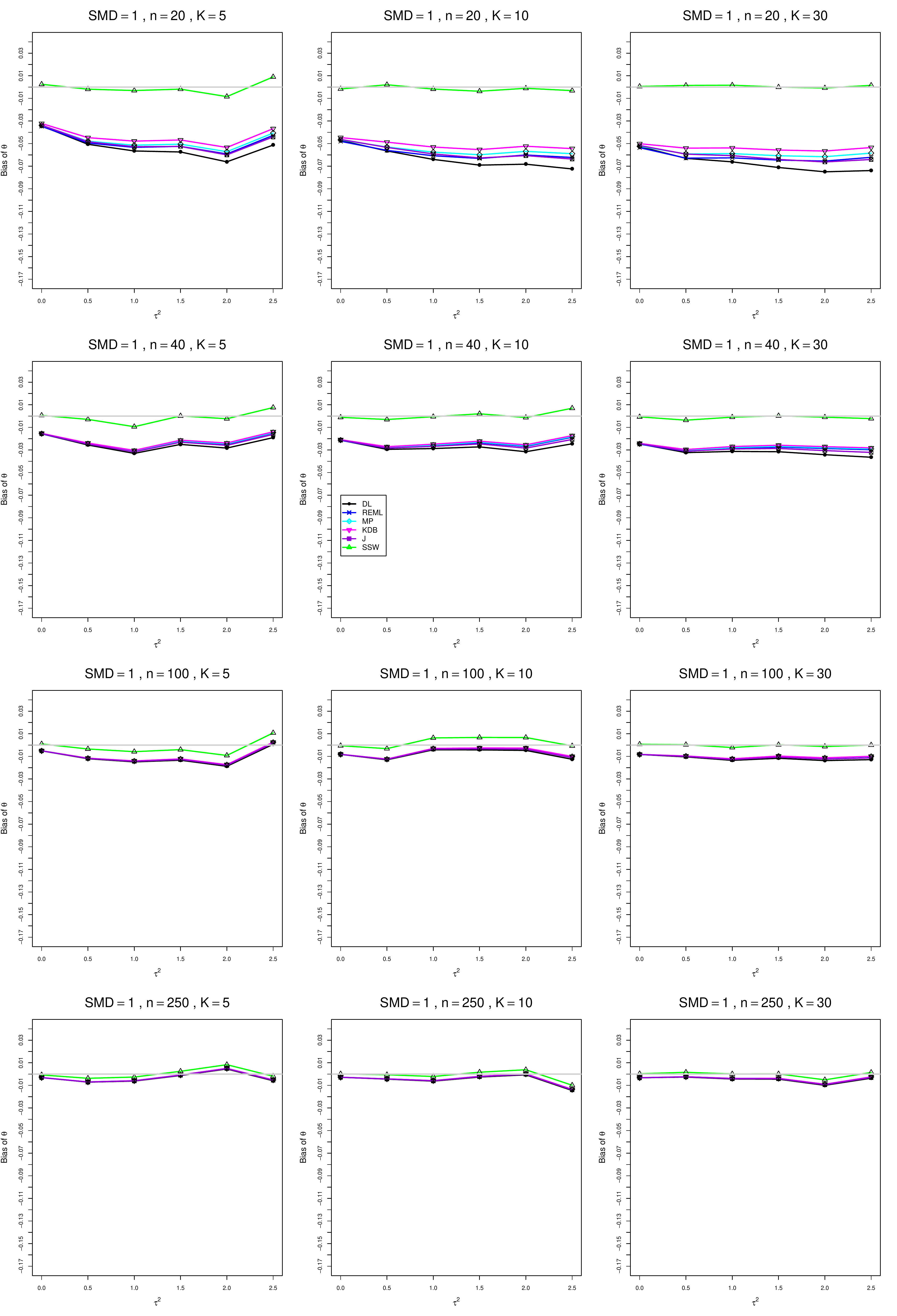}
	\caption{Bias of inverse-variance estimator of $\delta=1$, for $q=0.75$, $n=20,\;40,\;100,\;250$.
		\label{BiasThetaSMD1q75}}
\end{figure}

\begin{figure}[t]\centering
	\includegraphics[scale=0.35]{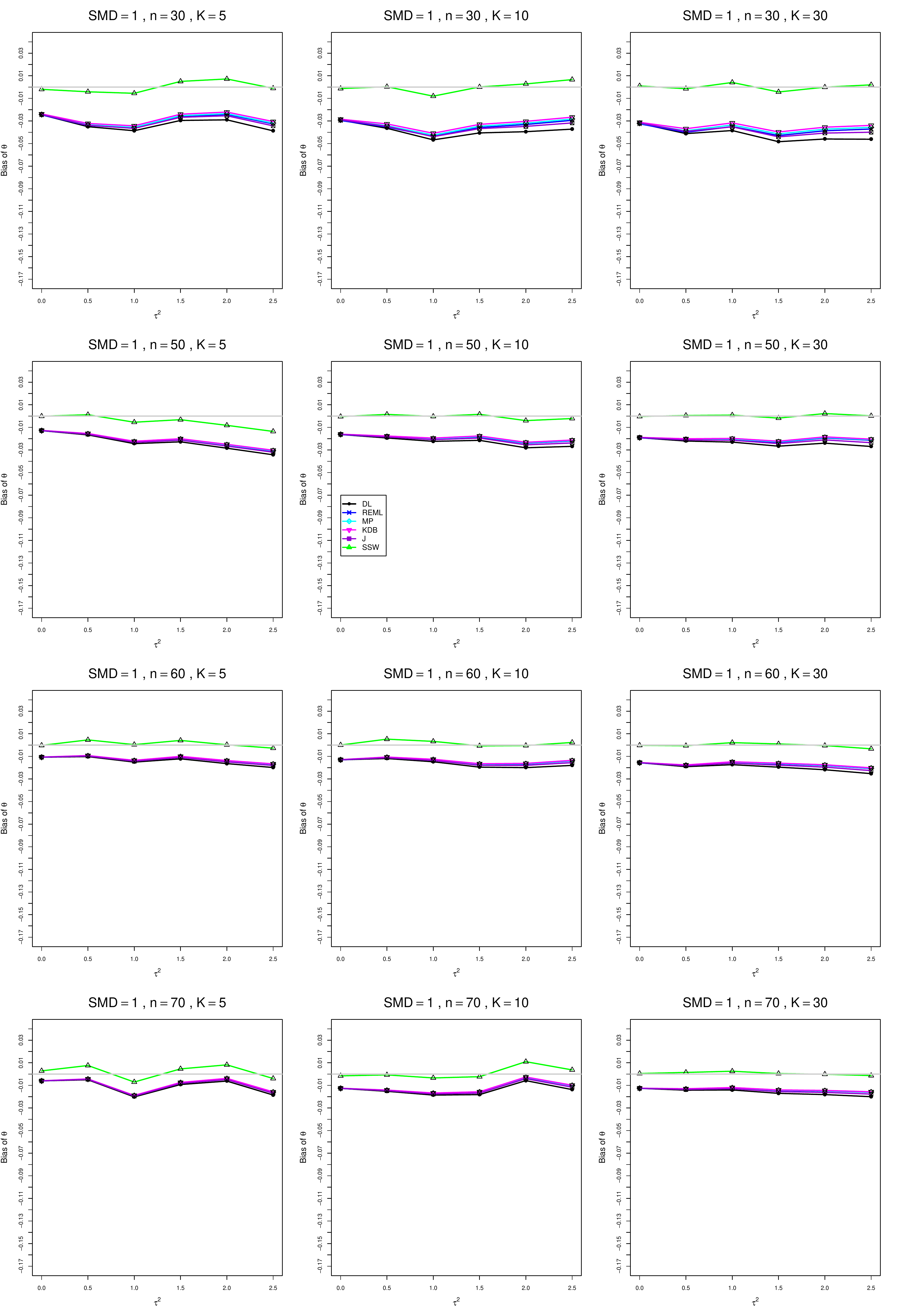}
	\caption{Bias of inverse-variance estimator of $\delta=1$, for $q=0.75$, $n=30,\;50,\;60,\;70$.
		\label{BiasThetaSMD1q75small}}
\end{figure}

\begin{figure}[t]\centering
	\includegraphics[scale=0.35]{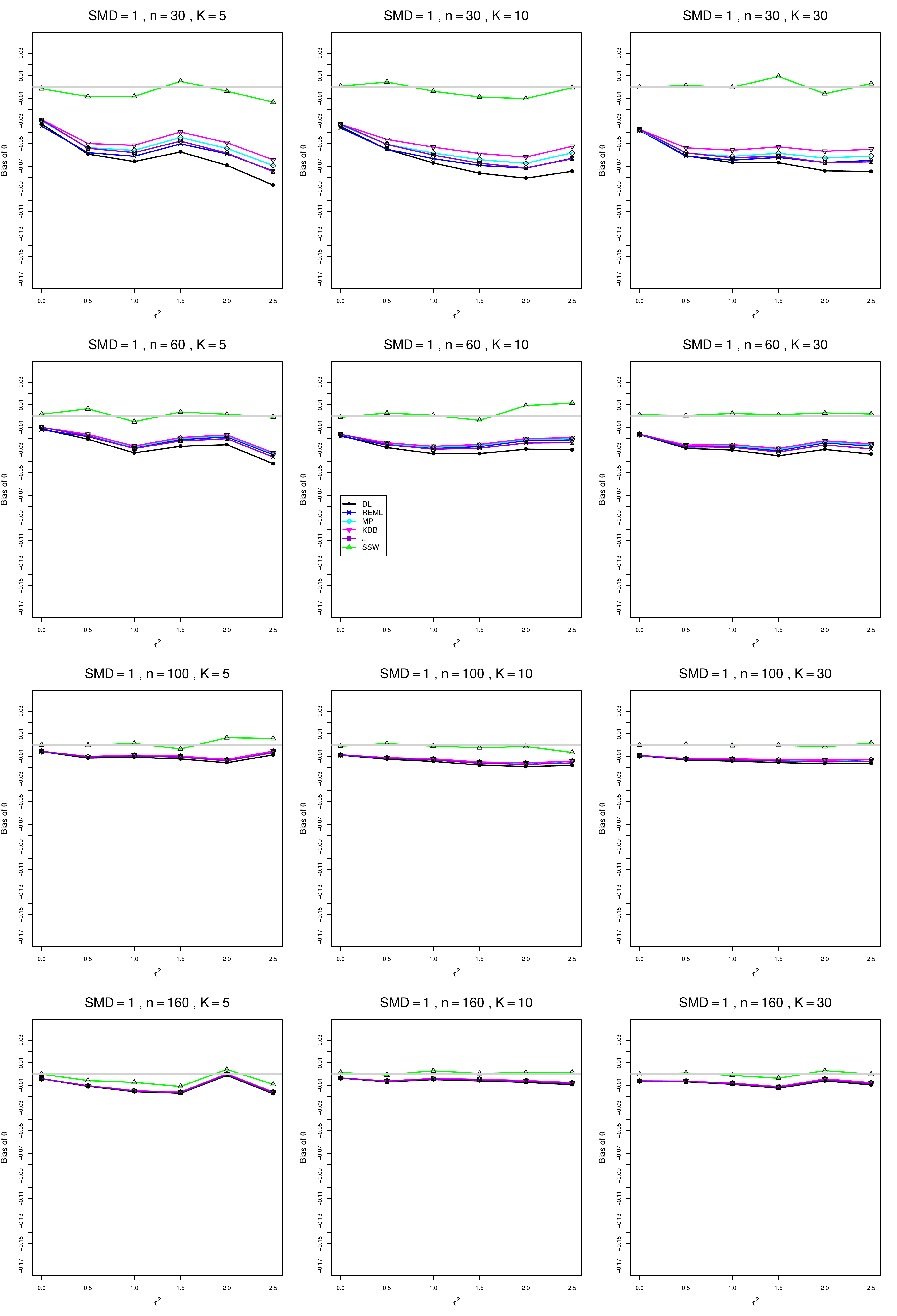}
	\caption{Bias of inverse-variance estimator of $\delta=1$, for  $q=0.75$, unequal sample sizes with $\bar{n}=30,\; 60,\;100,\;160$.
		\label{BiasThetaSMD1q75unequal}}
\end{figure}

\begin{figure}[t]\centering
	\includegraphics[scale=0.35]{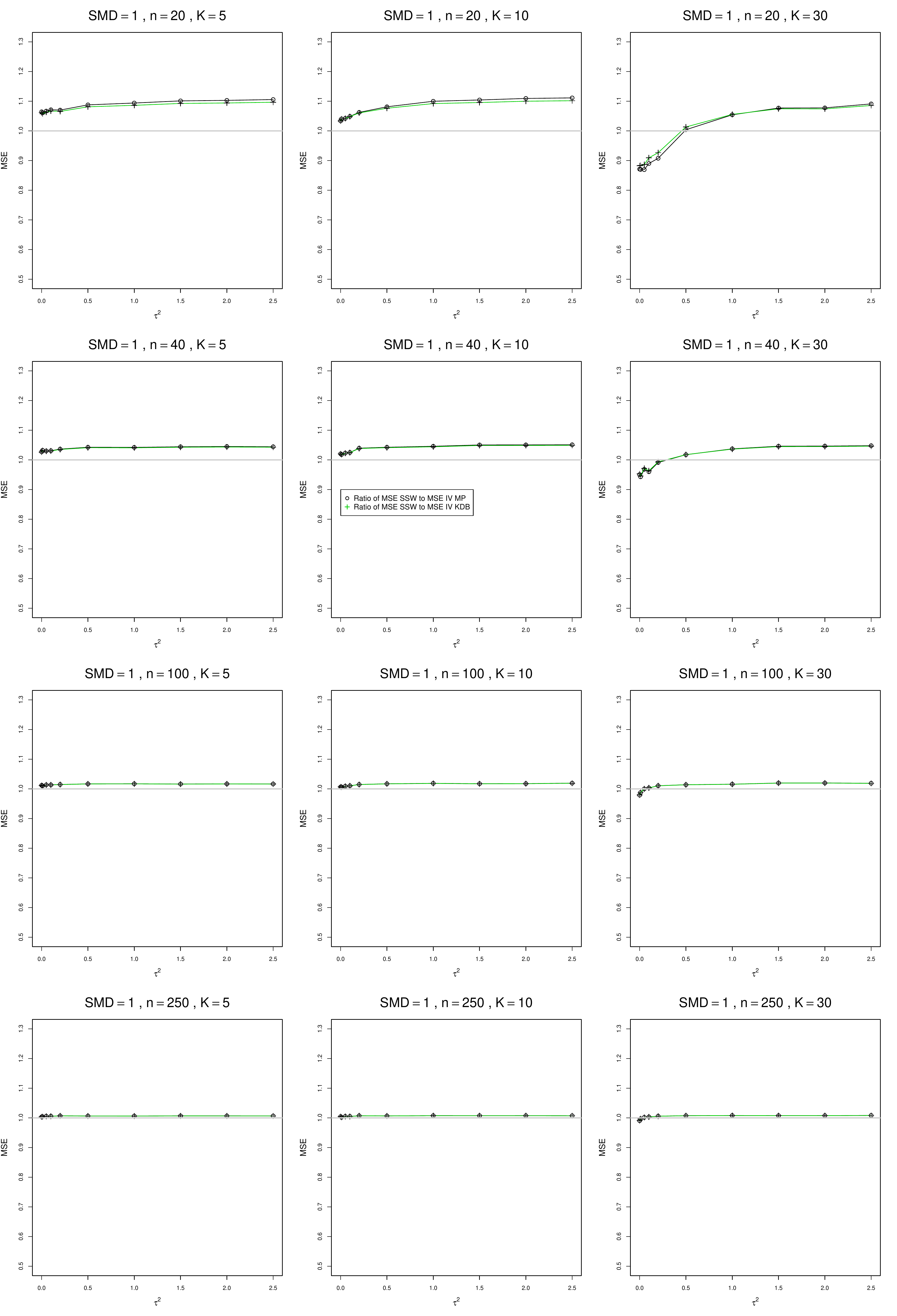}
	\caption{Ratio of mean squared errors of the fixed-weights to mean squared errors of inverse-variance estimator for $\delta=1$, for $q=0.75$, $n=20,\;40,\;100,\;250$.
		\label{RatioOfMSEwithSMD1q075fromMPandCMP}}
\end{figure}

\begin{figure}[t]\centering
	\includegraphics[scale=0.35]{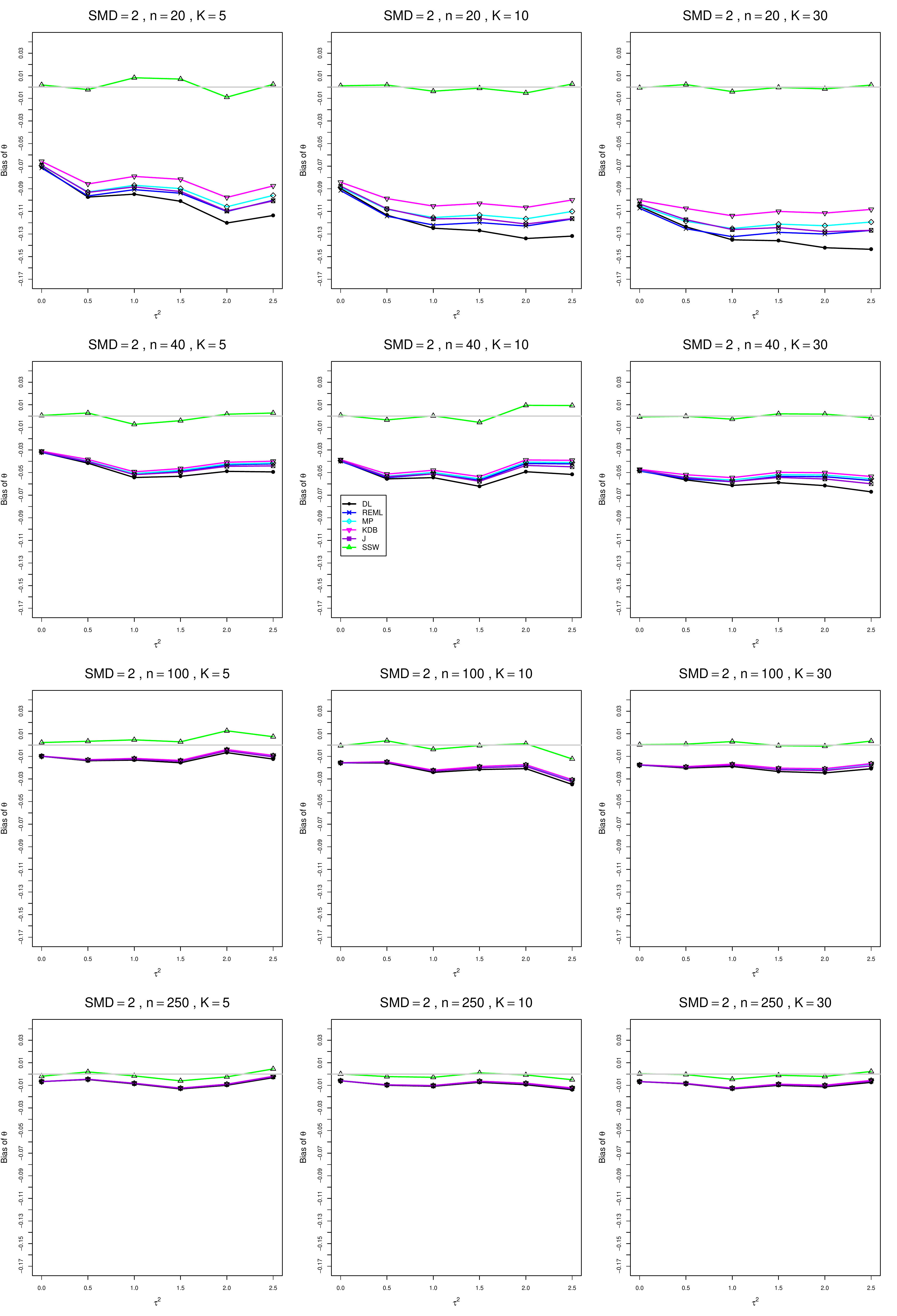}
	\caption{Bias of inverse-variance estimator of $\delta=2$, for $q=0.75$, $n=20,\;40,\;100,\;250$.
		\label{BiasThetaSMD2q75}}
\end{figure}

\begin{figure}[t]\centering
	\includegraphics[scale=0.35]{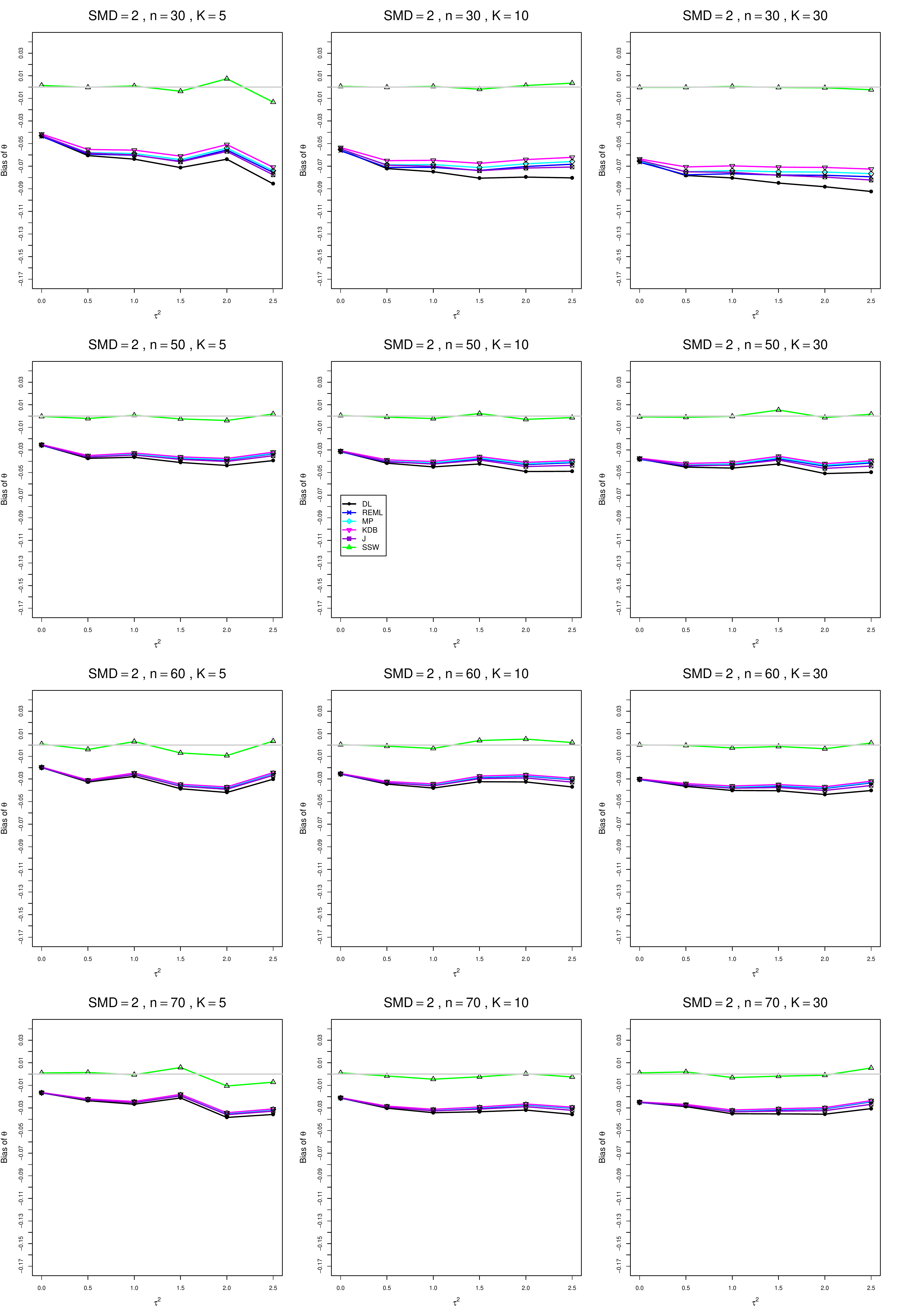}
	\caption{Bias of inverse-variance estimator of $\delta=2$, for $q=0.75$, $n=30,\;50,\;60,\;70$.
		\label{BiasThetaSMD2q75small}}
\end{figure}

\begin{figure}[t]\centering
	\includegraphics[scale=0.35]{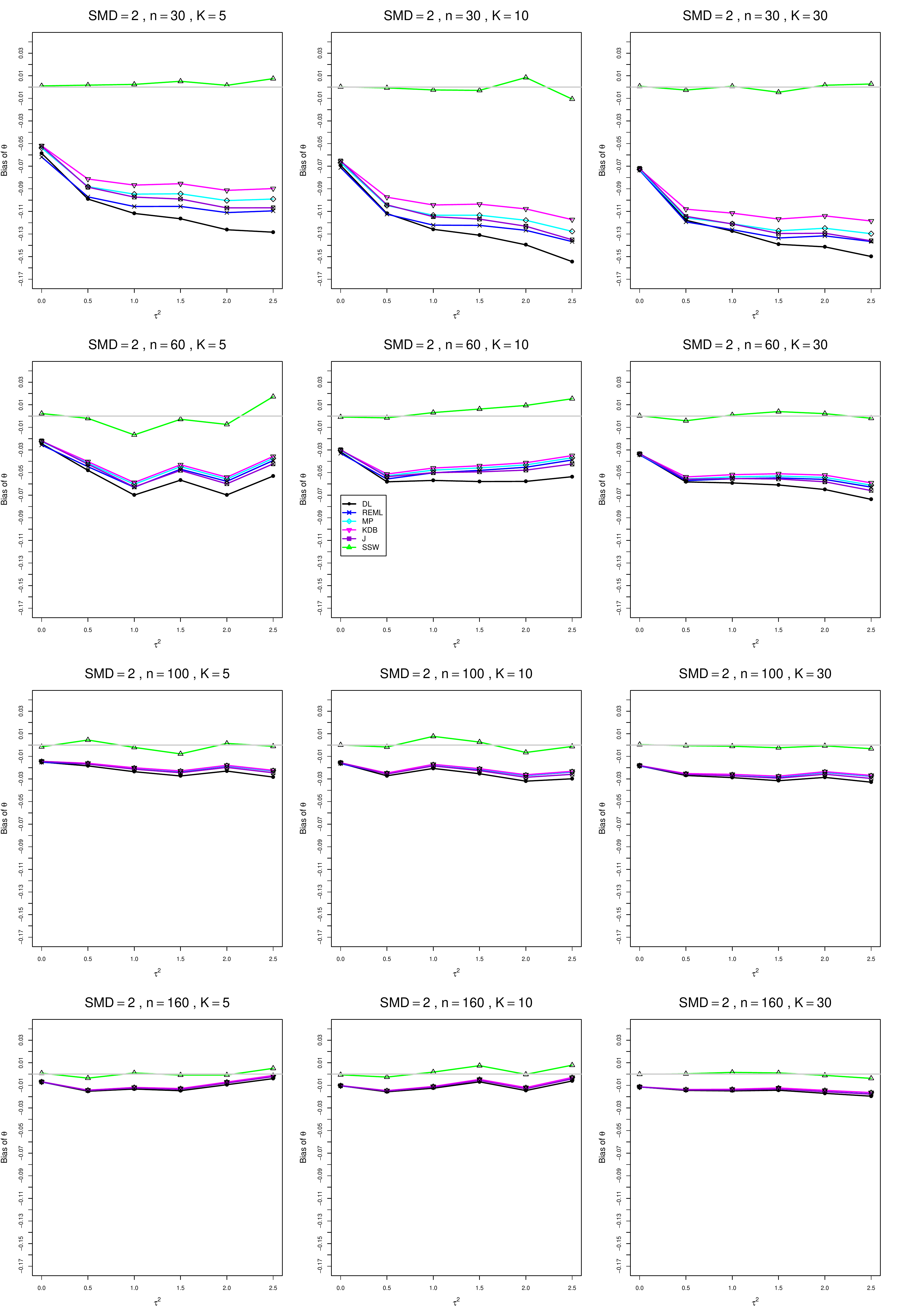}
	\caption{Bias of inverse-variance estimator of $\delta=2$, for  $q=0.75$, unequal sample sizes with $\bar{n}=30,\; 60,\;100,\;160$.
		\label{BiasThetaSMD2q75unequal}}
\end{figure}

\begin{figure}[t]\centering
	\includegraphics[scale=0.35]{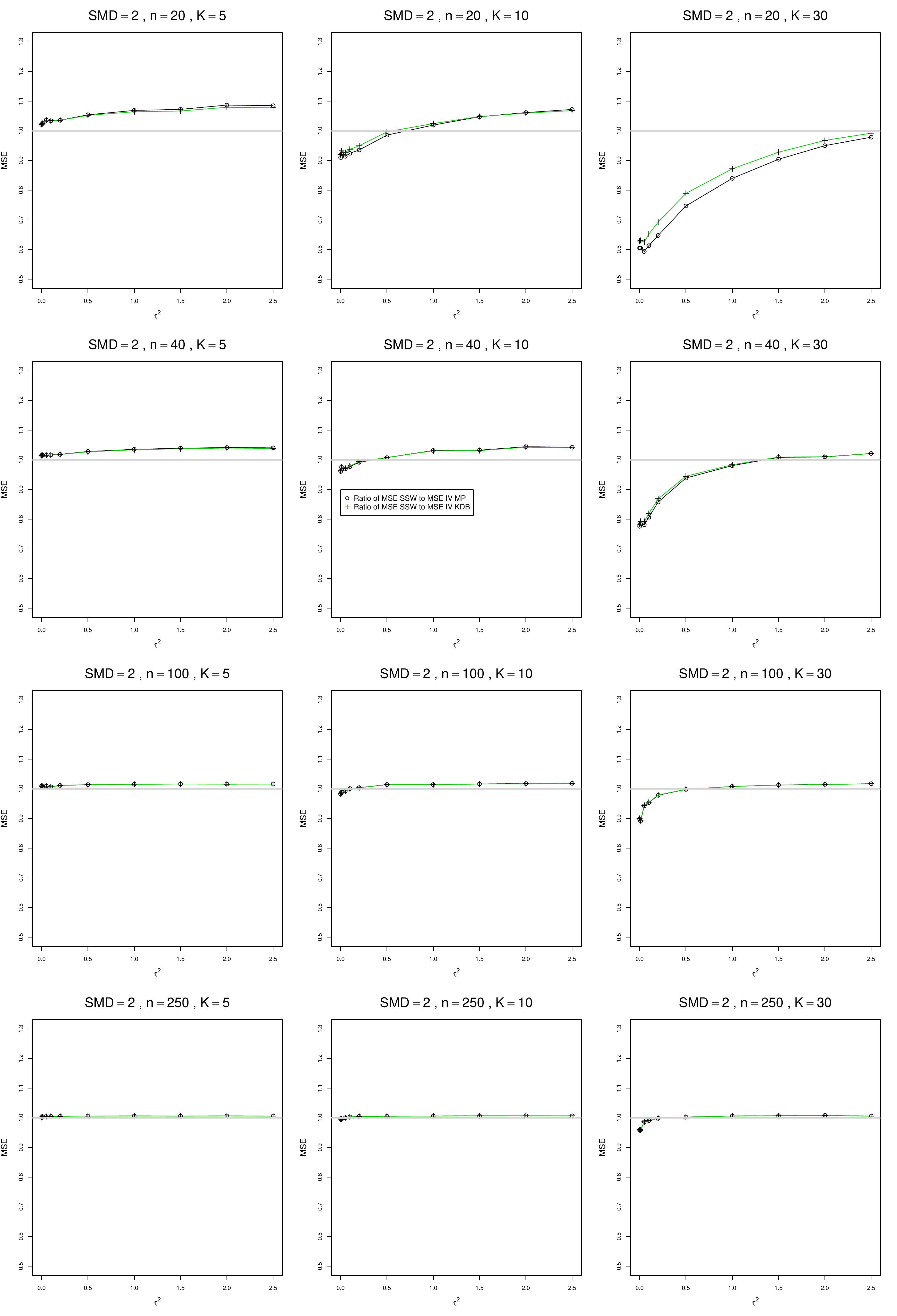}
	\caption{Ratio of mean squared errors of the fixed-weights to mean squared errors of inverse-variance estimator for $\delta=2$, for $q=0.75$, $n=20,\;40,\;100,\;250$. 
		\label{RatioOfMSEwithSMD2q075fromMPandCMP}}
\end{figure}

\clearpage
\setcounter{figure}{0}
\renewcommand{\thefigure}{B2.\arabic{figure}}
\section*{B2. Coverage of interval estimators $\hat{\delta}$ for $\hat{\tau}^2=0.0(0.5)2.5$.}
For coverage of $\hat{\delta}$, each figure corresponds to a value of $\delta$ (= 0, 0.5, 1, 1.5, 2 , 2.5), a value of q (= .5, .75), and a set of values of n (= 20, 40, 100, 250 or 30, 50, 60, 70) or $\bar{n}$ (= 30, 60, 100, 160).\\
Each figure contains a panel (with $\tau^2$ on the horizontal axis) for each combination of n (or $\bar{n}$) and $K (=5, 10, 30)$.\\
The interval estimators of $\delta$ are the companions to the inverse-variance-weighted point estimators
\begin{itemize}
	\item DL (DerSimonian-Laird)
	\item REML (restricted maximum likelihood)
	\item MP (Mandel-Paule)
	\item KDB (improved moment estimator based on Kulinskaya, Dollinger and  Bj{\o}rkest{\o}l (2011))
	\item J (Jackson)
\end{itemize}
and
\begin{itemize}
	\item HKSJ (Hartung-Knapp-Sidik-Jonkman)
	\item HKSJ KDB (HKSJ with KDB estimator of $\tau^2$)
	\item SSW (SSW as center and half-width equal to critical value from $t_{K-1}$
\end{itemize}
times estimated standard deviation of SSW with $\hat{\tau}^2$ = $\hat{\tau}^2_{KDB}$.

\begin{figure}[t]\centering
	\includegraphics[scale=0.35]{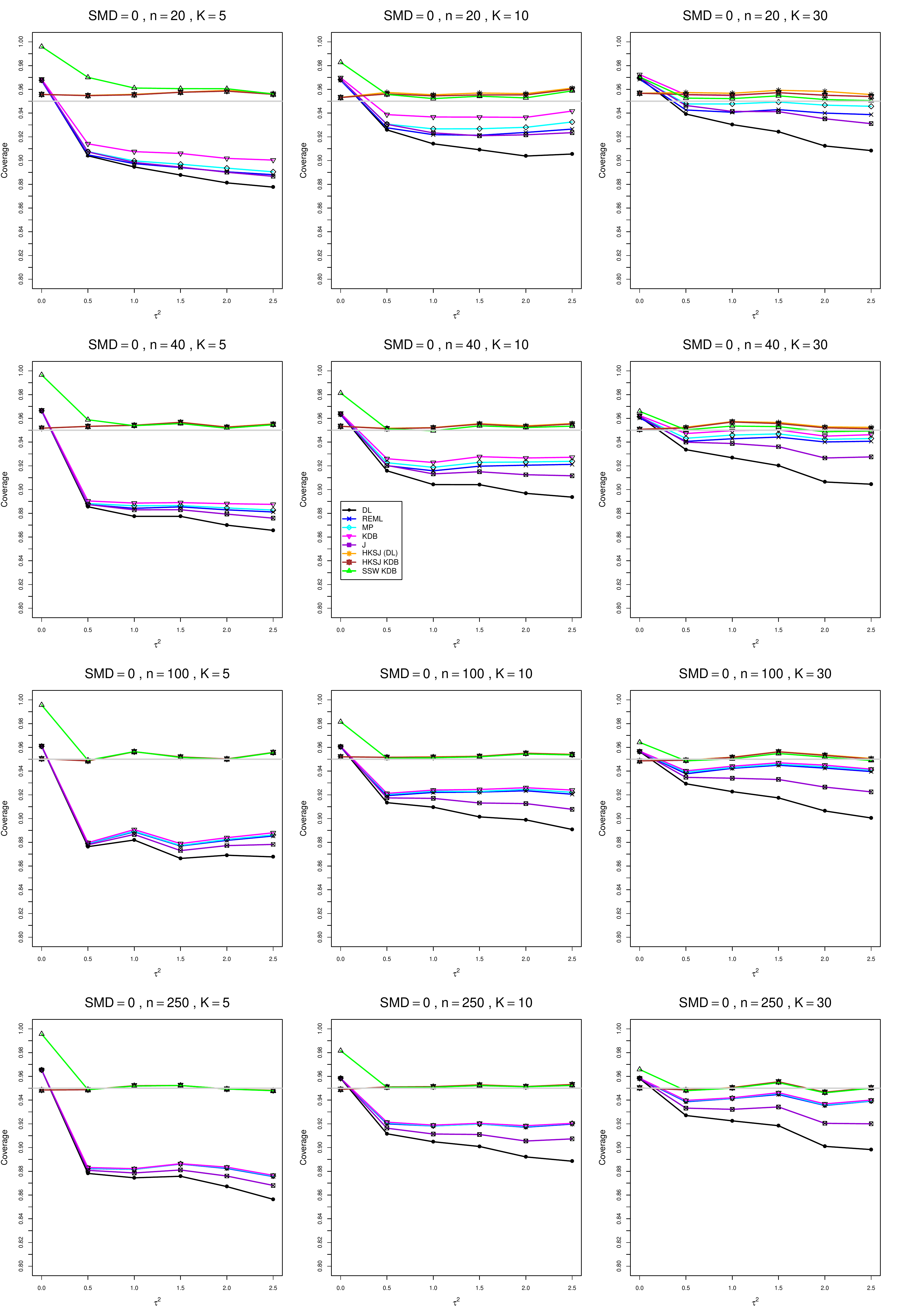}
	\caption{Coverage at  the nominal confidence level of $0.95$ of the  $\delta=0$,  for $q=0.5$, $n=20,\;40,\;100,\;250$.
		\label{CovThetaSMD0}}
\end{figure}

\begin{figure}[t]\centering
	\includegraphics[scale=0.35]{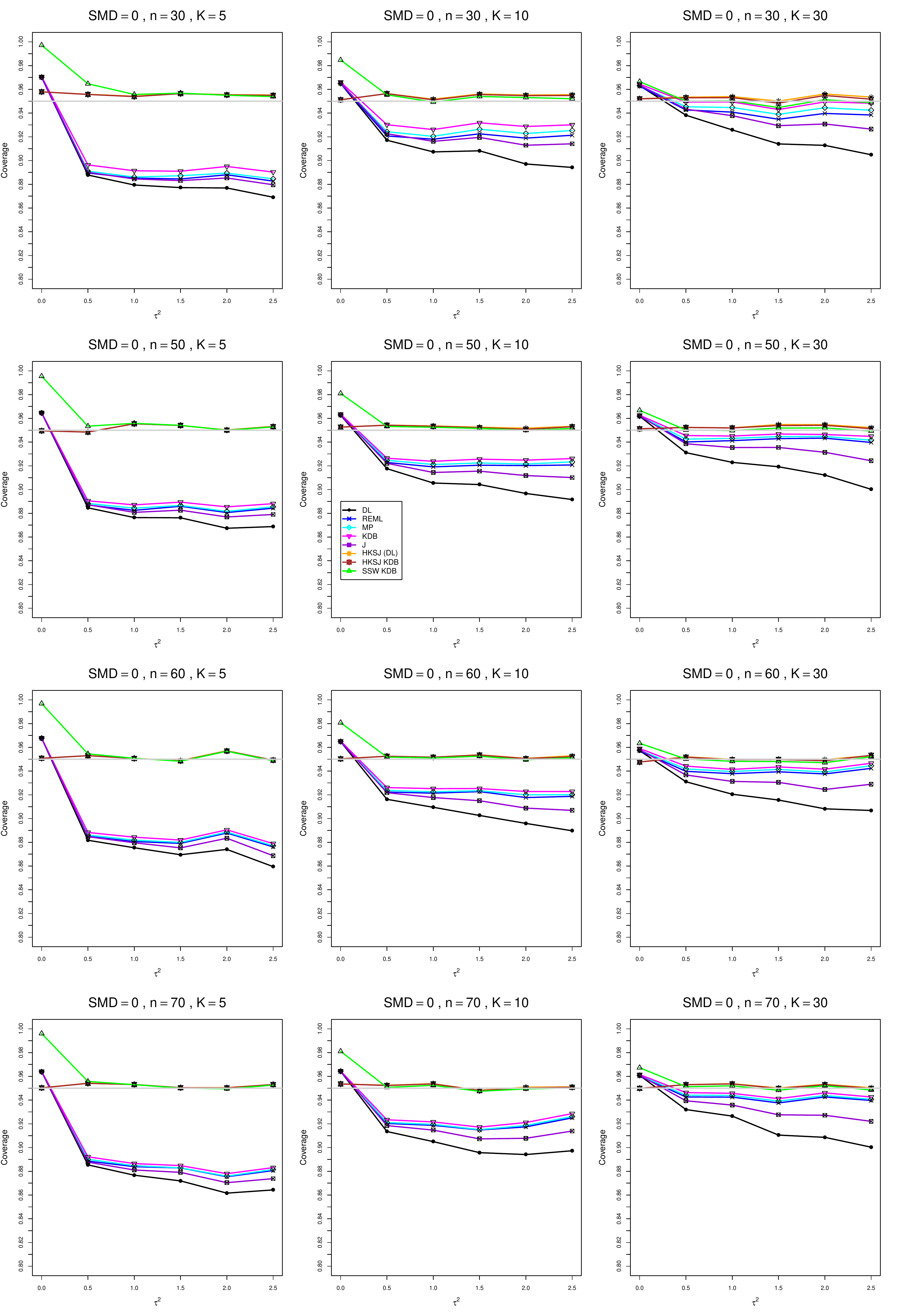}
	\caption{Coverage at  the nominal confidence level of $0.95$ of the  $\delta=0$,  for $q=0.5$, $n=30,\;50,\;60,\;70$.
		\label{CovThetaSMD0small}}
\end{figure}

\begin{figure}[t]\centering
	\includegraphics[scale=0.35]{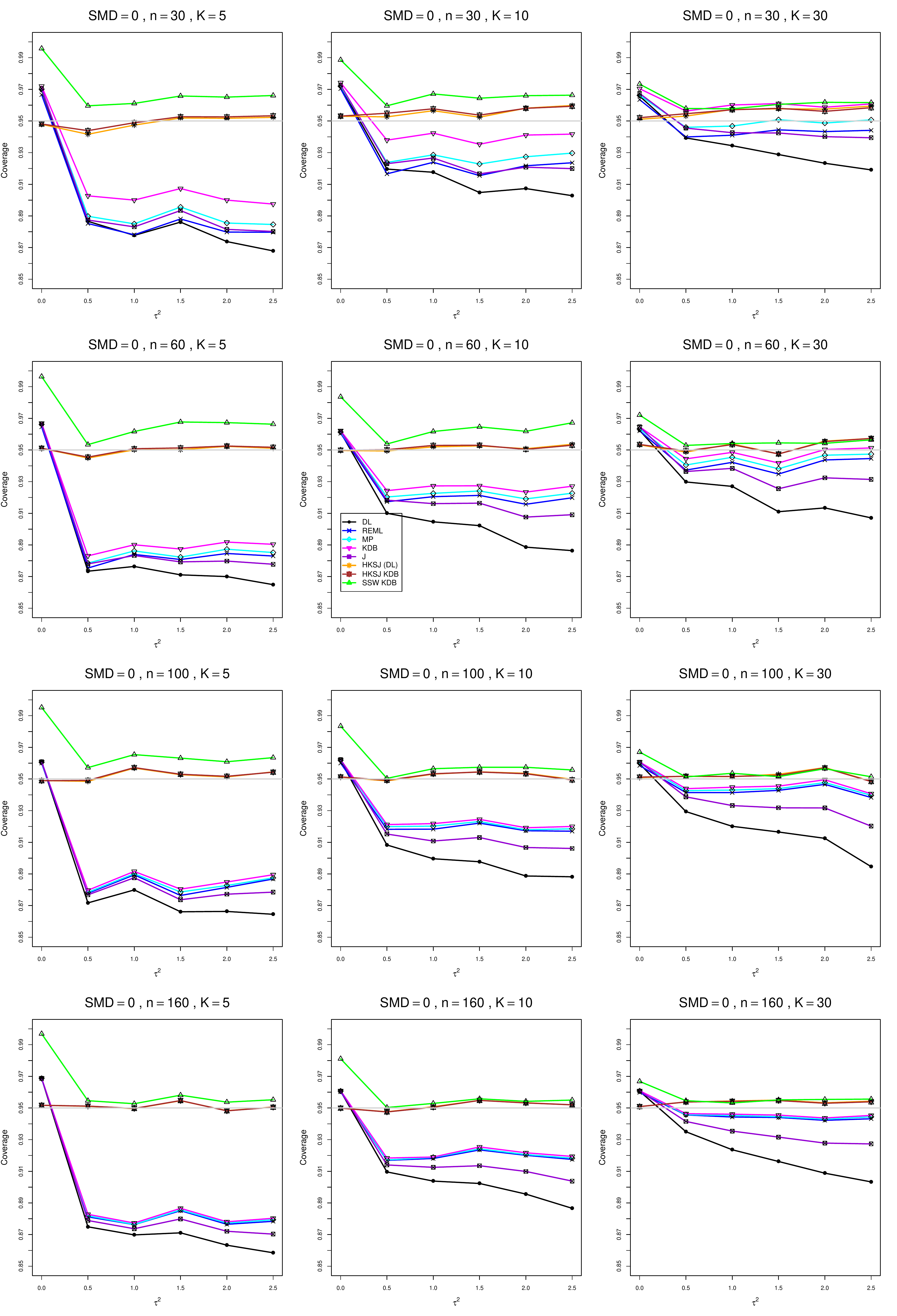}
	\caption{Coverage at  the nominal confidence level of $0.95$ of the $\delta=0$, for $q=0.5$,  unequal sample sizes with
		$\bar{n}=30,\; 60,\;100,\;160$.
		\label{CovThetaSMD0unequal}}
\end{figure}

\begin{figure}[t]\centering
	\includegraphics[scale=0.35]{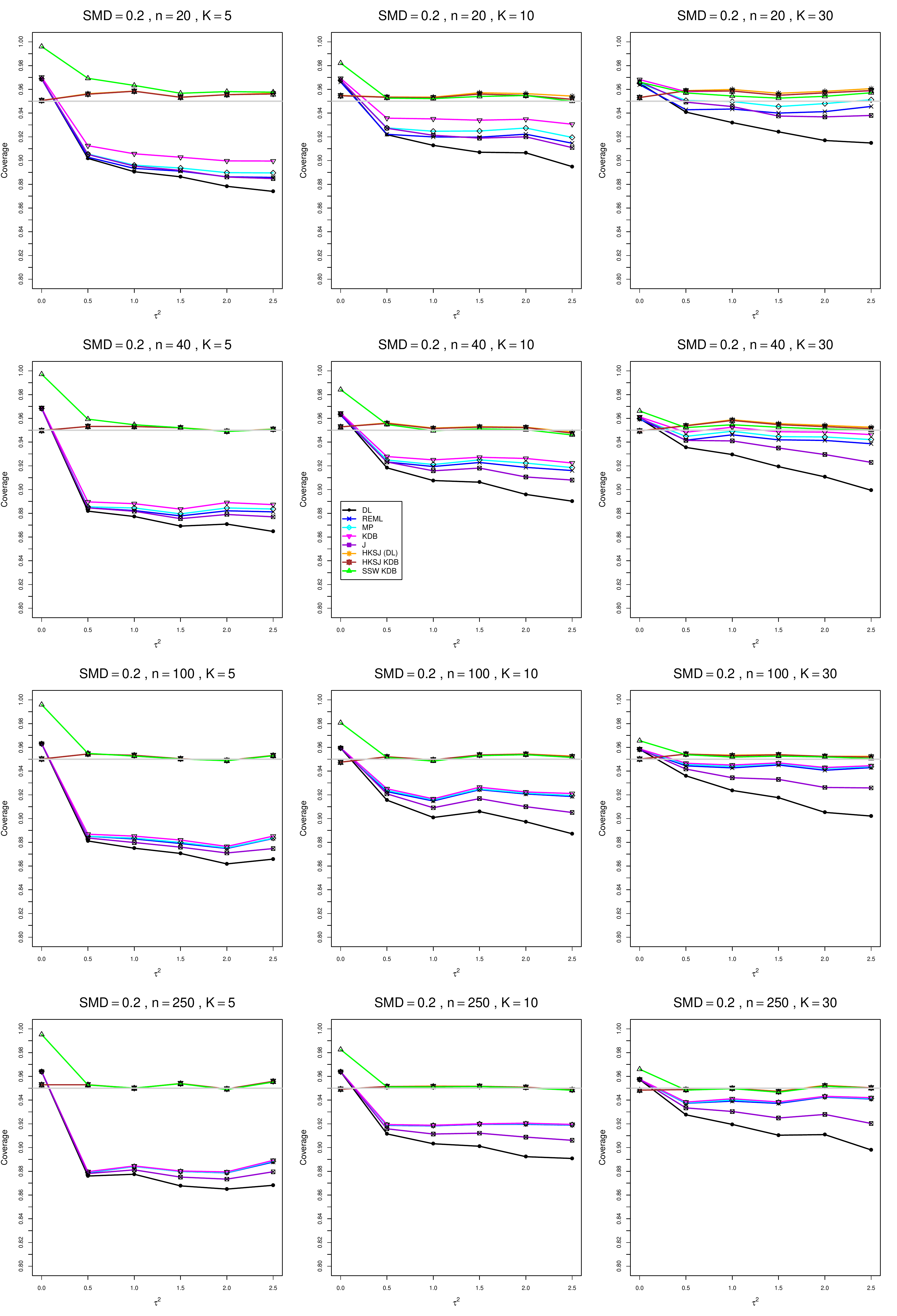}
	\caption{Coverage at  the nominal confidence level of $0.95$ of the  $\delta=0.2$,  for $q=0.5$, $n=20,\;40,\;100,\;250$.
		\label{CovThetaSMD02}}
\end{figure}

\begin{figure}[t]\centering
	\includegraphics[scale=0.35]{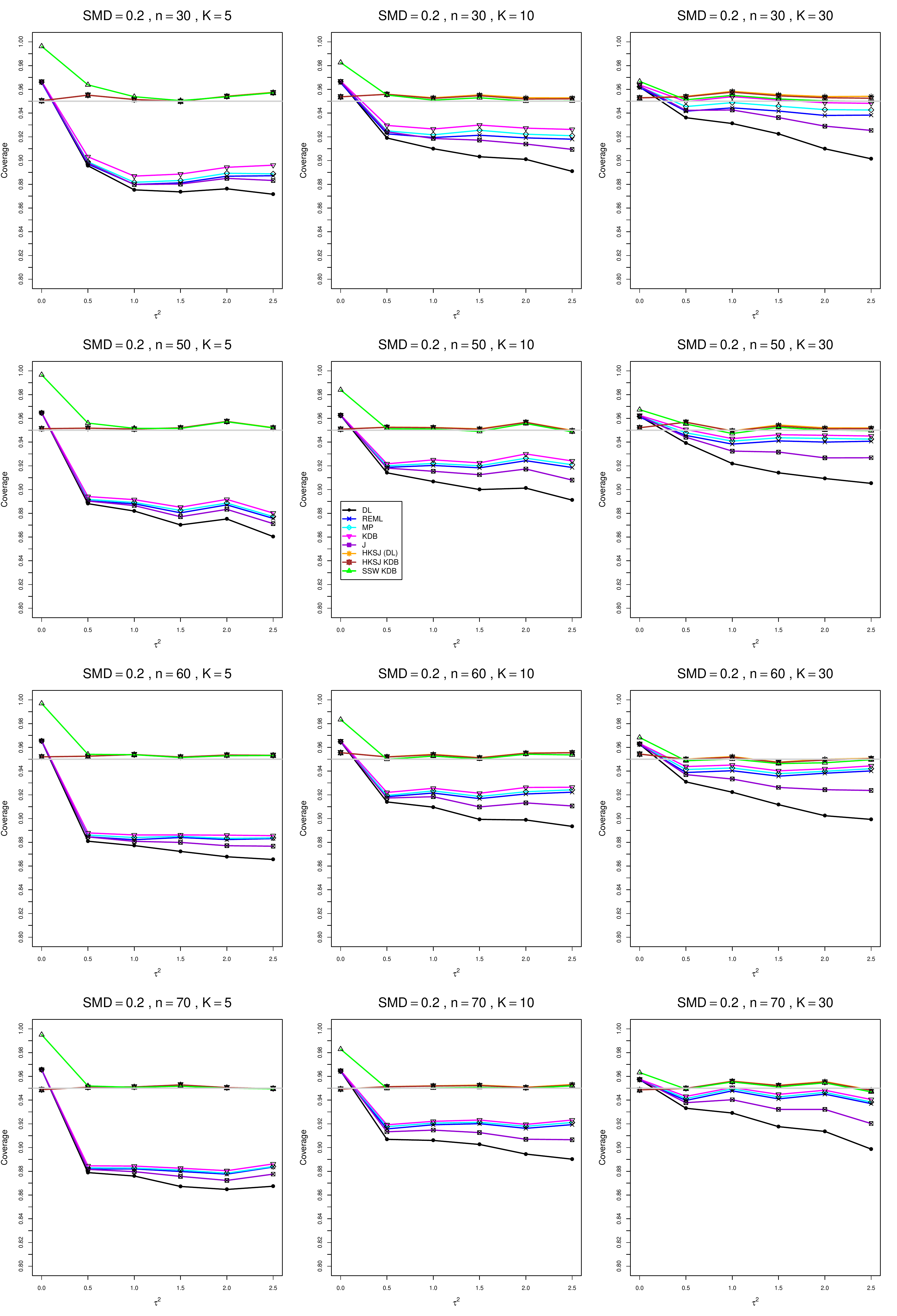}
	\caption{Coverage at  the nominal confidence level of $0.95$ of the  $\delta=0.2$,  for $q=0.5$, $n=30,\;50,\;60,\;70$.
		\label{CovThetaSMD02small}}
\end{figure}

\begin{figure}[t]\centering
	\includegraphics[scale=0.35]{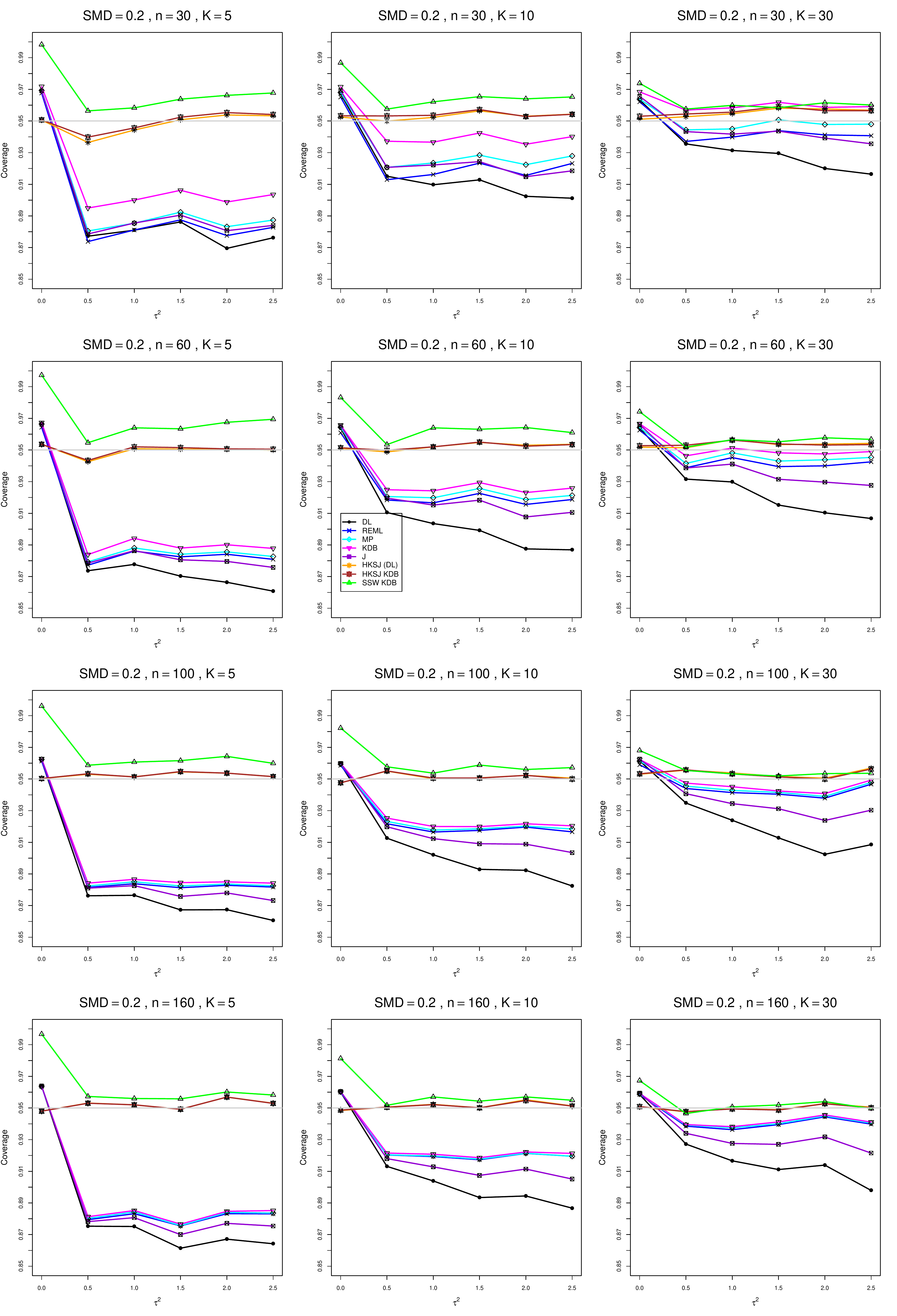}
	\caption{Coverage at  the nominal confidence level of $0.95$ of the $\delta=0.2$, for $q=0.5$,  unequal sample sizes with
		$\bar{n}=30,\; 60,\;100,\;160$.
		\label{CovThetaSMD02unequal}}
\end{figure}

\begin{figure}[t]\centering
	\includegraphics[scale=0.35]{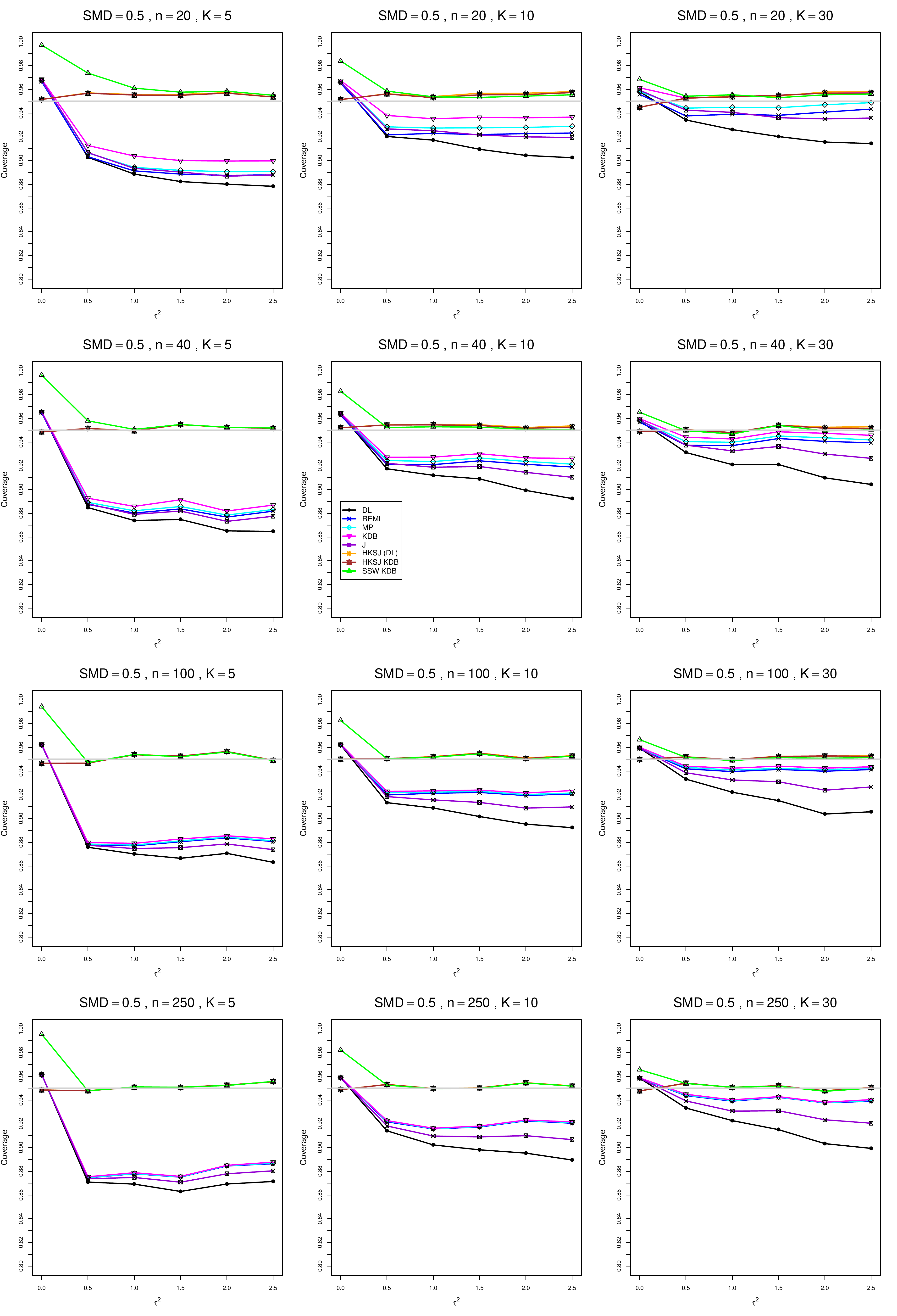}
	\caption{Coverage at  the nominal confidence level of $0.95$ of the  $\delta=0.5$,  for $q=0.5$, $n=20,\;40,\;100,\;250$.
		\label{CovThetaSMD05}}
\end{figure}

\begin{figure}[t]\centering
	\includegraphics[scale=0.35]{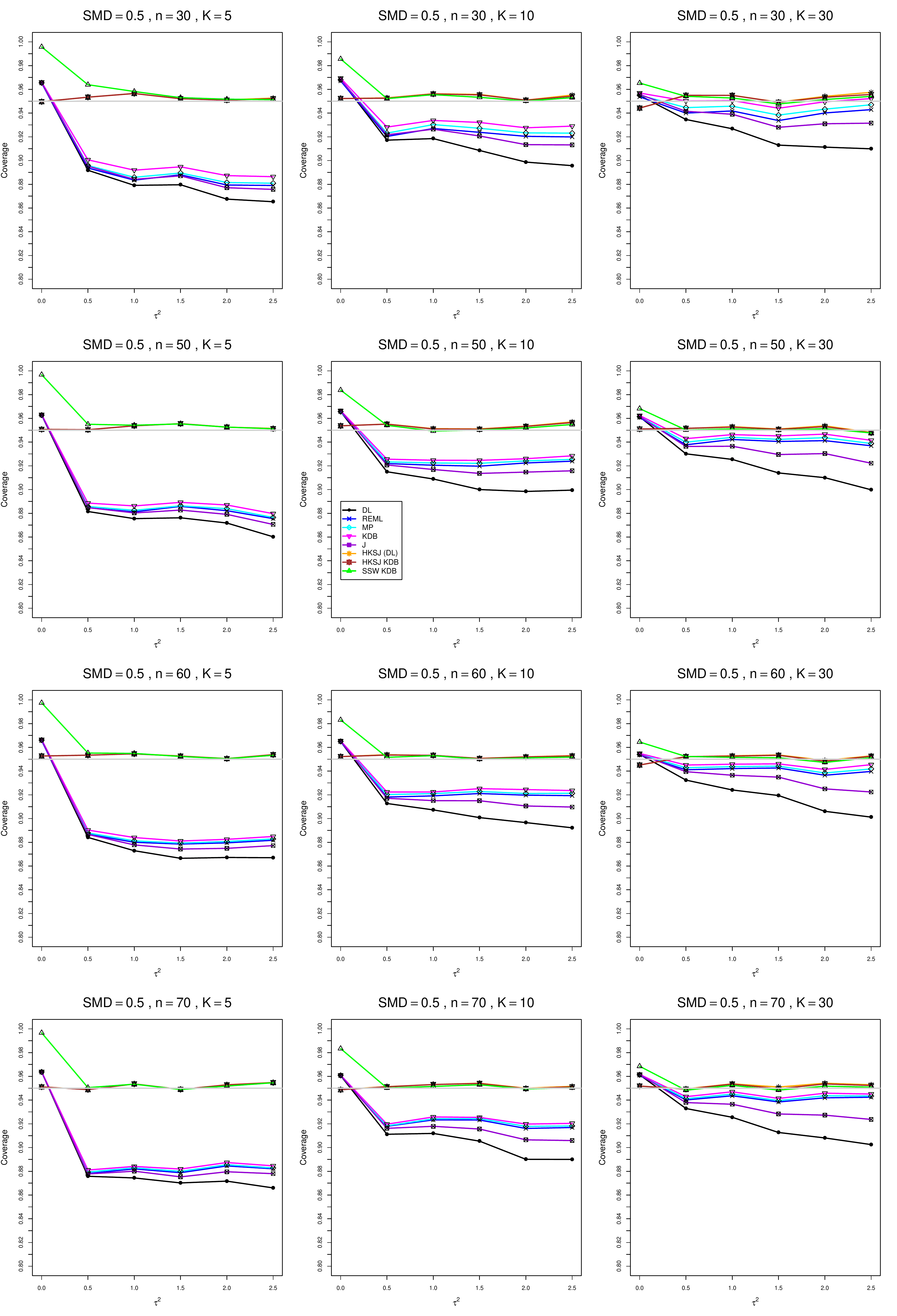}
	\caption{Coverage at  the nominal confidence level of $0.95$ of the  $\delta=0.5$,  for $q=0.5$, $n=30,\;50,\;60,\;70$.
		\label{CovThetaSMD05small}}
\end{figure}

\begin{figure}[t]\centering
	\includegraphics[scale=0.35]{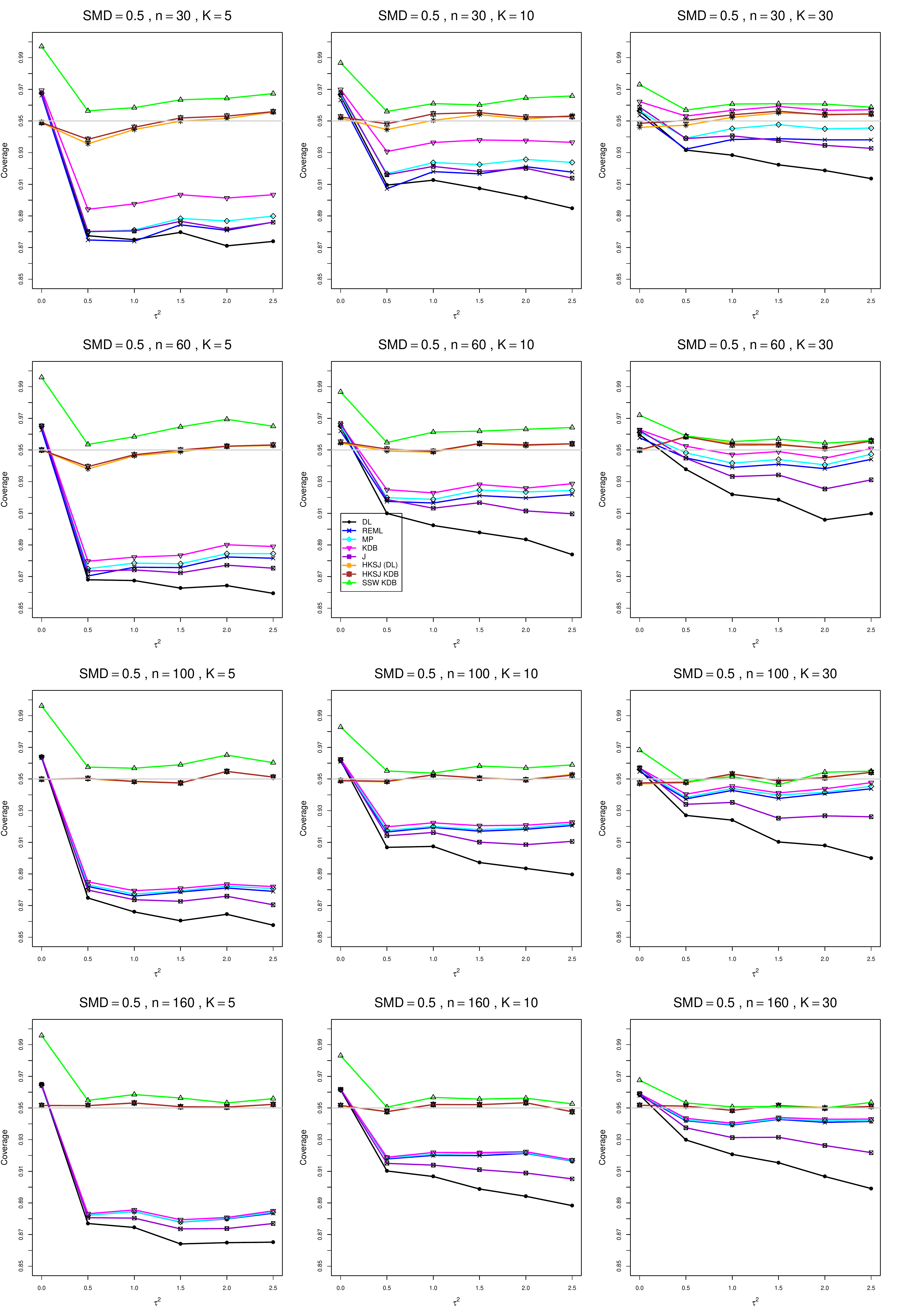}
	\caption{Coverage at  the nominal confidence level of $0.95$ of the $\delta=0.5$, for $q=0.5$,  unequal sample sizes with
		$\bar{n}=30,\; 60,\;100,\;160$.
		\label{CovThetaSMD05unequal}}
\end{figure}

\begin{figure}[t]\centering
	\includegraphics[scale=0.35]{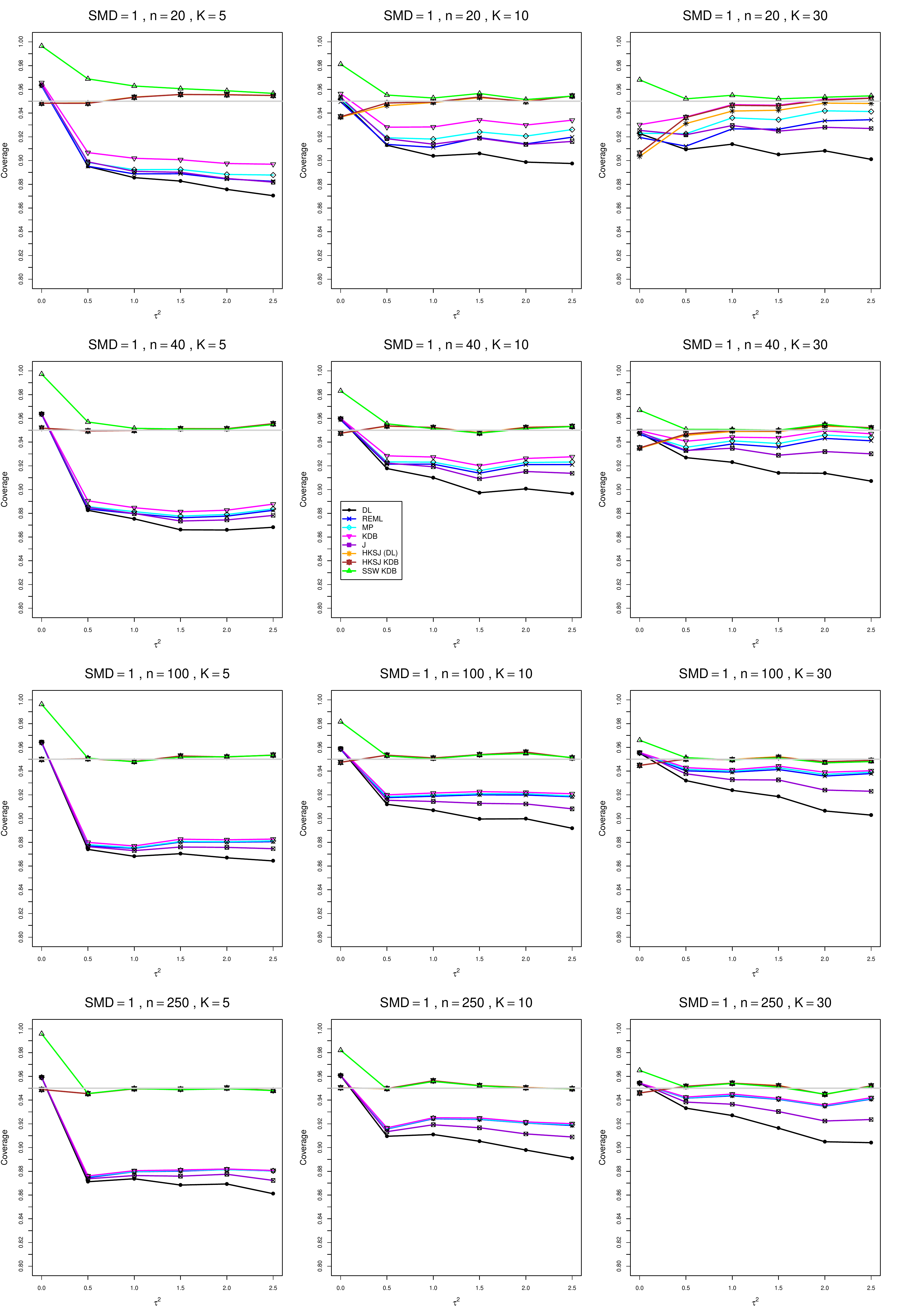}
	\caption{Coverage at  the nominal confidence level of $0.95$ of the  $\delta=1$,  for $q=0.5$, $n=20,\;40,\;100,\;250$.
		\label{CovThetaSMD1}}
\end{figure}

\begin{figure}[t]\centering
	\includegraphics[scale=0.35]{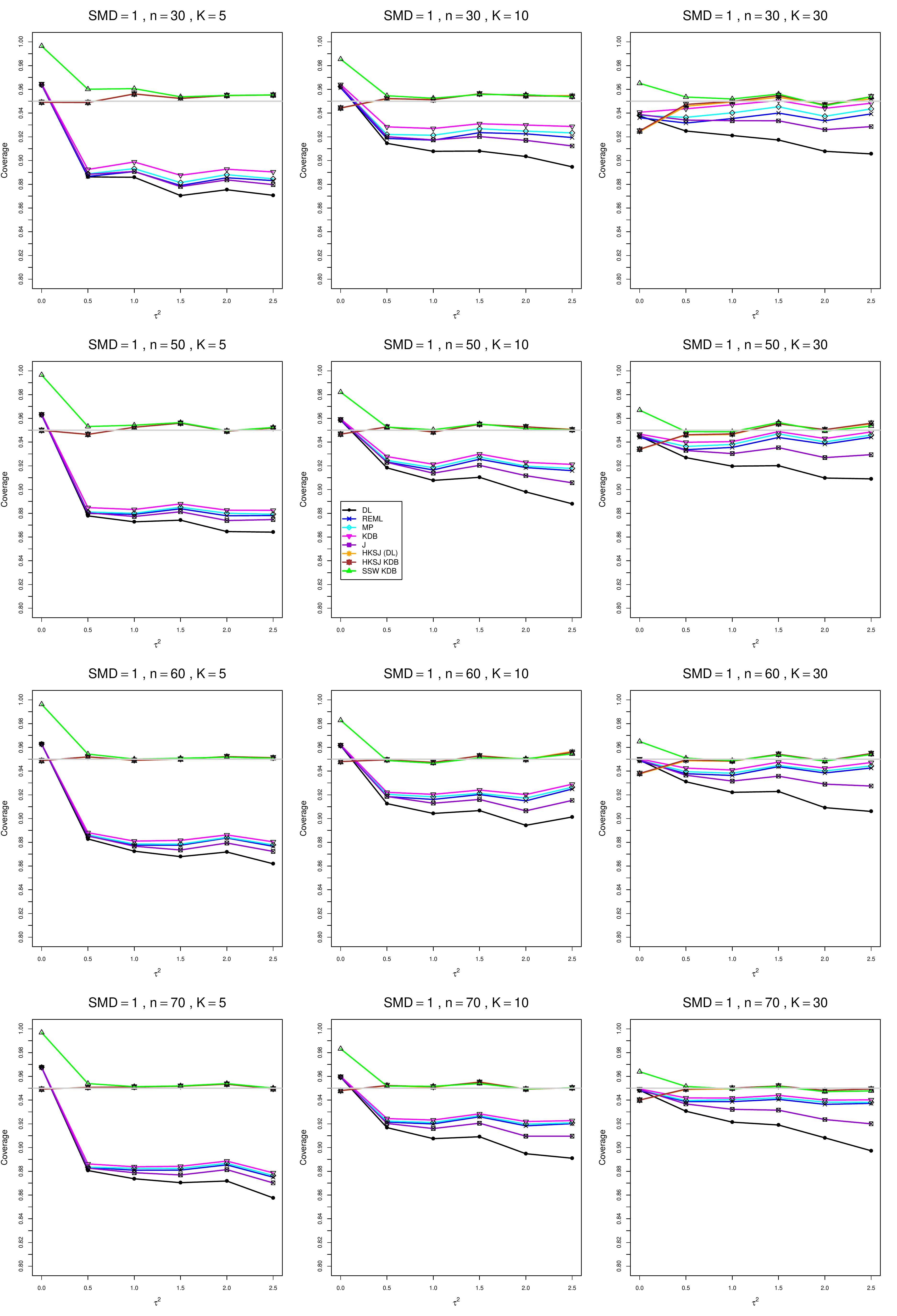}
	\caption{Coverage at  the nominal confidence level of $0.95$ of the  $\delta=1$,  for $q=0.5$, $n=30,\;50,\;60,\;70$.
		\label{CovThetaSMD1small}}
\end{figure}

\begin{figure}[t]\centering
	\includegraphics[scale=0.35]{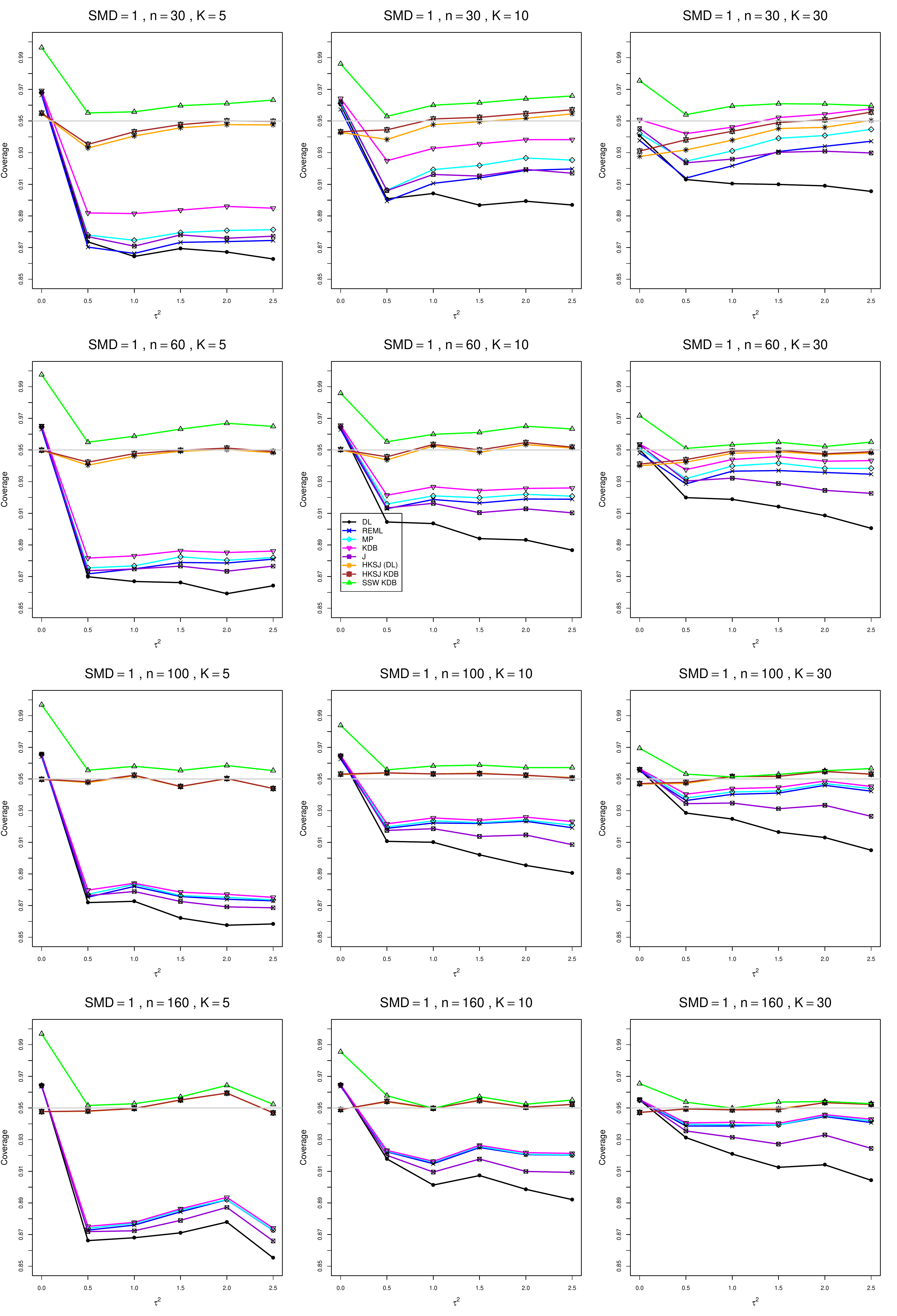}
	\caption{Coverage at  the nominal confidence level of $0.95$ of the $\delta=1$, for $q=0.5$,  unequal sample sizes with
		$\bar{n}=30,\; 60,\;100,\;160$.
		\label{CovThetaSMD1unequal}}
\end{figure}

\begin{figure}[t]\centering
	\includegraphics[scale=0.35]{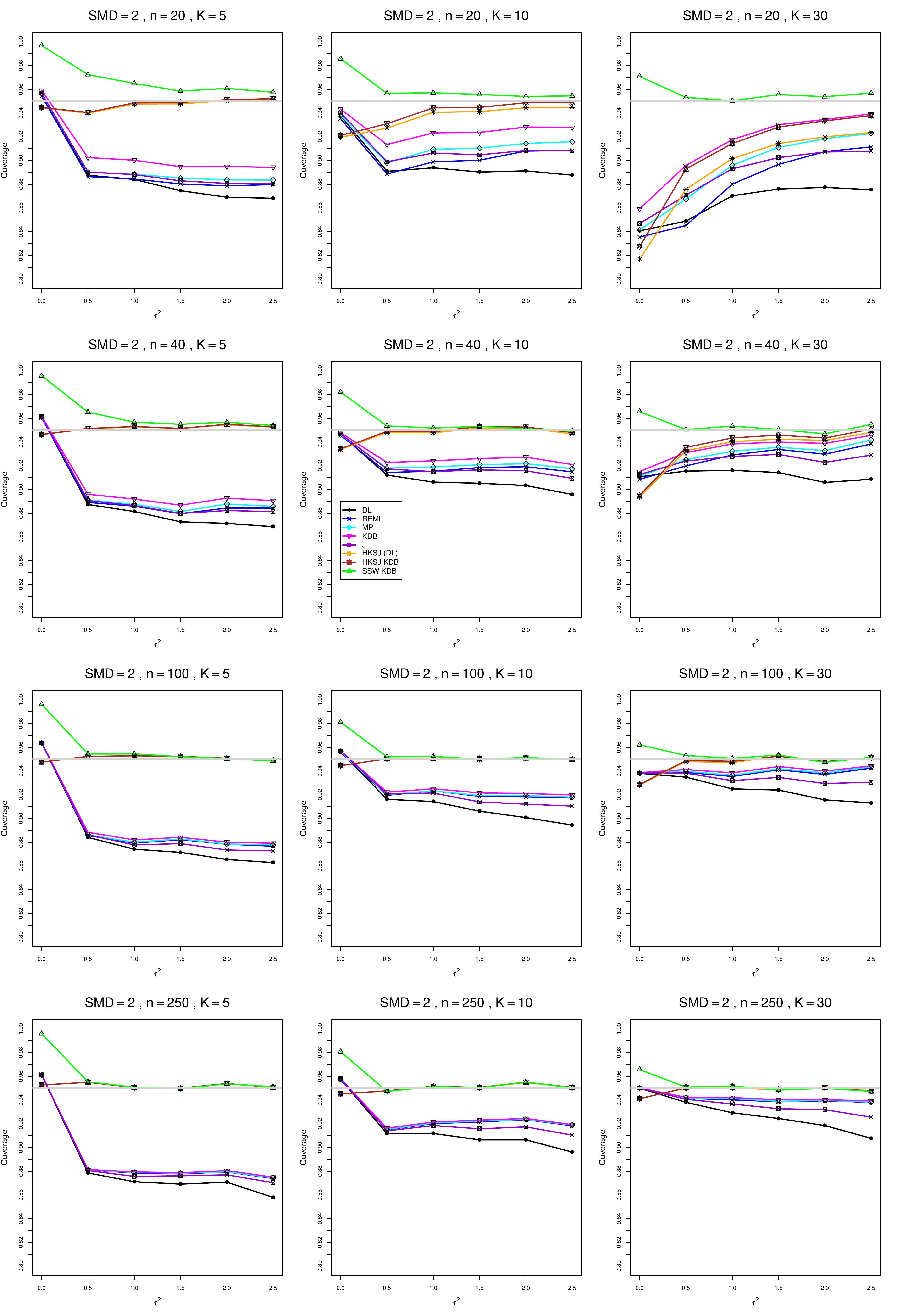}
	\caption{Coverage at  the nominal confidence level of $0.95$ of the  $\delta=2$,  for $q=0.5$, $n=20,\;40,\;100,\;250$.
		\label{CovThetaSMD2}}
\end{figure}

\begin{figure}[t]\centering
	\includegraphics[scale=0.35]{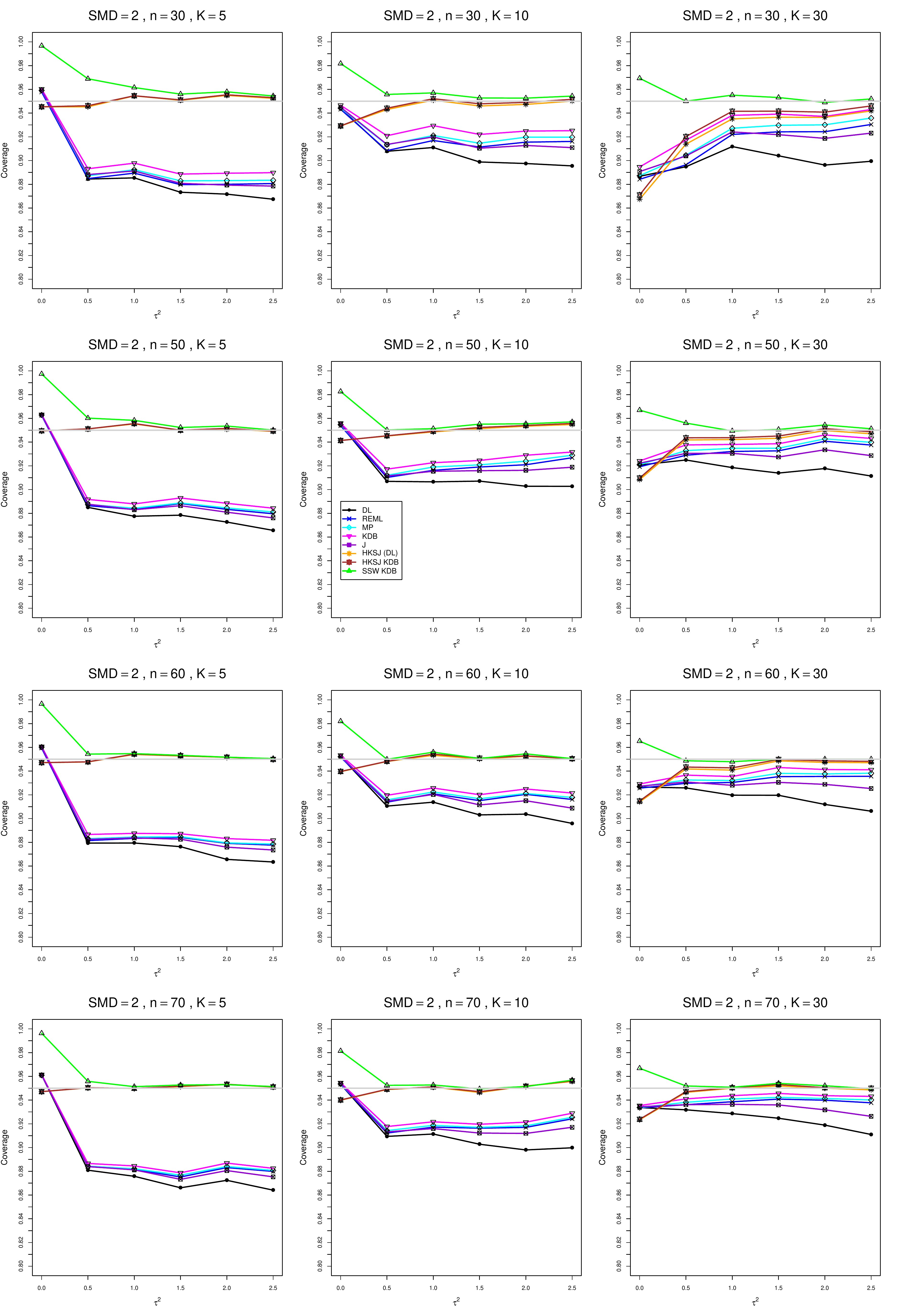}
	\caption{Coverage at  the nominal confidence level of $0.95$ of the  $\delta=2$,  for $q=0.5$, $n=30,\;50,\;60,\;70$.
		\label{CovThetaSMD2small}}
\end{figure}

\begin{figure}[t]\centering
	\includegraphics[scale=0.35]{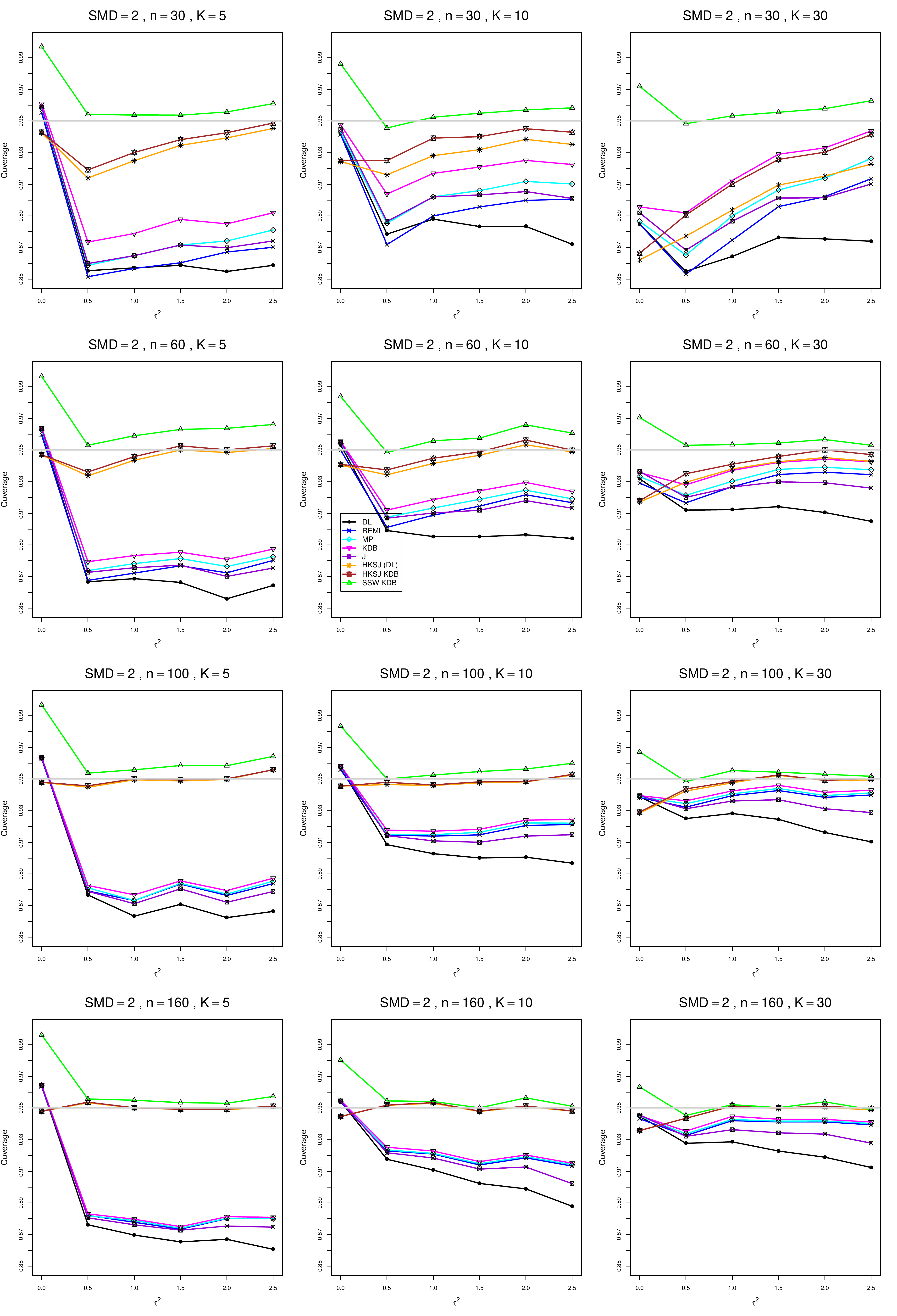}
	\caption{Coverage at  the nominal confidence level of $0.95$ of the $\delta=2$, for $q=0.5$,  unequal sample sizes with
		$\bar{n}=30,\; 60,\;100,\;160$.
		\label{CovThetaSMD2unequal}}
\end{figure}

\begin{figure}[t]\centering
	\includegraphics[scale=0.35]{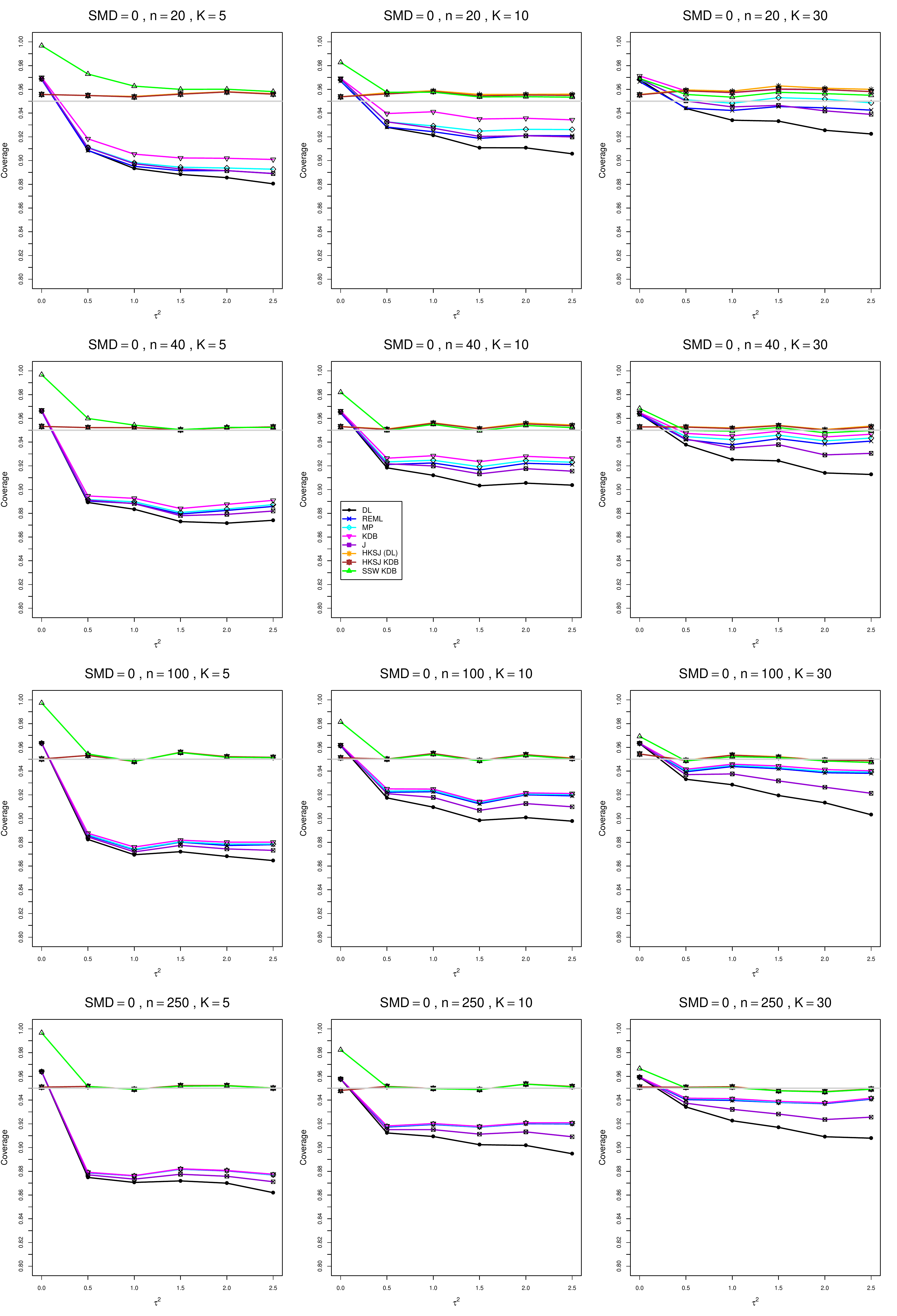}
	\caption{Coverage at  the nominal confidence level of $0.95$ of the  $\delta=0$,  for $q=0.75$, $n=20,\;40,\;100,\;250$.
		\label{CovThetaSMD0q75}}
\end{figure}

\begin{figure}[t]\centering
	\includegraphics[scale=0.35]{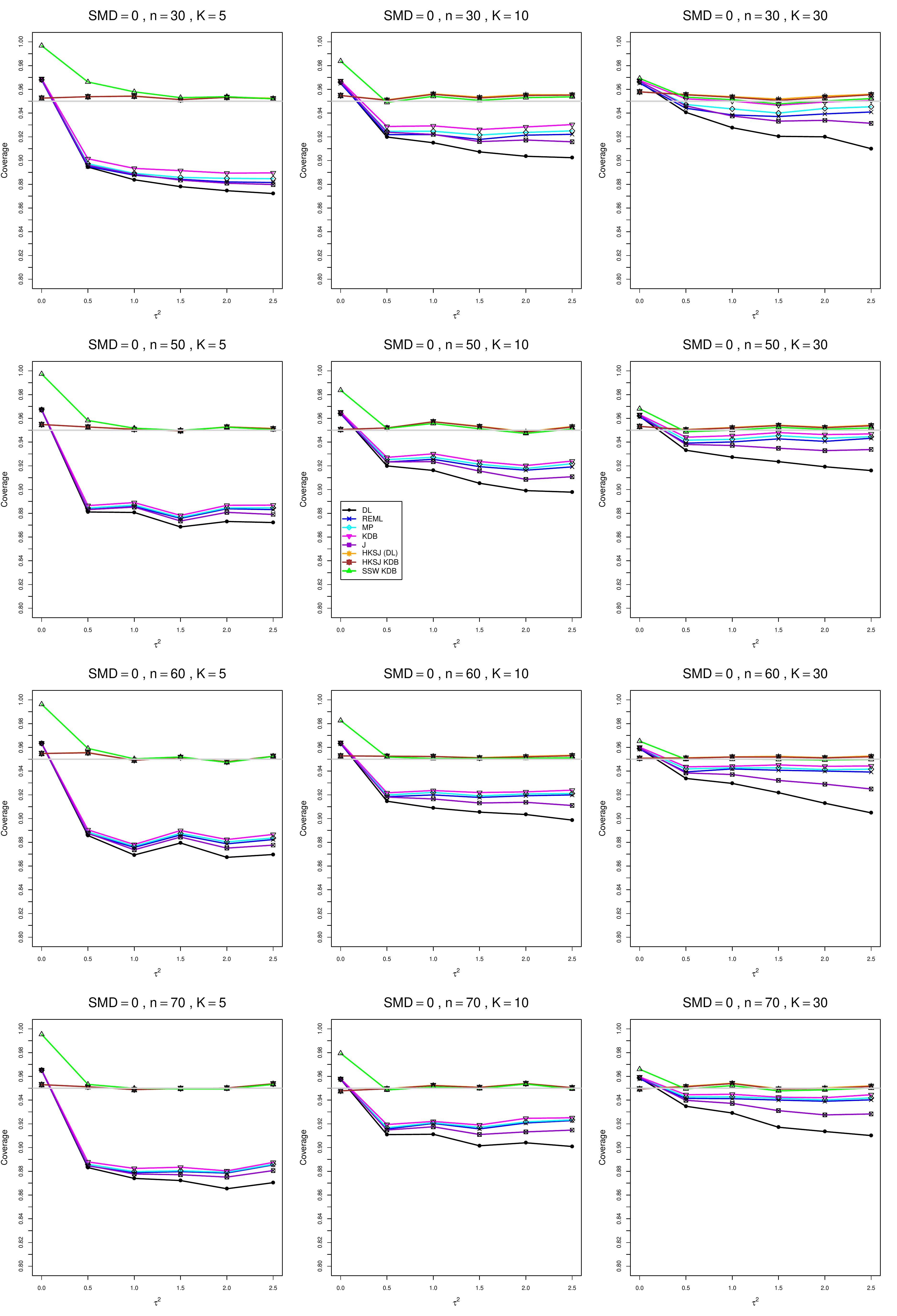}
	\caption{Coverage at  the nominal confidence level of $0.95$ of the  $\delta=0$,  for $q=0.75$, $n=30,\;50,\;60,\;70$.
		\label{CovThetaSMD0smallq75}}
\end{figure}

\begin{figure}[t]\centering
	\includegraphics[scale=0.35]{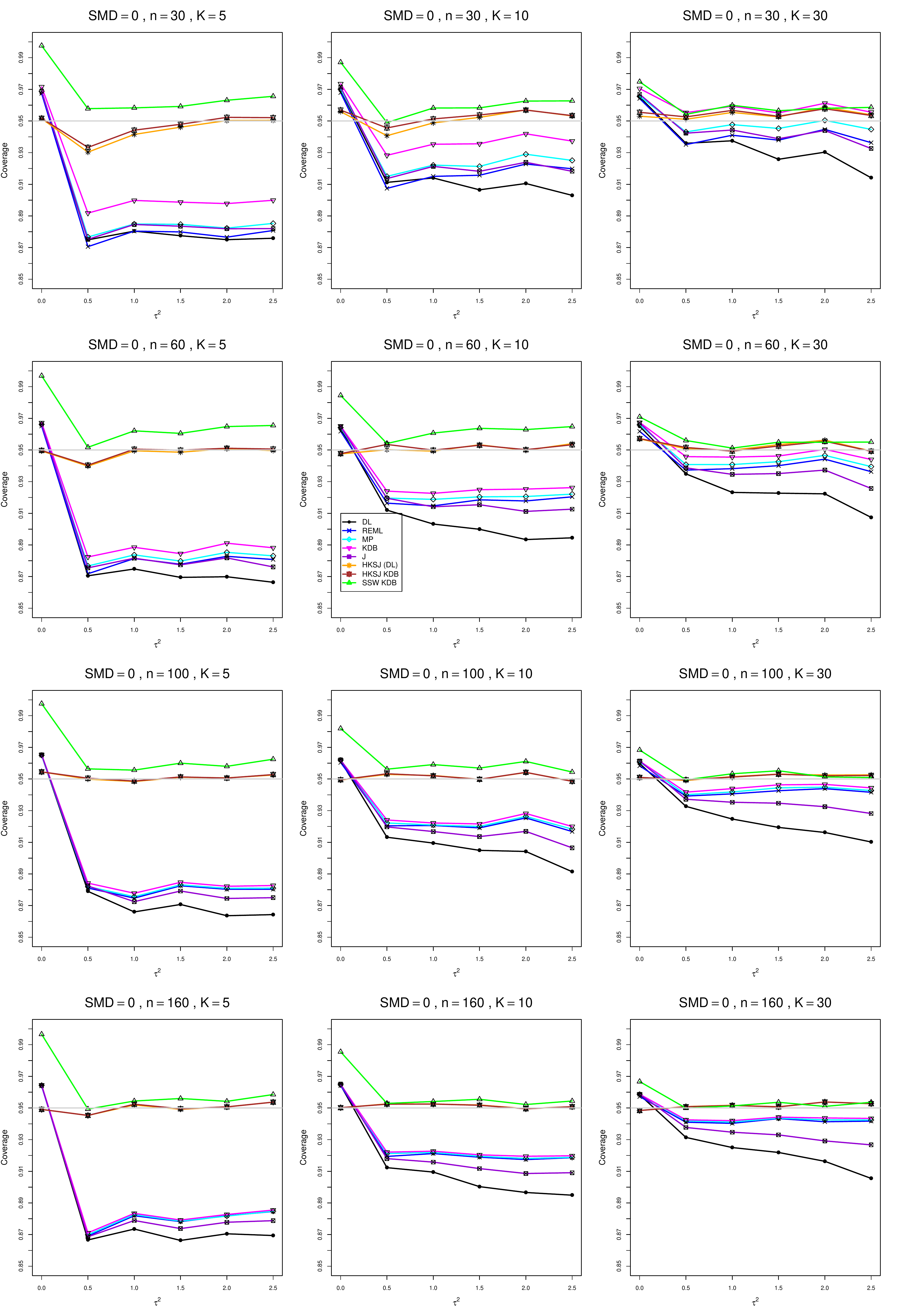}
	\caption{Coverage at  the nominal confidence level of $0.95$ of the $\delta=0$, for $q=0.75$,  unequal sample sizes with
		$\bar{n}=30,\; 60,\;100,\;160$.
		\label{CovThetaSMD0q75unequal}}
\end{figure}
\clearpage
\begin{figure}[t]\centering
	\includegraphics[scale=0.35]{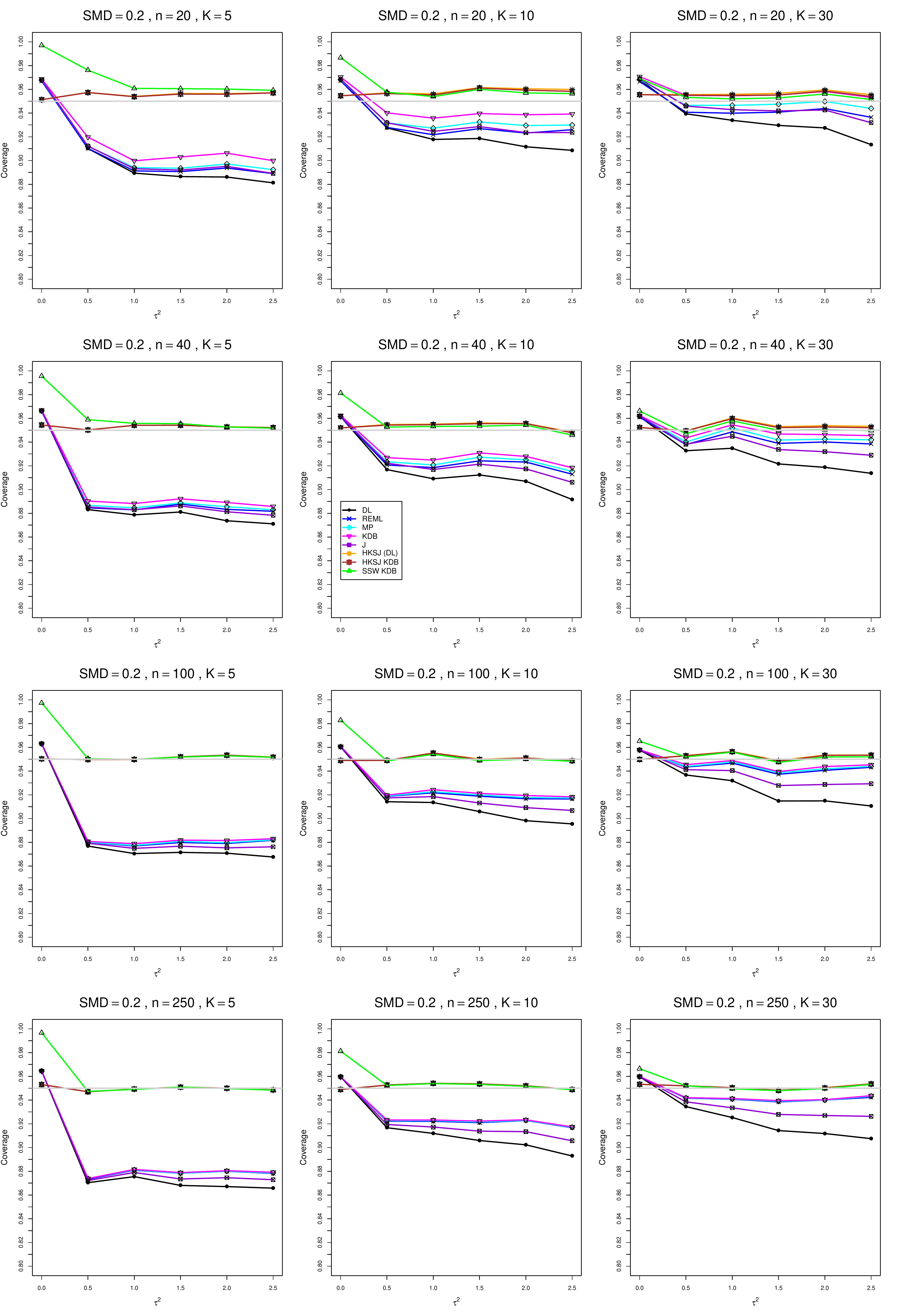}
	\caption{Coverage at  the nominal confidence level of $0.95$ of the  $\delta=0.2$,  for $q=0.75$, $n=20,\;40,\;100,\;250$.
		\label{CovThetaSMD02q75}}
\end{figure}

\begin{figure}[t]\centering
	\includegraphics[scale=0.35]{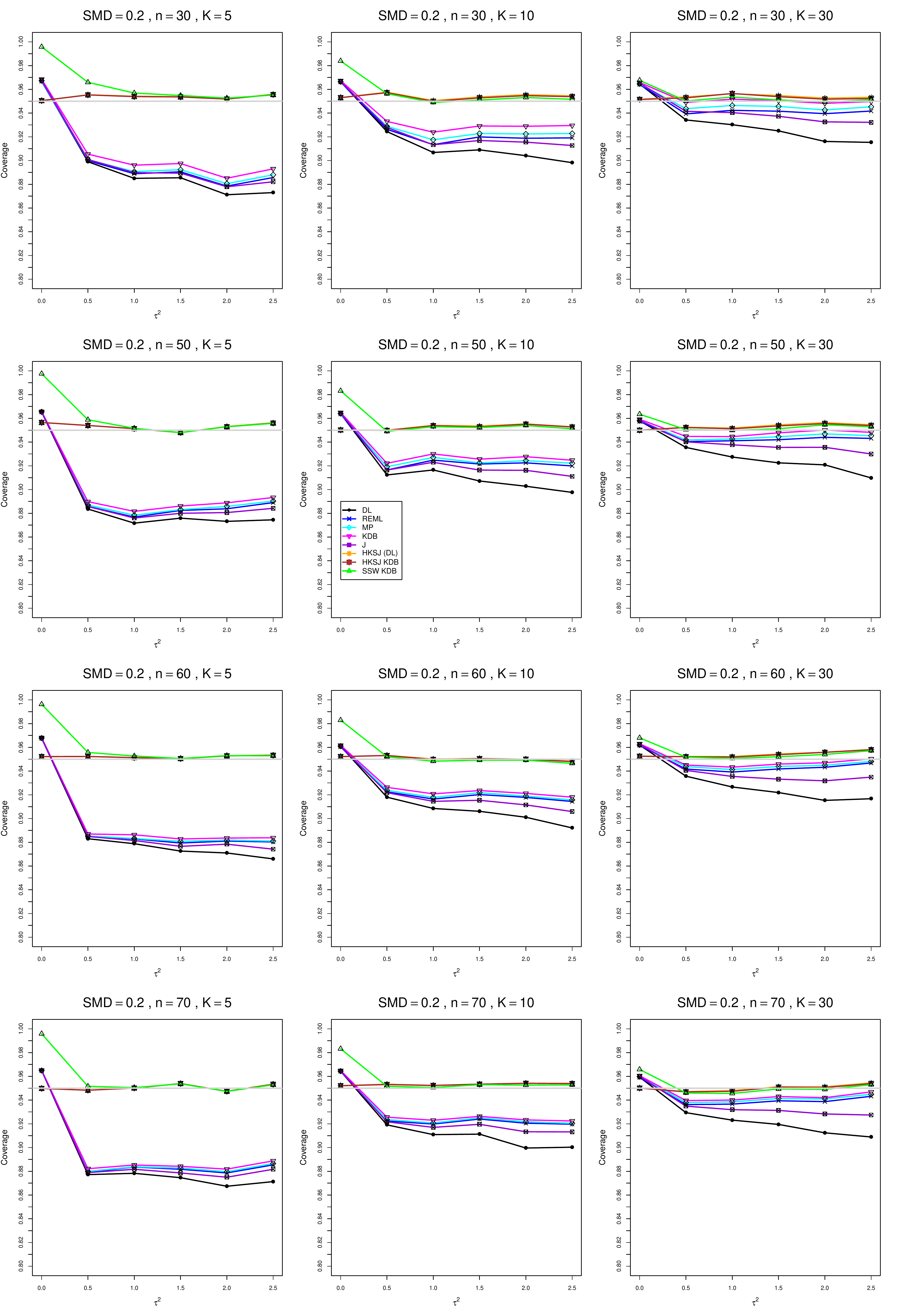}
	\caption{Coverage at  the nominal confidence level of $0.95$ of the  $\delta=0.2$,  for $q=0.75$, $n=30,\;50,\;60,\;70$.
		\label{CovThetaSMD02smallq75}}
\end{figure}

\begin{figure}[t]\centering
	\includegraphics[scale=0.35]{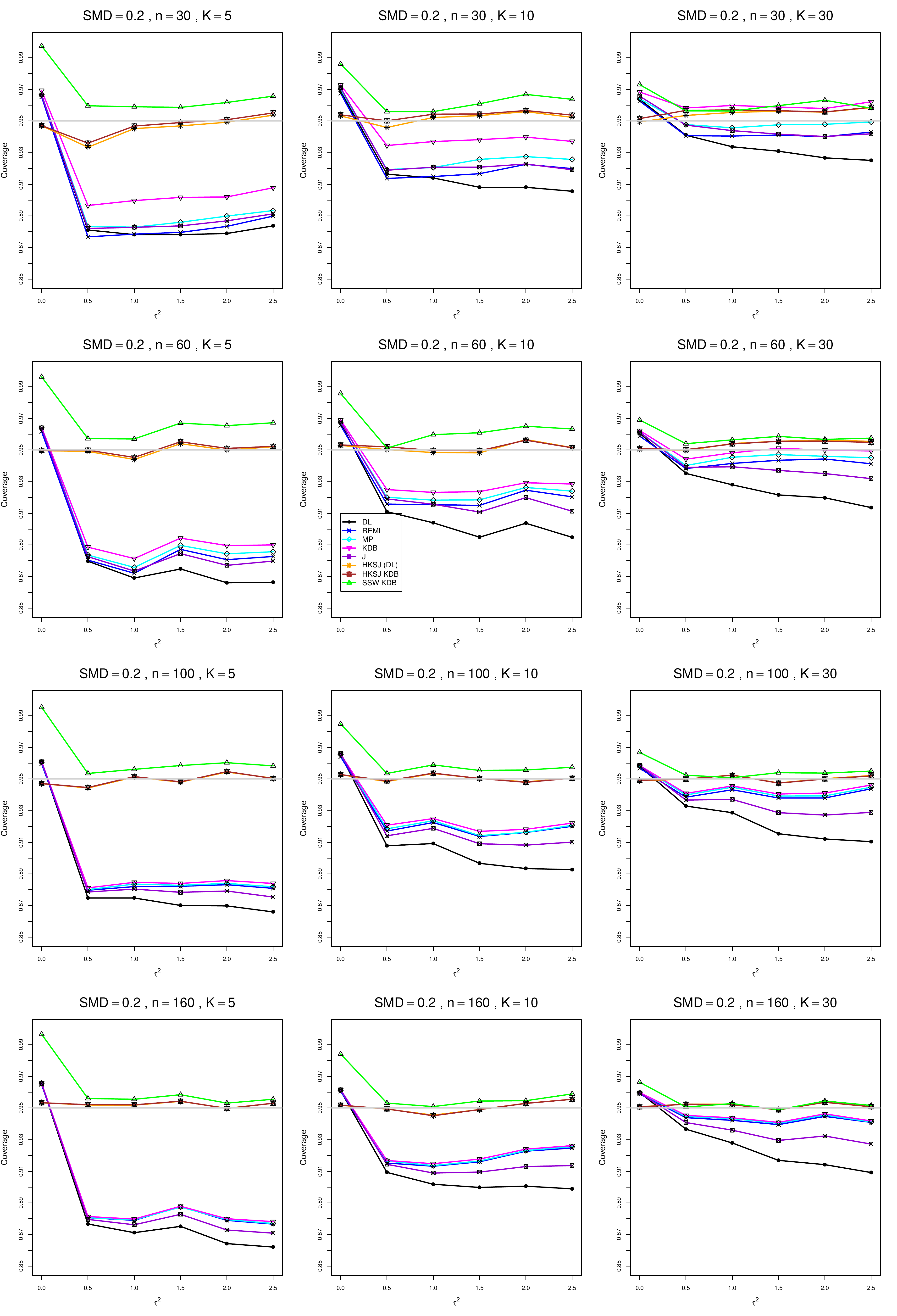}
	\caption{Coverage at  the nominal confidence level of $0.95$ of the $\delta=0.2$, for $q=0.75$,  unequal sample sizes with
		$\bar{n}=30,\; 60,\;100,\;160$.
		\label{CovThetaSMD02q75unequal}}
\end{figure}

\begin{figure}[t]\centering
	\includegraphics[scale=0.35]{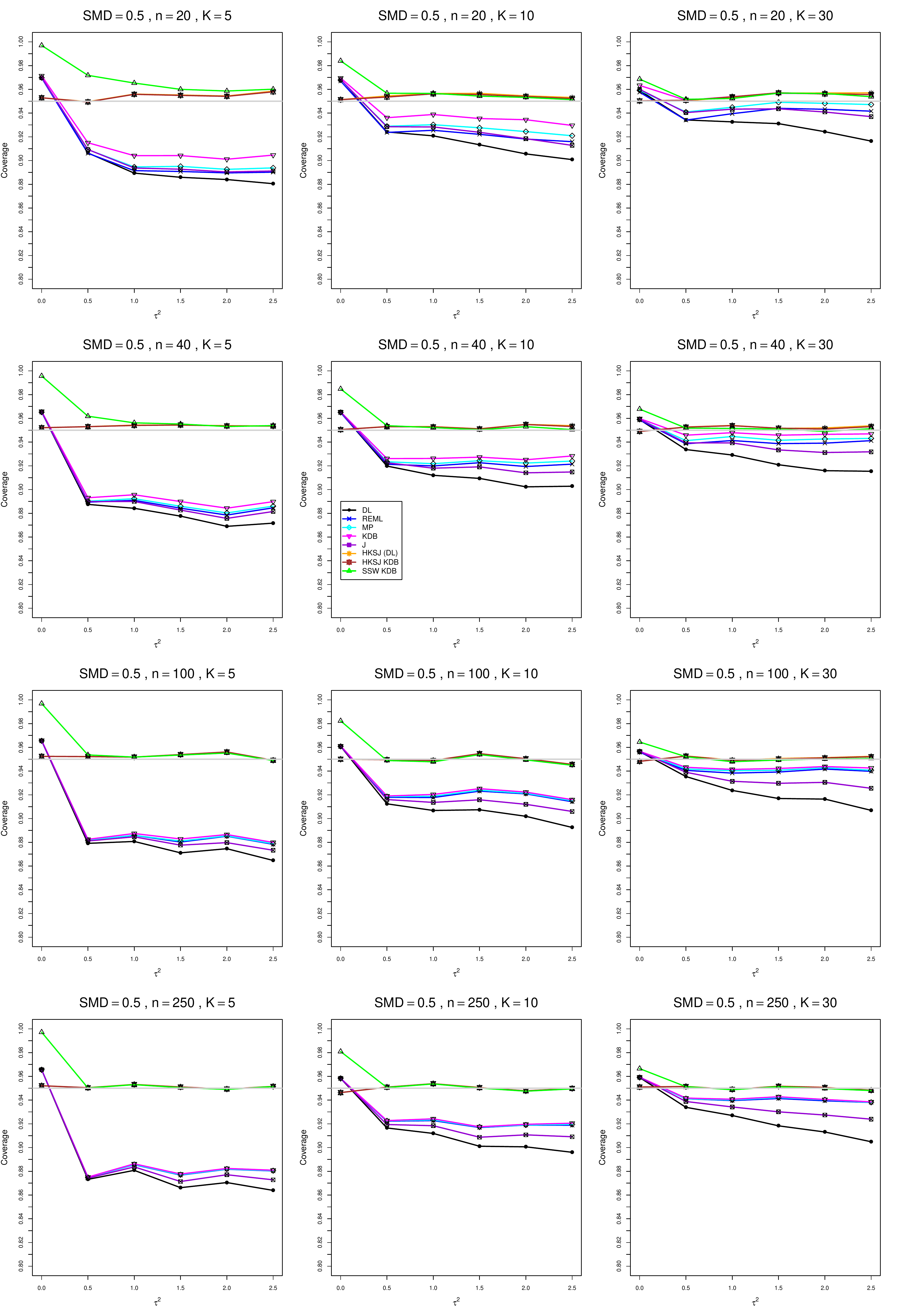}
	\caption{Coverage at  the nominal confidence level of $0.95$ of the  $\delta=0.5$,  for $q=0.75$, $n=20,\;40,\;100,\;250$.
		\label{CovThetaSMD05q75}}
\end{figure}

\begin{figure}[t]\centering
	\includegraphics[scale=0.35]{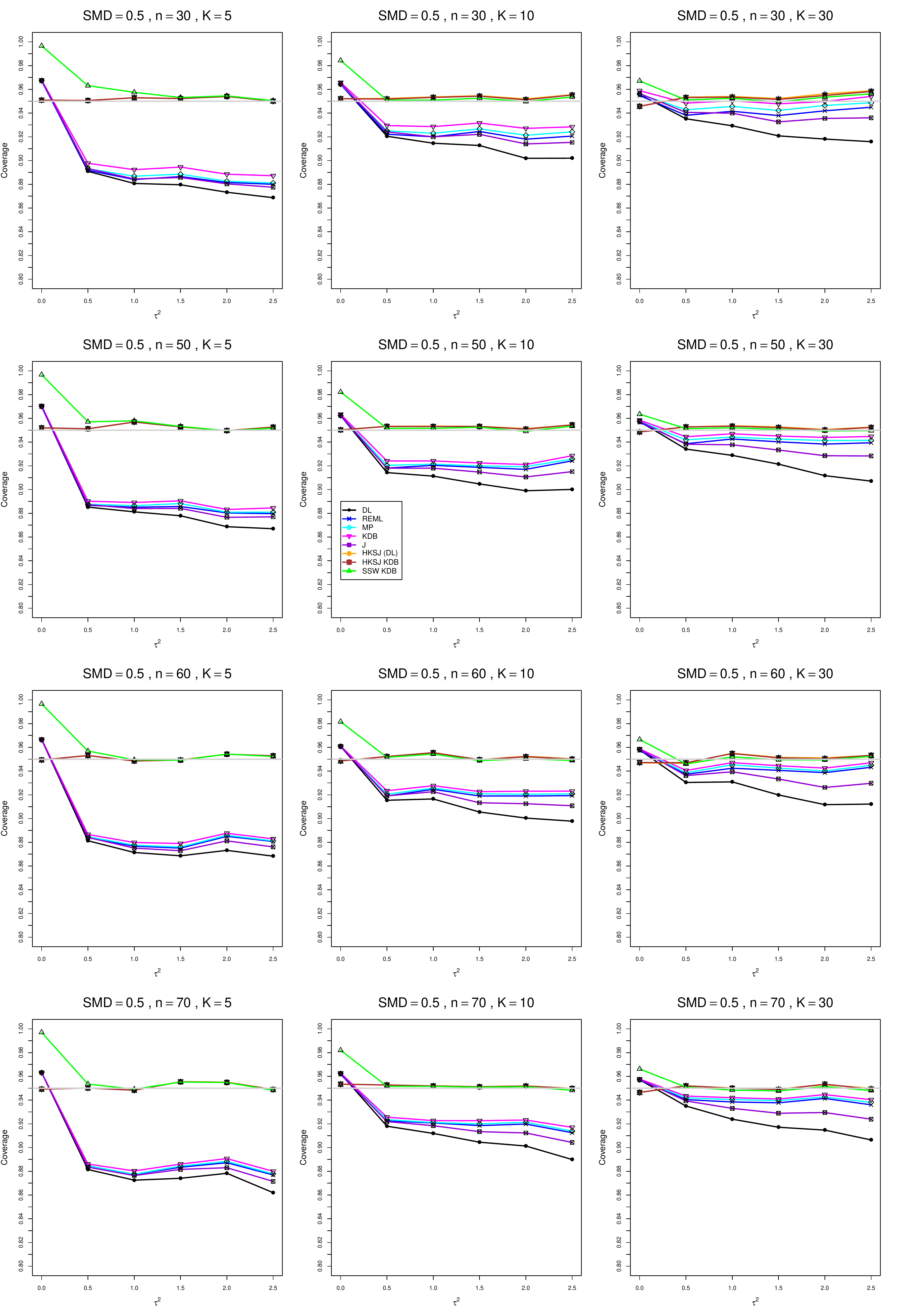}
	\caption{Coverage at  the nominal confidence level of $0.95$ of the  $\delta=0.5$,  for $q=0.75$, $n=30,\;50,\;60,\;70$.
		\label{CovThetaSMD075small}}
\end{figure}

\begin{figure}[t]\centering
	\includegraphics[scale=0.35]{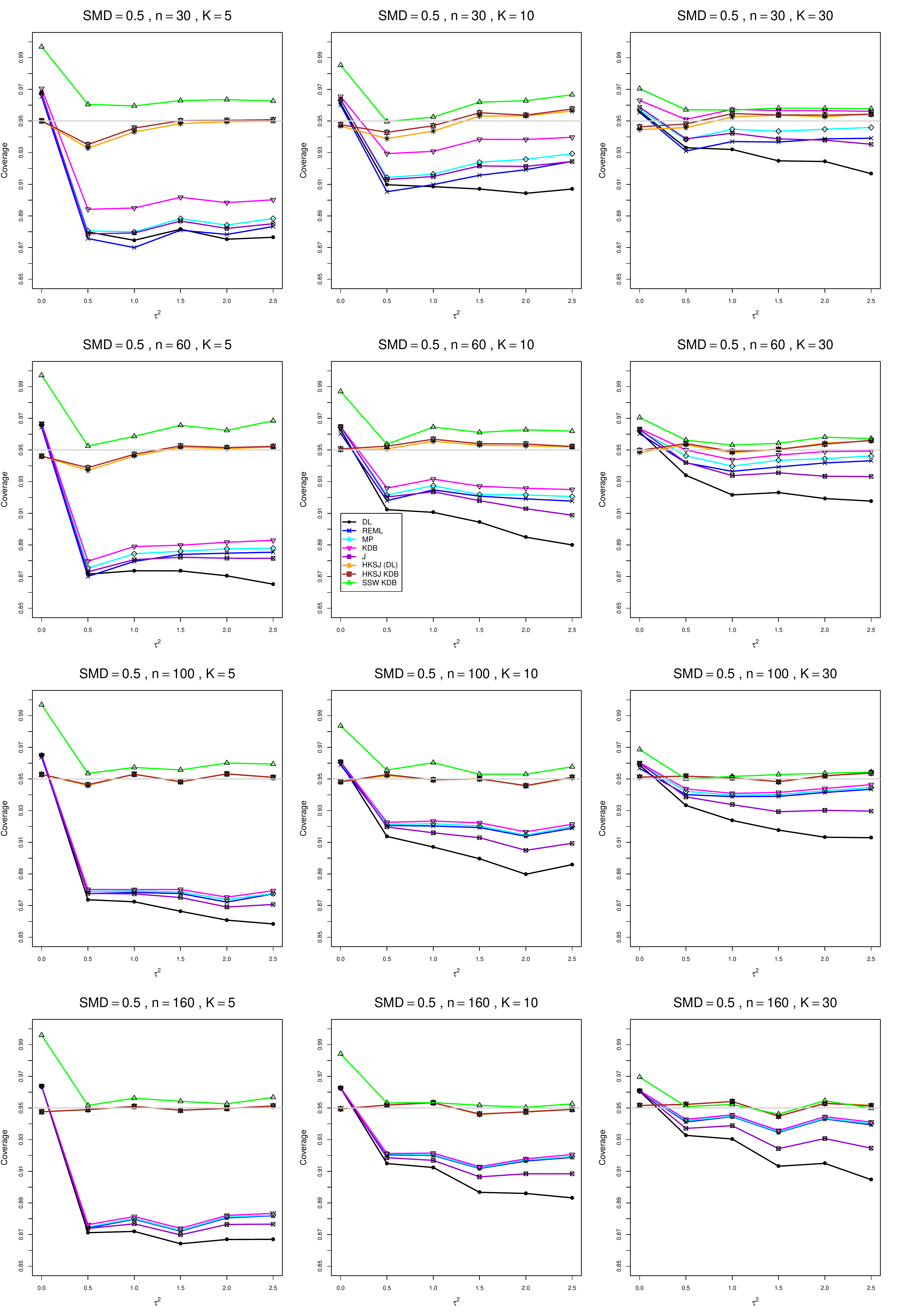}
	\caption{Coverage at  the nominal confidence level of $0.95$ of the $\delta=0.5$, for $q=0.75$,  unequal sample sizes with
		$\bar{n}=30,\; 60,\;100,\;160$.
		\label{CovThetaSMD05q75unequal}}
\end{figure}

\begin{figure}[t]\centering
	\includegraphics[scale=0.35]{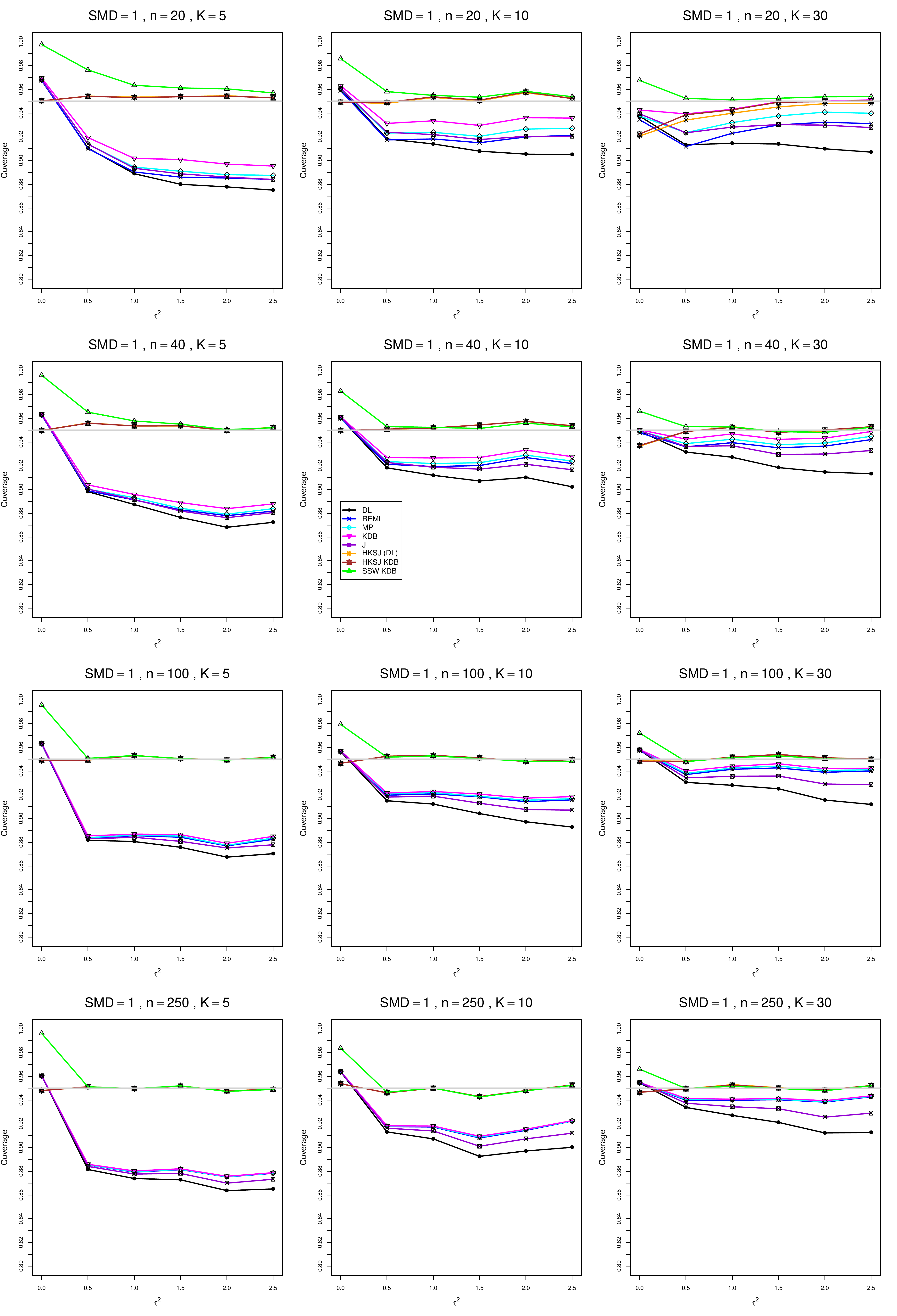}
	\caption{Coverage at  the nominal confidence level of $0.95$ of the  $\delta=1$,  for $q=0.75$, $n=20,\;40,\;100,\;250$.
		\label{CovThetaSMD1q75}}
\end{figure}

\begin{figure}[t]\centering
	\includegraphics[scale=0.35]{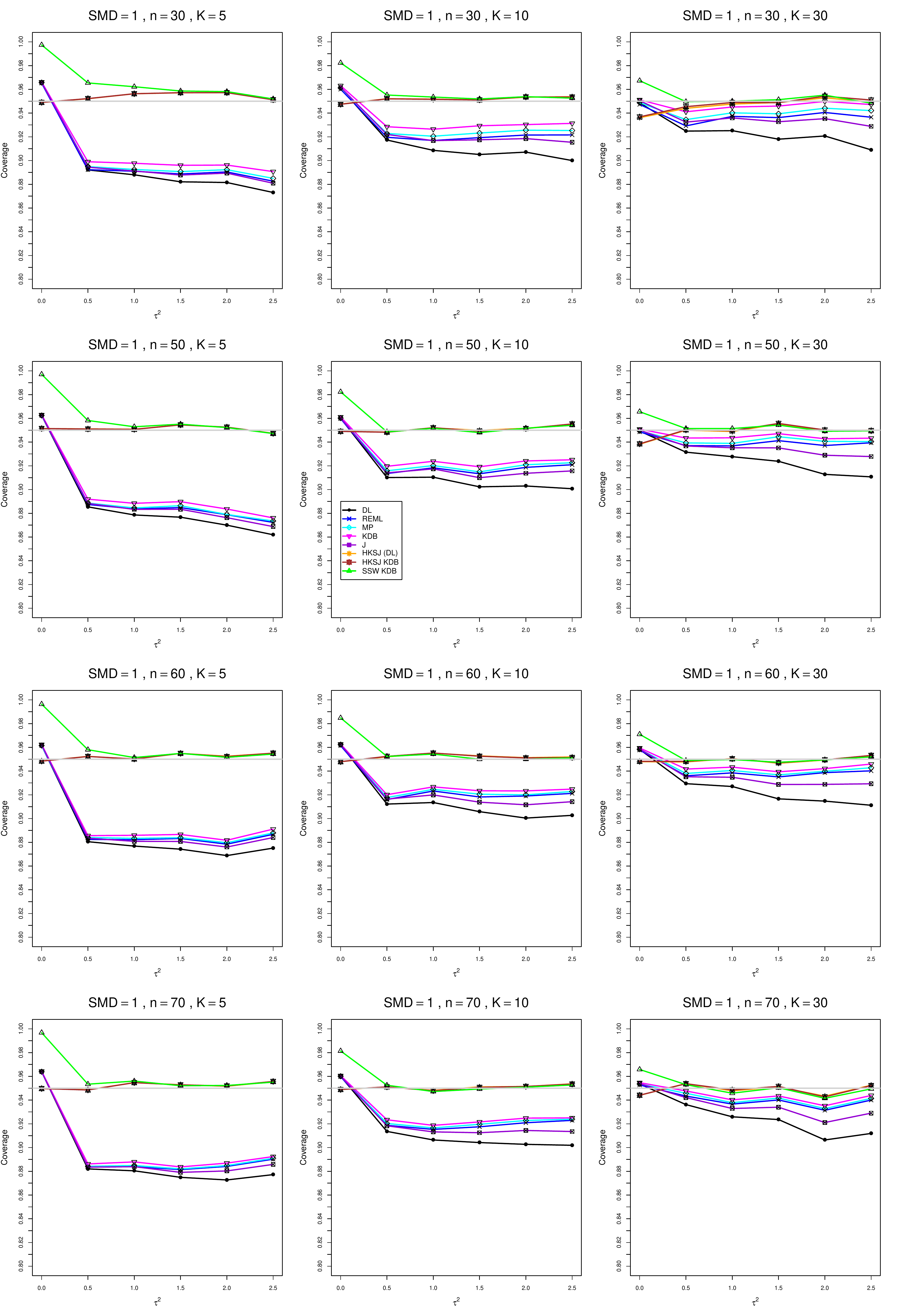}
	\caption{Coverage at  the nominal confidence level of $0.95$ of the  $\delta=1$,  for $q=0.75$, $n=30,\;50,\;60,\;70$.
		\label{CovThetaSMD1smallq75}}
\end{figure}

\begin{figure}[t]\centering
	\includegraphics[scale=0.35]{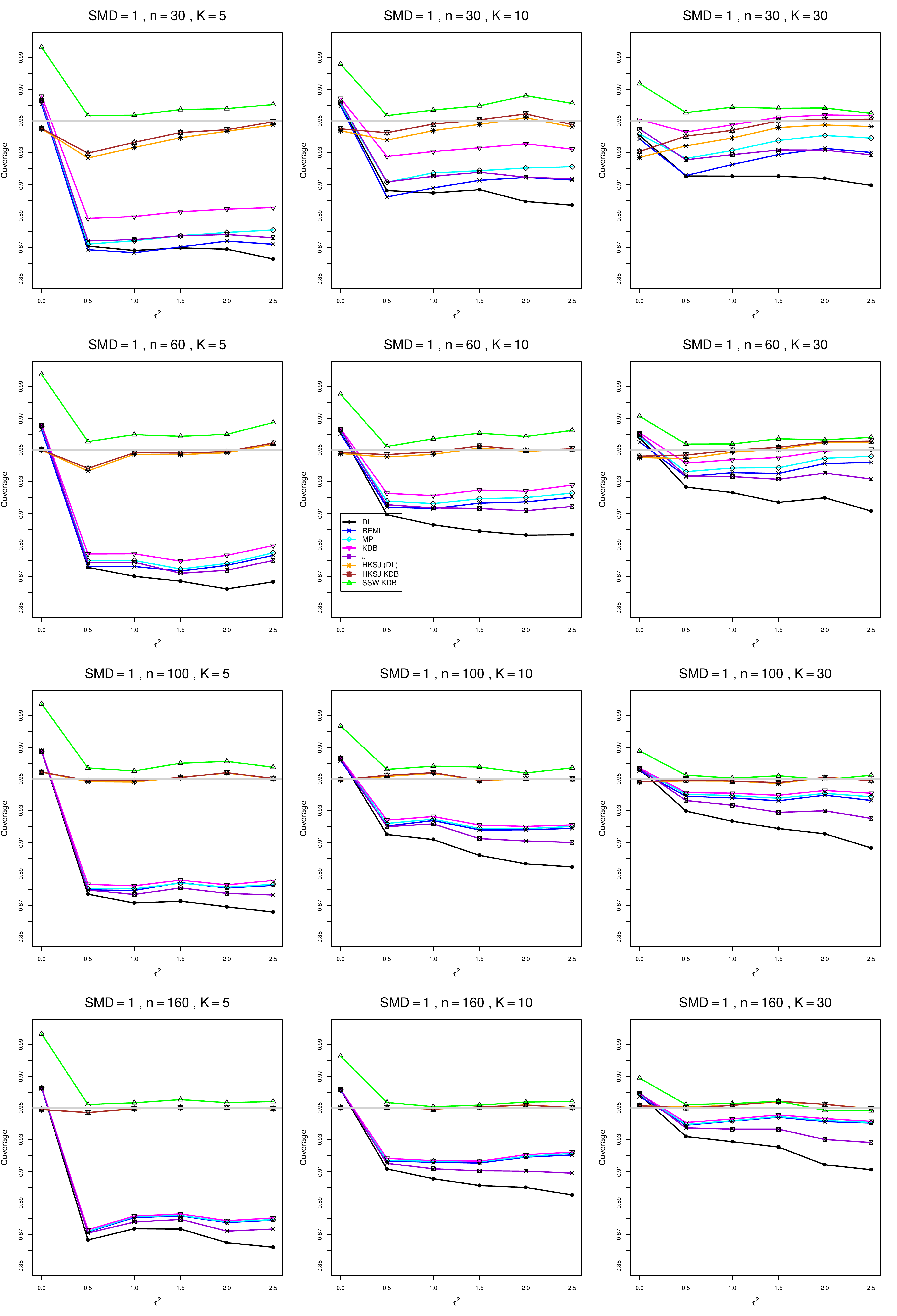}
	\caption{Coverage at  the nominal confidence level of $0.95$ of the $\delta=1$, for $q=0.75$,  unequal sample sizes with
		$\bar{n}=30,\; 60,\;100,\;160$.
		\label{CovThetaSMD1q75unequal}}
\end{figure}

\begin{figure}[t]\centering
	\includegraphics[scale=0.35]{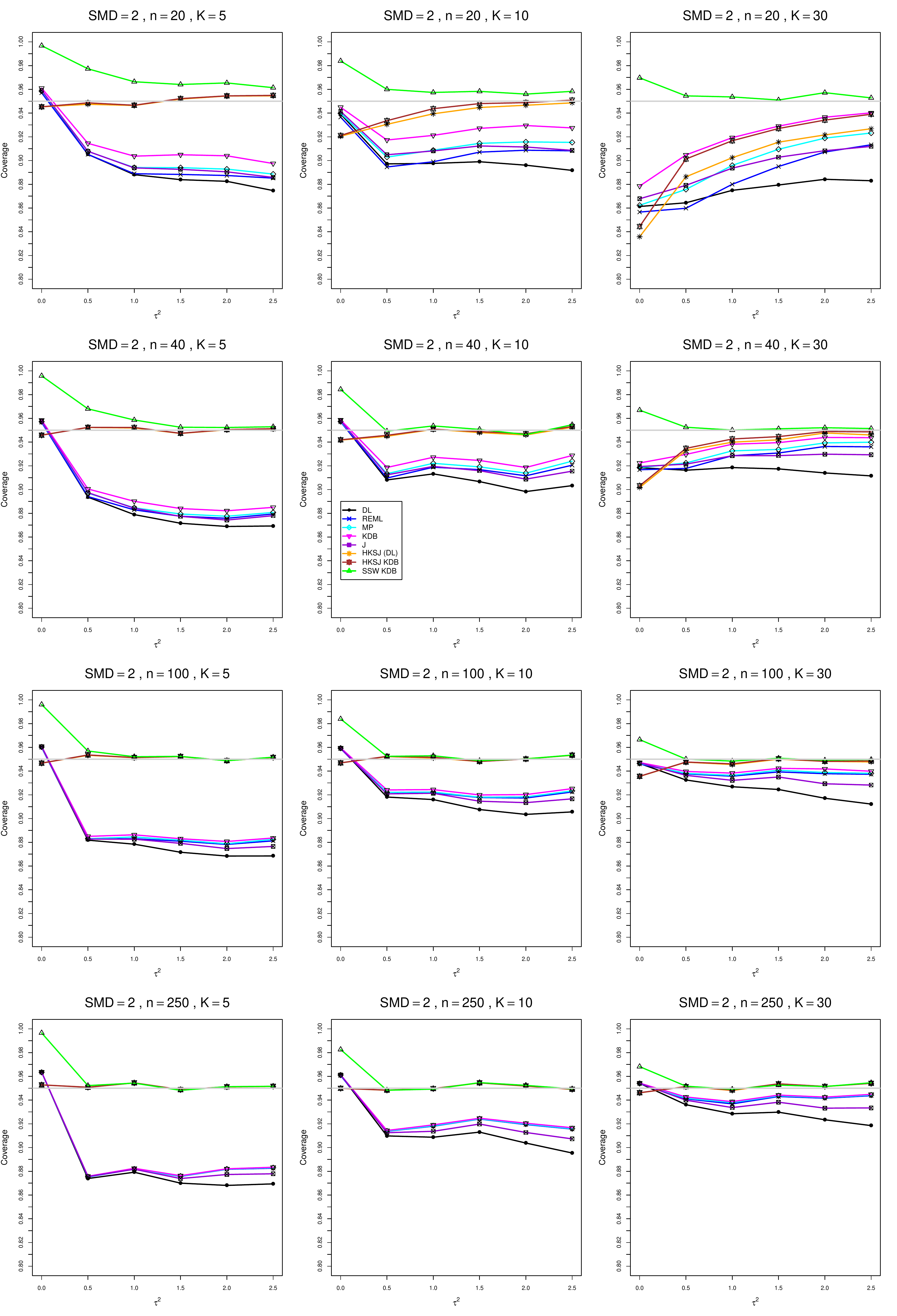}
	\caption{Coverage at  the nominal confidence level of $0.95$ of the  $\delta=2$,  for $q=0.75$, $n=20,\;40,\;100,\;250$.
		\label{CovThetaSMD2q75}}
\end{figure}

\begin{figure}[t]\centering
	\includegraphics[scale=0.35]{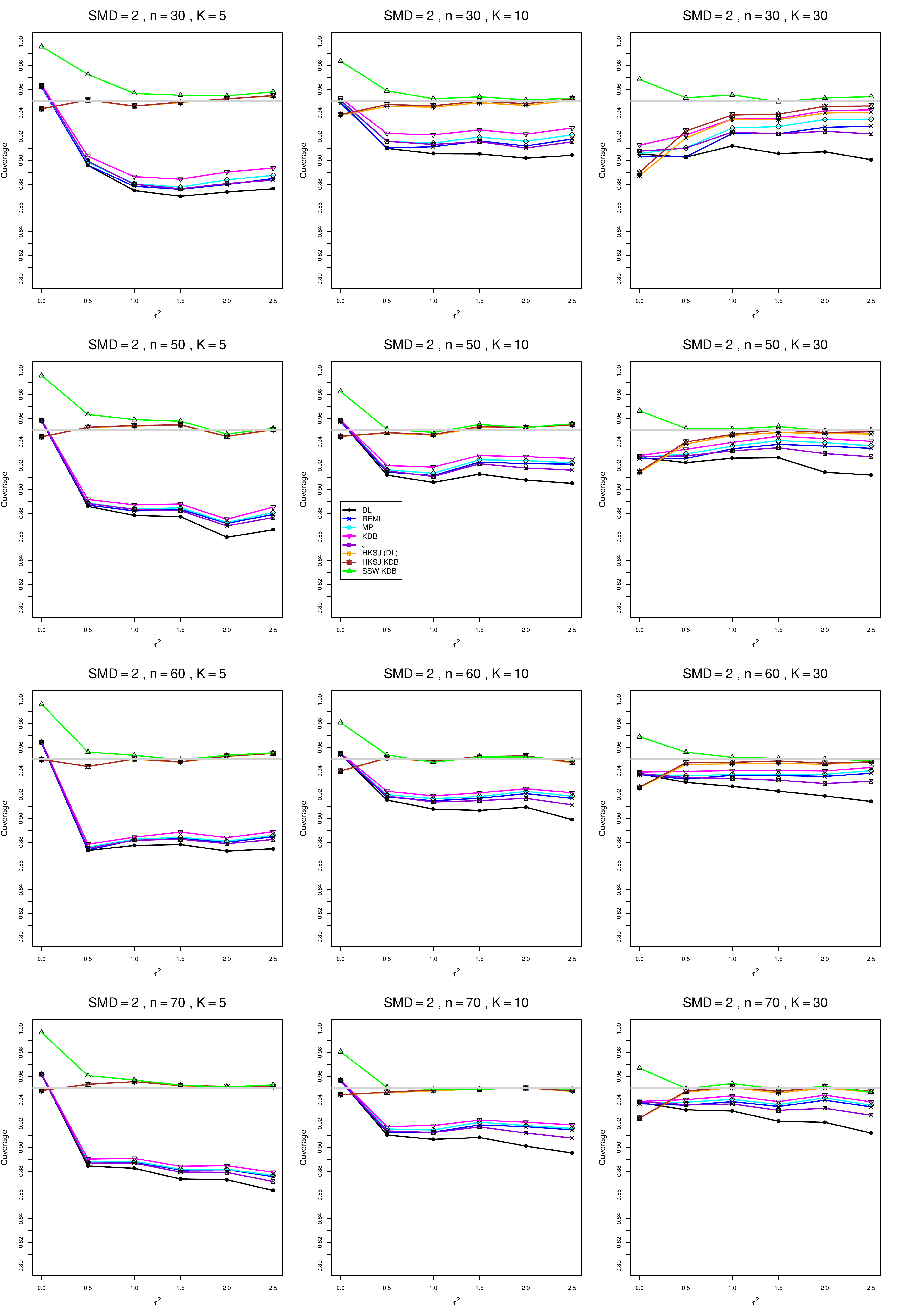}
	\caption{Coverage at  the nominal confidence level of $0.95$ of the  $\delta=2$,  for $q=0.75$, $n=30,\;50,\;60,\;70$.
		\label{CovThetaSMD2smallq75}}
\end{figure}

\begin{figure}[t]\centering
	\includegraphics[scale=0.35]{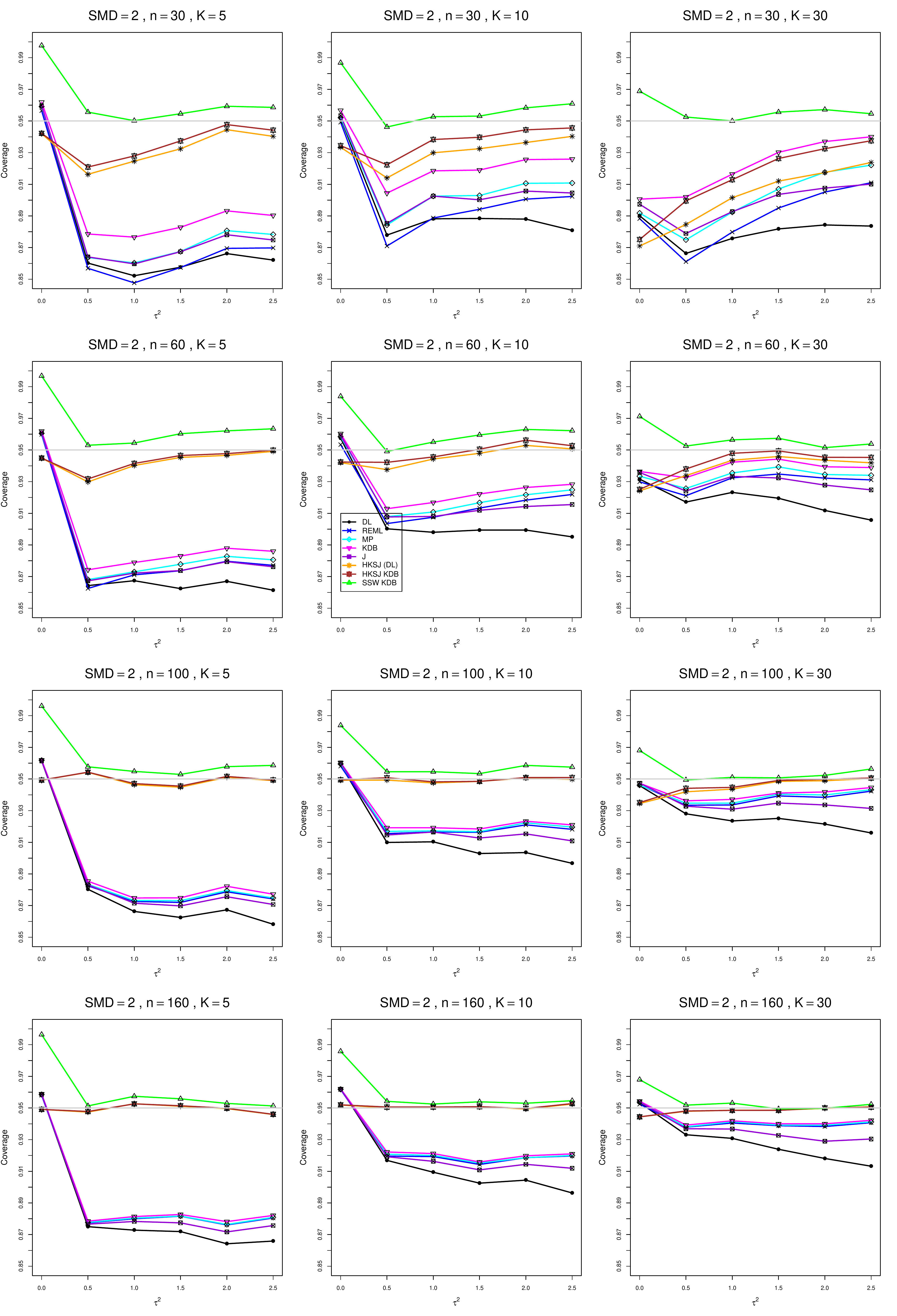}
	\caption{Coverage at  the nominal confidence level of $0.95$ of the $\delta=2$, for $q=0.75$,  unequal sample sizes with
		$\bar{n}=30,\; 60,\;100,\;160$ .
		\label{CovThetaSMD2q75unequal}}
\end{figure}
\end{document}